%% file: arxiv_main.tex
\pdfoutput=1
\documentclass[amsfonts, amssymb, amsmath, reprint, showkeys, nofootinbib, floatfix, superscriptaddress]{revtex4-2}
\usepackage[english]{babel}
\usepackage[utf8]{inputenc}
\usepackage[colorinlistoftodos, color=green!40, prependcaption]{todonotes}
\input{arxiv_preamble} 
\usepackage{hyperref}
\addto{\captionsenglish}{} % Supplementary Material}}

\begin{document}
\title{Excited state, non-adiabatic dynamics of large photoswitchable molecules using a chemically transferable machine learning potential}

\author{Simon Axelrod}
    % \email[Email: ]{saxelrod@g.harvard.edu }
    \affiliation{Department of Chemistry and Chemical Biology,
    Harvard University, Cambridge, MA, 02138}
    \affiliation{Department of Materials Science and Engineering, Massachusetts Institute of Technology, Cambridge, MA, 02139}

\author{Eugene Shakhnovich}
    % \email[]{shakhnovich@chemistry.harvard.edu}
    \affiliation{Department of Chemistry and Chemical Biology, 
    Harvard University, Cambridge, MA, 02138}
  
\author{Rafael Gómez-Bombarelli}
    \email[Corresponding author: ]{rafagb@mit.edu}
    \affiliation{Department of Materials Science and Engineering, Massachusetts Institute of Technology, Cambridge, MA, 02139}

\date{\today} % Leave empty to omit a date

\begin{abstract}
Light-induced chemical processes are ubiquitous in nature and have widespread technological applications. For example, photoisomerization can allow a drug with a photo-switchable scaffold such as azobenzene to be activated with light. In principle, photoswitches with desired photophysical properties like high isomerization quantum yields can be identified through virtual screening with reactive simulations. In practice, these simulations are rarely used for screening, since they require hundreds of trajectories and expensive quantum chemical methods to account for non-adiabatic excited state effects. Here we introduce a \textit{diabatic artificial neural network} (DANN) based on diabatic states to accelerate such simulations for azobenzene derivatives. The network is six orders of magnitude faster than the quantum chemistry method used for training. DANN is transferable to azobenzene molecules outside the training set, predicting quantum yields for unseen species that are correlated with experiment. We use the model to virtually screen 3,100 hypothetical molecules, and identify novel species with extremely high predicted quantum yields. The model predictions are confirmed using high-accuracy non-adiabatic dynamics. Our results pave the way for fast and accurate virtual screening of photoactive compounds.

\end{abstract}

% \keywords{first keyword, second keyword, third keyword}

\maketitle

\input{sections/section01.tex} 
% \vspace{0.5cm}

% \bibliographystyle{unsrtnat}
% \bibliography{main}

\clearpage
\footnotesize
\input{sections/methods.tex}

\input{sections/acknowledgements.tex}

 \renewcommand{\thetable}{S\arabic{table}}
 \renewcommand{\theequation}{S\arabic{equation}}
 \renewcommand{\thefigure}{S\arabic{figure}}
 \setcounter{equation}{0}
 \setcounter{table}{0}
 \setcounter{figure}{0}

 \clearpage
 \onecolumngrid
 \normalsize
\parbox[t]{\textwidth}{%
\begin{center}
  \large \textbf{Supplementary Information}\par
\end{center}
}
 \input{sections/sm}

 \bibliographystyle{unsrtnat}
 \bibliography{main}

\end{document}

%% file: arxiv_preamble.tex
\usepackage{amsthm}
\usepackage{mathtools}
\usepackage{physics}
\usepackage{xcolor}
\usepackage{graphicx}
\usepackage[left=23mm,right=22.5mm,top=35mm,columnsep=15pt]{geometry} 
\usepackage{adjustbox}
\usepackage{placeins}
\usepackage[T1]{fontenc}
\usepackage{lipsum}
\usepackage{csquotes}
\usepackage{scrextend}
\usepackage{booktabs}
\usepackage{multibib}
\usepackage{multirow} 
\usepackage{xltabular}
\usepackage{epstopdf}
\usepackage{xr} % For references to external documents
\usepackage{scrextend}

\usepackage{titlesec}
\usepackage{tikz}

\titleformat{\subsubsection}[runin]{\bfseries}{}{}{}[.]

\makeatletter
\newcommand\footnoteref[1]{\protected@xdef\@thefnmark{\ref{#1}}\@footnotemark}
\makeatother

%% file: sections/section01.tex
Light is a powerful tool for manipulating molecular systems. It can be controlled with high spatial, spectral and temporal precision to facilitate a variety of processes, including  energy transfer, intermolecular reactions, and photoisomerization \cite{evans2013applied}. These processes are used in areas as diverse as synthesis, energy storage, display technology, biological imaging, diagnostics and medicine \cite{kolpak2011azobenzene, evans2013applied, mai2020molecular}. Photoactive drugs, for instance, are photoswitchable compounds whose bioactivity can be toggled through light-induced isomerization. Precise spatiotemporal control of bioactivity allows photoactive drugs to be delivered in high doses with minimal off-target activity and side effects. Such therapeutics are a promising path for the treatment of cancer,  neurodegenerative diseases, bacterial infections, diabetes, and blindness \cite{broichhagen2015roadmap, lerch2016emerging}. 

Theory plays a key role in explaining and predicting photochemistry because empirical heuristics learned from thermally activated ground state processes typically do not apply to excited states \cite{mai2020molecular}. Computer simulations based on quantum mechanics can achieve impressive accuracy in the prediction of experimental observables. These include the isomerization efficiency and absorption spectrum of photoswitchable compounds \cite{mai2019unconventional, yu2020nonadiabatic}, which are key quantities in the design of photoactive drugs.

However, \textit{ab initio} methods in photochemistry are severely limited by their computational cost \cite{bannwarth2020hole}. In order to gather meaningful statistics for one molecule, hundreds of replicate simulations are needed, each of which involves thousands of electronic structure calculations performed in series with sub-femtosecond timesteps. The individual quantum chemical calculations are particularly demanding, requiring excited state gradients and some treatment of multireference effects. In some cases, both the ground- and excited-state gradients are required at each time step \cite{tully1998mixed, shalashilin2009quantum, ben2000ab}. Using \textit{ab initio} methods to compute photochemical properties of tens or hundreds molecules is impractical, and photodynamic simulations have not yet been used for large-scale virtual screening.

Among the most accurate and expensive electronic structure methods are multi-configuration perturbation techniques \cite{shiozaki2011communication, olsen1988determinant, ma2011generalized, li2011strong, ivanic2003direct}, but their cost and requirement for manual active space selection limit their use in virtual screening. The photochemistry community has made exciting developments over several years to overcome both of these hurdles. For example, reduced scaling techniques \cite{song2020reduced, song2021reduced} and graphics processing units \cite{seritan2020terachem} can significantly accelerate multi-reference calculations. The density matrix renormalization group (DMRG) \cite{marti2011new, sharma2012spin} and multi-reference density functional theory (DFT) methods \cite{hermes2019multiconfigurational, marian2019dft, li2014multiconfiguration, gagliardi2017multiconfiguration} have expanded the size of systems that can be treated with high accuracy. DMRG has also been used to automate the selection of active spaces for multi-reference methods \cite{stein2016automated, stein2019autocas}. Less accurate but more affordable black-box methods include spin-flip time-dependent DFT (SF-TDDFT) and hole-hole Tamm-Dancoff DFT \cite{yu2020ab}, among others \cite{filatov2015spin, lee2018eliminating}. Despite these developments, the cost of non-adiabatic simulations remains high. As discussed below, even SF-TDDFT is prohibitively expensive for virtual screening. Semi-empirical methods \cite{cusati2012semiempirical, inamori2020spin, de2019simplified} are currently the only affordable approach for large-scale screening. They provide qualitatively correct results across many systems, but are ultimately bounded by their approximations, with average energy errors of 15 kcal/mol \cite{inamori2020spin}.   % Other approaches focus on hardware and software development \cite{yu2020ab}, but cannot match the speed of semi-empirical methods. % For example, semi-empirical Hamiltonians have been successfully parameterized for specific chemical families \cite{cusati2012semiempirical}. Broadly-parameterized tight-binding approximations to spin-flip time-dependent density functional theory (TDDFT) have also been recently developed \cite{inamori2020spin, de2019simplified}. These methods benefit from being physics-driven, providing qualitatively correct results across many systems. However, their accuracy is ultimately bounded, because the result of a calculation depends on integrals that are approximated with simple functional forms. Indeed, energies predicted by tight-binding spin-flip TDDFT are typically in error by 15 kcal/mol \cite{inamori2020spin}. Other approaches focus on hardware and software development to accelerate existing methods. One example is the use of graphics processing units (GPUs) \cite{ufimtsev2008quantum, fales2020efficient, seritan2020terachem, snyder2016gpu, yu2020nonadiabatic, yu2020ab}. This approach can provide accurate results with improved speed, but cannot match the speed of semi-empirical methods.

A different approach is to use data-driven models in place of quantum chemistry (QC) calculations. Machine learning (ML) models trained on quantum chemical data can now routinely predict ground state energies and forces with sub-chemical accuracy \cite{qiao2020orbnet, schutt2021equivariant, unke2021spookynet}, and take only milliseconds to make predictions. These models have been successfully used in a variety of ground state simulations \cite{ang2021active, wang2020active, schutt2021equivariant}. %, including complex molecular reactions \cite{ang2021active, young2021transferable}, simulated binding of solvate ionic liquids \cite{wang2020active}, inter-dimensional transformations \cite{xie2021bayesian}, and ring polymer dynamics for molecular spectra \cite{schutt2021equivariant}. 
They have also been used to accelerate non-adiabatic simulations in a number of model systems \cite{chen2018deep, dral2018nonadiabatic, hu2018inclusion, li2021automatic, westermayr2019machine, westermayr2020combining, westermayr2020machine}. % Most recently, the SchNet deep learning architecture \cite{schutt2017schnet, schutt2018schnet} was incorporated into the SHARC platform, giving rise to the SchNarc program to accelerate photochemistry simulations \cite{westermayr2020combining}. 
However, excited state ML has not yet offered affordable photodynamics for hundreds of molecules of realistic size, which is the ultimate goal for predictive simulation in photopharmacology. % To our knowledge, the largest system analyzed thus far has been hexafluoro-2-butene \cite{li2021automatic}, which has 12 atoms and no rotatable bonds.  %Machine learning becomes increasingly difficult as the dimensionality of the system increases, particularly if there are rotatable bonds, meaning that these proofs of concept do not necessarily generalize to realistic systems. Indeed, as described below, surface hopping is controlled by conical intersection regions that are extremely difficult to learn in large systems. Moreover, it is in such large systems that machine learning could deliver the most significant computational savings. Developing methods that scale to large systems is therefore highly desirable, but so far unachieved. 
Further, no excited-state interatomic potentials have been developed that are transferable to different compounds. They therefore require thousands of QC calculations for every new species to serve as training data. % Training a model for one molecule can reduce computational cost by orders of magnitude, but still requires thousands of quantum chemistry calculations for every new species.
 % Previous calculations for different systems cannot be used to speed up future simulations. To scale photochemistry simulations to hundreds or thousands of different species, their computational demand must be reduced closer to that of a classical force field. This can only be achieved with chemically transferable models. Yet this is a well-known limitation of machine learning models. Since they are data-driven rather than physics-based, they often cannot extrapolate to new molecules. This is even more challenging for excited states, as they can be reactive in ways that depend on the specific excitation and the system. This makes transferability even more difficult than for ground states.

Here we make significant progress toward affordable, large-scale photochemical simulations and virtual screening with ML. To develop a transferable potential we focus on molecules from the same chemical family, studying derivatives of azobenzene, a prototypical photoswitch. The derivatives studied here contain up to 100 atoms, making them the largest systems fit with excited-state ML potentials to date. Combining an equivariant neural network \cite{schutt2021equivariant} and a physics-informed diabatic model, together with data generated by combinatorial exploration of chemical space, and configurational sampling through active learning, we produce a model that is transferable to large, unseen derivatives of azobenzene. This yields computational savings in excess of six orders of magnitude. Predicted isomerization quantum yields of unseen species are well-correlated with experimental values. The model is used to predict the quantum yield for over 3,100 hypothetical species, revealing rare molecules with extremely high \textit{cis}-to-\textit{trans} and \textit{trans}-to-\textit{cis} quantum yields.

\section*{Results}
\subsubsection*{Azobenzene photoswitches}
This work focuses on the photoswitching of azobenzene derivatives, but the methods are general and can be applied to other chemistries and other excited state processes. Azobenzene derivatives can exist as \textit{cis} and \textit{trans} conformers. % These isomers correspond to two benzene rings on the same or opposite sides of the N=N double bond, respectively. 
The conformations are local minima in the ground state, but not in the excited state. Photoexcitation of either can therefore induce isomerization into the other (see the potential energy schematics in Figs. \ref{fig:diabat}(a) and \ref{fig:schematics}(b)). A key experimental observable is the quantum yield, defined as the probability that excitation leads to isomerization. The yield depends critically on the dynamics near conical intersections (CIs), configurations in which the excitation energy is zero. In these regions the electrons can return to the ground state with non-zero probability.

Many approaches have been developed over several decades to model such non-adiabatic transitions. These include \textit{ab initio} multiple spawning \cite{ben2000ab} and cloning \cite{makhov2014ab}; Ehrenfest dynamics \cite{tully1998mixed, shalashilin2009quantum}; coherent switching with decay of mixing \cite{zhu2004coherent}; the variantional multi-configurational Gaussian method \cite{richings2015quantum}; exact factorization \cite{abedi2010exact, abedi2010exact, abedi2014mixed, min2017ab, curchod2017dynamics, ha2018surface}; the multi-configuration time-dependent Hartree (MCTDH) method \cite{beck2000multiconfiguration, wang2003multilayer}; Gaussian MCTDH \cite{burghardt1999approaches};
and trajectory surface hopping \cite{tully1990molecular}. A recent review of these methods can be found in Ref. \cite{mai2020molecular}. Surface hopping is a popular approach because of its simplicity and efficiency. In this method, independent trajectories are simulated with stochastic hops between potential energy surfaces (PESs). % The nuclei are propagated classically on the excited state surface before the hop, and classically on the ground state surface afterward. 
Depending on the curvature of the PESs and the location of the hop, a trajectory can end in the original isomer or in a new isomer (Figs. \ref{fig:diabat}(a) and \ref{fig:schematics}(b)). The quantum yield is the proportion of trajectories that end in a new isomer. % Like the other observables mentioned above, the yield depends critically on the dynamics near the CI.
Our goal is to predict the quantum yield of azobenzene derivatives after excitation from the singlet ground state ($S_0$) to the first singlet excited state ($S_1$). This can be accomplished with the surface hopping approach described above, using a fast surrogate ML model to generate the PESs. The impact of considering only the first excited state is discussed in Supplementary Sec. \ref*{sm_sec:sources_of_error}.

 %, corresponds to the symmetry-forbidden $n\pi^*$ transition \cite{bandara2012photoisomerization}.

\subsubsection*{ML architecture and training}
\begin{figure}[t!]
    \centering
    \includegraphics[width=0.45\textwidth]{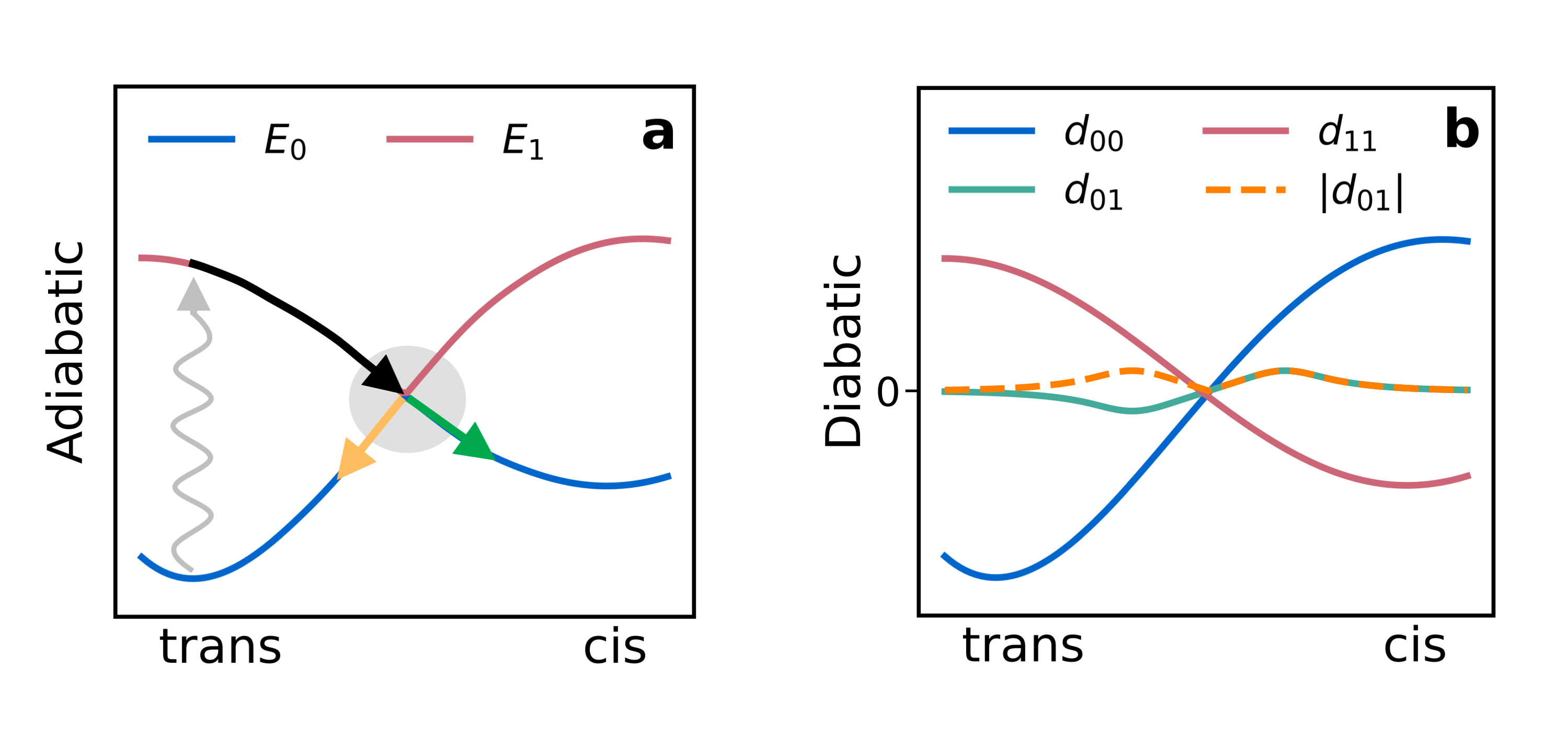}
    \caption{Depiction of the potential energy surfaces in azobenzene derivatives. (a) $S_0$ and $S_1$ adiabatic energies, with the CI region shaded in gray. Initial excitation is shown with a vertical zigzag line. Trajectories prior to hopping are shown in black. Reactive and unreactive trajectories after hopping are shown in green and yellow, respectively. (b) Diabatic energies $d_{nm} \equiv (\mathbf{H}_d)_{nm}$. The diagonal diabatic elements cross and become re-ordered along the isomerization coordinate. A CI occurs when the diagonal diabatic elements cross and the off-diagonal element becomes zero.}
    \label{fig:diabat}
\end{figure}
\begin{figure*}[t!]
    \centering
    \includegraphics[width=1.0\textwidth]{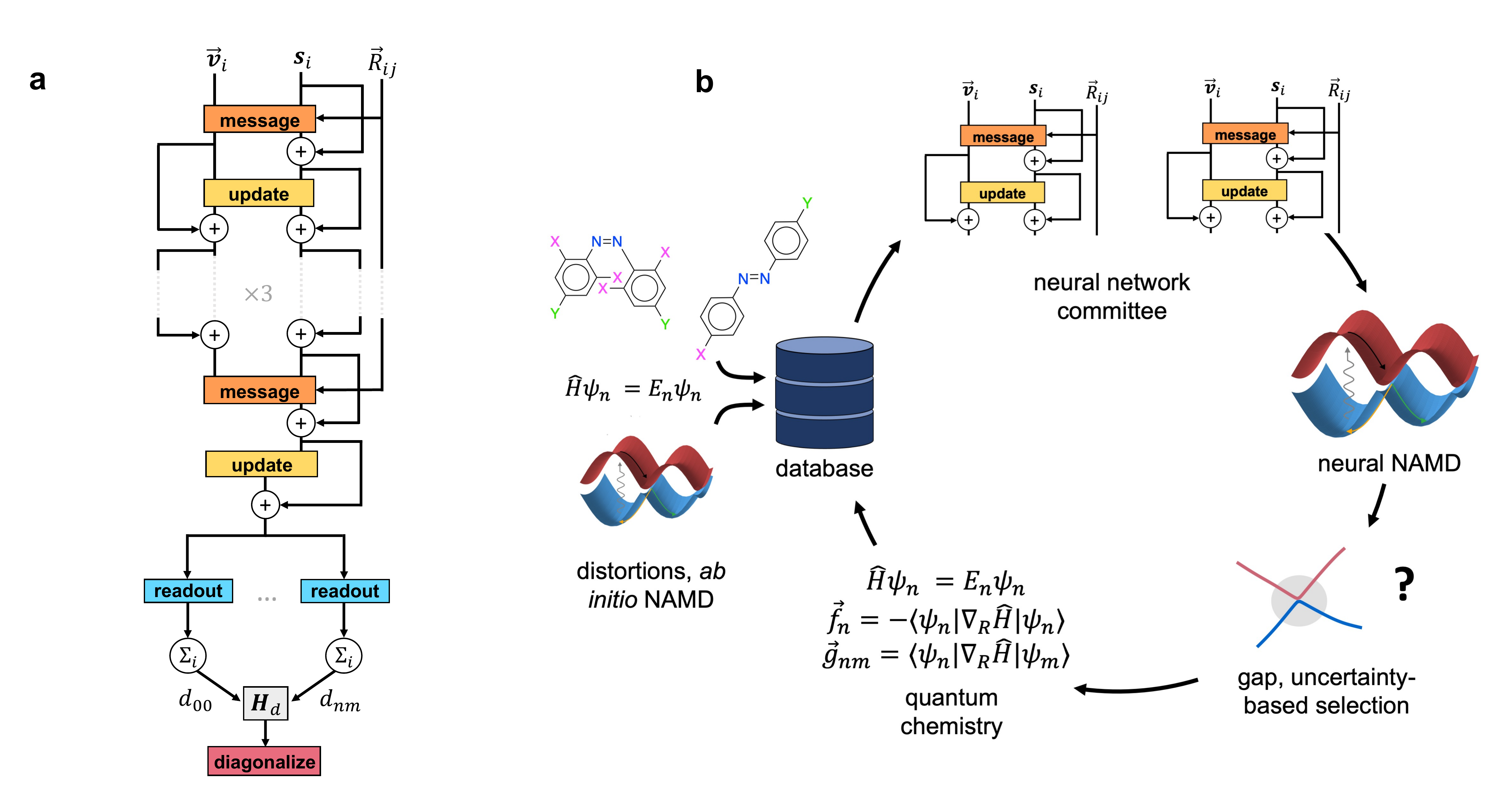}
    \caption{
    (a) Schematic of the DANN architecture, which is based on the PaiNN model. Scalar atomic features $\mathbf{s}_i$ and vectorial atomic features $\vec{\mathbf{v}}_i$ are updated through messages from neighboring atoms. The $\mathbf{s}_i$ are then mapped to atomic energies, which are summed to produce the diabatic Hamiltonian $\mathbf{H}_d$. The diabatic matrix is diagonalized to produce adiabatic quantities.  (b) Schematic of the active learning loop. Geometries and QC data are first generated through \textit{ab initio} NAMD, normal mode sampling, and inversion/rotation about the central N=N double bond. Two neural networks are then trained on the data and used to perform DANN-NAMD. Newly generated geometries with high committee variance and/or low predicted gaps receive QC calculations. The new calculations are added to the training data, the networks are retrained, and the cycle is repeated until convergence.
    % Schematic of the DANN architecture. (a) Network overview. $\Sigma_i$ denotes a sum over atoms and $||$ denotes concatenation. Note that $d_{nm} = d_{mn}$, so the model only generates $M_{\mathrm{d}}(M_{\mathrm{d}} + 1)/2$ outputs. (b) Attention pooling block for generating the molecular feature vector, with feature dimensions shown in gray. $\circ$ means element-wise multiplication, $\Sigma_F$ is a sum over features in each atom, and $\mathrm{softmax}_i$ is a softmax function applied over all atoms. $\mathbf{K}$ is a learnable feature vector and the $\mathbf{W}$ are learnable matrices. (c) Embedding block for converting $d_{11} - d_{00}$ into a vector of length 128. (d) Readout block. Dimensions on the left are for the readout of $d_{nn}$, and dimensions on the right are for the $d_{nm}$. 
    }
    \label{fig:schematics}
\end{figure*}
Our model is based on the PaiNN neural network \cite{schutt2021equivariant}, which uses equivariant message-passing to predict molecular properties. In this approach, an initial feature vector is generated for each atom using its atomic number. The vector is then updated through a set of neural network operations involving ``messages'', which incorporate the distance, orientation, and features of atoms within a cutoff distance. A series of updates leads to information being aggregated from increasingly distant atoms. Once the updates are complete, the atomic features are mapped to molecular energies using a neural network. 

This architecture can be used to predict energies and, through automatic differentiation, the forces for each state. However, models that predict adiabatic energies have a basic shortcoming for non-adiabatic molecular dynamics (NAMD). Since surface hopping is largely controlled by the energy gap when it is close to zero, small errors in the energies can lead to exponentially large errors in the hopping probability \cite{zhu1992two, zhu1993two}. This in turn can cause large errors in observable quantities like the quantum yield. This point is discussed in further detail in Supplementary Sec. \ref*{sm_sec:extended_methods}\ref*{sm_subsec:relevance_gap}. Further, since CIs are non-differentiable cusps in the energy gap, they are difficult to fit with neural networks. For $N$ atoms in a molecule, the network must predict two different energies that are exactly equal in $3N-8$ dimensions. We found this to be particularly challenging for \textit{trans} species that are outside the training set. As shown in Supplementary Sec. \ref*{sm_sec:compare_hop}, small errors in the gap lead to the incorrect prediction that many species never hop to the ground state.

\renewcommand{\arraystretch}{1.3}
\begin{table*}[t]
\centering
    \begin{tabular*}{\textwidth}{c  |c@{\extracolsep{\fill}}|ccccccccc}
        \toprule[1.1pt]
        &  & $E_0$ & $E_1$ & $\Delta E_{01}$ & $(\Delta E_{01})_{\mathrm{small}}\footnote{For these $R^2$ calculations, we computed the total sum of squares using $\mathrm{mean}\{ \Delta E_{01} \}$ instead of $\mathrm{mean}\{(\Delta E_{01})_{\mathrm{small}}\}$. The mean predictor should not know \textit{a priori} which gaps are small, and hence should predict the mean of all gaps.}$ & $\vec{F}_0$ & $\vec{F}_1$ & $\vec{g}_{01}$ \\ 
        %   \cline{1-1} \cline{3-9}
        \hline
        \multirow{2}{*}{Seen species \ } & \ MAE ($\downarrow$) \ & 0.86 & 1.01  & 0.75 & 0.47 & 1.00  & 1.17 & 0.87  \\
        & $R^2$ ($\uparrow$)  & 1.00 & 1.00 & 1.00 & 0.97 & 0.99 & 0.99 & 0.84 \\
        \hline
        \multirow{2}{*}{Unseen species \ } & MAE ($\downarrow$) &  3.06 & 3.77 & 1.89 & 0.97 & 1.72  & 2.31 & 1.36 \\
        & $R^2$ ($\uparrow$) & 0.99 & 0.98 & 0.98 & 0.95 & 0.97 & 0.86 & 0.50 \\
        \bottomrule[1.1pt]
        % \end{tabular}
    \end{tabular*}
    \caption{MAE and coefficient of determination ($R^2$) of the DANN model for various quantities. Units are kcal/mol for energies and kcal/mol/\AA \ for forces and force couplings. $E_i$ are energies, $\vec{F}_i$ are forces, $\Delta E_{01}$ is the energy gap, and $\vec{g}_{01}$ is the force NACV. $(\Delta E_{01})_{\mathrm{small}}$ denotes the energy gap when it is under 4.6 kcal/mol (0.2 eV).}
    \label{tab:accuracy}
\end{table*}
To remedy this issue we introduce a model based on diabatic states, which we call DANN (\textit{diabatic artificial neural network}; Fig. \ref{fig:schematics}(a)). The approach builds on previous work using neural networks for diabatization \cite{shu2020diabatization, williams2018neural, guan2019representation}. Much of the previous work could only be used for specific system types, such as semi-rigid molecules \cite{williams2018neural} and coupled monomers, and is thus not applicable to azobenzene. None of the methods have been used for large systems with significant conformational changes \cite{shu2020diabatization, guan2019representation}, such as azobenzene derivatives. Further, our work uses diabatization to ease the fitting of adiabatic states across chemical space. In particular, it addresses the issue of gap overestimation near conical intersections of unseen species, as described in Supplementary Secs. \ref*{sm_sec:ablation} and \ref*{sm_sec:compare_hop}. Our work uses diabatization to address this problem, whereas previous work developed diabatic states because of their favorable theoretical properties. We also note that gap overestimation in unseen species is both a newly-identified and newly-addressed problem, as previous work in ML-NAMD focused on single species only \cite{chen2018deep, dral2018nonadiabatic, hu2018inclusion, li2021automatic, westermayr2019machine, westermayr2020combining, westermayr2020machine}.

The diabatic energies form a non-diagonal Hamiltonian matrix, $\mathbf{H}_d$, which is diagonalized to yield adiabatic energies. When a 2$\times$2 sub-block of $\mathbf{H}_d$ has diagonal elements that cross, and off-diagonal elements that pass through zero, a CI cusp is generated (Fig. \ref{fig:diabat}). The diabatic energies that generate the cusp are smooth, which makes them easier to fit with an interpolating function than the adiabatic energies. In the DANN architecture, smoothness is imposed through a loss function related to the non-adiabatic coupling vector (NACV). The loss minimizes the value that the NACV takes when it is rotated from the adiabatic basis (Eq. (\ref{eq:g_nm})) into the diabatic basis. The NACV measures the change in overlap between two wavefunctions after a small nuclear displacement. If the NACV between two states is zero, then their wavefunctions must change slowly in response to a nuclear perturbation. Therefore, their energies cannot form the cusp in Fig. \ref{fig:diabat}(a), and must instead resemble the smooth energies in Fig. \ref{fig:diabat}(b).

The DANN model was trained on SF-TDDFT \cite{shao2003spin} calculations for 567,037 geometries, using the 6-31G* basis \cite{francl1982self} and BHHLYP \cite{becke1993new} exchange-correlation functional. Unlike traditional TDDFT \cite{levine2006conical}, SF-TDDFT provides an accurate description of the CI region \cite{filatov2013assessment}, and, unlike multi-reference methods, is fairly fast and requires no manual parameter selection. The configurations were sampled from 8,269 azobenzene derivatives, of which 164 were taken from the experimental literature. The remaining molecules were generated from combinatorial substitution using common literature patterns (Supplementary Tables \ref*{sm_tab:motifs} and \ref*{sm_tab:substituents}). 

The data generation process is shown in Fig. \ref{fig:schematics}. Initial data was generated through \textit{ab initio} NAMD with 164 species from the literature, together with normal-mode sampling and distortions of the combinatorial species to near-CI regions. The remaining data was generated through active learning. In each cycle we trained a committee of models, used one model to perform NN-NAMD, and used the committee variance and energy gap to choose NAMD geometries for new quantum chemistry calculations. The cycle was repeated five times in total; further details can be found in the Methods section.

\subsubsection*{Validation}
To test whether the model could reproduce experimental results for unseen molecules, we evaluated it on species that were outside the training set. The test set contained 40 species (20 \textit{cis}/\textit{trans} pairs), including 33 with experimental $S_1$ quantum yields in non-polar solution. Non-polar solution was chosen because it is the closest to the gas-phase conditions simulated here. Solvent effects can be easily incorporated into the model through transfer learning to implicit solvent calculations. Previously this was shown to require new calculations for only a small proportion of the training set \cite{ang2021active}.

The performance of the model is summarized in Table \ref{tab:accuracy}. Statistics are shown for both seen and unseen species. The former contains species that are in the training set, but geometries that are outside of it. The geometries were selected with the balanced sampling criteria described in Supplementary Sec. \ref*{sm_sec:balanced_sampling}. Geometries from unseen species were generated with DANN-NAMD using the final trained model. Half of the DANN-NAMD geometries were selected randomly from the full trajectory and half by proximity to a CI (Supplementary Eq. (\ref*{sm_eq:p_zn})). 100 configurations were chosen for each molecule. 

\begin{figure*}[t]
    \centering
    \includegraphics[width=\textwidth]{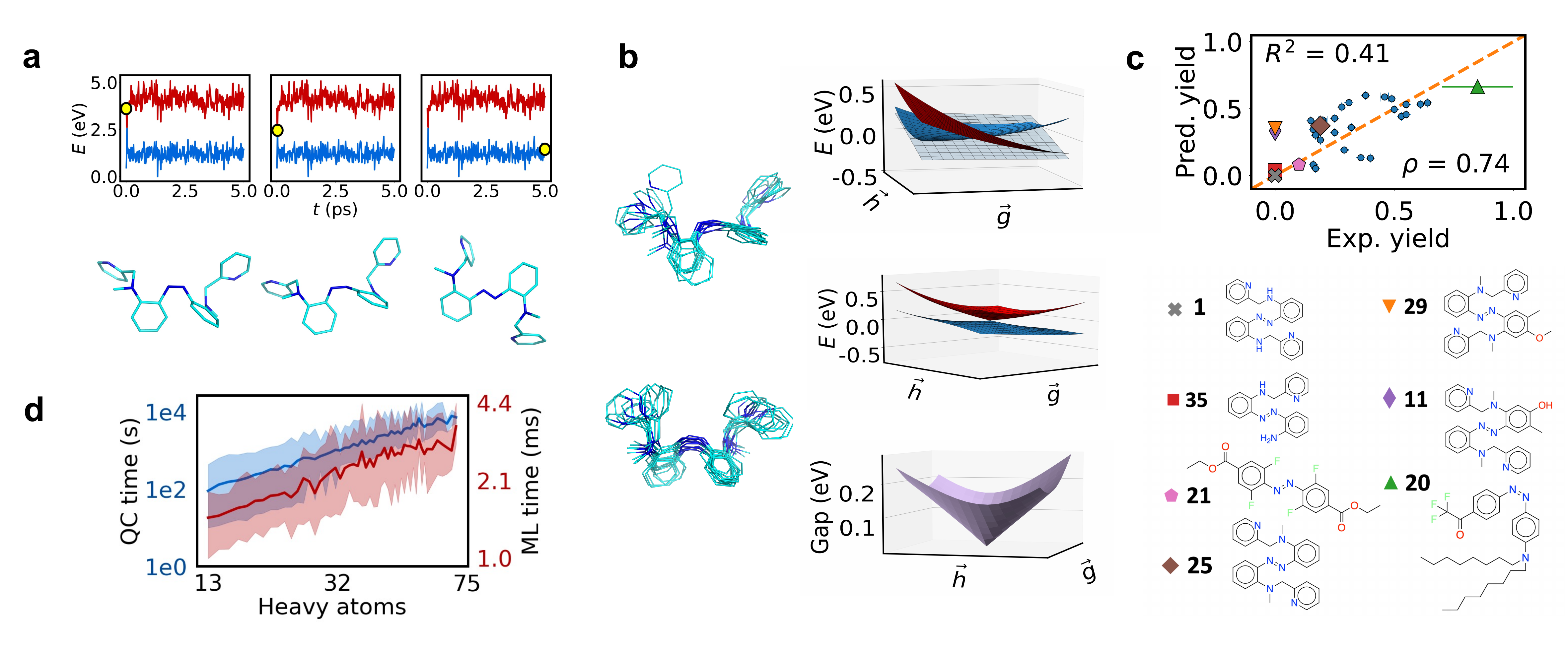}
    \caption{(a) Selected trajectory frames for a molecule outside the training set. The top panels show the $S_0$ and $S_1$ energy as a function of time. A yellow dot indicates the time at which the snapshot below was taken. (b) Left: Overlay of selected hopping geometries from reactive (top) and unreactive (bottom) trajectories. Right: PES as a function of branching plane coordinates at one of the reactive hopping geometries. Diabatic energies, adiabatic energies, and adiabatic gaps are shown from top to bottom. The diabatic coupling is shown in gray. (c) Predicted vs. experimental quantum yield for 33 species outside the training set. The $R^2$ value and Spearman rank correlation $\rho$ are both shown. Color-coded data points are defined below. (d) Node time for QC and ML calculations.}
    \label{fig:results}
\end{figure*}
For species in the training set, all quantities are accurate to within approximately 1 kcal/mol(/\AA). Apart from the NACV, all quantities have $R^2$ correlation coefficients close to 1. The $R^2$ of the NACV is 0.84. This may be somewhat low because diabatization cannot remove the curl component of the NACV in the diabatic basis \cite{mead1982conditions, baer1992study}. This would also explain the low $R^2$ value for the NACV in Ref. \cite{li2021automatic}, which computed it as the gradient of a scalar. For molecules outside the training set, all quantities apart from the energies have an error below 3 kcal/mol(/\AA). The energy gaps and ground state forces have $R^2$ correlation coefficients near 1. The gap error of 1.89 kcal/mol should be contrasted with the error of 15 kcal/mol in Ref. \cite{inamori2020spin}, which applied semi-empirical methods to azobenzene. The errors in the excited state forces are slightly larger, but still quite low. The correlation coefficient for the force NACV $g_{01}$ is rather poor. As described in Supplementary Sec. \ref*{sm_sec:compare_hop}, the yields of \textit{trans} derivatives are better correlated with experiment when using Zhu-Nakamura surface hopping than Tully's method. The latter uses the NACV and the former does not, so part of the difference may be explained by the high error in the force NACV. Nevertheless, there is still reasonable agreement between Tully's method and experiment, suggesting that errors in the force NACV do not spoil the dynamics.

Figure \ref{fig:results}(a) shows snapshots from an example DANN-NAMD trajectory, and panel (b) shows random samples of the hopping geometries. Reactive hopping geometries are shown on top, and non-reactive ones are shown below. The molecule is the (aminomethyl)pyridine derivative \textbf{26}, with the species numbering given in Supplementary Tables \ref*{sm_tab:test_lit_specs} and \ref*{sm_tab:train_lit_specs}. The overlays show \textit{cis}-\textit{trans} isomerization proceeding through inversion-assisted rotation, consistent with previous work \cite{toniolo2005simulation}. The dominant motion is rotation, with the CNNC dihedral angle increasing in magnitude from $ -10^{\circ}$ at equilibrium to $ -86^{\circ}$ at the hopping points. Significant changes also occur in the CNN and NNC angles, with each transitioning from $ 123^{\circ}$ to either $ 113^{\circ}$ or $135^{\circ}$. 

The predicted PES in the branching space $(\vec{g}, \vec{h})$ is shown beside the geometries. $\vec{g}$ is the direction of the force coupling and $\vec{h} \propto \nabla_{R} (\Delta E_{01})$ is the direction of the gap gradient. Each vector was computed with automatic differentiation using Eq. (\ref{eq:hd_nacv}).   The diabatic energies, adiabatic energies, and gap are shown from top to bottom. We see that the model generates a true CI, in which the $S_0$ and $S_1$ energies are exactly equal. Further, the degeneracy is lifted in both the $\vec{g}$- and $\vec{h}$-directions, so that the $S_1$ energy and gap each form a characteristic cone. These hallmarks of CIs are built into the model because the adiabatic energies are eigenvalues of a diabatic matrix. For example, the cone emerges from the fact that $d_{11} - d_{00}$ and $d_{01}$ each pass linearly through zero in different directions \cite{koppel2001construction}.
% \cite{higgins1961some, herzberg1963intersection, carrington1972triatomic, higgins1975intersection, davidson1977global, koppel1984multimode, koppel2001construction}.

Figure \ref{fig:results}(c) indicates that the predicted and experimental quantum yields of unseen species are correlated. The yields are for the 33 \textit{cis} and \textit{trans} species with experimental data in Supplementary Table \ref{sm_tab:test_lit_specs}. The $R^2$ value is 0.42, and the Spearman rank correlation coefficient $\rho$ is 0.74. While the $R^2$ value is somewhat low, the Spearman rank correlation is high. The Spearman coefficient measures the accuracy with which the model ranks species by quantum yield. $\rho$ only compares orderings, while $R^2$ compares the model error to the error of a mean predictor. This means that $\rho$ is a more forgiving metric, and also a more relevant metric for virtual screening. Since \textit{cis} isomers have yields 2 to 3 times higher than \textit{trans} isomers, the high value of $\rho$ means that the model properly separates the isomers into low- and high-yield groups.

Further, as shown in Supplementary Figs. \ref*{sm_fig:yields_zn_diabat} and \ref*{sm_fig:yields_tully_diabat}, the model produces meaningful rankings among \textit{trans} species. The correlation coefficients are $\rho=0.32$ using Tully's method \cite{tully1990molecular} and $\rho=0.57$ using the Zhu-Nakamura approach \cite{yue2018performance}. The model is largely able to differentiate between high- and low-yield \textit{trans} derivatives. Several such molecules are of interest. They are color-coded in the plots, with the legend given below. A full list of predictions is given in Supplementary Table \ref*{sm_tab:test_lit_specs}. We see, for example, that the (aminomethyl)pyridine derivatives \textbf{1} and \textbf{35} are both predicted to have near-zero yields. These species do not isomerize from \textit{trans} to \textit{cis}, because strong N-H hydrogen bonds lock the planar \textit{trans} conformation in place \cite{bandara2010proof}. Replacing the NH group in \textbf{1} with $\mathrm{N-CH_3}$ gives species \textbf{25}. This molecule isomerizes because there is no hydrogen bonding. This, too, is predicted by the model. Further, the hepta-tert-butyl derivative \textbf{17} has an experimental and predicted yield of zero. This is likely because of steric interactions among the bulky tert-butyl groups. While able to account for these two different mechanisms, the model fails to predict the subtle electronic effects in species \textbf{11} and \textbf{29}. Resonance interactions between oxygen lone pairs and the azo group modify the PES, such that there is no rotational CI \cite{bandara2011short}. There is instead a concerted inversion CI, which occurs too early along the path between \textit{trans} and \textit{cis} to allow for isomerization. The changes in the PES may either be too small or too specific to the substituents for the model to predict without fine tuning. Finally, derivatives with high yields are partly distinguished from those with low but non-zero yields. An example is \textbf{21}, whose experimental yield of 10\% is half that of \textit{trans}-azobenzene. The model properly identifies this molecule as having a low yield, but also mistakenly does the same for several high-yield species. The accuracy for unseen species could always be improved with transfer learning, in which the model is fine-tuned with a small number of calculations from a single molecule (discussed below). This would increase the computational cost, but would still be orders of magnitude less expensive than \textit{ab initio} NAMD. 

While meaningful correlations are produced for \textit{trans} species, the same is not true of \textit{cis} molecules ($\rho=0.02$). This may be because there are no \textit{cis} derivatives with zero yield. Nevertheless, the model properly identifies \textbf{20} as having the highest yield. Further, it does not mistakenly assign a zero yield to any derivative. This is noteworthy because, as shown in Fig. \ref{fig:screening}(a) and (b), several hypothetical \textit{cis} species are predicted to have zero yield. Synthesis of non-switching \textit{cis} derivatives and comparison to predictions could therefore be of interest in the future.

Overall, we observe moderate correlation between predicted and experimental yields. The Spearman correlation is high when including both isomers, moderate for \textit{trans} isomers, and low for \textit{cis} isomers. The $R^2$ value, a measure of numerical error compared to that of a mean predictor, is moderate when including both isomers and near-zero when separating them. Indeed, the MAEs of the mean predictor are 9.5\%, 10.3\%, and 17.7\% for \textit{trans}, \textit{cis}, and all species, respectively. The model MAEs before (after) subtracting the mean signed error are 14.4\% (13.5\%), 11.5\% (11.2\%) and 13.2\% (13.0\%). In addition to model error, sources of error include inaccuracies in SF-TDDFT, approximations in surface hopping, solvent effects, and experimental uncertainty. These are discussed in depth in Supplementary Sec. \ref*{sm_sec:sources_of_error}. Each source of error affects both $R^2$ and $\rho$, but is expected to have a larger effect on $R^2$. The rank correlation with experiment is encouraging given the difficulty of the task, as captured by the sensitivity of the yield to model errors in the PES \cite{yue2018performance}, and given the sources of error outside the model. Further, as discussed below, DANN provides an excellent starting point for fine-tuned, molecule-specific models that can be used for high-accuracy simulations of single species.  

Figure \ref{fig:results}(d) shows that DANN-NAMD is extremely fast. The plot shows the node time, defined as $t_{\mathrm{calc}} / n_{\mathrm{calc}}$, where $t_{\mathrm{calc}}$ is the calculation time per geometry, and $n_{\mathrm{calc}}$ is the number of parallel calculations that can be performed on a single node. We see that ML speeds up calculations by five to six orders of magnitude. The direct comparison of the pre-trained model node times and QC node times is appropriate because the model generalizes to unseen species. This means that it incurs no extra QC cost for any future simulations. The minimum speedup corresponds to the smallest molecules (14 heavy atoms or 24 total atoms), and the maximum to the largest molecules (70 heavy atoms or 99 total atoms). This reflects the different scaling of the QC and ML calculations. Empirically we see that DANN scales as $N^{0.49}$ for $N$ heavy atoms, while SF-TDDFT scales as $N^{2.8}$. These values come from fitting the timings to $t = A \cdot N^x$, where $t$ is the computational time, $A$ and $x$ are fitted constants, and $N$ is the number of heavy atoms. DANN's apparent sub-linear scaling is an artifact of diagonalizing $\mathbf{H}_d$; when the diagonalization is removed, the scaling becomes linear. This is the expected scaling for a message-passing neural network with a fixed cutoff radius. Evidently diagonalizing $\mathbf{H}_d$ introduces a large overhead with weak dependence on system size. Nevertheless, we see that DANN is still quite fast.

% , though DFT scaling can be made linear for large enough systems \cite{mohr2015accurate}. 

\subsubsection*{Virtual screening}
Having shown that the model is fast and generalizes in the chemical and configurational space of azobenzenes, we next used it for virtual screening of hypothetical compounds. We first retrained the network on all available data, including species that were originally held out, for a total of 631,367 geometries in the training set. We then predicted the quantum yields of 3,100 combinatorial species generated through literature-informed substitution patterns, as in Ref. \cite{gomez2016design}. This screen served two purposes. The first was to gather statistics about the distribution of photophysical properties of azobenzenes at a scale not accessible to experiments or traditional simulations. The second was to identify molecules with rare desirable properties. In particular, we sought to find molecules with high $c\xrightarrow{}t$ or $t\xrightarrow{}c$ quantum yields and redshifted absorption spectra. The former is important because increasing the ratio $\mathrm{QY}_{a \xrightarrow{} b} \hspace{0.05cm}  / \hspace{0.05cm} \mathrm{QY}_{b \xrightarrow{} a}$, where QY is the quantum yield, can lead to more complete \textit{a}$ \xrightarrow{} $\textit{b} transformation under steady state illumination. This is critical for precise spatial control of drug activity when the two isomers have different biological effects \cite{velema2014photopharmacology}. Redshifting is a crucial requirement for photo-active drugs, since human tissue is transparent only in the near-IR \cite{velema2014photopharmacology}. % We identified several promising candidates with intriguing physical properties and confirmed our predictions with \textit{ab initio} NAMD (ab-NAMD).  

\begin{figure*}[t]
    \centering
    \includegraphics[width=\textwidth]{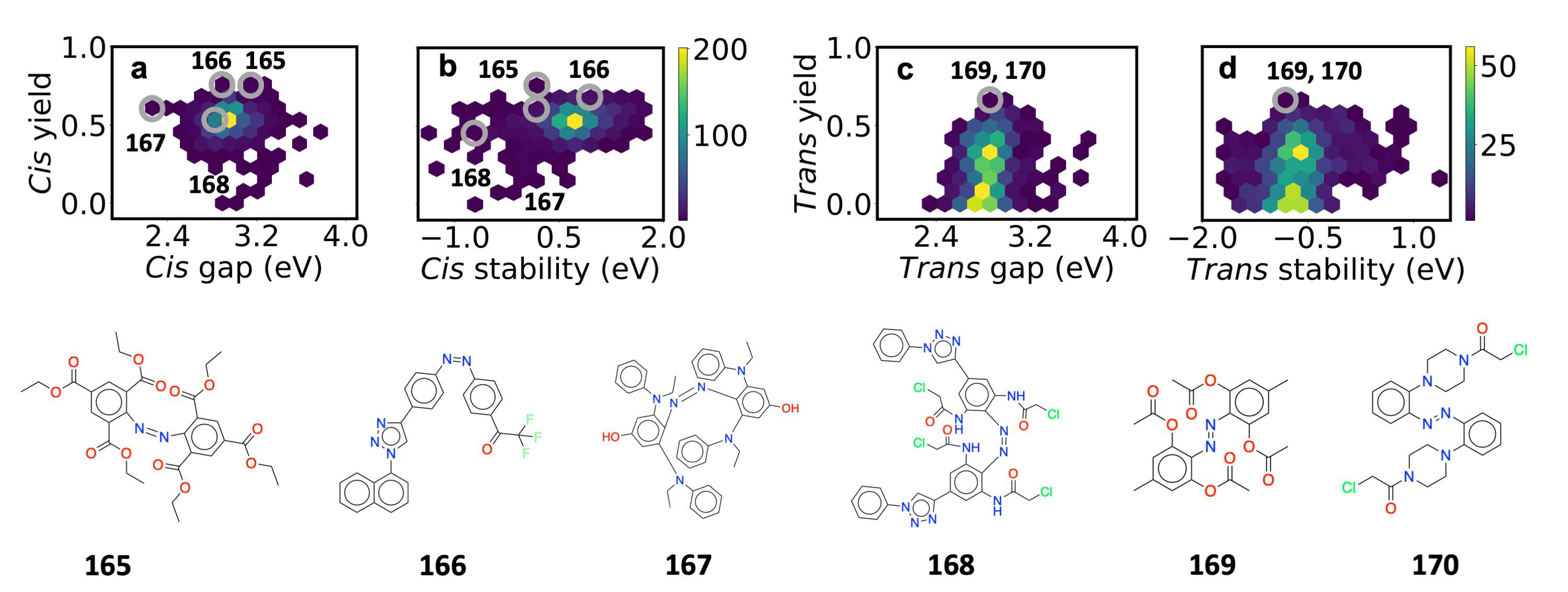}
    \caption{Results of virtual screening. Species of interest are circled in gray and shown below the plots. (a) Predicted yield vs. excitation energy for \textit{cis} derivatives. (b) Predicted yield vs. stability for \textit{cis} derivatives. (c)-(d) As in (a)-(b), but for \textit{trans} derivatives.  }
    \label{fig:screening}
\end{figure*}

The results are shown in Fig. \ref{fig:screening}. Panels (a) and (c) show the predicted yield vs. mean gap. For each species we averaged the gap over the configurations sampled during neural network ground state MD. The thermal averaging led to a typical blueshift of 0.2-0.3 eV relative to the gaps of single equilibrium geometries. The mean excitation energies are 2.95 eV for \textit{cis} derivatives and 2.84 eV for \textit{trans} species; the gaps are 2.98 eV and 2.97 eV for the respective unsubstituted compounds. The average gaps and their differences are similar to experimental measurements for azobenzene \cite{bandara2012photoisomerization}. The average $c\to t$ and $t \to c$ yields are 54\% and 24\%, respectively, while those of the unsubstituted species are 59\% and 37\%. These are consistent with experimental results in non-polar solution, for which the base compound has yields of 44-55\% and 23-28\% \cite{bandara2012photoisomerization}; the former is closer to 55\% on average. However, the yield of the base \textit{trans} compound is overestimated with respect to both theory and experiment \cite{bandara2012photoisomerization, yue2018performance, yu2020nonadiabatic}. The mean (median) proportion of trajectories ending in the ground state after 2 ps are 92\% (100\%) for \textit{cis} species and 31\% (17\%) for \textit{trans} species. The standard deviations are 25\% and 30\%, respectively.

\begin{figure*}[t]
    \centering
    \includegraphics[width=\textwidth]{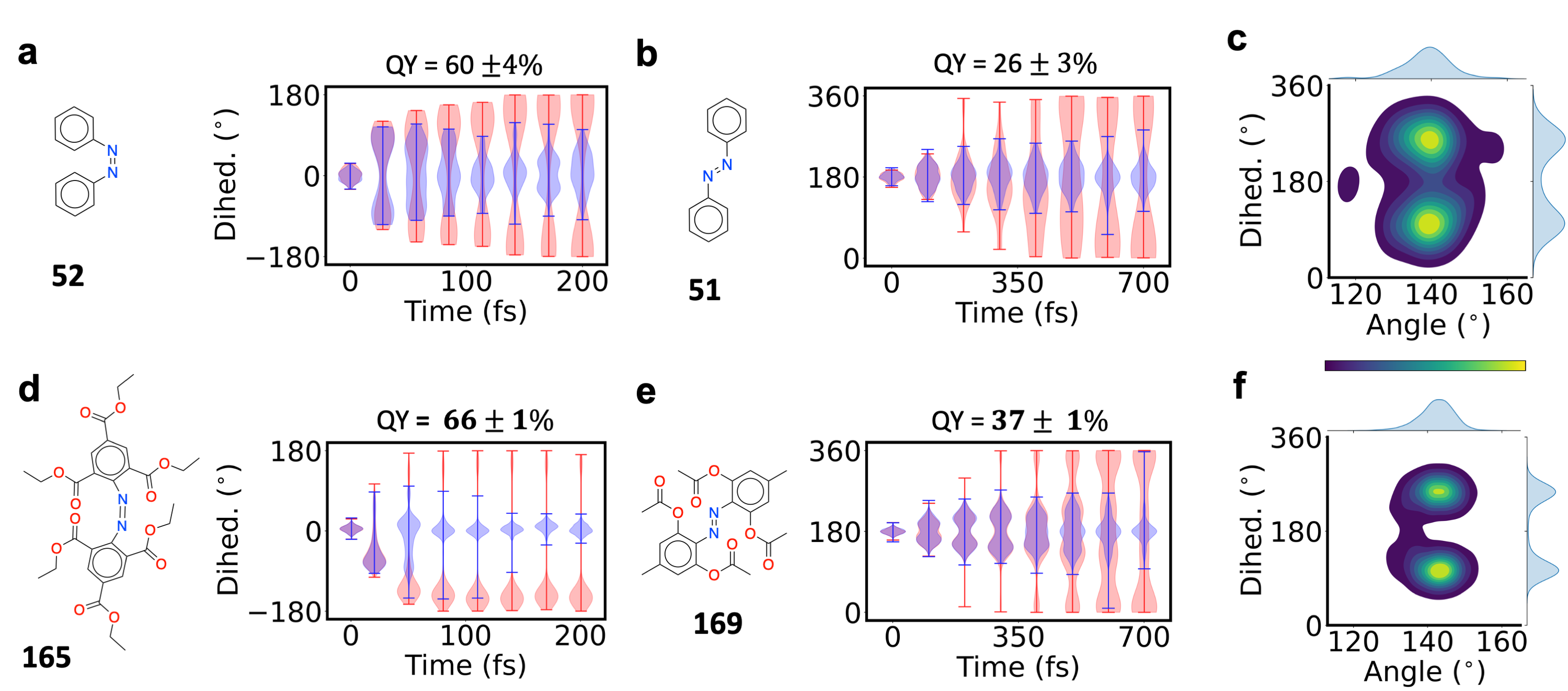}
    \caption{Visualization of high-accuracy non-adiabatic dynamics for several compounds of interest. (a)-(b), (d)-(e) Violin plots showing the CNNC dihedral angle vs. time. Reactive and non-reactive NAMD trajectories are shown in red and blue, respectively. The violin width at a given dihedral angle indicates the density of trajectories with that angle. The yield of each compound is shown above the plots. For ease of visualization we have used the range [$-180$, $180$] for \textit{cis} dihedral angles and [$0$, $360$] for \textit{trans} dihedral angles. (c)  Distribution of hopping geometries for \textit{trans} azobenzene. (f) As in (c), but for the derivative \textbf{169}. The density is visualized with kernel density estimation as a function of the CNNC dihedral and $\mathrm{max}(\alpha_{\mathrm{CNN}}, \alpha_{\mathrm{NNC}})$, where $\alpha$ is an angle. Yellow corresponds to the highest density and blue to the lowest. The marginal distributions over single coordinates are shown above and to the right of each plot. }
    \label{fig:screen_dynamics}
\end{figure*}
Panels (b) and (d) show the yield plotted against the isomeric stability, defined as $E_{{trans}} - E_{{cis}}$ for \textit{trans} isomers and $E_{{cis}} - E_{{trans}}$ for \textit{cis} isomers. The energy $E$ is the median value of the configurations sampled in the ground state; we used the median to reduce the effect of outlier geometries. On average the \textit{trans} isomers are more stable than the \textit{cis} isomers by 0.66 eV (15.3 kcal/mol), which is similar to experimental values over 10 kcal/mol for azobenzene \cite{dias1992enthalpies}. The stability is of interest for three reasons. First, a large absolute value indicates that one isomer is dominant at room temperature. This is essential for photoactive drugs, and is the case for regular azobenzene. Second, an inverted stability, in which \textit{cis} is more stable than \textit{trans}, allows for stronger absorption at longer wavelengths. This is because the dipole-forbidden $n-\pi^*$ ($S_1$) transition is significantly stronger for \textit{cis} than for \textit{trans} \cite{bandara2012photoisomerization}. Third, in photopharmacology, one often wants to deliver a drug in inactive form, and activate it with light in a localized region. If \textit{trans} happens to be active and \textit{cis} inactive, then localized activation is only possible if \textit{cis} is more stable.

Several species of interest are shown in Fig. \ref{fig:screening}. The molecules \textbf{165} and \textbf{166} have predicted  $c\to t$ yields of 75 $\pm$ 6\% and 72 $\pm$ 6\%, respectively, which are well above the \textit{cis} average of 55\%. The species \textbf{169} and \textbf{170} have predicted  $t\to c$  yields of 66 $\pm$ 7\% and 63 $\pm$ 10\%, respectively, which are three times the average \textit{trans} yield. Molecule \textbf{167} is highly redshifted, with a mean predicted gap of 2.26 eV (548 nm), and a standard deviation of 0.87 eV. QC calculations on the geometries sampled with DANN gave a gap of 2.26 $\pm$ 0.61 eV, in good agreement with predictions. The mean gap is lower than the median of 2.52 eV, which reflects the presence of several ultra-low gap structures. The low gap and large variance mean that \textbf{167} may be able to absorb in the near IR. The redshifting is likely because of the six electron donating groups, which increase the HOMO energy, together with the crowding of the four \textit{ortho} substituents. The latter distorts the molecule, leading to twisted configurations with smaller gaps. Finally, species \textbf{168} is more stable in \textit{cis} form than \textit{trans} form. The predicted \textit{cis} stability is $-0.79$ eV ($-18$ kcal/mol), in good agreement with the QC prediction of $-0.92$ eV ($-21$ kcal/mol). As mentioned above, this inverted stability can be a desirable property for photopharmacology.

To validate the yield results, we performed DANN-NAMD using highly accurate species-specific models. As described in Supplementary Sec. \ref*{sm_sec:tl}, we generated a model for each species by refining the base network with data from that species alone. The data was generated through several active learning cycles, resulting in 1,200 to 2,500 training geometries for each compound. We used this approach in place of \textit{ab initio} NAMD because of the latter's prohibitive cost for large molecules. The QC computational cost for fine-tuning was at most 3\% of that of an \textit{ab initio} simulation, and hence far less demanding. The average gradient calculation for a molecule with 50 atoms took 58 minutes for two surfaces using 8 cores, and the average NACV calculation took 55 minutes. Fine-tuning with 2,000 geometries for a medium-sized molecule would thus take 30,000 core hours. For \textit{ab initio} NAMD, a conservative estimate of 100 trajectories run for 1 ps each, with only one gradient computed per frame, would take 780,000 core hours.

We also computed the yields of \textit{cis} and \textit{trans} azobenzene for comparison. For these species we used full \textit{ab initio} simulations, because of the central role of the unsubstituted compound as a reference point and because simulations were fairly affordable for such small molecules. Issues with spin contamination also hampered the fine-tuning process for these compounds (see Supplementary Sec. \ref*{sm_sec:tl}).

Initially we generated refined models for species \textbf{165}, \textbf{167}, \textbf{169} and \textbf{170}. It became clear early on that only \textbf{165} and \textbf{169} had high yields, and so we focused on those molecules thereafter. Using molecule-specific models, we computed the quantum yields of \textbf{165} and \textbf{169} to be $66 \pm 1\%$ and $37 \pm 1\%$, respectively. The computed yields for \textit{cis} and \textit{trans} azobenzene are $60 \pm 4\%$ and $26 \pm 3\%$, respectively, which are in excellent agreement with experiment \cite{bandara2012photoisomerization}. Both of the new molecules have higher quantum yields than the associated base compounds. The improvement is particularly large for species \textbf{169}: its yield is 11 points higher than \textit{trans} azobenzene, which is a relative enhancement of 42 percent. We argue below that that this significant increase has an intuitive physical explanation.

The dynamics of the four molecules are shown in Fig.  \ref{fig:screen_dynamics}. Panels (a) and (b) show the CNNC dihedral angle vs. time for azobenzene, and panels (d) and (e) show the same for the derivatives. Both the substituted and unsubstituted \textit{cis} isomers rapidly proceed through a rotational CI, but the derivative rotates much more quickly. Indeed, we see that the isomerization of the derivative is complete within 75 fs, while the base compound takes nearly 130 fs. The excited state lifetimes are $34.2 \pm 0.3$ fs and $63 \pm 3$ fs for the derivative and base compound, respectively, indicating that the former reaches the CI earlier than the latter. These observations may explain the enhanced yield, since a higher rotational velocity leads to more efficient isomerization \cite{yu2014trajectory}. We also note that the derivative rotates in only the counter-clockwise direction, while \textit{cis} azobenzene rotates in both directions, but this is not expected to affect the yield. 

The two \textit{trans} molecules behave in qualitatively different ways. In \textit{trans} azobenzene, the distribution of dihedral angles slowly widens with time (Fig. \ref{fig:screen_dynamics}(b)). This is consistent with a rotational barrier \cite{yu2020nonadiabatic, yue2018performance}, as different trajectories overcome the barrier at different times, and so the torsion angle becomes uniformly distributed. Additionally, as seen in the marginal dihedral distribution of Fig. \ref{fig:screen_dynamics}(c), many of the geometries hop near 180$^{\circ}$. This agrees with Ref. \cite{yu2020nonadiabatic}, which identified a non-reactive planar CI and a reactive twisted CI as the main hopping points for \textit{trans} azobenzene. The non-reactive CI leads exclusively back to \textit{trans}, while the reactive CI leads to \textit{cis} and \textit{trans} in different proportions. Using the method described in Supplementary Sec. \ref*{sm_subsec:ci_pathways}, we found that 26\% of the trajectories proceed through the planar CI and 74\% through the rotational CI. This is the same distribution reported in Ref. \cite{yue2018performance}. Approximately 36\% of the rotational trajectories generate \textit{cis} azobenzene, giving an overall yield of 26\%. This is in good agreement with previous computational and experimental values \cite{yu2020nonadiabatic}. % This is similar to Ref. \cite{yu2020nonadiabatic}, which found that 40\% of the trajectories pass through the planar CI, leading to a somewhat lower yield of 25\%.

By contrast, nearly all trajectories of \textbf{169}, including non-reactive trajectories, rotate significantly. This can be seen in the marginal dihedral distribution in Fig. \ref{fig:screen_dynamics}(f), in which the hops are tightly localized around $180 \pm 77^{\circ}$. Only 5\% of the trajectories hop at the planar CI, which is five times lower than the base compound. Additionally, the yield of the rotational trajectories increases from 36\% to 40\%. The inhibition of the planar CI pathway, together with the enhancement of the rotational yield, leads to an overall yield increase from 26\% to 37\%. While the enhanced reactive yield does not have a simple explanation, the reason for the planar pathway inhibition can be clearly seen in Fig. \ref{fig:screen_dynamics}(e). Whereas the rotation of \textbf{51} is stochastic, leading to a uniform distribution of angles, the rotation of \textbf{169} is initially concerted. Nearly all trajectories rotate in unison to a dihedral angle of $180 \pm 45^{\circ}$ at 300 fs. Past 300 fs, hopping begins and the trajectories separate from each other. Hence they proceed through the rotational reactive CI, and become distributed between $0^{\circ}$ and $360^{\circ}$ after hopping. The planar non-reactive CI is avoided because of the molecule's initial rotation. This explanation is consistent with the presence of bulky \textit{ortho} groups, which twist the equilibrium structure and hence weaken the N=N double bond. This lowers the excited state barrier to rotation, which leads to an initial torsion and hence increases the yield. % Lastly, we note that this concerted motion reduces the excited state lifetime from \textbf{update!} 1 ps to 400 fs. Such rapid internal conversion is quite desirable since it decreases the probability of intersystem crossing. Triplet states generated through intersystem crossing can react with oxygen to form toxic singlet oxygen \cite{broichhagen2015roadmap}, which makes fast excited state dynamics essential to photopharmacology.

\section*{Discussion}
The DANN model shows high accuracy and transferability among azobenzene derivatives. One limitation is that the unseen species contained functional groups that were present to some degree in the training set. Model performance was generally higher for more highly represented functional groups, though some groups were highly represented yet difficult to fit, while others were weakly represented and well-fit (Supplementary Sec. \ref{sm_sec:func_groups}). Moreover, the model cannot be applied to other chemical families without additional training data. Further, as shown in Supplementary Sec. \ref*{sm_sec:compare_hop}, it substantially overestimates the excited state lifetime for a number of \textit{trans} derivatives. On the other hand, semi-empirical methods provide qualitatively correct predictions across a variety of chemistries, but cannot match DANN's in-domain accuracy, and cannot be improved with more reference data. Adding features from semi-empirical calculations, as done in the OrbNet model \cite{qiao2020multi}, may therefore prove useful in the future. Recent developments accounting for non-local effects and spin states have improved neural network transferability \cite{unke2021spookynet}, and could also be beneficial for excited states. The model could be further improved with high-accuracy multi-reference calculations, solvent effects, and the inclusion of the bright $S_2$ state. The use of spin-complete methods in particular is of crucial importance, since spin contamination prevented fine-tuning the model for the base compounds. It may also have affected the accuracy of the DANN model in general. Thus spin-complete, affordable alternatives are of particular interest \cite{yu2020ab}.  % SF-TDDFT is accurate enough that the model should not require much extra data to do so. The density matrix renormalization group and/or multi-reference methods combined with DFT should be affordable for a small portion of the dataset \cite{marti2011new, sharma2012spin, li2014multiconfiguration}. This transfer learning strategy can also be applied to calculations in different solvents using implicit solvent models \cite{tomasi2005quantum}. Inclusion of the bright $S_2$ state may also be of interest. 
Active learning could be accelerated through differentiable sampling with adversarial uncertainty attacks \cite{schwalbe2021differentiable}, which would improve the excited state lifetimes. Transfer learning could also be used to improve performance for specific molecules. Only a small number of \textit{ab initio} calculations would be required to fine-tune the model for an individual species. 

Diabatization may also prove to be useful for reactive ground states. Reaction barriers can often be understood as transitions from one diabatic state to another \cite{van2010diabatic}. The diabatic basis %  has been used to analyze proton transfer \cite{hammes1994proton}, $\mathrm{S_N}1$ reactions \cite{kim1992theoretical}, and $\mathrm{S_N}2$ reactions \cite{mo2000ab}, and 
may make reactive surfaces easier to fit with neural networks.

In conclusion, we have introduced a diabatic multi-state neural network potential trained on over 630,000 geometries at the SF-TDDFT BHHLYP/6-31G* level of theory, covering over 8,000 unique azobenzene molecules. We used DANN-NAMD to predict the isomerization quantum yields of derivatives outside the training set, and the results were well-correlated with experiment. We also identified several hypothetical compounds with high quantum yields, redshifted excitation energies, and inverted stabilities. The network architecture, diabatization approach, and chemical and configurational diversity of the training data allowed us to produce a robust and transferable potential. The model can be applied off-the-shelf to new molecules, producing results that replicate those of SF-TDDFT at orders of magnitude lower computational cost. \\

%% file: sections/methods.tex
\section*{Methods}
\subsubsection*{Network and training}

As explained in Supplementary Sec. \ref*{sm_sec:extended_methods}, a unique challenge for non-adiabatic simulations is their sensitivity to the energy difference between states. Using a typical neural network to predict energies and forces for NAMD leads to some molecules becoming incorrectly trapped in the excited state. This is partly caused by overestimation of the gap and/or an incorrectly shaped PES in the vicinity of the CI. To address this issue we introduce an architecture based on \textit{diabatic} states, whose smooth variation leads to more accurate neural network fitting (Fig. \ref{fig:diabat}(b)). 

In general diabatic states must satisfy \cite{schuurman2007vibronic}
\begin{align}
& \big( \mathbf{U}^{\dagger}\big[ \nabla_{R}  \mathbf{H}_{d}  \big] \mathbf{U} \big)_{nm} = \begin{cases} -\vec{f}_{n}, \text{ if } n = m  \\ \vec{g}_{nm}, \text{ if } n \neq m. \end{cases} \label{eq:hd_nacv}
\end{align}
where $\nabla_R$ is the gradient with respect to $\vec{R}$, $\mathbf{U}$ diagonalizes the diabatic Hamiltonian through
\begin{align}
& \big( \mathbf{U}^{\dagger} \mathbf{H}_{d} \mathbf{U} \big)_{nm} = E_n \hspace{0.05cm} \delta_{nm}, \label{eq:u_def}
\end{align}
and $\vec{f}_n = -\nabla_{R} E_n$ is the adiabatic force for the $n^{\mathrm{th}}$ state. The dependence on $\vec{R}$ has been suppressed for ease of notation. $\vec{g}_{nm}$ is the force coupling,
\begin{alignat}{2}
 \vec{g}_{nm}(\vec{R}) & = \left\langle \psi_n(\vec{r}; \vec{R}) \left\vert \nabla_R \hat{H}(\vec{r}, \vec{R})  \right\vert \psi_m(\vec{r}; \vec{R})  \right\rangle \nonumber \\
& = (E_m(\vec{R}) - E_n(\vec{R})) \ \vec{k}_{nm}(\vec{R}) \label{eq:g_nm},
\end{alignat}
where $\hat{H}(\vec{r}, \vec{R)}$ is the clamped nucleus Hamiltonian, $\psi_n(\vec{r}; \vec{R})$ is the $n^{\mathrm{th}}$ adiabatic wavefunction, and the matrix element is an integral over the electronic degrees of freedom $\vec{r}$. The vector $\vec{k}_{nm}(\vec{R})$ is the derivative coupling:
\begin{align}
\vec{k}_{nm}(\vec{R}) = \left\langle \psi_n(\vec{r}; \vec{R}) \Big\vert  \nabla_R \psi_m(\vec{r}; \vec{R}) \right\rangle \label{eq:k_nm}
\end{align}
Combined with the following reference geometry conditions (Supplementary Sec. \ref*{sm_sec:extended_methods}),
\begin{align}
 (E_0, \ E_1)  = \begin{cases} (d_{00}, \ d_{11})  , & \text{ if }\vec{R} \in \text{\textit{trans}} \\ (d_{22}, \ d_{00}) , & \text{ if }\vec{R} \in \text{\textit{cis}}, \end{cases} \label{eq:ref_geom_azo}
\end{align}
we arrive at three sets of constraints, Eqs. (\ref{eq:hd_nacv}), (\ref{eq:u_def}), and (\ref{eq:ref_geom_azo}). In principle only Eqs. (\ref{eq:hd_nacv}) and (\ref{eq:u_def}) are required for the states to be diabatic. However, we found the reference loss to provide a minor improvement in the gap near CIs (Supplementary Table \ref*{sm_tab:ablated}).

We use a neural network to map the nuclear positions $\vec{R}_i$ and charges $Z_{i}$ to the diabatic matrix elements $d_{nm}$, and a loss function to impose Eqs. (\ref{eq:hd_nacv}), (\ref{eq:u_def}) and (\ref{eq:ref_geom_azo}). The adiabatic energies $E_n$ are generated by diagonalizing $\mathbf{H}_{d}$, and the forces and couplings by applying Eq. (\ref{eq:hd_nacv}) and using automatic differentiation. The design of the network is shown schematically in Fig. \ref{fig:schematics}(a). The general form of the diabatic loss function is
\begin{align}
\mathcal{L} & = \mathcal{L}_{\mathrm{core}} + \mathcal{L}_{\mathrm{ref}} + \mathcal{L}_{\mathrm{nacv}}. \label{eq:overall_loss}
\end{align}
Here $\mathcal{L}_{\mathrm{core}}$ penalizes errors in the adiabatic energies, forces, and gaps, % , while $\mathcal{L}_{\text{small}}$ penalizes errors in gaps under 0.2 eV (for example, the shaded region in Fig. \ref{fig:diabat}(b)). 
$\mathcal{L}_{\mathrm{ref}}$ imposes Eq. (\ref{eq:ref_geom_azo}) and $\mathcal{L}_{\mathrm{nacv}}$ imposes Eq. (\ref{eq:hd_nacv}) for $n \neq m$.
% The parameters $\alpha$, $\beta$, $\gamma$ and $\delta$ are loss trade-offs that depend on the training epoch $\tau$. In the adiabatic model both $\gamma$ and $\delta$ and are zero. 
The terms are defined explicitly in Supplementary Eqs. (\ref*{sm_eq:core_loss})-(\ref*{sm_eq:nacv_loss}).

For the network itself we adopt the PaiNN equivariant architecture \cite{schutt2021equivariant}. In this approach a set of scalar and vector features for each atom are iteratively updated through a series of convolutions (Fig. \ref{fig:schematics}(a)). In the message block, the features of each atom gather information from atoms within a cutoff distance, using the interatomic displacements.
The scalar and vector features for each atom are then mixed in the update phase. Hyperparameters can be found in Supplementary Table \ref*{sm_tab:hyperparams}. Most were taken from Ref. \cite{schutt2021equivariant}, but some were modified based on experiments with azobenzene geometries. Further details of the PaiNN model can be found in Ref. \cite{schutt2021equivariant}.  Once the elements of $\mathbf{H}_{d}$ are generated, the diabatic matrix is diagonalized to yield the transformation matrix $\mathbf{U}$ and the adiabatic energies $E_n$. The vector quantities $\vec{f}_n$ and $\vec{g}_{nm}$ are given by Eq. (\ref{eq:hd_nacv}). When non-adiabatic couplings are not required, the $\vec{f}_{n}$ can be calculated by directly differentiating the $E_n$. This is more efficient than Eq. (\ref{eq:hd_nacv}), since it requires only $M_{\mathrm{ad}}=2 < M_{\mathrm{d}}(M_{\mathrm{d}}+1)/2=6 $ gradient calculations. This approach was used for NAMD runs, which required only diabatic energies, adiabatic energies, and adiabatic forces, while Eq. (\ref{eq:hd_nacv}) was used for training. % 

\subsubsection*{Data generation and active learning}

Data was generated in two different ways. First, we searched the literature for azobenzene derivatives that had been synthesized and tested experimentally. This yielded 164 species (82 \textit{cis} and 82 \textit{trans}). For these species we performed \textit{ab initio} NAMD, yielding geometries that densely sampled configurational space. Second, to enhance chemical diversity, we generated nearly 10,000 species through combinatorial azobenzene substitution. This was done using 48 common literature substituents and four common substitution patterns (Supplementary Tables \ref*{sm_tab:motifs} and \ref*{sm_tab:substituents}). We then performed geometry optimizations, normal mode sampling, and inversion/rotation about the central N=N bond to generate configurations. QC calculations were performed on 25,212 combinatorial geometries. All calculations were performed with Q-Chem 5.3 \cite{shao2015advances}, using SF-TDDFT \cite{shao2003spin} with the BHHLYP functional \cite{becke1993new} and 6-31G* basis \cite{francl1982self}.

Two neural networks were trained on the initial data and used to perform DANN-NAMD. Initial positions and velocities for DANN-NAMD were generated from classical MD with the Nos\'{e}-Hoover thermostat \cite{nose1984unified, hoover1985canonical}. The initial trajectories were unstable because the networks had not been trained on high-energy configurations. To address this issue we used active learning \cite{ang2021active, wang2020active} to iteratively improve the network predictions (Fig. \ref{fig:schematics}(b)). After each trajectory we performed new QC calculations on a sample of the generated geometries. For all but the last two rounds of active learning, geometries were selected according to the variance in predictions of two different networks, where the networks were initialized with different parameters and trained with different random batches. In the last two rounds, half the geometries were selected by network variance, and half by proximity to a CI. Further details are given in Supplementary Sec. \ref*{sm_sec:active_learning}. The new data was then added to the training set and used to retrain the networks. The cycle was repeated three times with all species and another two times with azobenzene alone. In all, we computed ground-state gradients, excited-state gradients, and NACVs with SF-TDDFT for 641,367 geometries. 96\% of the geometries were from the 164 literature species. 88\% were generated through \textit{ab initio} NAMD and 8\% through active learning. The remaining 4\% were from the combinatorial species. 1.5\% were generated through geometry optimizations, 1.5\% through inversion/rotation, and 1\% through normal-mode sampling.

We initially set out to train a model using energies and forces alone. Since analytic NACVs are unavailable for many \textit{ab initio} methods, an adiabatic architecture could have been used with a wider variety of methods. NACVs also add computational overhead, and so generating training data for an adiabatic model would have taken less time. To this end we initially used the Zhu-Nakamura (ZN) surface hopping method \cite{yu2014trajectory}, which only requires adiabatic energies and forces. However, the issues with adiabatic models described in Supplementary Sec. \ref*{sm_sec:compare_hop} led us to develop the diabatic approach. Since diabatic states can be used with any surface hopping method, we used the diabatic model to perform Tully's fewest switches (FS) surface hopping \cite{tully1990molecular} after the last round of active learning. All results in the main text use the FS method. A comparison of FS and ZN results is given in Supplementary Sec. \ref*{sm_sec:compare_hop}.

%% file: sections/acknowledgements.tex
\section*{Data availability} 
\label{sec:data_avail}
The quantum chemistry data is available at \url{https://doi.org/10.18126/unc8-336t}. A detailed description of how to load and interpret the data is given in the README file. Source data of experimental and predicted quantum yields are provided with this paper.

\section*{Code availability} \label{sec:code_avail}
Trained models and computer code are available in the Neural Force Field repository at \url{https://github.com/learningmatter-mit/NeuralForceField}. Notebook tutorials explain how to train the models and perform DANN-NAMD. 

\section*{Acknowledgements} \label{sec:acknowledgements}
The authors thank Wujie Wang, Daniel Schwalbe-Koda, Shi Jun Ang (MIT), Kristof Sch\"utt, and Oliver Unke (Technische Universit\"at Berlin) for scientific discussions and access to computer code. Harvard Cannon cluster, MIT Engaging cluster, and MIT Lincoln Lab Supercloud cluster at MGHPCC are gratefully acknowledged for computational
resources and support. Financial support from DARPA (Award HR00111920025) and MIT-IBM Watson AI Lab is acknowledged.

\section*{Author Contributions} \label{sec:contrib}
S.A. conceived the project, and developed the methodology with R.G.-B. and E.S. S.A. performed the calculations under the guidance of R.G.-B. S.A. wrote the first draft of the manuscript, and all authors contributed to the final version.

\section*{Competing interests} \label{sec:competing}
The authors declare no competing interests.

%% file: sections/sm.tex
% \parbox[t]{\textwidth}{%
% \begin{center}
%   \large \textbf{Supplementary Information}\par
% \end{center}
% }
% \newcommand{\red}[1]{\textcolor{red}{#1}}
% \newcommand{\blue}[1]{\textcolor{blue}{#1}}

\tableofcontents

\section{Extended methods}
\label{sm_sec:extended_methods}
\subsection{Relevance of adiabatic gap}
\label{sm_subsec:relevance_gap}
Accurate prediction of the energy gap is crucial for generating reliable NAMD results. The gap controls the hopping rate, and in turn observables like the photoisomerization quantum yield, excited state lifetime and time-resolved photoelectron spectrum \cite{hudock2007ab}. To understand the importance of the gap, consider that in most approaches, the hopping rate is determined by the derivative form of the NACV \cite{tully1990molecular, mai2018nonadiabatic}. The derivative coupling between states $n$ and $m$ can be written as $\vec{g}_{nm} / \Delta E_{nm}$, where $\Delta E_{nm}$ is the energy gap and $\vec{g}_{nm}$ is the force coupling \cite{schuurman2007vibronic, herbert2016beyond}. The gap in the denominator leads to singular derivative coupling at CIs, and therefore to a guaranteed hop. The coupling can alternatively be obtained from the gap alone from through its first and second derivatives \cite{westermayr2020combining}. The energy difference also features prominently in the Zhu-Nakamura method, in which the hopping rate is approximately exponential in the square of the gap \cite{yu2014trajectory}. Therefore, it is crucial to accurately predict the energy gap in any NAMD simulation.

\subsection{Diabatization}
The adiabatic energies of a system, $\{ E_n(\vec{R}) \}$, are the energies produced by a QC calculation. They depend on the nuclear coordinates $\vec{R}$, and form the usual Born-Oppenheimer PESs. In the adiabatic basis the nuclear kinetic energy operator, related to $\nabla_{R}^2$, is non-diagonal. Its off-diagonal elements are related to the NACVs, which generate transitions between adiabatic PESs \cite{martinez1997molecular}. The derivative form of the NACV also becomes singular at CIs, which is an undesirable numerical property. The diabatic basis is designed to remove this singularity \cite{kryachko2000diabatic}. The diabatic Hamiltonian $\mathbf{H}_d(\vec{R})$ is a rotation of the adiabatic energies into a new basis, given by $\mathbf{H}_d(\vec{R}) = \mathbf{U}(\vec{R}) \hspace{0.05cm} \mathrm{diag}(\{ E_n(\vec{R}) \} ) \hspace{0.05cm} \mathbf{U}^{\dagger}(\vec{R})$. Here $\mathrm{diag}$ denotes a diagonal matrix, and $\mathbf{U}(\vec{R})$ is a rotation matrix that depends on the nuclear coordinates (see Eq. (\ref*{eq:u_def})). $\mathbf{H}_d$ can also be viewed as the clamped nucleus Hamiltonian expressed in the basis of diabatic wave functions. These wave functions are given by $\psi_{\mathrm{d}, n}(\vec{r}; \vec{R})= \sum_m U_{nm} (\vec{R}) \hspace*{0.05cm} \psi_{\mathrm{ad}, m}(\vec{r}; \vec{R})$, where $\psi_{\mathrm{ad}, m}$ is the $m^{\mathrm{th}}$ adiabatic wave function and $\vec{r}$ denotes the electronic coordinates.

The diabatic states are defined such that the nuclear kinetic energy is approximately diagonal \cite{kryachko2000diabatic}. The states are instead coupled through off-diagonal elements in the potential energy matrix, known as diabatic couplings. The diabatic couplings are smooth functions of the nuclear coordinates. The diagonal elements are also smooth, maintaining their orbital character and switching energy ordering through a CI (Fig. \ref*{fig:diabat}(b)). % Indeed, this uniformity property may be viewed as an alternative definition of diabatic states, and forms the basis of orbital-based \cite{nakamura2001direct, nakamura2002direct, nakamura2003extension, yang2013direct, li2015model} and property-based \cite{subotnik2008constructing, subotnik2010predicting, ou2013electronic, hoyer2014diabatization, hoyer2016dq} diabatization methods. 
In many applications, diabatic states are preferred over adiabatic ones because the singular coupling is removed. Here we prefer them because diabatic energies are easier to fit than adiabatic energies. That is, even if one is only interested in adiabatic energies and not NACVs, it is easier to learn the diabatic energies and then diagonalize $\mathbf{H}_d$ than to learn the adiabatic energies directly. This is because the diabatic states possess no CI cusps or avoided crossings (Fig. \ref*{fig:diabat}), and are thus more easily fit by interpolating functions such as neural networks. As discussed below, diabatic states improve the model accuracy for species outside the training set.

While many diabatization methods exist, the most common ones cannot be straightforwardly applied to the current problem. Property-based methods were developed for charge-transfer type problems \cite{hoyer2016dq}, while orbital-based approaches \cite{nakamura2001direct} are not implemented for TDDFT. %Such methods are not suitable for the generation of initials training data, since they scale steeply with the system size and require user expertise for the selection of diabatic orbitals. 
Diabatic models that are parameterized by electronic structure data \cite{plasser2019highly, schuurman2007vibronic, williams2018neural} are not designed for systems undergoing large distortions. Approaches that solve for the adiabatic-to-diabatic transformation matrix \cite{guan2017construction} are difficult to implement in practice, because the matrix varies rapidly near a CI. % Further, in this approach one needs non-adiabatic couplings with unambiguous phases. For large systems the only viable method of phase correction is to compute the wavefunction overlap between sequential frames in an \textit{ab initio} trajectory \cite{mai2018nonadiabatic}. This precludes the use of active learning \cite{smith2018less, ang2021active, wang2020active} to generate new geometries.

Recent work introduced neural network diabatization based on reference geometries \cite{shu2020diabatization, shu2021permutationally}. In this procedure one assumes that the diabatic Hamiltonian is diagonal at a set of known reference geometries. At such geometries the elements of $\mathbf{H}_d$ are equal to the adiabatic energies, but possibly reordered. For example, for two states and two reference geometries, one would have $\mathbf{H}_d = \mathrm{diag}(E_0, E_1)$ at the first geometry and $\mathbf{H}_d = \mathrm{diag}(E_1, E_0)$ at the second (Fig. \ref*{fig:diabat}(b)). A neural network is then trained to produce $\mathbf{H}_d$, such that its eigenvalues are always equal to the adiabatic energies, and its elements are as above at the reference geometries. These two constraints generate $\mathbf{H}_d$ at all intermediate geometries.

This method was successfully applied to thiophenol dissociation, yielding results consistent with the fourfold way \cite{shu2020diabatization}. However, as shown by the NACV error in Table \ref{sm_tab:ablated}, this method alone does not generate true diabatic states for azobenzene. To understand why, consider that near a CI, the true diabatic coupling $d_{01}$ must closely resemble the coupling shown in Fig. \ref*{fig:diabat}(b) ($d_{nm}$ is shorthand for $(\mathbf{H}_d)_{nm}$). This is because $d_{01}$ must be linear in displacements about a CI, and quadratic only for Renner–Teller type intersections \cite{koppel2001construction}. However, for two diabatic states, only the square of the diabatic coupling $\vert d_{01} \vert^2$ enters into the expression for the adiabatic energies. The model might then generate an off-diagonal element similar to $|d_{01}|$. This error would incur no penalty on the network, because the diagonal elements would properly switch ordering and the adiabatic energies would be correct. Hence the model would not generate smooth diabatic states. 

To remedy this issue we used the combined loss described in the Methods section. The NACV component of this loss was also used in Ref. \cite{guan2019representation}. We note that the number of diabatic states and choice of orderings in Eq. (\ref*{eq:ref_geom_azo}) may depend on the system. For azobenzene we were interested in fitting $M_{\mathrm{ad}}=2$ adiabatic states, and in general one should use $M_{\mathrm{d}} \geq M_{\mathrm{ad}}$ diabatic states. In this work chose $M_{\mathrm{d}}=3$ because it significantly improved on the results of $M_{\mathrm{d}} = 2$. In fact, with $M_{\mathrm{d}}=2$, the $R^2$ correlation of the force NACV was negative. The orderings in Eq. (\ref*{eq:ref_geom_azo}) were chosen by taking a small sample of azobenzene configurations, training several small models with different diabatic orderings, and picking the one with the best results. Thus no system knowledge is required to choose the orderings. The only system knowledge required is the set of reference geometries\footnote{Typically the reference geometries are reactants and products. For reactions in which the product is not known \textit{a priori}, one might use the following approach. First, train an adiabatic network on a single molecule. We have found that adiabatic models match or outperform diabatic ones for single species. Then use the model to simulate dynamics. By using active learning to improve the model's coverage of configurational space, the simulations can eventually discover the product. If derivatives of the molecule are expected to have similar reactions, then their products are also now known. With knowledge of the reactants and products, and hence the reference geometries, one can now build a diabatic model. If this is not possible, then one can train the model without $\mathcal{L}_{\mathrm{ref}}$. Table \ref{sm_tab:ablated} shows that this will likely decrease the model performance near CIs by a small amount.}, but the reference loss can be omitted with negligible impact on model performance.

The diabatic model leads to true CIs, where the ground and excited state energies are exactly equal. To see why, consider Fig. \ref*{fig:results}(b), which shows the two diagonal diabatic elements (red and blue) and the off-diagonal coupling (light gray). The off-diagonal element passes linearly through zero in the $\vec{h}$ direction. The diagonal elements cross in the $\vec{g}$ direction, and so $\Delta \equiv d_{11} - d_{00}$ passes linearly through zero. Therefore, one can begin at a geometry for which $\Delta = 0$, and move in the $\vec{h}$ direction until $d_{01}=0$ without changing $\Delta$. The final geometry will therefore be a CI. The fact that both $d_{01}$ and $\Delta$ are locally linear, and hence must pass through zero, is known theoretically \cite{koppel2001construction} and properly predicted by the model. 

\subsection{Network loss}
The neural network loss terms are defined as:
\begin{alignat}{2}
& \mathcal{L}_{\mathrm{core}} = && \sum_{n}  \rho_{E_n}  \cdot \mathrm{mse} \big(E_n \big) + \rho_{f_n} \cdot \mathrm{mse} \big(\vec{f}_n\big)  \nonumber \\
 &  && \hspace*{-0.35cm} + \sum_{n > m} \rho_{\Delta E_{nm}}  \cdot \mathrm{mse}\big( \Delta E_{nm} \big)    \label{sm_eq:core_loss} \\
% & \mathcal{L}_{\text{small}} =  && \sum_{n > m} \rho_{\mathrm{small}} \cdot  \mathrm{mse}\big( \Delta E^{^{\mathrm{small}}}_{nm} \big)  \\ 
& \mathcal{L}_{\mathrm{ref}} =  && \sum_{n} \rho_{\mathrm{ref}} \cdot 
\mathrm{mse} \big( d_{nn}(\vec{R}) \big) \big\vert_{\vec{R} \in \{\text{\textit{cis}, \textit{trans}}\} } \label{sm_eq:ref_loss} \\
& \mathcal{L}_{\mathrm{nacv}}  =  && \sum_{n > m } \rho_{\mathrm{nacv}} \cdot \mathrm{mse}\big( A_{n}\big( \vec{R} \big) A_m \big( \vec{R} \big) \ \vec{g}_{nm}(\vec{R}) \big), \label{sm_eq:nacv_loss}
\end{alignat}
where $\Delta E_{nm} = E_n - E_m$ is the energy gap. Each loss function is a sum over mean squared error (MSE) terms scaled by different weights $\rho$. For scalar molecular quantities $X$ the loss is given by $\mathrm{MSE}(X) = \sum_{j=1}^{M} (X_j - \hat{X}_j)^2 / M$, where $\hat{X}$ is the predicted quantity and the sum is over $M$ geometries in a batch. For atomic vectorial quantities the mean is additionally over the $3 L_j$ vector components, where $L_j$ is the number of atoms in the $j^{\mathrm{th}}$ geometry. 

The loss term $\mathcal{L}_{\mathrm{core}}$ contains the usual energy and force losses, plus an additional term for gap errors. While the gap term was not used in previous work, we found it to be crucial for the systems studied here. Indeed, without this loss term, one would expect the gap MAE in adiabatic models to be the geometric sum of the energy MAEs. This is what we found when using $\mathcal{L}_{\mathrm{core}}$ without the gap loss. Table \ref{sm_tab:ablated} shows that, when using the full core loss, the gap MAE is actually lower than each individual energy MAE. 

The reference loss, $\mathcal{L}_{\mathrm{ref}}$, is a sum over geometries $\vec{R}$ which are considered to be \textit{cis} or \textit{trans}. At these geometries the target $d_{nn}$ are given by Eq. (\ref*{eq:ref_geom_azo}). A geometry is considered to be \textit{cis} or \textit{trans} if its central CNNC atoms deviate from those of the equilibrium structure by <0.15 \AA. The distance is computed as the root-mean-square deviation (RMSD) among the atoms after alignment. 

The NACV loss imposes diabaticity. It involves the force NACV $\vec{g}_{nm}$, and a phase correction $A_{n}(\vec{R}) = \pm 1$. The phase correction is chosen separately for each geometry to minimize the error between the predicted and target force coupling. This factor is necessary because each adiabatic wavefunction can have an arbitrary sign \cite{richter2011sharc}. The signs cancel for diagonal terms like $E_n$ and $\vec{f}_n$, but not for off-diagonal terms like $\vec{g}_{nm}$. The $A_{n}$ account for these arbitrary sign changes.

\subsection{Data generation}
\label{subsec:data_gen}

To train a useful model one must generate reliable QC data. TDDFT \cite{runge1984density} typically offers a good compromise between speed and accuracy for modeling excited states. However, it has known instabilities near CIs \cite{levine2006conical}, and, as a result of the Brillouin theorem, produces the wrong branching space dimensionality for $S_0/S_1$ intersections \cite{gozem2014shape}. These issues, which can be traced back to the single-reference description of the excitation, can be partially alleviated with SF-TDDFT \cite{shao2003spin}. In this approach the excitation is performed with respect to a high-spin reference state. This yields some transitions that have double-excitation character with respect to the singlet ground state. Here we used SF-TDDFT with the BHHLYP functional \cite{becke1993new} and the 6-31G* basis \cite{francl1982self}. SF-TDDFT/BHHLYP is well-benchmarked for CIs in a number of molecules \cite{filatov2013assessment, nikiforov2014assessment}. Because SF is not spin complete, the singlet states must be identified based on their square spins $\langle S^2 \rangle$. We used the approach of Ref. \cite{yue2018performance}, which identifies singlets as the two states with the lowest $\langle S^2 \rangle$ from the three excitations of lowest energy. Calculations were performed with the Q-Chem package \cite{shao2015advances}.

Near-CI regions of the combinatorial species were sampled by setting the central CNNC dihedral angle of relaxed geometries to $\pm$90 degrees and/or the CNN/NNC angles to $180$ degrees. The other internal coordinates were not changed. The former corresponds to the rotation pathway and the latter to inversion. This led to 11 possible combinations of rotation and inversion, including pure rotation, pure inversion, concerted inversion, and inversion-assisted rotation \cite{bandara2012photoisomerization}.

\section{Model accuracy and ablation studies}
\label{sm_sec:ablation}
\begin{table}[h]
\centering
    \begin{tabular*}{\textwidth}{c @{\extracolsep{\fill}} |c|c|c|c|c|c|c|c}
          \toprule[1.1pt]
          & $E_0$ & $E_1$ & $\Delta E_{01}$ & $(\Delta E)_{\mathrm{small}}$ & sgn$(\Delta E)_{\mathrm{small}} $ & $\vec{F}_0$ & $\vec{F}_1$ & $\vec{g}_{01}$ \\  
        %   \cline{1-1} \cline{3-9}
        %  \midrule[1.1pt]
          \hline
          DANN & 3.06 & 3.77 & 1.89 & \textbf{0.97} & -0.29 & 1.72  & 2.31 & \textbf{1.36} \\
          $-\mathcal{L}_{\mathrm{nacv}}$ & 2.31 & \textbf{2.49} & \textbf{1.65} & 1.22 & 0.52 & 1.72 & 2.39 & 2.21 \\
          $-\mathcal{L}_{\mathrm{ref}}$ & 2.21 & 2.58 & 1.68 & 1.06 & 0.16 & 1.72 & 2.32 & 1.37 \\ 
          adiabat \ & \textbf{2.11} & 3.24 & 2.43 & 1.24 & 0.60 & 1.69 & 2.31 & --- 
          \\ 
          adiabat (st. 1) \ & 3.28 & 2.99 & 1.88 & 1.49 & 0.95 & \textbf{1.67} & \textbf{2.30} & ---  \\
          \ median \  & 16.86 & 19.04 & 22.80 & 36.04 & -36.04 & 17.27 & 17.18 & 2.05  \\ 
          \bottomrule[1.1pt]
    \end{tabular*}
    % \footnotetext[1]{Computed using the Hessian of the gap \cite{westermayr2020combining} for $\Delta E_{01} < 0.5$ eV, and set to zero otherwise.}
    \caption{Performance of different ablated models. Each column, apart from $\mathrm{sgn} (\Delta E)_{\mathrm{small}}$, shows the MAE of a different quantity. $\mathrm{sgn} (\Delta E)_{\mathrm{small}}$ is the mean signed error of $(\Delta E)_{\mathrm{small}}$, given by mean\{$( E)_{\mathrm{small}}^{\mathrm{pred}} - (\Delta E)_{\mathrm{small}}^{\mathrm{target}}$\}. ``st. 1'' indicates the first stage of training of the adiabatic model, as described in Supplementary Sec. \ref{sm_sec:training}. The best score in each category is shown in bold.}
    \label{sm_tab:ablated}
\end{table}

\begin{table*}[t]
\centering
    \begin{tabular*}{\textwidth}{c@{\extracolsep{\fill}}|ccccccccc}
        \toprule[1.1pt]
        &  $E_0$ & $E_1$ & $\Delta E_{01}$ & $(\Delta E_{01})_{\mathrm{small}}$ & $\vec{F}_0$ & $\vec{F}_1$ & $\vec{g}_{01}$ \\ 
        \hline
        \ MAE ($\downarrow$) \ & 0.75 & 0.96 & 0.72 & 0.40 & 0.94 & 1.13 & 0.87   \\
        $R^2$ ($\uparrow$)  & 1.00 & 0.99 & 0.99 & 0.97 & 1.00 & 0.99 & 0.88 \\
        \bottomrule[1.1pt]
    \end{tabular*}
    \caption{Test set MAE for the DANN model trained on all species.}
    \label{sm_tab:all_specs}
\end{table*}

Here we compare the DANN model to a model trained without a NACV loss (``$-\mathcal{L}_{\mathrm{nacv}}$''), a model trained without a reference geometry loss (``$-\mathcal{L}_{\mathrm{ref}}$''), an adiabatic model, and a median predictor. The latter predicts the median ground- and excited-state energy for each species, and the median value among all species for other quantities. The MAEs for unseen species are compared in Table \ref{sm_tab:ablated}. Half the geometries were sampled randomly and half by proximity to a CI, as described in Supplementary Sec. \ref{sm_sec:training}. Of particular interest are $(\Delta E)_{\mathrm{small}}$, the MAE of the gap when it is under 0.2 eV, and $\mathrm{sgn}(\Delta E)_{\mathrm{small}}$, the average overestimation of small gaps by the model. DANN actually \textit{underestimates} the small gaps, indicating that it does not ``miss'' conical intersections. Using an adiabatic model or removing the NACV loss increases both the error and overestimation of $(\Delta E)_{\mathrm{small}}$. The same is true of removing $\mathcal{L}_{\mathrm{ref}}$, but the effect is smaller. While the changes in $(\Delta E)_{\mathrm{small}}$ appear minor, we show in Supplementary Sec. \ref{sm_sec:compare_hop} that they lead to noticeable quantum yield differences for a number of species.

Also of note is the difference in NACV error among the different models. Removing $\mathcal{L}_{\mathrm{nacv}}$ substantially increases the NACV error, even when $\mathcal{L}_{\mathrm{ref}}$ is used. Since $\mathcal{L}_{\mathrm{nacv}}$ imposes diabaticity, this shows that the reference loss alone does not give accurate diabatic states. We also see that the adiabatic model is much worse at predicting $\Delta E$ away from the CI than the diabatic model. This is because of the large loss weight used for $(\Delta E)_{\mathrm{small}}$ in the second stage of adiabatic training (see Supplementary Sec. \ref{sm_sec:training}). Indeed, the results after the first stage are much better for $\Delta E$, but far worse for $(\Delta E)_{\mathrm{small}}$. Finally, the models without reference or NACV losses have far lower energy errors than DANN. It may be possible to decrease DANN's energy errors by increasing the weight of the energy loss.

A key benefit of the diabatic model is that it accurately predicts the gap near CIs. It also produces the NACVs, adiabatic energies, and forces all with one model. Without the diabatic model, one might learn the force coupling as a separate property. This has been done in other work by taking the gradient of a learnable scalar \cite{westermayr2020combining, li2021automatic}. However, one would still have to divide by the adiabatic gap to obtain the derivative coupling. Thus gap prediction errors from an adiabatic model would still be problematic. A separate option would be to compute the NACV through the Hessian of the gap \cite{westermayr2020combining}. This is quite slow, and would have similar problems with the gap.

Table \ref{sm_tab:all_specs} shows the test set statistics for the DANN model trained on all species. This model was used for virtual screening, as described in the main text. Here the test set simply consists of 5,000 held-out geometries. Unlike in Table \ref{sm_tab:ablated}, the test set contains mostly seen species, and the geometries were not generated from DANN-NAMD with the trained model. The results are quite similar to the ``seen species'' rows in Table \ref*{tab:accuracy}.

\section{Experimental data}
\label{sm_sec:parsing_exp}

We performed an extensive literature search to find quantum yields of azobenzene derivatives. To best compare to the vacuum conditions simulated here, we chose quantum yields measured in non-polar solvents. When multiple results were available in different non-polar solvents, we computed the prediction error relative to the mean experimental value. To compare to the simulated $S_1$ dynamics, we selected yields measured at the peak of the highest-wavelength absorption band, which is a dipole-forbidden $n-\pi^*$ transition in unsubstituted azobenzene. All sources used here specifically reported yields at wavelengths close to the $n-\pi^*$ and $\pi-\pi^*$ absorption peaks, so it was straightforward to choose the appropriate wavelength. Lastly, we only selected yields at or around room temperature, since yields can have a strong temperature dependence \cite{malkin1962temperature}. 

% Add in references to all the reported yields in the main text -- should give them the credit

\section{Sources of error}
\label{sm_sec:sources_of_error}

\subsection{Experimental error}
\label{sm_subsec:exp_error}
There are two main sources of uncertainty in the experimental quantum yield results. The first is the error in the absorption coefficients $\varepsilon$ of the two isomers. The absorption strengths are used in standard rate equations to compute the yield; see, for example, Refs. \cite{rau1988photoisomerization, mohr2004photochemistry} and the supplementary of Refs. \cite{knie2014ortho, moreno2015two}. The error can be traced back to the overlapping absorption spectra of the two isomers, which must be disentangled. The magnitude of the error depends on the method used to compute $\varepsilon$, which itself is related to the year of publication. For example, Ref. \cite{rau1988photoisomerization}, published in 1988, indicated that the absorption coefficients were the primary source of error, but did not give an estimate for their uncertainty. They isolated the different isomers using chromatography and computed their absorption coefficients separately. When the isomers could not be isolated in sufficient amounts, the authors used the time-dependent absorption approach of Ref. \cite{rau1984further}. This method had relative errors of of $\pm$25\% for $n-\pi^*$ quantum yields \cite{rau1984further} (i.e, yield $\to \mathrm{yield} \times (1 \pm 0.25)$). Ref. \cite{mohr2004photochemistry}, published in 2004, reported a yield of $0.7$ to $1.0$ for compound \textbf{20}, a large range stemming from the overlap of the isomers' spectra and the method used to separate them  \cite{fischer1967calculation}. More modern methods have lower errors; for example, Ref. \cite{moreno2015two}, published in 2015, reported a relative error of only $\pm$10\% in the supplementary.

The second important source of error is the overlap of the $n-\pi^*$ and $\pi-\pi^*$ absorption bands. The more strongly the two bands overlap, the higher the error of using only the $S_1$ state in the dynamics. % Hence the prediction error can actually depend on the experimental band overlaps. 
Moreover, many experiments do not irradiate at the precise maximum of the $n-\pi^*$ band. Often a representative $n-\pi^*$ or $\pi-\pi^*$ wavelength is chosen, and then used for each derivative \cite{knie2014ortho, malkin1962temperature}, even though the derivatives have different absorption peaks. % The magnitude of this uncertainty depends on how far the irradiation wavelengths are from the peak wavelength.

To get an idea of the range of errors this could introduce, consider the results of Ref. \cite{zimmerman1958photochemical} from 1958. This work computed the quantum yield of unsubstituted azobenzene over several wavelengths. The \textit{trans} to \textit{cis} yield was measured as 21\% at 405 nm and 27\% at 436 nm ($10^{-3}$ M solution). The difference is because 405 nm light excites more of the $\pi-\pi^*$ band (313 nm) than 436 nm light. 405 nm excitation leads to a lower yield because the $\pi-\pi^*$ yield is only 11\%. The difference is even more pronounced for the \textit{cis} quantum yield, which drops from 55\% to 40\% moving from 436 nm to 546 nm. This result does not have a simple explanation, as there is no lower energy band beyond 436 nm. In principle, the error due to overlapping bands could be mitigated with an explicit dipole-electric field coupling term \cite{marquetand2011nonadiabatic}, which would excite multiple states in different proportions. % We can estimate the error of only using the $S_1$ state in the dynamics by comparing the $S_1$ and $S_2$ yields. 
For derivatives with completely overlapping $n-\pi*$ and $\pi-\pi*$ bands, the $S_1$ approximation would incur a maximum error of the $S_2$ yield minus the $S_1$ yield. This is because the $S_2$ state is much brighter than $S_1$, and would therefore dominate the experimental yield. The yield difference is about 10 percentage points for both \textit{cis} and \textit{trans} unsubstituted azobenzene \cite{bortolus1979cis, ronayette1974isomerisation}. Therefore, in the worst case scenario, we would expect an error of about 10 percentage points from the $S_1$ approximation. The error would likely be closer to the 6\% reported in Ref. \cite{zimmerman1958photochemical} at wavelengths of moderate overlap.

We can further quantify these errors by computing the range of results from different studies.  Measurements of the $t \to c$ azobenzene yield range from 20\% to 28\% between 1979 and 1987 \cite{bortolus1979cis, rau1984further, bortolus1987cis}, though this could also be related to solvent effects (Sec. \ref{sm_subsec:comp_errors}).The yields of compounds \textbf{9} and \textbf{10} range from 16\% to 24\% and 44\% to 50\%, respectively, with references from 1962 and 1988 \cite{malkin1962temperature, rau1988photoisomerization}. Aggregating the results of all three compounds gives an average yield range of 7.3 percentage points, and an average relative error of $\pm 14\%$.

A final source of uncertainty is specific to Refs. \cite{bandara2010proof, bandara2011short}. Several species were reported to have zero yield over a large range of irradiation wavelengths. However, as noted in the footnote to Table \ref{sm_tab:test_lit_specs}, the dipole-allowed $S_2$ transition was highly redshifted and thus overlapped strongly with the $S_1$ transition. The absorbance changes after irradiation were quite small, indicating a near-zero yield, but could have been due to the $S_1$ transition. Hence it is possible that the $S_1$ yield is not precisely zero.

\subsection{Computational error}
\label{sm_subsec:comp_errors}

There are several sources of computational error. The first is error in the PES, which can be decomposed into error from SF-TDDFT and error from the model. While it is intractable to perform SF-TDDFT for all species with experimental data, our results for unsubstituted azobenzene suggest that it is rather accurate. As reported in the main text, we computed yields of 60 ± 4\% and 26 ± 3\% for $c \to t$ and $t \to c$, respectively. Experimental measurements between 1979 to 1987 gave $t\to c$ yields between 20\% and 28\% \cite{bortolus1979cis, rau1984further, bortolus1987cis, ronayette1974isomerisation}. Experimental measurements in 1974 and 1979 gave $c\to t$ yields of 55\% and 56\% \cite{bortolus1979cis, ronayette1974isomerisation}, while a measurement in 1958 gave 48\% $\pm$ 5\% \cite{zimmerman1958photochemical}. All measurements were performed in non-polar solvents. The yields computed with the original model were 59\% and 37\% for $c \to t$ and $t \to c$, respectively. Hence the main source of error seems to be the model, rather than SF-TDDFT.

Solvent effects may also affect the yield. These effects can be decomposed into a systematic term and a non-systematic term. It has been argued that non-polar solvents systematically reduce the quantum yield relative to vacuum \cite{tiberio2010does}. However, careful analysis of the experimental data reported in \cite{granucci2007excited} and cited in \cite{tiberio2010does} does not support this conclusion. Indeed, given the good agreement between non-polar experimental results and both SF-TDDFT  and hh-TDA DFT \cite{yu2020nonadiabatic}, systematic effects of non-polar solvents are likely to be small. Solvent-specific effects are part of the range of reported experimental values, since different works often use different non-polar solvents. Typical ranges were discussed in Sec. \ref{sm_subsec:exp_error}.
 
A final source of error is the approximations in surface hopping. A large body of literature has examined surface hopping's accuracy; see, for example, Refs. \cite{landry2012recover, chen2016accuracy, zobel2021surface, ibele2021comparing}. In this work, we can roughly evaluate its performance by comparing its results to those of \textit{ab initio} multiple spawning (AIMS) \cite{yu2020nonadiabatic}. AIMS is a fully quantum mechanical method and can thus be treated as a benchmark. We computed the $t \to c$ yield as 26 $\pm$ 3\% using surface hopping with SF-TDDFT, and Ref. \cite{yu2020nonadiabatic} computed the yield as 24\% $\pm$ 6\% using AIMS with hh-TDA DFT. While there may be some error cancellation from the different quantum chemistry methods, the good agreement is still encouraging.

\section{Influence of different functional groups}
\label{sm_sec:func_groups}

Here we discuss how the model transferability depends on the functional groups in a compound. To answer this question, we analyzed each of the 40 unseen species in the test set, and computed the model error for each geometry. As described in the main text, the geometries were taken from DANN-NAMD with the trained model; half were selected by proximity to a CI and half selected randomly. For each functional group, we aggregated the errors of all geometries that contained the group, and then computed the mean. 

\begin{figure}[t]
    \centering
    \includegraphics[width=\textwidth]{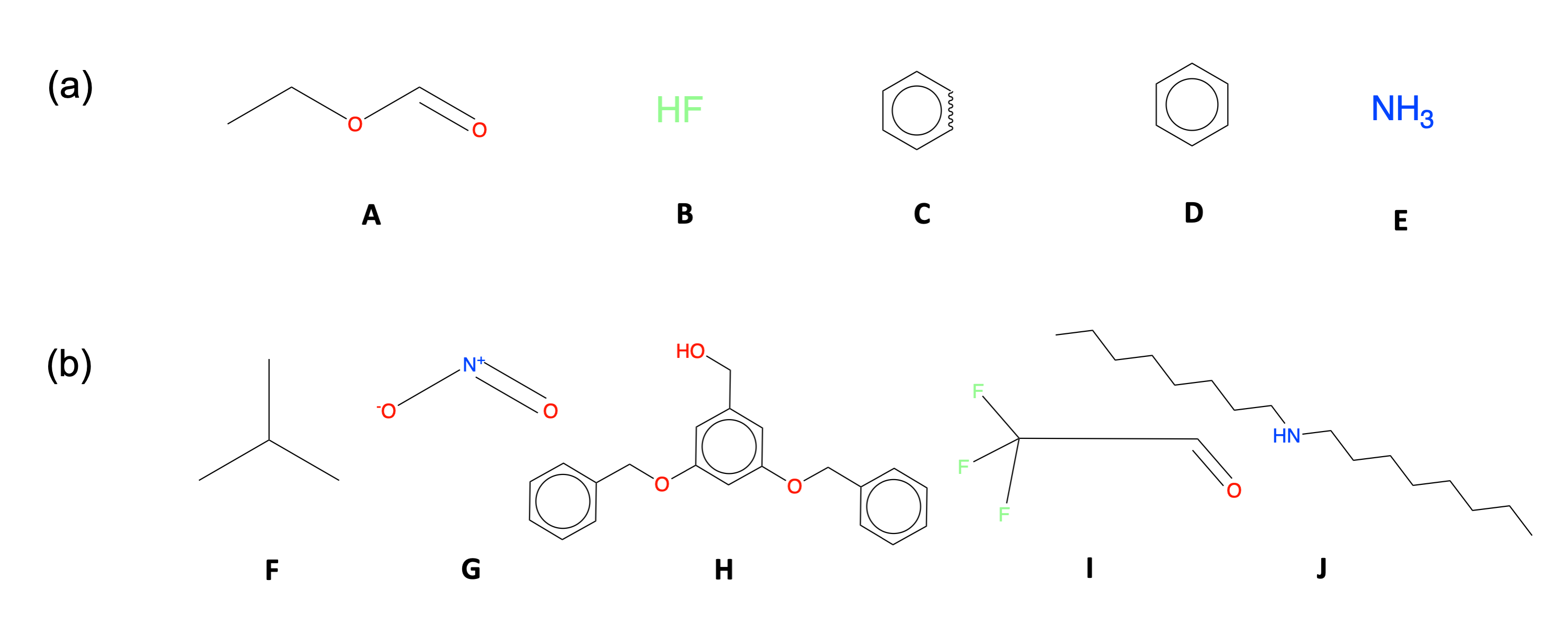}
    \caption{Functional groups in the test set of unseen species. (a) The five functional groups with the lowest gap error. Error increases from left to right. \textbf{C} is fused to a benzene ring in azobenzene. (b) The five groups with the highest gap error. Error decreases from left to right.}
    \label{sm_fig:func_groups}
\end{figure}

The functional groups with the lowest gap errors are shown in Fig. \ref{sm_fig:func_groups}(a), and those with the highest are shown in panel (b). Table \ref{sm_tab:func_groups} shows the number of times that each functional group appears in the training set, together with the errors for all properties. Of the groups with the lowest errors, we see that both \textbf{B} and \textbf{C} are well-represented in the training set, which may explain their accurate results. The other groups each have about 1,000 samples in the training set. This is smaller than that of \textbf{B} and \textbf{C}, but not negligible. The compounds are also rather simple, which may partly explain their low error.

Of the groups with the highest gap error, only the nitro group \textbf{G} is well-represented in the training set. There are almost 10,000 training geometries with nitro groups, yet the error is still quite high. This may be because $\mathrm{NO_2}$ is a strong electron-withdrawing group, and can thus have a significant effect on the electronic structure of azobenzene. Of the remaining substituents, both \textbf{H} and \textbf{J} are rather complicated, and each has only about 200 samples in the training set. Groups \textbf{F} and \textbf{I} both have about 1,000 samples in the training set, and are thus moderately represented. \textbf{I} may have large errors because of the electron-withdrawing effects of the three fluorines, or because it is found together with \textbf{J} in compound \textbf{20}. \textbf{F}, the \textit{tert}-butyl group, likely has large errors simply because it is bulky and thus leads to distorted geometries.

\begin{figure}[t]
    \centering
    \includegraphics[width=0.9\textwidth]{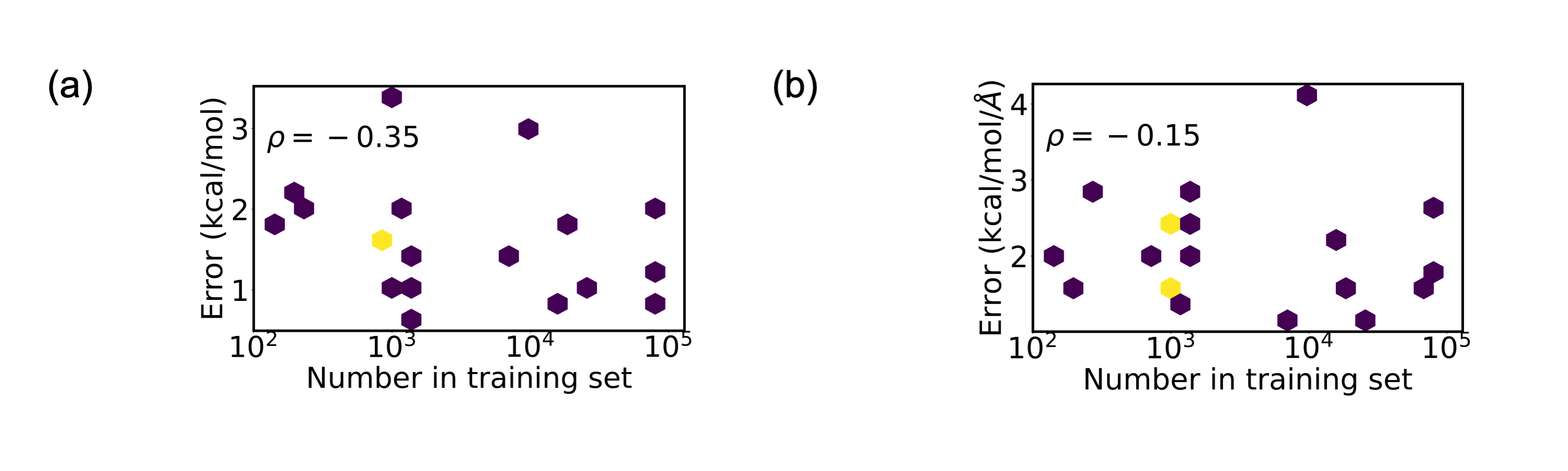}
    \caption{Test set error by functional group, plotted against the number of times the group appears in the training set. The Spearman rank correlation $\rho$ is also shown. (a) Gap error. (b) Excited state force error.}
    \label{sm_fig:func_group_err_plots}
\end{figure}

We see that the transferability of the model depends on how well-represented a functional group is in the training set, and how complicated its electronic effects are. The former is supported by Fig. \ref{sm_fig:func_group_err_plots}, which shows that the functional group error is anti-correlated with its prevalence in the training set. However, the effect is stronger for the gap than the excited state forces ($\rho = -0.35$ and $\rho=-0.15$, respectively), and neither fully explains the error. The remaining error can best be explained by analyzing the groups themselves.


\begin{table}[h]
    \centering
    \begin{tabular*}{\textwidth}{c @{\extracolsep{\fill}} |c|c|c|c|c|c|c|c|c}
          \toprule[1.1pt]
          & Num. train & $E_0$ & $E_1$ & $\Delta E_{01}$ & $(\Delta E)_{\mathrm{small}}$ & sgn$(\Delta E)_{\mathrm{small}} $ & $\vec{F}_0$ & $\vec{F}_1$ & $\vec{g}_{01}$ \\  
          \hline
          \ \textbf{A} \  & 1,219 & 1.05 & 0.94 & 0.63 & 0.61 & -0.04 &  1.24 & 1.32 & 1.19  \\
          \ \textbf{B} \ & 71,940 & 0.95 & 1.02 & 0.81 & 0.52 & -0.01 & 1.33 & 1.66 & 1.62  \\
          \ \textbf{C} \ & 17,624 & 1.08 & 1.14 & 0.85 & 0.37 & 0.12 &  1.24 & 1.52 & 1.62\\
          \ \textbf{D} \ & 994 & 0.73 & 1.00 & 0.91 & 0.44 & -0.07 & 1.18 & 1.70 & 1.54 \\
          \ \textbf{E} \ & 1,154 & 1.09 & 1.21 & 0.95 & 0.72 & -0.04 & 1.61 & 2.42 & 1.41 \\
          \hline 
          \ \textbf{F} \ & 979 & 2.28 & 3.36 & 3.39 & 1.14 & -0.31 & 1.53 & 1.53 & 1.96 \\
          \ \textbf{G} \ & 9,578 & 0.94 & 3.49 & 2.90 & 0.68 & -0.16 & 1.59 &  4.11 & 1.83 \\
          \ \textbf{H} \ & 200 & 1.62 & 3.17 & 2.18 & 0.44 & -0.13 & 0.99 & 1.67 &  0.68 \\
          \ \textbf{I} \ & 1,225 & 1.33 & 2.90 & 2.10 & 0.78 & -0.17 & 1.63 & 2.90 & 0.91  \\
          \ \textbf{J} \ & 236 & 1.33 & 2.90 & 2.10 & 0.78 & -0.17 & 1.63 & 2.90 & 0.91  \\
          \bottomrule[1.1pt]
    \end{tabular*}
    \caption{Test set error by functional group. The groups are divided into those with the lowest gap errors (top) and those with the highest (bottom). ``Num. train'' refers to the number of geometries in the training set that contain the functional group. While each of the functional groups is contained in the training set, the precise combinations and arrangements of groups (and hence the molecular graphs) are unique to the test set.}
    \label{sm_tab:func_groups}
\end{table}

\section{Spin contamination}
\label{sm_sec:spin_contam}
\begin{figure}[t]
    \centering
    \includegraphics[width=\textwidth]{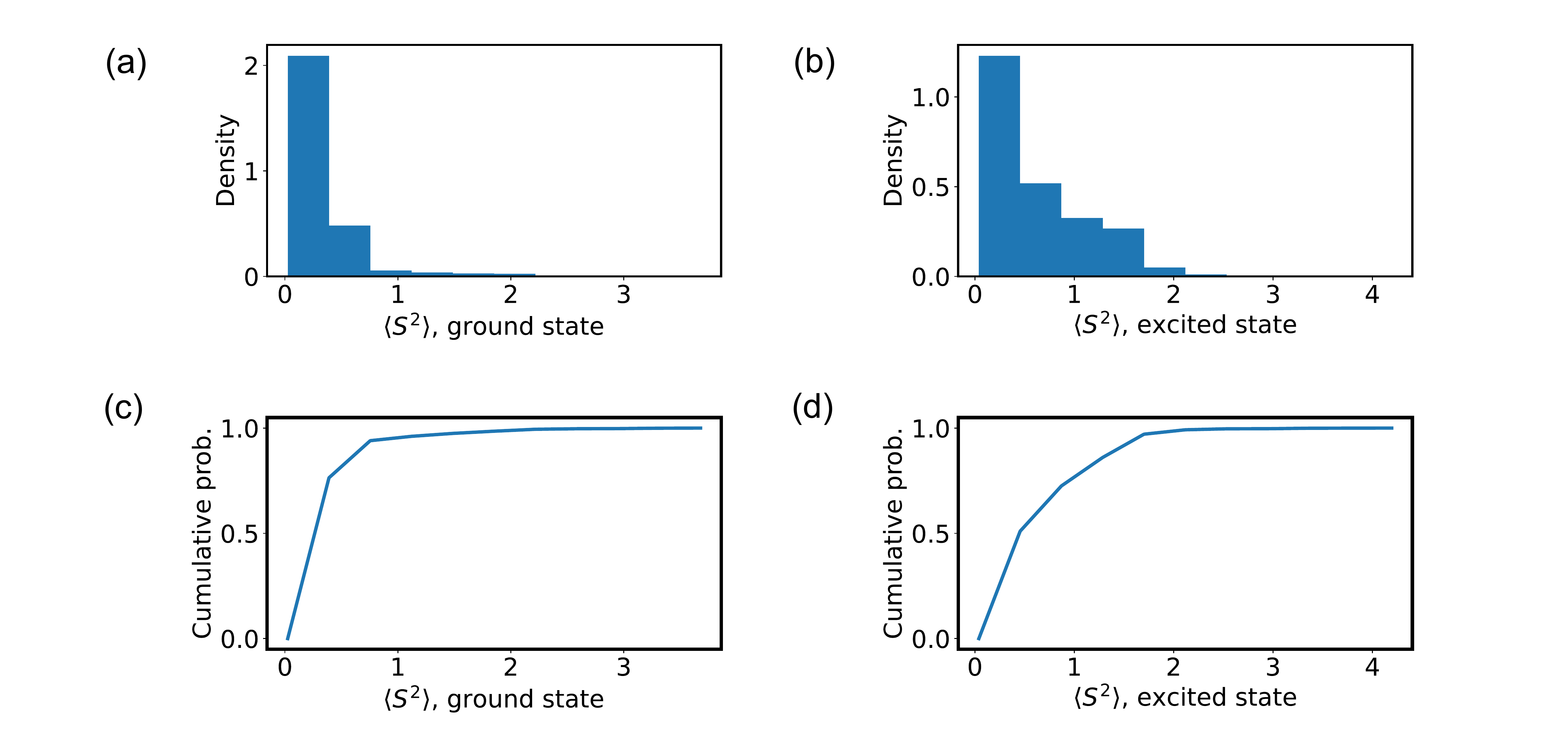}
    \caption{Analysis of spin contamination in the training set. (a) Distribution of $\langle S^2 \rangle$ in the ground state. (b) As in (a), but for the excited state. (c) Cumulative probability as a function of $\langle S^2 \rangle$ in the ground state. (d) As in (c), but for the excited state.}
    \label{sm_fig:spin_contam}
\end{figure}
Here we discuss the amount of spin contamination in the training set. Figure \ref{sm_fig:spin_contam} shows the square spin, $\langle S^2 \rangle$, for both the ground and excited states. The spin contamination is quite low for the ground state, with $\langle S^2 \rangle$ under 1.0 for 96\% of geometries. It is much higher for the excited state, with 76\%, 82\%, and 93\% of geometries having $\langle S^2 \rangle$ under 1.0, 1.2, and 1.4, respectively. Most of the geometries with excited-state $\langle S^2 \rangle$ $\gtrapprox 1.2$ were near the non-reactive $S_1/S_2$ CI, characterized by near-planarity and CNN angles near 108$^\circ$ \cite{yu2020nonadiabatic}. The $S_0/S_1$ CIs, on the other hand, did not have severe spin contamination. The spin-contamination near the $S_1/S_2$ CI led to difficulty in fine-tuning the model for unsubstituted azobenzene. We also visually inspected a sample of geometries with extreme spin contamination, $\langle S^2 \rangle > 1.8$, and found that nearly all had broken apart. We kept these geometries in the training set so that the model could learn the high energy of bond breaking.

\section{Surface hopping results with different models}
\label{sm_sec:compare_hop}
\begin{figure}[t]
    \centering
    \includegraphics[width=0.8\textwidth]{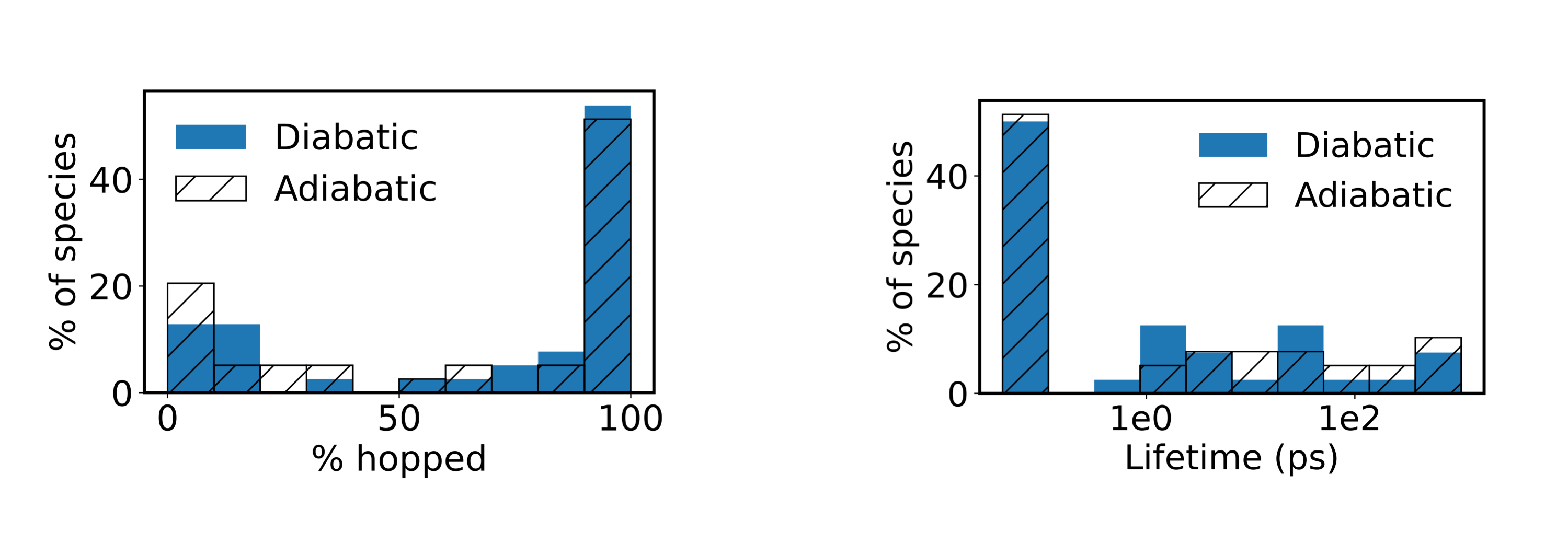}
    \caption{Comparison of ZN hopping statistics in diabatic and adiabatic models. (a) Proportion of trajectories in each species that hopped to the ground state. (b) Excited state lifetime for each species.}
    \label{sm_fig:compare_hop}
\end{figure}

Figure \ref{sm_fig:compare_hop} compares ZN hopping statistics of adiabatic and diabatic models for unseen species. Panel (a) shows the distribution of hopping percentages among species. The hopping percentage is defined as the percent of trajectories for a given species that end in the ground state. The $y$-axis is the percent of all species that correspond to each bin. Panel (b) has an analogous plot, but with the hopping percentage replaced by the lifetime. The lifetime was estimated from an exponential fit of $S_1$ population, $p =  \mathrm{exp}(-(t-t_{\mathrm{on}})/\tau) \ \Theta(t-t_{\mathrm{on}}) + \Theta(t_{\mathrm{on}} - t )$. Here $p  \in  [0, 1]$ is the $S_1$ population, $t$ is the time, $t_{\mathrm{on}}$ is the fitted turn-on time, $\tau$ is the fitted lifetime, and $\Theta$ is the Heaviside step function. Trajectories that did not contain any hops were assigned the maximum lifetime of all the other trajectories. 

The \textit{cis} derivatives have lifetimes around 50 fs, consistent with computational results for \textit{cis} azobenzene \cite{yue2018performance}. Nearly 100\% of all \textit{cis} trajectories ended in the ground state. The \textit{trans} derivatives have a wide distribution of lifetimes. Some are between 1 and 2 ps, which is similar to \textit{trans} azobenzene \cite{yue2018performance}. Others are in the range of tens to hundreds of ps, which reflects the high proportion of trajectories that never returned to the ground state. These are very likely incorrect. They may be because of barriers between $S_1$ minima and $S_0/S_1$ CIs, which are known to exist for \textit{trans} azobenzene \cite{yu2020nonadiabatic}. Relatively small errors in the barrier may lead to large over-estimations of the lifetime. 

Excited state barriers make it even more important to have accurate PESs near CIs. Since trajectories spend little time near crossing regions, trapping becomes even more severe when the gap near CIs is overestimated. The diabatic model helps to address this problem: in each plot we can see that the diabatic model leads to more hopping. For example, using the adiabatic model, 21\% of species have hopping percentages under 10\%. This number is reduced to 13\% for the diabatic model.

\begin{figure}[t]
\includegraphics[width=\textwidth]{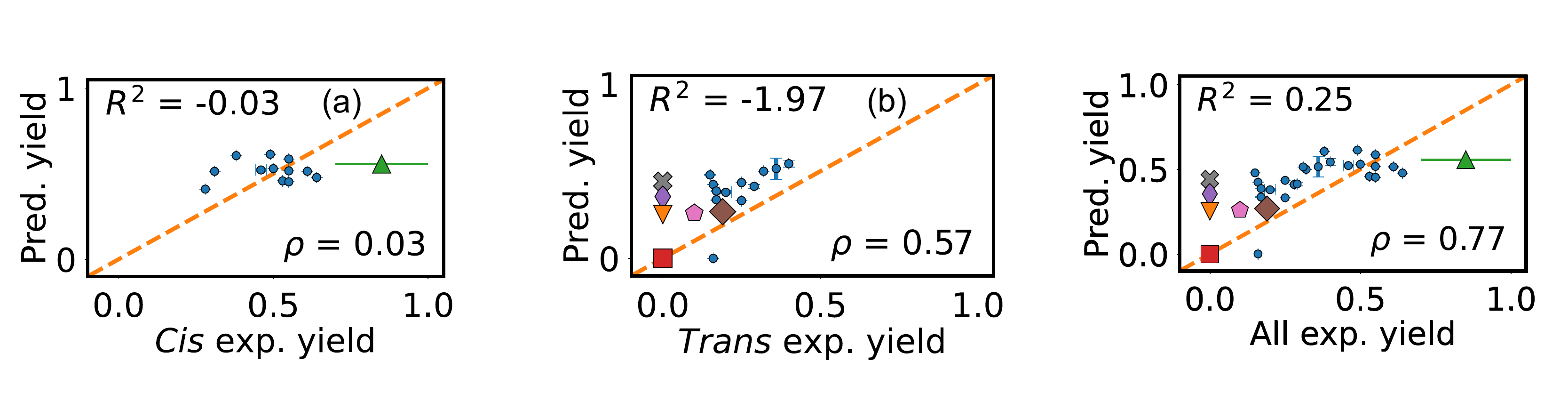}
\caption{Experimental vs. predicted ZN yields using the diabatic model. (a) \textit{Cis} isomers. (b) \textit{Trans} isomers. (c) All species.}
\label{sm_fig:yields_zn_diabat}
\end{figure}

\begin{figure}[t]
\includegraphics[width=\textwidth]{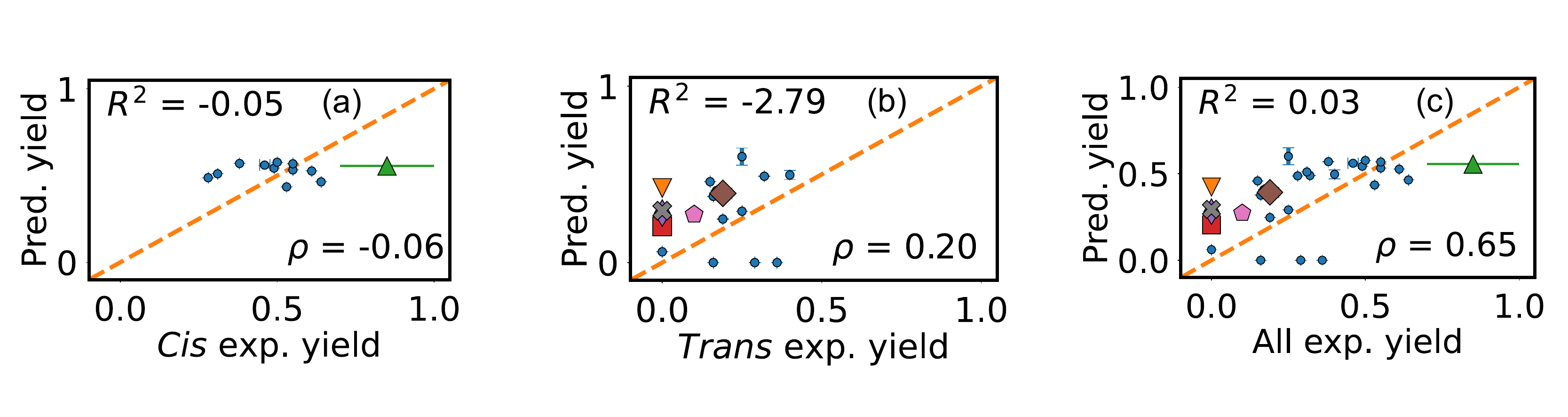}
\caption{Experimental vs. predicted ZN yields using the adiabatic model. (a) \textit{Cis} isomers. (b) \textit{Trans} isomers. (c) All species.}
\label{sm_fig:yields_zn_adiabat}
\end{figure}

\begin{figure}[t]
\includegraphics[width=\textwidth]{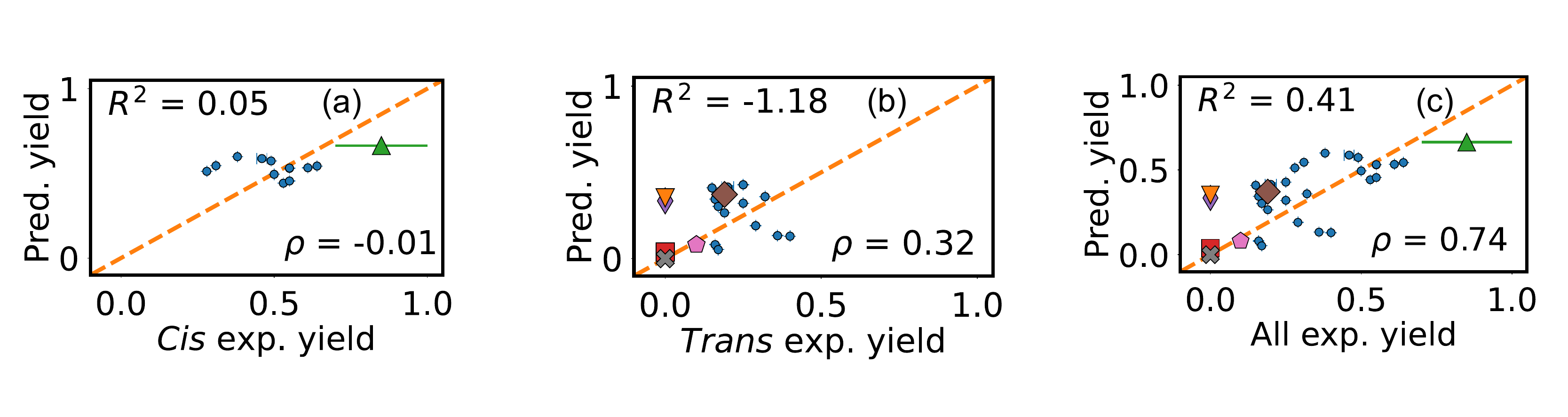}
\caption{Experimental vs. predicted FS yields using the diabatic model. (a) \textit{Cis} isomers. (b) \textit{Trans} isomers. (c) All species. }
\label{sm_fig:yields_tully_diabat}
\end{figure}

The diabatic model also improves the quantum yield. Figure \ref{sm_fig:yields_zn_diabat} shows the ZN quantum yields with the diabatic model, and Fig. \ref{sm_fig:yields_tully_diabat} shows the diabatic FS yields. The Spearman rank correlation for the \textit{trans} species is fairly high with both the ZN and FS methods, with the model accurately predicting low yields for a number of species. Figure \ref{sm_fig:yields_zn_adiabat} shows the ZN yield with the adiabatic model. The correlation among all species is reasonable, but for \textit{trans} species is rather low. This is because of three molecules with zero predicted yield, all of which became trapped in the excited state. The diabatic model only predicts zero yield for one of these. Further, the diabatic model properly predicts zero yield for species \textbf{35}, while the adiabatic model does not. For these reasons the diabatic model has a fairly high correlation with experiment. Still, it is clear that excited state trapping of \textit{trans} species is not a fully solved problem. Preferential sampling of excited state barriers may help to further address this issue in the future.

\section{Architecture}
\begin{table}[t]
\centering
\begin{tabular}{c|c|c}
     \hline
     Hyperparameter & Meaning & Value or name \\
     \hline
     $F$ & dimension of hidden atomic features & 128 \\
     $n_{\mathrm{conv}}$ & number of convolutions & 5 \\
    $n_{\mathrm{RBF}}$ & number of radial basis functions (RBF) & 20 \\
    $R_{\mathrm{cut}}$ & cutoff distance for convolutions & 5.0 \AA \\
    activation function & activation used in message-passing and readout & Swish \\
    learnable $k$ & whether $k$ parameters in RBF are learnable & true \\
    skip & output is sum of per-convolution outputs & true only for adiabatic \\
    % $F_{\mathrm{embed}}$ & dimension of the $d_{nn} - d_{mm}$ embedding & 128 \\
    % pooling activation & activation applied to attention-weighted features & softmax \\
    $M_{\mathrm{d}}$ & number of diabatic states & 3 \\
    $M_{\mathrm{ad}}$ & number of ground truth adiabatic states & 2 \\
    \hline
\end{tabular}
\caption{Model hyperparameters}
\label{sm_tab:hyperparams}
\end{table}

All models were implemented in PyTorch \cite{NEURIPS2019_9015}. The model hyperparameters are given in Table \ref{sm_tab:hyperparams}, and an in-depth explanation of the PaiNN parameters can be found in Ref. \cite{schutt2021equivariant}. Note that we used five convolutions instead of the three used originally, as this substantially improved model performance. % The attention pooling and diabatic parameters apply only to the diabatic model, but the remaining parameters apply to both models.  
Following DimeNet \cite{klicpera2020directional}, we allowed the $k$ values in the radial basis functions to be updated during training. For the adiabatic model, we predicted each property as a sum over per-convolution properties, which was also used in DimeNet. In particular, each convolution had a readout network to convert the atomic features to an output. The final property was obtained by summing each of these outputs.

Several variations on the architecture were tested. For example, we trained both two- and three-state diabatic models on 5,000 azobenzene configurations. We found that adding a third diabatic state significantly decreased the error for all properties. We then trained models using all possible three-state reference orderings and chose the best one. We also experimented with intensive pooling for off-diagonal energies (see Supplementary Sec. \ref{sm_sec:intensive}). In particular, we generated a molecular fingerprint through an attention-weighted average of atomic fingerprints. We then mapped this fingerprint to the $d_{nm}$ for $n \neq m$. Even though the $d_{nm}$ are intensive for $n\neq m$, this approach did not improve model predictions.

We also experimented with several adiabatic models. The model in the main text predicted $E_0$ and $E_1$ directly. This approach was used in previous work for single-molecule non-adiabatic dynamics \cite{chen2018deep, dral2018nonadiabatic, hu2018inclusion, li2021automatic, westermayr2019machine, westermayr2020combining, westermayr2020machine} and for the prediction of absorption spectra across chemical space \cite{westermayr2020deep}. We also examined three models that predicted $E_0$ directly and $E_1$ as the sum of $E_0$ and a learned gap. We tried three different pooling methods for the learned gap $\Delta E$: taking the mean over atomwise gaps, taking an attention-weighted average over atomwise gaps, and applying a dense readout network to an attention-weighted sum of atomic fingerprints. All approaches performed quite poorly compared to learning $E_0$ and $E_1$ separately as summed atomwise energies. This is problematic, because only adiabatic models that predict $E_1$ as $E_0 + \Delta E$ can guarantee the positivity of the gap (e.g. by squaring $\Delta E$ or applying a softplus function). Hence the adiabatic model in the main text sometimes generated predictions with $E_1 < E_0$. This is impossible in the diabatic model by construction.

\section{Training}
\label{sm_sec:training}
\begin{table}[t]
\centering
\scalebox{1}{
    \begin{tabular}{cccccccccc}
        \toprule[1.1pt]
        $\rho_{E_0}$ &  $\rho_{E_1}$ & $\rho_{\Delta  E_{01}}$ & $\rho_{f_0}$ & $\rho_{f_1}$ & $\rho_{\mathrm{ref}}$ & $\rho_{\mathrm{nacv}}$   \\ 
         \hline 
         0.2 & 0.1 & 0.5  & 1 & 1 & 0.01 & 1 \\
        \bottomrule[1.1pt]
    \end{tabular}
    }
    \caption{Loss parameters used to train all diabatic models. }
    \label{sm_tab:fixed_loss}
\end{table}
\begin{table}[t]
\centering
\scalebox{1}{
    \begin{tabular}{l|ccccccccccc}
        \toprule[1.1pt]
         & lr & $\mathrm{lr}_{\mathrm{min}}$ & $\rho_{\Delta  E_{01}}$ &  $\rho_{\mathrm{small}}$   \\ 
         \hline 
         Stage 1 &  $10^{-4}$ & $10^{-5}$ & 0.5 & 0  \\
         Stage 2 & $10^{-5}$ & $10^{-6}$  & 1.0 & 100  \\
        \bottomrule[1.1pt]
    \end{tabular}
    }
    \caption{Variable loss parameters used for training the adiabatic model. lr is the starting learning rate and $\mathrm{lr}_{\mathrm{min}}$ is the minimum learning rate, at which point training is stopped. }
    \label{sm_tab:var_loss}
\end{table}
The training set contained 562,037 geometries from 8,197 species. 5,000 geometries from 308 species were used for validation. The remaining 74,322 geometries from 40 species were held out for testing, so that the predicted yields of unseen species could be compared with experiment.  A different random seed was used to determine the training and validation splits for each committee model, and also to initialize the different models. After training, we ran FS DANN-NAMD on 40 holdout species using the trained diabatic model. For each species we selected 50 geometries randomly and 50 by CI proximity (Eq. (\ref{sm_eq:p_zn})), for a total of 4,000 geometries. These geometries were used as the test set, giving the ``unseen'' statistics in Table \ref*{tab:accuracy}. The DANN-NAMD geometries from the diabatic network were used for \textit{all} models, including the adiabatic and ablated ones in Table \ref{sm_tab:ablated}. The ``seen'' statistics were generated using the validation set. For all statistics, a phase correction was applied to $\vec{g}_{01}$ to minimize the prediction error, as in Eq. (\ref{sm_eq:nacv_loss}).

The model for screening new azobenzene derivatives was trained on all species. Inclusion of the holdout species provided an additional 74,322 geometries. We used 631,367 geometries for training, 5,000 for validation, and 5,000 for testing. This corresponded to 8,215 training species, 332 validation species, and 303 test species.

Training was performed over energies and forces/force couplings in units of kcal/mol and kcal/mol/\AA, respectively. Per-species reference energies were subtracted from each energy. These were obtained by summing atomic reference energies, computed using multi-variable linear regression from (atom type, count) to relaxed geometry energy. Configurations with 10-$\sigma$ energy and force outliers were removed prior to training. Those with forces or energies $\geq$450 kcal/mol(/\AA) from the mean were also removed. We found that more stringent removal of outliers led to less stable trajectories. For example, removing 3-$\sigma$ outliers and maximal energies/forces of $\geq$300 kcal/mol(/\AA) led to unstable ground state trajectories for six of the 40 unseen species. The 10-$\sigma$ and 450 kcal/mol(/\AA) criteria led to no diverging ground state trajectories. A small proportion of excited state trajectories were still unstable. We discarded all excited state trajectories that produced NaN geometries or energies, which were at most a few percent of the trajectories for a given species.

Models were trained with the Adam algorithm using a batch size of 60. Geometries were sampled for each batch using Eq. (\ref{sm_eq:p_sample}), as described below. The loss was given by Eq. (\ref*{eq:overall_loss}), using parameters in Table \ref{sm_tab:fixed_loss}. Note that the range of NACVs is approximately 10 times smaller than the range of forces. This means that $\rho_{\mathrm{nacv}}=1$, $\rho_{f_i}=1$ gives much higher weight to the forces. We experimented with a range of NACV coefficients between 0.1 and 10, and found that $\rho_{\mathrm{nacv}} = 1$ gave the best performance.

The learning rate was initialized to $10^{-4}$ and reduced by a factor of two if the validation loss had not improved in 10 epochs.  %Training was restarted using the next loss function once the learning rate fell below its minimum value.  The model with the best validation loss was used to initialize the next sequence. The final model was selected as the one with the lowest validation loss, according to the final loss function. 
The final model was selected as the one with the lowest validation loss. Training was performed on a single 32 GB Nvidia Volta V100 GPU, and took 13 days to complete.

For diabatic models, the training was stopped once the learning rate reached $10^{-5}$. Adiabatic models were trained in a two-step process, using different loss functions in each stage. Each step ended once the learning rate fell below a certain value. The following loss function was used with different loss trade-offs at different stages:
\begin{align}
& \mathcal{L}  = \mathcal{L}_{\mathrm{core}} + \mathcal{L}_{\mathrm{small}}, \nonumber \\
& \mathcal{L}_{\text{small}} =  \sum_{n > m} \rho_{\mathrm{small}} \cdot  \mathrm{mse}\big( \Delta E^{^{\mathrm{small}}}_{nm} \big),
\end{align}
where $\mathcal{L}_{\text{small}}$ penalizes errors in gaps under 0.2 eV. The parameters for each stage are given in Table \ref{sm_tab:var_loss}, and the $\rho_{E}$ and $\rho_f$ coefficients are the same as in Table \ref{sm_tab:fixed_loss}. The first stage emphasized energy gaps and gradients, while the second stage emphasized small gaps to fine-tune the model near conical intersections. We also experimented with scheduled training and using $\mathcal{L}_{\mathrm{small}}$ for diabatic models, but did not find any improvements. 

\section{Balanced sampling}
\label{sm_sec:balanced_sampling}
A custom data sampler was used during training because the dataset was imbalanced in the following ways. First, the combinatorial species only had a few geometries each, and so would be very rarely sampled during training. Second, there were more equilibrium \textit{trans} geometries than \textit{cis} geometries. Third, there were more equilibrium geometries in general than near-CI configurations. Our sampler addressed these imbalances by giving higher sampling probability to underrepresented species and/or configurations. 

\begin{figure*}[t]
    \centering
    \includegraphics[width=1.0\textwidth]{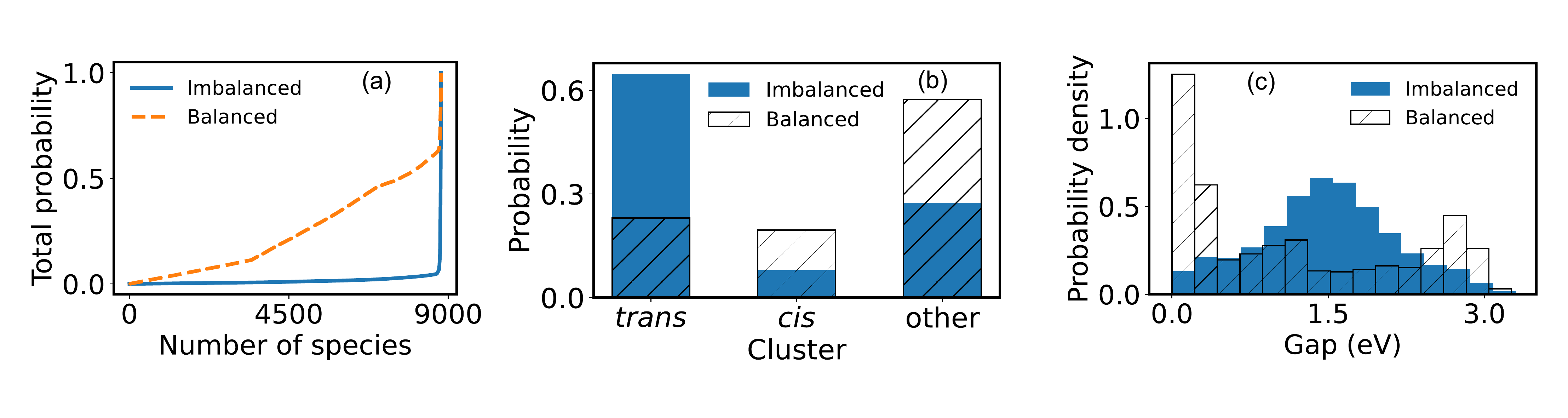}
    \caption{Sampling probabilities for different geometries in the training set, with and without the custom sampler. (a) Cumulative probability $p(i)$ of sampling any species $\leq i$. The species are ordered from fewest to most geometries. (b) Probability of sampling a geometry in each of the three clusters. (c) Probability density of sampling a geometry with a given energy gap. }
    \label{sm_fig:sampling}
\end{figure*}

To explain the sampling procedure, let us define two types of sampling weights. Sampling weights that are balanced by species are denoted by $w$, and those that are not are denoted by $v$. For example, the weights $w_{\mathrm{cluster}}(g_{i, A})$ and $v_{\mathrm{cluster}}(g_{i, A})$ are both assigned to the $i^{\mathrm{th}}$ geometry in species $A$, denoted $g_{i, A}$, based on the cluster of configurations that it belongs to. This cluster is denoted $k$, where $k \in$ (\textit{cis}, \textit{trans}, other). For a geometry $g_{i, A}$ in cluster $k$, the weights are given as follows:
\begin{alignat}{2}
& w_{\mathrm{cluster}}(g_{i,A}) =  \left( \frac{1 }{n_{\mathrm{spec}}  \cdot n_{\mathrm{clusters} }} \right) \left( \frac{1}{n_{k, A}}  \right) \nonumber    \\ 
&  v_{\mathrm{cluster}}(g_i, A) = \left( \frac{1 }{n_{\mathrm{geoms}} \cdot n_{\mathrm{clusters}}  }\right) \left( \frac{n_A}{n_{k, A}} \right)  \label{sm_eq:cluster_balance}
\end{alignat}
Here $n_{k, A}$ is the number of geometries of species $A$ in cluster $k$, and $n_A$ is the number of geometries in species $A$. $n_{\mathrm{spec}}$ is the total number of species, $n_{\mathrm{clusters}}$ is the total number of clusters, and $n_{\mathrm{geoms}}$ is the total number of geometries. The above definitions ensure that there is an equal probability $P_k$ of sampling any cluster $k$:
\begin{align}
& \text{Using }w: \ P_k = \sum_A \sum_{g_{i, A} \in \ k} w_{\mathrm{cluster}}(g_{i, A}) = \sum_A \frac{1}{n_{\mathrm{spec}} \cdot n_{\mathrm{clusters}}  } = \frac{1}{n_{\mathrm{clusters}} } \\ 
& \text{Using }v: \ P_k = \sum_A \sum_{g_{i, A} \in \ k} v_{\mathrm{cluster}}(g_{i, A}) = \sum_A \frac{n_A}{n_{\mathrm{geoms}} \cdot n_{\mathrm{clusters}} } = \frac{1}{n_{\mathrm{clusters}} } 
% =  \left( \frac{n_{k, A} \cdot  n_{\mathrm{spec}} \cdot  n_{k, A}^{-1}}{n_{\mathrm{spec}} \cdot \sum_j n_{j, A}^{-1} }\right)   =  \frac{1}{ \sum_j n_{j, A}^{-1} } = \mathrm{const} \\
% & \text{Using }v: \ P_k = \sum_A \sum_{g_i \in \ k, A} v_{\mathrm{cluster}}(g_i) =  \left( \frac{n_{k, A} \cdot  n_{\mathrm{spec}} \cdot  n_{k, A}^{-1}}{n_{\mathrm{spec}} \cdot \sum_{j, B} n_{j, B}^{-1} }\right)   
\end{align}
Here we have used the fact that $\sum_{g_i \in \ k, A} \ 1 = n_{k, A}$, $\sum_{A} \ 1 = n_{\mathrm{spec}}$, and $\sum_A n_A = n_{\mathrm{geoms}}$. The difference between the two weights is that $w$ leads to balanced sampling among species, while $v$ does not:
\begin{align}
& \text{Using }w: \ P_A = \sum_k \sum_{g_{i, A} \in k} w_{\mathrm{cluster}}(g_{i, A}) = \sum_k \ \frac{1}{n_{\mathrm{spec}} \cdot n_{\mathrm{clusters}}  } = \frac{1}{n_{\mathrm{spec}}} \\
& \text{Using }v: \ P_A = \sum_k \sum_{g_{i, A} \in k} w_{\mathrm{cluster}}(g_{i, A}) = \sum_k \ \frac{n_A}{n_{\mathrm{geoms}} \cdot n_{\mathrm{clusters}}  } = \frac{n_A}{n_{\mathrm{geoms}}}.
\end{align}
Here we have used the fact that $\sum_k \ 1 = n_{\mathrm{cluster}}$. We see that $w$ gives equal weights for each species, whereas $v$ gives weights proportional to the number of geometries in that species.

Similarly to the cluster weights, we define Zhu-Nakamura weights $w_{\mathrm{ZN}}$ and $v_{\mathrm{ZN}}$, such that geometries with a higher hopping probability $p_{\mathrm{ZN}}$ are sampled more often. The corresponding expressions are:
\begin{alignat}{2}
& w_{\mathrm{ZN}}(g_{i, A}) = \left(\frac{ 1}{n_{\mathrm{spec}} \cdot \sum_j p_{\mathrm{ZN}}(g_{j, A})} \right) p_{\mathrm{ZN}}(g_{i, A}) \\
& v_{\mathrm{ZN}}(g_{i, A}) = \left( \frac{1 }{\sum_{j, B} p_{\mathrm{ZN}}(g_{j, B})} \right) p_{\mathrm{ZN}}(g_{i, A}).
\end{alignat}
The overall sampling weight for a given geometry is determined by each $w$ and $v$, together with two user-defined weights: $P_{\mathrm{spec}} \in [0, 1]$, the importance of species balance, and $P_{\mathrm{ZN}} \in [0, 1]$, the importance of sampling geometries with high hopping rates. $P_{\mathrm{cluster}} = 1 - P_{\mathrm{ZN}}$ is the importance of sampling different clusters in a balanced way. The final sampling weight is then given by
\begin{alignat}{2}
p(g_{i, A}) & =  P_{\mathrm{ZN}} \left[ P_{\mathrm{spec}} \cdot w_{\mathrm{ZN}}(g_{i, A}) + (1 - P_{\mathrm{spec}}) \cdot v_{\mathrm{ZN}}(g_{i, A}) \right] \nonumber \\
& % \quad & + & 
+ P_{\mathrm{cluster}} \left[ P_{\mathrm{spec}} \cdot w_{\mathrm{cluster}}(g_{i, A}) + (1 - P_{\mathrm{spec}}) \cdot v_{\mathrm{cluster}}(g_{i, A}) \right]. \label{sm_eq:p_sample}
\end{alignat}
In applying Eq. (\ref{sm_eq:p_sample}) we used $P_{\mathrm{spec}} = 0.6$ and $P_{\mathrm{cluster}} = 0.5$. We defined a geometry as \textit{cis} or \textit{trans} if its RMSD with respect to the corresponding reference structure was $\leq 0.25$ \AA, and ``other'' otherwise. In contrast to diabatic reference geometries, the RMSD was computed using all atoms, not just the central CNNC atoms. Note that while our clustering approach was specific to \textit{cis}-\textit{trans} isomerization, many black-box clustering methods exist, such as hierarchical clustering \cite{schwalbe2021differentiable}. Any of these methods could have been used in place of our user-defined clusters.

In principle $p_{\mathrm{ZN}}$ should depend on trajectory-specific factors such as velocity, and is therefore not a function of geometry alone. However, during simulations, hops almost always occurred below 0.5 eV. Further, since $p_{\mathrm{ZN}}$ is approximately exponential in the square of the gap, we approximated it with
\begin{align}
p_{\mathrm{ZN}}(\Delta E) \approx \mathrm{exp}(-\Delta E ^2 / (2 \Delta E_0^2)), \label{sm_eq:p_zn}
\end{align}
where $\Delta E$ is the energy gap and $\Delta E_0 = 0.15$ eV. This choice of $\Delta E_0$ gave $p(\Delta E = 0.5 \text{ eV}) \approx 3\times 10^{-3}$. This meant that geometries were usually selected only if their gap was under 0.5 eV.

The sampling probabilities for the training set are shown in Fig. \ref{sm_fig:sampling}. Panel (a) shows the cumulative probability of sampling different species, with the compounds ordered from fewest to most geometries. Without the custom sampler the probability was quite small for most molecules. This was because the majority were combinatorially generated and had few geometries. 90\% of the probability was contained in the last 41 species. This probability was reduced to 35\% when using the balanced sampler. Panel (b) shows that most geometries were \textit{trans} isomers and few were \textit{cis}. The custom sampler gave approximately equal probability to the two isomers, and the highest probability to the \textit{other} group. Most geometries in this group had small gaps, and hence were highly weighted by $w_{\mathrm{ZN}}$ and $v_{\mathrm{ZN}}$. This is demonstrated in panel (c), which shows that near-CI geometries had the highest probabilities when using the balanced sampler.

\section{Dynamics}
\label{sm_sec:dynamics}
Simulations were performed in two stages. First the starting geometries and velocities were generated for each NAMD trajectory. For \textit{ab initio} simulations we optimized the ground state geometry of each species and computed its normal modes. These modes were used to sample geometries and velocities from the harmonic oscillator Wigner distribution at temperature $T$ = 300 K \cite{barbatti2016effects}. For neural trajectories we ran classical MD for 15 ps with the Nos\'{e}-Hoover thermostat \cite{nose1984unified, hoover1985canonical}, and selected random samples to start the NAMD trajectories. Classical MD was implemented with ASE \cite{ase-paper}. The effective mass $Q$ was set to $(3 N - 6) \cdot \tau^2  k_{\mathrm{B}} T $, where $N$ is the number of atoms, $k_{\mathrm{B}}$ is Boltzmann's constant, $T = 300$ K, and $\tau = 25$ fs is the relaxation time. Initial atomic velocities were sampled from a Maxwell-Boltzmann distribution at 300 K. Optimized geometries were used for the initial atomic positions. The total linear and angular momenta of the system were set to zero, and the target kinetic energy was set to $ (3N -6) \cdot k_{\mathrm{B}} T/ 2$. The time step of the simulation was set to 0.5 fs, and the neighbor list of the system was updated every 10 time steps (5 fs). All pairs of atoms within with 7 \AA \ were considered neighbors. In each step the distance between all pairs of neighbors was computed, and only pairs within 5 \AA \ of each other were used in the model. This procedure meant that atoms entering the 5 \AA \ cutoff between neighbor list updates would not be missed (i.e., an extra 2 \AA \ ``skin'' was addded). All ground state and excited state trajectories in this work were propagated with the velocity Verlet algorithm \cite{verlet1967computer}. Frames for NAMD were taken only after the first 1 ps of ground state MD to allow for equilibration.

Next we ran ZN dynamics \cite{yu2014trajectory, yue2017benchmark, yue2018performance} using $N_{\mathrm{trj}}$ trajectories, where $N_{\mathrm{trj}}$ was 10 during the active learning cycle and 500 during final inference. %The ZN method uses a form of local diabatization to infer diabatic quantities from adiabatic ones, and these values determine the hopping rate \cite{yu2014trajectory, yue2017benchmark, yue2018performance}. However, since our model is already diabatic, we used the diabatic parameters themselves.
Hops were restricted to gaps $\leq 0.5$ eV to avoid unphysical transitions, the time step was set to 0.5 fs \cite{yu2014trajectory, yue2017benchmark, yue2018performance}, and the neighbor list was again updated every 10 steps. Excited state trajectories were run for 1.5 ps during active learning and 5 ps for final inference. All other details of the implementation can be found in Refs. \cite{yu2014trajectory, yue2017benchmark, yue2018performance}. For each species we performed inference in parallel over 100-250 geometries at a time, depending on the number of atoms in the molecule. This was done by batching together geometries from 100-250 trajectories and evaluating the model on the batch. The batch size depended on the size of the molecule, which determined the GPU memory consumption. FS simulations were also performed for final inference. The ZN parameters above were used for FS DANN-NAMD of molecules in the holdout set. For virtual screening we used 100 trajectories and 2 ps of excited state dynamics. Species with a hopping rate under 10\%, or with molecular graphs that changed during ground state dynamics, were deemed unreliable. These molecules were excluded from Fig. \ref*{fig:screening}, but not from the average hopping percentage reported in the main text. % We repeated the simulations for the top three candidates using 500 trajectories, 15 ps of ground state dynamics (1 ps of equilibration), and 2 ps of excited state dynamics. NAMD for each species was performed on a single 32 GB Nvidia Volta V100 GPU.

The quantum yield was calculated as $Y = n_R / n_T$, where $n_R$ is the number of reactive trajectories and $n_T$ is the total number of trajectories. The uncertainty was computed as the standard deviation of 1,000 bootstrapped samples. To identify reactivity, we computed the RMSD between the CNNC atoms in the final frame and the corresponding atoms in the optimized \textit{cis}/\textit{trans} geometries. A trajectory was considered reactive if it started as \textit{cis} and ended closer to \textit{trans}, or vice-versa. Trajectories that ended in the excited state were excluded from $n_R$ and $n_T$. The yield was reported as $0 \pm 0$ if all trajectories ended in the excited state.

\section{Fewest switches implementation}
\label{sm_sec:tully}
Tully's surface hopping propagates the nuclei on one electronic surface at a time, called the active surface. The expansion coefficients of the electronic wavefunction are also propagated, and are used to make stochastic changes in the active surface. The coefficient vector $\mathbf{c}$, expressed in a basis denoted by ``rep'', evolves as \cite{mai2015general}
\begin{align}
  \frac{d}{d t} \mathbf{c}^{\mathrm{rep}} = - \frac{i}{\hbar} [\mathbf{H}^{\mathrm{rep}} - i \mathbf{T}^{\mathrm{rep}}] \mathbf{c}^{\mathrm{rep}}. \label{sm_eq:c_eqn}
\end{align}
The Hamiltonian matrix is $\mathbf{H}^{\mathrm{rep}} = \bra{\psi^{\mathrm{rep}}_{n} } H \ket{\psi^{\mathrm{rep}}_{m} }$, and the coupling term is
\begin{align}
    (\mathbf{T}^{\mathrm{rep}})_{nm} = \left\langle  \psi^{\mathrm{rep}}_{n}(\vec{R}(t)) \left\vert \frac{\partial}{\partial t} \right\vert \psi^{\mathrm{rep}}_{m} (\vec{R}(t))  \right\rangle   = \vec{v} \cdot \vec{k}^{\mathrm{rep}}_{nm}, \label{sm_eq:T}
\end{align}
where $\vec{v}$ is the classical velocity of the nuclei. In principle the nuclei can be propagated on a PES in any electronic basis, and hops can be performed between states in that basis. For example, nuclei could be propagated on diabatic PESs and hops could occur between diabatic states. However, it is well-established that surface hopping in the adiabatic basis gives the best results \cite{richter2011sharc}. Hence the nuclei should be propagated in the adiabatic basis, and hops should be decided using $\mathbf{c}^{\mathrm{ad}}$. 

While the adiabatic basis should be used for hopping, $\mathbf{c}$ can be \textit{propagated} in the diabatic basis, and then transformed into the adiabatic basis for all subsequent manipulations (e.g. deciding on hops, adding decoherence, etc.). Indeed, using Eq. (\ref{sm_eq:c_eqn}) in the adiabatic basis can require small time steps for so-called trivial crossings, where derivative NACVs are quite narrow and large \cite{plasser2012surface}. Since global diabatic states are almost never available in \textit{ab initio} simulations, a local diabatization method is often used to propagate $\mathbf{c}$ \cite{plasser2012surface, mai2018nonadiabatic}. This method is very stable and can be used with a fairly large time step.  % In this approach, one uses the overlap of adiabatic wavefunctions in successive steps to generate a diabatic basis in each step. The transformation minimizes the time derivative in Eq. (\ref{sm_eq:T}) in each step. It is \textit{local} because the time derivative depends on both the velocity and geometry, and so the transformation is unique to each trajectory. 
In this work we have a \textit{global} diabatic basis that can replace local diabatization. Following Refs. \cite{plasser2012surface, mai2015general}, we then propagate $\mathbf{c}$ as follows:
\begin{align}
    & \mathbf{c}^{\mathrm{ad}}(t + \Delta t) = \mathbf{P}^{\mathrm{ad}}(t, t+ \Delta t) \hspace*{0.05cm} \mathbf{c}^{\mathrm{ad}}(t), \label{sm_eq:p_rep} \\
    & \mathbf{P}^{\mathrm{ad}}(t, t+ \Delta t) = \mathbf{U}^{\dagger}(t + \Delta t) \hspace*{0.05cm} \mathbf{P}^{\mathrm{d}}(t, t+ \Delta t)  \label{sm_eq:p_di_to_ad} \hspace*{0.05cm} \mathbf{U}(t) \\
    &  \mathbf{P}^{\mathrm{d}}(t, t+ \Delta t)  = \Pi_{k=1}^{K}  \ \mathrm{exp} \left[ -i \mathbf{H}^{\mathrm{d}}_k \Delta t / K \right]  \label{sm_eq:p_d} \\
    & \mathbf{H}_k^{\mathrm{d}} = \mathbf{H}^{\mathrm{d}}(t) + \frac{k}{K} \left[ \mathbf{H}^{\mathrm{d}}(t + \Delta t) - \mathbf{H}^{\mathrm{d}}(t) \right]. \label{sm_eq:p_interp}
\end{align}
Here $\mathbf{P}^{\mathrm{ad}}(t, t + \Delta t)$ is the propagator of $\mathbf{c}^{\mathrm{ad}}$ from $t$ to $t + \Delta t$ (Eq. (\ref{sm_eq:p_rep})), and $\Delta t$ is the timestep. The adiabatic propagator is a transformation of the diabatic propagator into the adiabatic basis (Eq. (\ref{sm_eq:p_di_to_ad})), using the transformation matrix $\mathbf{U}$ (Eq. (\ref*{eq:u_def})). The diabatic propagator is a matrix product of $K$ sub-propagators (Eq. (\ref{sm_eq:p_d})), where $K$ is the number of electronic substeps for each nuclear step. ``exp'' denotes a matrix exponential, not element-wise exponentiation. The $k^{\mathrm{th}}$ sub-propagator uses a linear interpolation between $\mathbf{H}^{d}(t)$ and $\mathbf{H}^{d}(t + \Delta t)$ to approximate $\mathbf{H}_d(t + k \Delta t / K)$ (Eq. (\ref{sm_eq:p_interp})). The diabatic Hamiltonian is produced by the neural network model. The probability of hopping from state $n$ to $m$ is \cite{mai2015general}
\begin{align}
    p_{n \to m} = \left(1 - \frac{\vert c^{\mathrm{ad}}_n(t + \Delta t) \vert^2}{ \vert c^{\mathrm{ad}}_n(t) \vert^2} \right) \frac{\mathrm{Re} \left[ c_m^{\mathrm{ad}}(t + \Delta t) \left(P^{\mathrm{ad}}_{mn}\right)^* \left( c^{\mathrm{ad}}_n(t)\right)^* \right] }{\vert c^{\mathrm{ad}}_n(t) \vert^2  - \mathrm{Re}\left[ c^{\mathrm{ad}}_n (t + \Delta t) \left(P^{\mathrm{ad}}_{nn}\right)^*  \left( c^{\mathrm{ad}}_n(t)\right)^* \right] }, \label{sm_eq:p_hop}
\end{align}
where $\mathrm{Re}(x)$ is the real part of $x$. Note that while our approach follows that of SHARC \cite{mai2015general}, we have written our own code and made it publicly available \cite{nff}. Our repository also contains code for surface-hopping with the ZN method.

In this work we initialized $\mathbf{c}^{\mathrm{ad}}(t=0) = [0, 1, 0]$, used Eqs. (\ref{sm_eq:p_rep})-(\ref{sm_eq:p_interp}) to generate $\mathbf{c}^{\mathrm{ad}}(t + \Delta t)$, and computed the hopping probability with Eq. (\ref{sm_eq:p_hop}). $\mathbf{c}^{\mathrm{ad}}$ was expressed in a three-state basis because our model used three diabatic states, and hence $\mathbf{U}$ had three dimensions. Since the model was only trained on the first and second adiabatic energies, we set $p=0$ for all transitions to the third state. Hops to state $m$ were performed if \cite{mai2015general}
\begin{align}
    \sum_{i=1}^{m-1} p_{n \to i} < r \leq \sum_{i=1}^{m-1} p_{n \to i} + p_{n \to m},
\end{align}
where $0 \leq r \leq 1$ is a random number. The momentum was rescaled after each hop to conserve the total energy. The momentum was multiplied by a constant factor, rather than being re-scaled in the direction of the NACV, to avoid the overhead of a NACV calculation (see below) \cite{mai2018nonadiabatic}. If the factor was complex---a so-called ``frustrated'' hop---then no transition was made \cite{mai2018nonadiabatic}. 

We also tested propagation in the adiabatic basis. In this case we used the NACV to evaluate $\mathbf{T}$, and constructed $P^{\mathrm{ad}}$ from the interpolation of $\mathbf{H}^{\mathrm{ad}} -i \mathbf{T}$ \cite{mai2015general}. Both Eq. (\ref{sm_eq:p_hop}) and Tully's original hopping expression \cite{tully1990molecular} were used. The momentum was re-scaled in the direction of the NACV \cite{landry2012recover}. In all cases the results were very similar to those of the diabatic basis. Diabatic propagation was ultimately chosen because of its reported stability \cite{plasser2012surface}, and because of its computational efficiency. In particular, the diabatic propagation requires only diabatic energies and one adiabatic gradient. It therefore uses only one forward and one backward pass through the neural network. By contrast, constructing the NACVs requires gradients for each diabatic element (Eq. (\ref*{eq:hd_nacv})). For three diabatic states, this corresponds to one forward pass and six backward passes. While shared convolution layers and caching mean that $N$ gradients do not take $N\times$ as long as one gradient, we found that they still added significant time. Hence diabatic propagation was chosen as the most efficient method.

The decoherence correction of Ref. \cite{granucci2007critical} was used to counter the over-coherence of Tully's original method. A sign correction was also used to remove random sign changes in the eigenvectors of $\mathbf{H}_d$. The sign correction $A_m$ for the $m^{\mathrm{th}}$ eigenvector $\mathbf{v}_m$ was computed as
\begin{align}
&  A_m = \mathrm{sgn} (S_{n'm}) \\
& S_{nm} = \mathbf{v}_n^{\dagger}(t) \mathbf{v}_{m}(t + \Delta t) = [\mathbf{U}^{\dagger}(t) \mathbf{U}(t + \Delta t)]_{nm} \\
&  n' = \mathrm{argmax}_{n} \{ \vert S_{nm} \vert \}.
\end{align}
%State $m$ at time $t + \Delta t$ has the largest absolute overlap with state $n'$ at time $t$. Since these two states are the most similar, the phase correction for $\mathbf{v}_{m}(t + \Delta t)$ is decided using $\mathbf{v}_{n'}(t)$. 
The transformation matrix and NACVs were corrected through
\begin{align}
& U_{nm}(t + \Delta t) \to A_m U_{nm}(t + \Delta t) \\
& \vec{k}_{nm}(t + \Delta t) \to A_n A_m \hspace*{0.05cm} \vec{k}_{nm}(t + \Delta t) \\
& \vec{g}_{nm}(t + \Delta t) \to A_n A_m \hspace*{0.05cm} \vec{g}_{nm}(t + \Delta t).
\end{align}
Note also that $S_{nm} \approx \bra{\psi_n^{\mathrm{ad}}(t)} \ket{\psi_m^{\mathrm{ad}}(t + \Delta t)}$:
\begin{align}
\bra{\psi_n^{\mathrm{ad}}(t)} \ket{\psi_m^{\mathrm{ad}}(t + \Delta t)} & = \sum_{ij} U_{in}^*(t) U_{jm}(t + \Delta t) \bra{\psi_i^{\mathrm{d}}(t)} \ket{\psi_j^{\mathrm{d}}(t + \Delta t)} \nonumber \\
& \approx  \sum_{ij} U^{\dagger}_{ni}(t) U_{jm}(t + \Delta t) \bra{\psi_i^{\mathrm{d}}(t)} [(1 -  (\vec{v} \Delta t) \cdot \nabla_R) \ket{\psi_j^{\mathrm{d}}(t)}] \nonumber \\
& = \sum_{ij} U^{\dagger}_{ni}(t) U_{jm}(t + \Delta t) \hspace*{0.05cm} \bra{\psi_n^{\mathrm{d}}(t)} \ket{\psi_m^{\mathrm{d}}(t )}  \nonumber \\
& = [\mathbf{U}^{\dagger}(t) \mathbf{U}(t + \Delta t) ]_{nm} = S_{nm}.
\end{align}
Here we used the fact that there is no derivative coupling between diabatic states, and that the diabatic states at a given time are orthonormal (Eq. (\ref{sm_eq:orthonormal})).

All simulations were performed with a time step of 0.5 fs and $K=25$ substeps for electronic propagation. Unlike in ZN dynamics, hops were not restricted to gaps under 0.5 eV. We found that restricting hops improved the ZN results but hurt the FS results. For example, with no maximum gap, most neural ZN trajectories starting from \textit{trans}-azobenzene hopped at $\sim 1.5$ eV. This did not match the results of Ref. \cite{yue2018performance}, which used \textit{ab initio} ZN with the same level of DFT theory. After adding the restriction, most hops occurred at gaps under 0.1 eV, in agreement with Ref. \cite{yue2018performance}. By contrast, most FS hops occurred under 0.1 eV even without a maximum restriction, though some still occurred at large gaps. The lifetime and yield without a maximum also better matched previous calculations.

\section{Active learning}
\label{sm_sec:active_learning}
New geometries were selected for QC calculations using two different criteria. In the first three active learning loops, new geometries were chosen based on the prediction variance of two neural networks. Our aim was to select geometries with fairly uncertain predictions. We did not want geometries with extremely high uncertainty, as these usually corresponded to broken graphs that were outside the target learning space for the model. We therefore used a log-normal target distribution for the uncertainty. Target uncertainties were randomly sampled from this distribution, and geometries with the closest variance to the targets were selected. The log-normal probability of obtaining a sample $x$ is given by $P(x) = \mathrm{exp}\left[ -(\mathrm{ln}(x/s) -\mu )^2 / 2\sigma^2 \right]/(x \sigma \sqrt{2 \pi})$, where $s$, $\sigma$ and $\mu$ are positive numbers. Under this distribution, both completely certain and completely uncertain predictions have zero probability, with a peak for predictions with medium uncertainty. The decay of the probability at large uncertainty is slow, such that highly uncertain geometries can still be selected with reasonable probability. We set $\mu=0$, $\sigma=1$, and $s=7$, giving a target distribution with a mode of 2.5 kcal/mol(/\AA) and a mean of 11.5 kcal/mol(/\AA). Committee variances were computed for both forces and energies for each electronic state, and the largest variance was compared with the target variance from the distribution. For each species we selected 16 geometries from ground state MD and 33 from surface hopping. Only geometries from the 164 literature species were chosen. This led to approximately 8,000 new data points in each active learning cycle. Other, simpler methods of geometry selection are certainly possible; for example, random selection is also known to work quite well \cite{wang2020active, ang2021active}. Our approach successfully increased the model quality throughout active learning, but we did not thoroughly compare it to other methods. Such a comparison may be of interest in the future.

In the next two loops we used only azobenzene, with the goal of densely sampling the CI region. Half the geometries were selected randomly from the excited state dynamics, and half were selected based on the gap. Because the model overestimated the gap in uncertain regions, we did not want to simply select geometries with low predicted gap. Rather, in each trajectory we identified avoided crossing geometries, and assigned equal sampling probability to all geometries before and after that crossing, up to a maximum predicted gap of 1.7 eV. An avoided crossing geometry was defined as having a gap that was lower than in the previous and subsequent time step. In this way we aimed to identify regions that could have small gaps, even if the model did not predict them to be so.

\section{Validation}

\subsection{\textit{Ab initio} NAMD}
\textit{Ab initio} simulations were used for unsubstituted \textit{cis} and \textit{trans} azobenzene. Starting configurations and velocities were generated with \textit{ab initio} MD using Q-Chem to drive the dynamics. Ten MD simulations were performed for each species. Each simulation was initiated with a different geometry produced by ground-state neural network MD. The simulations were run for 10 ps each using a time step of 0.5 fs, with the the BP86 functional \cite{becke1988density, perdew1986density} and 6-31G* basis \cite{francl1982self}. Initial velocities were sampled from a thermal distribution at 300 K, with rotation and translation projected out. The nuclei were propagated with the Velocity Verlet algorithm. A Nos\'{e}-Hoover chain of length 5, timescale $\tau=25$ fs, and temperature $T=300$ K was used as a thermostat. The Fock extrapolation order was set to six, and twelve Fock matrices from previous steps were used in the extrapolation.

167 and 178 excited state trajectories were generated for \textit{trans} and \textit{cis} azobenzene, respectively. Each trajectory was initialized with a different set of coordinates and velocities, randomly sampled from the ten ground state simulations after 1 ps of equilibration. Spin-flip TDDFT was used with the BHHLYP functional \cite{becke1993new} and 6-31G* basis. DIIS with geometric direct minimization (GDM) \cite{van2002geometric} was used to improve SCF convergence. We found this to be critical: with the usual DIIS algorithm, the SCF cycle failed to converge for 36\% of trajectories. This usually occurred within the first 200 fs. 

The dynamics were run with in-house scripts, which can be found at \url{https://github.com/learningmatter-mit/NeuralForceField}. We propagated the elements of the electronic wavefunction in the adiabatic basis, and corrected the sign of the force NACV by minimizing its change from the previous step. The momentum was re-scaled in the direction of the NACV after a hop \cite{landry2012recover}. All other parameters were unchanged from the DANN-NAMD simulations. \textit{Cis} trajectories were propagated for 200 fs, and \textit{trans} trajectories for a maximum of 1.5 ps. For the former, we extended trajectories that had hopped within the last 50 fs, or not hopped at all, until at least 50 fs had passed since they hopped. For the latter, we stopped a trajectory if it had lasted at least 500 fs and had hopped more than 300 fs earlier. % \textit{Trans} trajectories were stopped once they hopped to the ground state and remained there for at least 200 fs. %The time step was set to 0.5 fs, and 20 electronic sub-steps were used for each nuclear step. Tully's FS surface hopping was used with the decoherence correction of Ref. \cite{landry2012recover}. State-tracking was used to account for spin contamination. The same Fock extrapolation parameters were used as above. 

\subsection{Transfer learning}
\label{sm_sec:tl}

To validate the yield results for substituted compounds, we performed DANN-NAMD for the top candidates using a set of highly accurate models. Each model was fine-tuned for a single species only, which allowed it to achieve high accuracy on that one molecule. This transfer learning strategy was used in place of \textit{ab initio} NAMD because the latter would be prohibitively slow for all but the smallest molecules. For example, consider molecule \textbf{169}, which has only 54 atoms. We found that a single gradient or NACV calculation for this species took approximately 50 minutes with 8 CPU cores. Since \textit{trans} derivatives must be simulated for at least 1 ps, and since the time step must be no larger than 0.5 fs, we would need to perform 2,000 QC calculations for each trajectory. Assuming parallel calculation of the NACVs and gradients at each step, an \textit{ab initio} simulation would take 70 days. % Further, this would require a significant amount of computational hardware. Assuming a modest 100 trajectories and 40 cores per node, together with parallel gradient and NACV calculations, we would need continuous access to 40 nodes in total. 
% While the QC calculations could be accelerated with GPU-based software \cite{ufimtsev2008quantum, yu2020ab}, our tests indicated that this would only reduce the wall time by a factor of five to ten. Therefore, we would still only be able to validate a small number of molecules. 
By contrast, the fine-tuning approach allows us to perform highly accurate simulations for tens to hundreds of species. 
% Indeed, the species-specific networks are so accurate that full \textit{ab initio} simulations are unnecessary. It would be better to simply test the top candidates experimentally than further validate them computationally.

% Second, according to the ``computational funnel'' paradigm, one should use more accurate and more resource-intensive methods for fewer candidates at each successive step in screening \cite{pyzer2015high}. The method accuracy and candidate pool should both be modified slowly, so that one does not eliminate too many candidates  The transferable model is six orders of magnitude faster than QC, and can screen thousands to tens of thousands of candidates; the species-specific model is three orders of magnitude faster, and can screen tens to hundreds of molecules; and QC can screen under ten molecules. The funnel paradigm is better followed if we move from the transferable model to species-specific models, as the drop-off in prediction quality and number of species is less extreme. This allows us to perform highly accurate simulations for a larger number of species. Indeed, the species-specific networks are so accurate that full \textit{ab initio} simulations are unnecessary. It would be better to simply test the top candidates experimentally than further validate them with computational methods.
\begin{table}[t]
\centering
\scalebox{1}{
    \begin{tabular*}{\textwidth}{c@{\extracolsep{\fill}}ccccccccccccccccccccc}
        \toprule[1.1pt]
         lr & $\mathrm{lr}_{\mathrm{min}}$ & patience & factor & batch size & $\rho_{E_0}$ &  $\rho_{E_1}$ & $\rho_{\Delta  E_{01}}$ & $\rho_{f_0}$ & $\rho_{f_1}$ & $\rho_{\mathrm{ref}}$ & $\rho_{\mathrm{nacv}}$ \\ 
         \hline 
          $10^{-4}$ & $10^{-5}$ & 50 & 0.5 & 1 & 0.3 & 0.3 & 0.5  & 1 & 1 & 0 & 1  \\
        \bottomrule[1.1pt]
    \end{tabular*}
    }
    \caption{Training parameters used for transfer learning. ``Patience'' refers to the number of epochs without an improvement in validation loss before the learning rate is reduced. ``Factor'' is the amount by which the learning rate is lowered. }
    \label{sm_tab:tl_params}
\end{table}
Each new model was refined from the original network using QC data from a single species. The initial training geometries were sampled from DANN-NAMD simulations with the original DANN model. A committee of three DANN models, each trained on the entire dataset, was used to select the initial geometries (see below). Three fine-tuned models were then trained. Each was initialized with a different random seed and trained with different data splits. 90\% of the data was used for training and 10\% for validation. The first model was subsequently used for DANN-NAMD. Geometries were selected from these simulations using the newly-trained committee, and each geometry received QC calculations. This data was added to the training set, each committee model was re-trained, and the cycle was repeated as in Fig. \ref*{fig:schematics}(b). After each cycle we evaluated the model accuracy using the geometries generated by DANN-NAMD.

Initially we selected only 50 geometries in each round of active learning. Once we narrowed our focus to two species, we sampled 500 geometries in each round. The geometries were sampled according to the following four strategies. One third was sampled randomly. One third was chosen by the prediction variance in the excited state forces. Those with the highest variance were selected. We used only the uncertainty in the forces, and not the energies, because the former is a better indicator of trajectory instability \cite{schwalbe2021differentiable}. One sixth was chosen by proximity to a CI (Eq. (\ref{sm_eq:p_zn}) with $\Delta E = 0.2$ eV). The final sixth was chosen to sample excited-state barriers. In particular, we sampled geometries with probability
\begin{align}
    p \propto \mathrm{exp} \left[ (E - E_{\mathrm{min}}) / (k_\mathrm{B} T) \right]. \label{sm_eq:p_barrier}
\end{align}
Here $E$ is the excited state energy, $E_{\mathrm{min}}$ is the minimum excited sate energy in the trajectory, and $T=300$ K. We only sampled configurations from 50 fs or later in the simulations. This was done to avoid the initial high-energy geometries encountered before relaxation. Equation (\ref{sm_eq:p_barrier}) is inversely proportional to the room-temperature Boltzmann probability, and thus assigns the highest weight to the highest-energy configurations.

Transfer learning has previously been used to account for solvent effects \cite{ang2021active} and to reach higher levels of QC theory \cite{smith2019outsmarting,ang2021active} with only a small portion of the training set. Typically the majority of the network weights are frozen during re-training. This leaves modifiable parameters in only the final few layers. However, we found that the best results were obtained without any parameter freezing. We therefore allowed all parameters to be modified during re-training. Further, we found it best to start with the normal learning rate rather than a reduced one. We also did not use a reference loss, as its effect on the original DANN results was minor, and possibly even harmful. The full set of training parameters can be found in Table \ref{sm_tab:tl_params}. All trajectory parameters were unchanged from the screening phase. For final inference we performed DANN-NAMD for 5 ps with 2,000 trajectories, using 500 ps of ground state MD.

\begin{table*}[t]
\centering
    \begin{tabular*}{\textwidth}{c@{\extracolsep{\fill}}|cccccccccc}
        \toprule[1.1pt]
        Sampled by \ & Metric &  $E_0$ & $E_1$ & $\Delta E_{01}$ & $\vec{F}_0$ & $\vec{F}_1$ & $\vec{g}_{01}$ \\ 
        \hline
        ZN & MAE ($\downarrow$)  & 0.48 & 0.42  &  0.44 & 0.83 & 0.82 & 0.52   \\
        & $R^2$ ($\uparrow$) & 1.00  & 1.00 & 0.96 & 0.99 & 0.99  & 0.86  \\
        \hline
        Barrier & MAE ($\downarrow$) & 0.62 & 0.93 & 0.86 & 0.67 & 0.82 & 0.66  \\
        & $R^2$ ($\uparrow$) & 0.99 & 0.99 & 0.99 & 1.00 & 0.99 & 0.88 \\
        \hline
        Random & MAE ($\downarrow$) & 0.74 & 0.80 & 0.86 & 0.71 & 0.77 & 0.89   \\
        & $R^2$ ($\uparrow$) & 0.99 & 0.99 & 0.98 & 1.00 & 1.00 & 0.88 \\
        \hline
        Uncertainty & MAE ($\downarrow$) & 0.74 & 1.26 & 1.09 & 0.83 & 1.36 & 1.04 \\
        & $R^2$ ($\uparrow$) & 0.99 & 0.99 & 0.99 & 0.99 & 0.97 & 0.73 \\
        \bottomrule[1.1pt]
    \end{tabular*}
    \caption{Test set statistics for the final TL model of species \textbf{165}. Results are divided by the method used to select samples. ``ZN'' uses Zhu-Nakamura gap-based sampling [Eq. (\ref{sm_eq:p_zn})] and ``barrier'' uses Eq. (\ref{sm_eq:p_barrier}). 1,244 geometries were used for fine-tuning.}
    \label{sm_tab:tl_accuracy_165}
\end{table*}

\begin{table*}[t]
\centering
    \begin{tabular*}{\textwidth}{c@{\extracolsep{\fill}}|cccccccccc}
        \toprule[1.1pt]
        Sampled by \ & Metric &  $E_0$ & $E_1$ & $\Delta E_{01}$ & $\vec{F}_0$ & $\vec{F}_1$ & $\vec{g}_{01}$ \\ 
        \hline
        ZN & MAE ($\downarrow$) & 0.67 & 0.54 & 0.39 & 0.87 & 0.84 & 0.64  \\
        & $R^2$ ($\uparrow$) & 0.98 & 0.98  & 0.99 & 0.99 & 0.99 & 0.82 \\
        \hline
        Barrier & MAE ($\downarrow$) & 0.89 & 0.74 & 0.41 & 0.61 & 0.57 & 0.51 \\
        & $R^2$ ($\uparrow$) & 0.99 & 0.95 & 1.00 & 1.00 & 1.00 & 0.98 \\
        \hline
        Random & MAE ($\downarrow$) & 0.58 & 0.66 & 0.51 & 0.67 & 0.70 & 0.78 \\
        & $R^2$ ($\uparrow$) & 0.99 & 1.00 & 1.00 & 1.00 & 1.00 & 0.96 \\
        \hline
        Uncertainty & MAE ($\downarrow$) & 0.72 & 1.66 & 1.90 & 0.83 & 1.53 & 1.01  \\
        & $R^2$ ($\uparrow$) & 0.99 & 0.93 & 0.97 & 0.99 & 0.95 & 0.92 \\
        \bottomrule[1.1pt]
    \end{tabular*}
    \caption{As in Table \ref{sm_tab:tl_accuracy_165}, but for species \textbf{169}. 2,445 geometries were used for fine-tuning.}
    \label{sm_tab:tl_accuracy_169}
\end{table*}
  
The accuracy of the fine-tuned networks is shown in Tables \ref{sm_tab:tl_accuracy_165} and \ref{sm_tab:tl_accuracy_169}. Statistics are shown for 500 geometries that were sampled from DANN-NAMD for each molecule using the final trained networks. The model errors are well below 1 kcal/mol(/\AA) for geometries sampled by all methods other than uncertainty. The error on the uncertainty-selected configurations is under 2 kcal/mol(/\AA) in all cases. Since these geometries are specifically chosen to have the highest error, and since their errors are still rather small, we can be confident in the accuracy of the models. We note, however, that the average error for the uncertain geometries depends on how many trajectories are run. Running more trajectories means sampling more configurations, and hence finding more geometries with high uncertainty. As mentioned above, we ran 100 trajectories for 5 ps each for both screening and transfer learning (2,000 trajectories were used for the final predictions). We saved frames every 15 fs, leading to 33,333 geometries in total. Since we picked the 167 most uncertain frames, our sample is roughly the same as choosing the top 0.5\% of all geometries with the highest error. 

One reason that we did not use transfer learning for the unsubstituted compounds was severe spin contamination. This is an artifact of unrestricted SF-TDDFT, and is a well-documented problem in NAMD for \textit{trans} azobenzene \cite{yue2018performance}. After the first round of active learning, we found that the force error for the unsubstituted models increased to 8 kcal/mol/\AA. This error did not drop significantly with more data, even though the new geometries were fairly close to the Franck-Condon region. This issue did not occur with any of the derivatives. We found that the error was correlated with $\langle S^2 \rangle$, and reasoned that our method of singlet selection (Sec. \ref{sm_sec:extended_methods}\ref{subsec:data_gen}) was likely choosing triplet excited states. This highlights the issues inherent in SF-TDDFT, and reinforces the need for low-cost, spin-complete alternatives \cite{filatov2015spin, lee2018eliminating, yu2020ab}.

Lastly, we note that both the \textit{ab initio} and transfer-learned \textit{cis} quantum yields were noticeably higher than in other SF-TDDFT studies \cite{yue2018performance}. This is likely because we used MD to initiate the trajectories, rather than normal-mode or Wigner sampling based on the harmonic approximation. Our experiments showed that normal-mode sampling led to decreased \textit{cis} yields, closer to those reported in Ref. \cite{yue2018performance}. Unlike the \textit{trans} isomer, \textit{cis} azobenzene is somewhat flexible, with significant torsions occurring during ground-state MD. This indicates that the harmonic approximation should be avoided when possible. Indeed, using MD ground state sampling together with FS dynamics for \textit{cis} azobenzene, we obtained a yield of 60 $\pm$ 4\%; this is in excellent agreement with experimental values in non-polar solution, which are close to 55\% on average \cite{bandara2010proof}. It is in much better agreement than the value of 34\% obtained with FS surface hopping in Ref. \cite{yue2018performance}. Since we used the same electronic structure method and the same surface hopping approach, we can be confident that the difference is mainly due to MD sampling.

\subsection{Conical intersection pathways}
\label{sm_subsec:ci_pathways}
We labeled the CI pathway of each \textit{trans} trajectory according to its last hopping geometry. Each geometry was compared to two reference planar CIs and two reference rotational CIs. The trajectory was labeled by the reference CI that was the closest to the hopping geometry. That is, if a trajectory hopped at a geometry closer to one of the rotational CIs, it was labeled a rotational trajectory, and similarly for the planar CI.

The reference CIs were chosen from the set of all hopping geometries in the trajectories. The two rotational CIs were those with dihedral angles closest to $90^{\circ}$ and $270^{\circ}$, respectively, and $\mathrm{max}(\alpha_{\mathrm{CNN}}, \alpha_{\mathrm{NNC}})$ closest to $138^{\circ}$. The two planar CIs were those with dihedral angles closest to $186^{\circ}$ and $174^{\circ}$, respectively, and both $\mathrm{max}(\alpha_{\mathrm{CNN}}$ and $\alpha_{\mathrm{NNC}})$ closest to $148^{\circ}$. The angles are those of the optimized CI geometries in Ref. \cite{yu2020nonadiabatic}. For the derivative \textbf{169}, we further optimized each of the reference structures to minimize the gap and hence obtain a true CI. For each trajectory we computed the RMSD between the CNNC atoms of the hopping geometry and the CNNC atoms of the reference CIs.

We additionally enforced that a hopping geometry could only be considered planar if $\vert 180 - \theta \vert \leq \theta_0 $, where $\theta$ is the CNNC dihedral angle and $\theta_0$ is a cutoff value. We chose $\theta_0 = 75^{\circ}$ for azobenzene and $65^{\circ}$ for \textbf{169}. Without this constraint, many of the hopping geometries labeled ``planar'' in \textbf{169} actually led to isomerization. This made a dihedral constraint necessary. On the other hand, when we only labeled all geometries with $\vert \theta - 180 \vert \leq 45^{\circ}$ as planar \cite{yu2020nonadiabatic}, we found through visual inspection that many non-reactive CI geometries were mislabeled as rotational. Using the minimal distance to a reference CI, together with a modest dihedral constraint, led to the most qualitatively reasonable results. We confirmed that none of the geometries labeled as ``planar CI'' with our metric led to switching, which further reinforced the soundness of our approach.

\section{Figure details}
\subsection{Computational speed-up}

Here we describe how the speeds of ML and QC calculations were computed. For QC, we first computed the run time of one gradient calculation on a single geometry, denoted $t_{\mathrm{calc}}$. The node time was then calculated as $t_{\mathrm{node}} = t_{\mathrm{calc}} / n_{\mathrm{calc}}$, with $n_{\mathrm{calc}} = (\text{cores per node}) / (\text{cores per job})$. We assumed 40 cores per node. All QC jobs were performed with 8 cores, and so $n_{\mathrm{calc}}$ was equal to 5.

For ML we performed a batched calculation on ten copies of each geometry, and computed one gradient. The node time was computed as $t_{\mathrm{node}} = t_{\mathrm{calc}} / n_{\mathrm{calc}}$, with $n_{\mathrm{calc}} = 10 \cdot (\text{total memory per gpu}) / (\text{memory used in calculation}) \cdot (\text{GPUs per node})$. We set $(\text{total memory per gpu}) = 32$ GB and $(\text{GPUs per node}) = 2$. We used a script that performed one batched calculation on ten copies of a random geometry, and re-ran it 7,000 times. For each iteration we used the \texttt{nvidia-smi} command with the PID of the current job to access the GPU memory. We re-ran the script many times, rather than running many calculations in one script, because \texttt{nvidia-smi} does not account for all freed memory until a job is finished. That is, after one calculation is finished, \texttt{nvidia-smi} shows that GPU memory is still being occupied by the job. This occurs even after all local variables are deleted. Hence running multiple calculations in one script would yield an overestimated GPU memory.

In the above calculation, we implicitly assumed that multiplying the batch size by $x$ would lead to an $x$-fold increase in memory. In practice we have found that the memory is increased by less than that. This means that the ML speedup in Fig. \ref*{fig:results} is a conservative estimate.

\subsection{Diabatic energies}
Since three diabatic states were used in the DANN model, and since all three were coupled near the CI, it would be incorrect to plot only two states in Fig. \ref*{fig:results}(b). We therefore applied a fixed rotation matrix, $\mathbf{U} \neq \mathbf{U}(\vec{R})$, to generate a new diabatic Hamiltonian with only two coupled states. The new diabatic Hamiltonian was given by $\mathbf{H}_d'(\vec{R}) = \mathbf{U}^{\dagger} \hspace{0.05cm} \mathbf{H}_d(\vec{R}) \mathbf{U}$. Note that any position-independent rotation matrix can be brought outside the gradient in Eq. (\ref*{eq:hd_nacv}), and so $\mathbf{H}_d'$ is still diabatic. The rotation matrix was chosen to diagonalize $\mathbf{H}_d$ at the CI, so that $\mathbf{U}^{\dagger} \hspace*{0.05cm} \mathbf{H}_d (\vec{R}_{\mathrm{CI}}) \hspace*{0.05cm} \mathbf{U} = \mathrm{diag}(\{ E \} )$. This led to $d_{00}' = d_{11}'$ and $d_{01}'=0$ at the CI. Hence the lowest two rotated states were the most important contributors to $E_0$ and $E_1$, and were therefore used in Fig. \ref*{fig:results}(b).

\section{Intensive and extensive quantities}
\label{sm_sec:intensive}
The DANN model predicts each property by summing over atomic contributions, just as in the PaiNN model. This guarantees size-extensivity. However, as explained below, the off-diagonal elements of $\mathbf{H}_d$ should be intensive, in the sense that atoms not involved in the excitation should not contribute to these values. Nonetheless, we found that summation for these terms gave better results than averaging. This was true even when using a learnable weighted average. Atom-wise summation can still generate accurate $d_{nm}$, because the readout network can simply map unimportant atoms to zero.

To demonstrate extensivity and intensivity, consider two uncoupled subsystems, $A$ and $B$. The total clamped nucleus Hamiltonian is $H(\vec{r}, \vec{R}) = H^A(\vec{r}, \vec{R}) + H^B(\vec{r}, \vec{R}) $. Let the excitation of interest be in subsystem $A$. In this case the adiabatic states involve only excitations in subsystem $A$, so that the $n^{\mathrm{th}}$ diabatic wave function is written as 
\begin{align}
\psi_{\mathrm{d}, n}(\vec{r}; \vec{R}) = \psi_{\mathrm{ad}, 0}^{B}(\vec{r}; \vec{R}) \sum_k U_{nk} \psi_{\mathrm{ad}, k}^{A}(\vec{r}; \vec{R}).
\end{align}
The diabatic state is a direct product of the ground state wave function of system $B$, $\psi_{\mathrm{ad}, 0}^{B}(\vec{r}; \vec{R})$, and a rotation of the adiabatic states of system A, $\psi_{\mathrm{ad}, k}^{A}(\vec{r}; \vec{R})$. The matrix elements of $\mathbf{H}_d$ are then given by
\begin{alignat}{2}
(\mathbf{H}_d)_{nm} & = \langle \psi_{\mathrm{d}, n} \vert H(\vec{r}, \vec{R}) \vert \psi_{\mathrm{d}, m} \rangle \nonumber \\
& = \sum_{kl} U_{nk}^* U_{ml} \langle \psi^{B}_{\mathrm{ad}, 0} \psi^{A}_{\mathrm{ad}, k}   \vert  \  H_A(\vec{r}, \vec{R})  + H_B(\vec{r}, \vec{R}) \  \vert \psi^{B}_{\mathrm{ad}, 0} \psi^{A}_{\mathrm{ad}, l}  \rangle  \nonumber \\
& = \sum_{kl} U_{nk}^* U_{ml} \big( E_{k, A} + E_B \big) \delta_{kl} \nonumber \\
& = \sum_{k} U^*_{in} U_{mk} \hspace*{0.05cm} E_{k, A} + E_B \big(\mathbf{U U^{\dagger}}\big)_{mn} \nonumber \\
& = \mathbf{H}^A_{\mathrm{d}, nm} + E_B \hspace*{0.05cm} \delta_{nm}. \label{sm_eq:intensive}
\end{alignat} 
Hence the diagonal elements of $\mathbf{H}_d$ each gain the adiabatic energy of $B$, while the off-diagonal elements remain the same. To understand this result physically, consider that adding a scalar multiplied by the identity yields eigenvalues that are each shifted by the scalar. This means that the eigenvalues of $\mathbf{H}_d$ are each shifted by $E_B$. Therefore the excitation energy is unchanged, which is the expected behavior upon adding an uncoupled system. Note also that extensivity is sometimes described as doubling the energy when the system size is doubled. However, this definition is too narrow, as it applies to only one adiabatic energy at a time. For example, if subsystem B is a copy of subsystem A, then $E_B = E_{0, A}$, meaning that $E_0 = E_{0, A} + E_{B} = 2 E_{0, A}$, and so the ground state energy is indeed doubled. However, $E_1 = E_{1, A} + E_B = 2 E_{1, A} - \mathrm{gap}$, where $\mathrm{gap} = E_{1, A} - E_{0, A}$. Therefore, the excited state energy is not doubled. A more general definition is that the energy of the second subsystem is added to each adiabatic energy.

This analysis shows that the off-diagonal elements of $\mathbf{H}_d$ are intensive. The adiabatic energy gap is also intensive. Intensive here means that adding a subsystem that does not participate in the excitation does not modify the quantity. Such properties can naturally be modeled with an attention mechanism \cite{bahdanau2014neural, kim2017structured, noam_lr, gat} that learns the importance of each atom to the excitation. For example, the excitation energy can be modeled as an attention-weighted sum over atomwise quantities. However, as discussed in Supplementary Sec. \ref{sm_sec:training}, this approach led to worse performance than simply predicting extensive energies for each state.

\section{Proof of diabaticity}
\label{sm_sec:proof}
Here we prove Eq. (\ref*{eq:hd_nacv}) in the main text. Using bra-ket notation for the wave functions, we define the diabatic states as a linear combination of adiabatic states,
\begin{align}
& \ket{\psi_{\text{d,}n}} = \sum_{m} \ket{\psi_{\text{ad,}m}} V^*_{nm}, \label{sm_eq:di_def}
\end{align}
where $\mathbf{V}$ is a unitary matrix. Multiplying each side by $V_{nn'}$ and summing over $n$ yields
\begin{align}
& \ket{\psi_{\text{ad,}n}} = \sum_{m} \ket{\psi_{\text{d,}m}} V_{mn},
\end{align}
where we have renamed the dummy indices, $n \to m$ and $n' \to n$. We have also used the fact that $(\mathbf{V}^{\dagger} \mathbf{V})_{nm} = \delta_{nm}$ for any unitary matrix, where $\delta_{nm}$ is the Kronecker delta. Given the connection between diabatic and adiabatic states, we can relate $\mathbf{V}$ to the derivative coupling $\vec{k}$:
\begin{align}
& \vec{k}_{nm} \equiv \bra{ \psi_{\text{ad,}n}}  \nabla_{R} \ket{ \psi_{\text{ad,}m} } \nonumber \\
& = \sum_{ij} V^*_{jn} \left( \nabla_R V_{im} \bra{\psi_{\mathrm{d}, j}}  \ket{\psi_{\mathrm{d, }i}} + V_{im} \bra{\psi_{\mathrm{d}, j}}  \nabla_R \ket{\psi_{\mathrm{d}, i}} \right) \nonumber \\
& = \sum_{i} V_{in}^* \nabla_R V_{im}. \label{sm_eq:deriv_coupling_diabat}
\end{align}
We have used the fact that, by definition, the derivative coupling between any two diabatic states is zero. We have also used the orthonormality of the diabatic states:
\begin{align}
    \bra{\psi_{\mathrm{d}, j}} \ket{\psi_{\mathrm{d}, i}} = \sum_{nm} V_{jn} V^*_{im} \bra{\psi_{\mathrm{ad}, n}} \ket{\psi_{\mathrm{ad}, m}} = \sum_{nm} V_{jn} V^*_{im} \delta_{nm} = \left(\mathbf{V} \mathbf{V}^{\dagger} \right)_{ij} = \delta_{ij}, \label{sm_eq:orthonormal}
\end{align}
which follows from the orthonormality of the adiabatic wave functions. Meanwhile, by construction, the Hamiltonian produced by the model has eigenvalues equal to the adiabatic energies:
\begin{align}
\left( \mathbf{H}_{\mathrm{d}} \right)_{nm} & = \left( \mathbf{U} \hspace*{0.05cm} \mathrm{diag} ( \{ E \} ) \hspace*{0.05cm} \mathbf{U}^{\dagger} \right)_{nm} \nonumber  \\
& = \sum_{ij} U_{ni} \hspace*{0.05cm} (E_i \delta_{ij}) \hspace*{0.05cm} U_{mj}^* \nonumber \\
&  = \left(\sum_{i} U_{ni} \bra{\psi_{\mathrm{ad}, i}} \right) \hat{H} \left( \sum_j U_{mj}^* \ket{\psi_{\mathrm{ad} , j}}  \right), \label{sm_eq:h_d_elec}
\end{align}
% \nonumber \\
% & =  \bra{\psi_{\mathrm{d, }n}} H \ket{\psi_{\mathrm{d, }m}}.
% \end{align}
where $\mathbf{U}$ is the unitary matrix that diagonalizes $\mathbf{H}_d$, $\hat{H}$ is the Hamiltonian operator, and we have used the relation $\bra{\psi_{\mathrm{ad}, j}} \hat{H} \ket{\psi_{\mathrm{ad}, i}} = E_i \bra{\psi_{\mathrm{ad}, j}} \ket{\psi_{\mathrm{ad}, i}} = E_i \delta_{ij}$. Comparing Eqs. (\ref{sm_eq:h_d_elec}) and (\ref{sm_eq:di_def}), we see that $\mathbf{H}_d$ is the representation of $\hat{H}$ in a different electronic basis. In particular, if we choose $\mathbf{U} = \mathbf{V}$, such that $\mathbf{U}$ satisfies Eq. (\ref{sm_eq:deriv_coupling_diabat}), then $\mathbf{H}_d$ is the representation of $\hat{H}$ in the diabatic basis.

The model could be trained directly with Eq. (\ref{sm_eq:deriv_coupling_diabat}). However, the equation is numerically ill-posed because $\vec{k}_{nm}$ diverges at conical intersections. It is preferable to work instead with the force coupling, $\vec{g}_{nm}$. We now show that Eq. (\ref*{eq:hd_nacv}), which is defined in terms of $\vec{g}_{nm}$, holds only if (\ref{sm_eq:deriv_coupling_diabat}) is satisfied. The left-hand side of Eq. (\ref*{eq:hd_nacv}) can be written as
\begin{align}
& \sum_{ij} U^*_{in} \nabla_R \left( \mathbf{U} \hspace*{0.05cm} \mathrm{diag} ( \{ E \} ) \hspace*{0.05cm}  \mathbf{U}^{\dagger} \right)_{ij} U_{jm}\nonumber \\
& = \sum_{ijk} U^*_{in} \left[ (\nabla_R E_k) \hspace*{0.05cm}  U_{ik} U^*_{jk} + E_k \hspace*{0.05cm}  (\nabla_R U_{ik}) U^*_{jk} + E_k \hspace*{0.05cm} U_{ik}  (\nabla_R U^*_{jk}) \right] U_{jm}.
\end{align}
The first term is
\begin{align}
& \sum_{k} (\mathbf{U}^{\dagger} \mathbf{U})_{nk} \hspace*{0.05cm} (\nabla_R E_k) \hspace*{0.05cm} (\mathbf{U}^{\dagger} \mathbf{U})_{km} = (\nabla_R E_n) \delta_{nm},
\end{align}
where we have used the fact that $(\mathbf{U}^{\dagger} \mathbf{U})_{nm} = \delta_{nm}$. Substituting Eq. (\ref{sm_eq:deriv_coupling_diabat}) with $\mathbf{V} = \mathbf{U}$, the second term is
\begin{align}
\sum_{k} \vec{k}_{nk} (\mathbf{U}^{\dagger} \mathbf{U})_{km} E_k = \vec{k}_{nm} E_m.
\end{align} Performing the same substitution for the third term gives
\begin{align}
    \sum_k \vec{k}^*_{mk} (\mathbf{U}^{\dagger}  \mathbf{U})_{nk}  E_k = -\vec{k}_{nm} E_n,
\end{align}
where we have used the anti-Hermitian property $\vec{k}_{mn}^* = -\vec{k}_{nm}$.
Adding the three terms gives
\begin{align}
\left( \mathbf{U}^{\dagger} (\nabla_R H_d) \mathbf{U}\right)_{nm} = \begin{cases} \nabla_R E_n, & \text{ if }n=m, \\ (E_m - E_n) \hspace*{0.05cm} \vec{k}_{nm}, & \text{ if }n \neq m. \end{cases}
\end{align}
Noting that $\vec{f}_n = -\nabla_R E_n$ and $\vec{g}_{nm} = (E_m - E_n) \hspace*{0.05cm} \vec{k}_{nm}$ gives Eq. (\ref*{eq:hd_nacv}) in the main text.
Hence Eq. (\ref*{eq:hd_nacv}) can only hold if Eq. (\ref{sm_eq:deriv_coupling_diabat}) is true. Therefore, enforcing Eq. (\ref*{eq:hd_nacv}) ensures that there is no derivative coupling between any pair of diabatic states. Note that in the main text we used three diabatic states, and trained only on energies and couplings between the first two adiabatic states. This still means that all three diabatic states are properly diabatic: if Eq. (\ref*{eq:hd_nacv}) is satisfied for even one pair of adiabatic states, then the derivative coupling must be zero between \textit{all} pairs of diabatic states.

% \section{Tully surface hopping}
% The simulation was performed at the SF-TDDFT/BHHLYP level of theory with the 6-31G* basis using Q-Chem. State-tracking was used to identify singlet states. Trajectories were initialized with Maxwell-Boltzmann velocities at 300 K and Wigner-sampled geometries from ground state normal modes. \textbf{100} independent trajectories were run for 1.5 ps each. 

\newpage
\section{Training species}
\label{sm_sec:train_specs}

Here we provide the motifs and substituents used for combinatorial molecule generation, together with a list of the literature species used for dense configurational sampling. Species with a net charge were excluded from training but are given here for completeness. In some cases only the \textit{cis} or \textit{trans} isomer was actually investigated experimentally, but in all cases we reference the publication for both isomers.

Many species in both the training and test set had experimental $S_1$ yields in non-polar solution. We did not put all such molecules in the test set, because this would mean losing hundreds of thousands of training geometries. Instead we used the 40 species with the fewest QC calculations.

\begin{table}[h]
\caption{Motifs used for combinatorial species generation. Examples of literature species for each motif are also given. The species numbers are those in Tables \ref{sm_tab:test_lit_specs} and \ref{sm_tab:train_lit_specs}. } 
\label{sm_tab:motifs}
\end{table}

\renewcommand\tabularxcolumn[1]{m{#1}}
\begin{xltabular}{\textwidth}{|>{\centering\arraybackslash}m{.3\linewidth}|>{\centering\arraybackslash}m{.6\linewidth}|}
\toprule
Graphs & Literature species \\

\toprule

\hline
\raisebox{-.45\height}{\includegraphics[height=0.10\textwidth,trim=0 -5 0 -5]{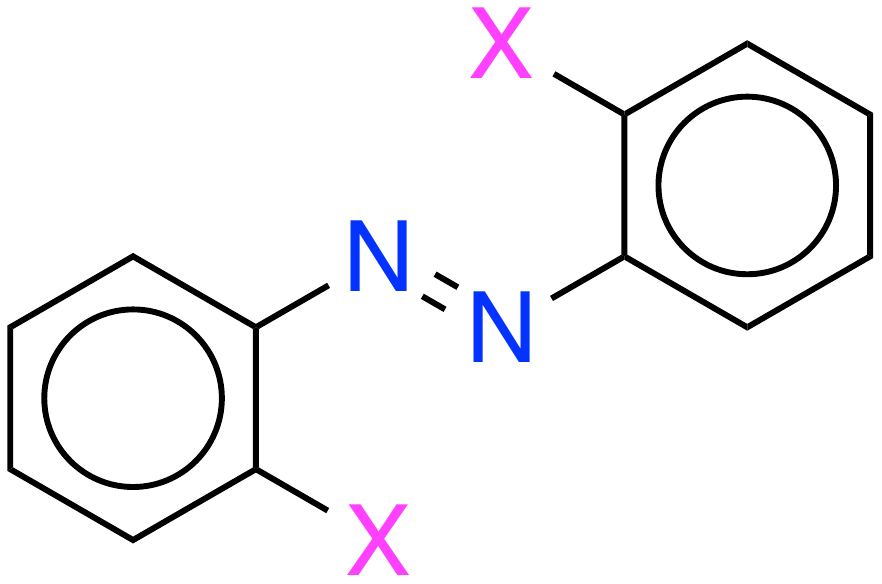}}  \raisebox{-.45\height}{\includegraphics[height=0.07\textwidth,trim=0 -5 0 -5]{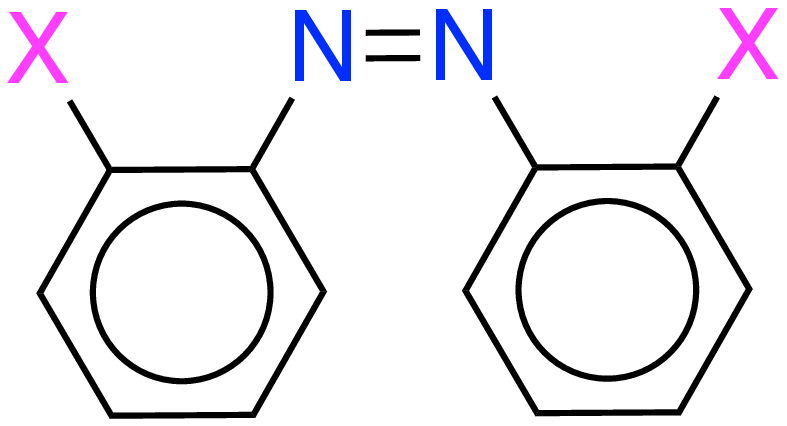}} &  1, 2, 25, 26, 63, 64, 125, 126 \\

\hline
\raisebox{-.45\height}{\includegraphics[height=0.1\textwidth,trim=0 -5 0 -5]{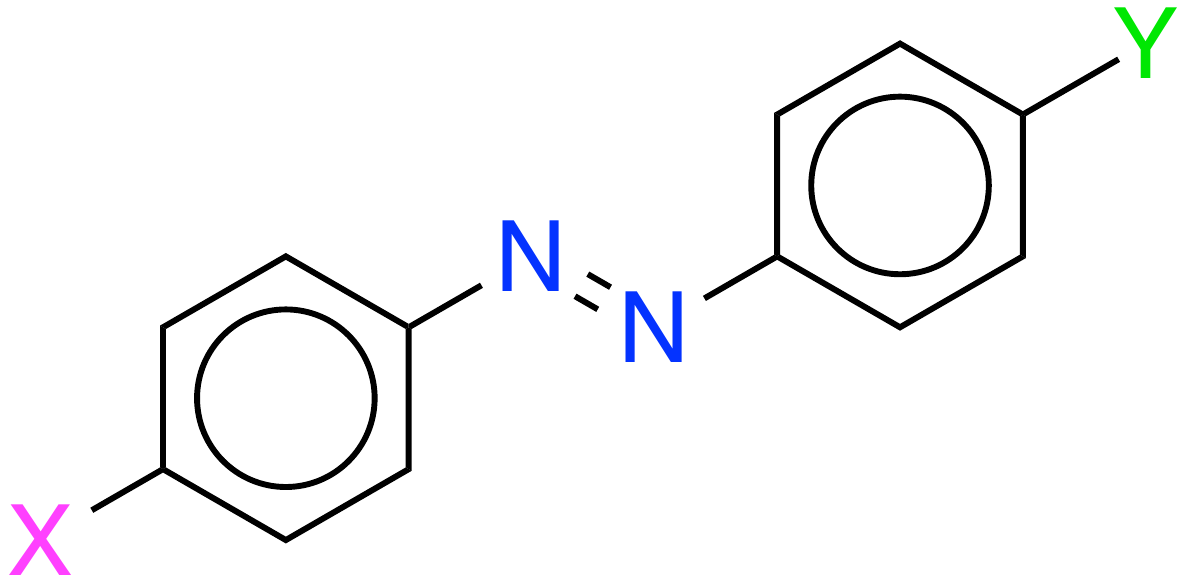}} \  \raisebox{-.45\height}{\includegraphics[height=0.1\textwidth,trim=0 -5 0 -5]{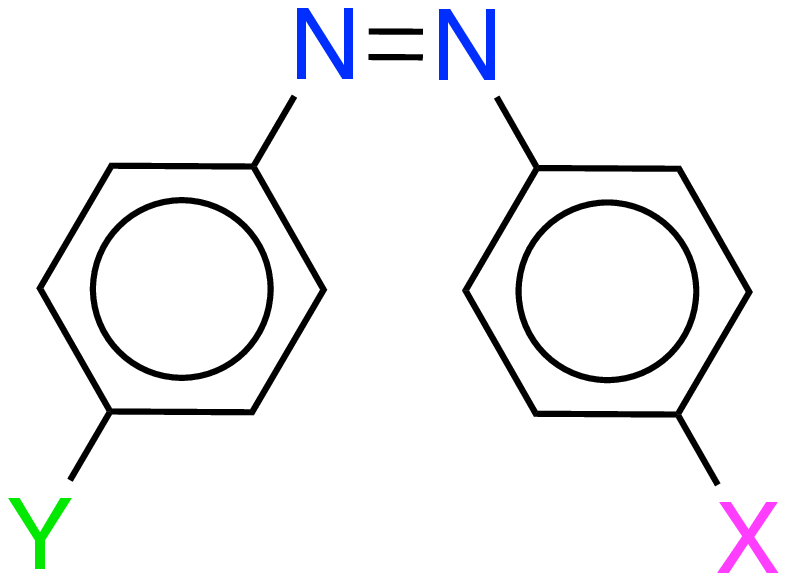}} &  19, 20, 27, 28, 37, 38, 39, 40, 61, 62, 69, 70, 73, 74, 75, 76, 83, 84, 97, 98, 115, 116, 117, 118, 119, 120, 121, 122, 123, 124, 127, 128, 129, 130, 147, 148, 155, 156, 157, 158, 163, 164  \\

\hline
\raisebox{-.45\height}{\includegraphics[height=0.10\textwidth,trim=0 -5 0 -5]{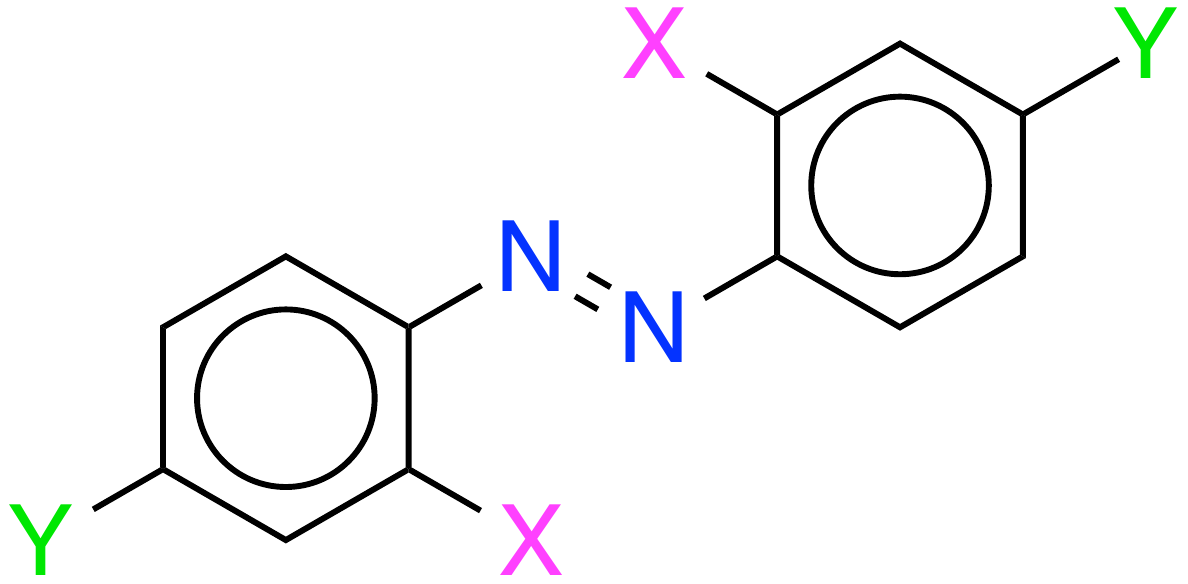}}  \raisebox{-.45\height}{\includegraphics[height=0.1\textwidth,trim=0 -5 0 -5]{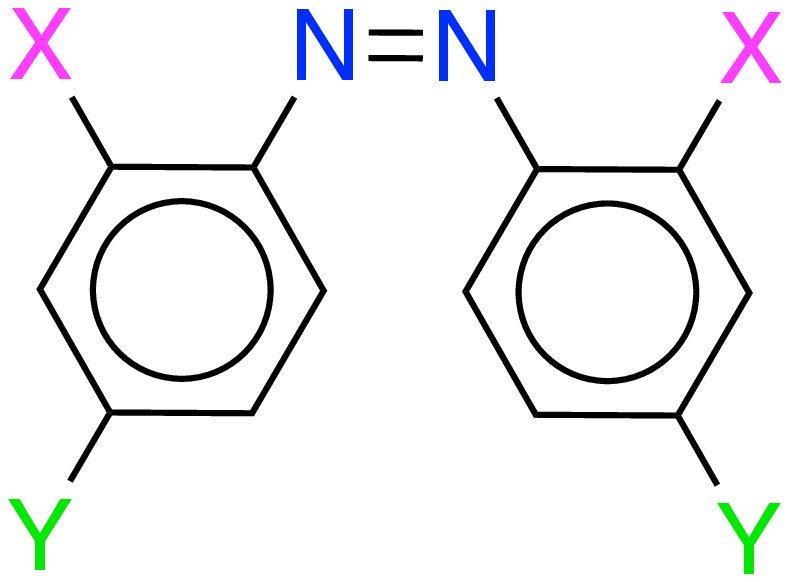}} & 67, 68, 87, 88, 99, 100, 101, 102, 103, 104, 105, 106, 107, 108 \\

\hline
\raisebox{-.45\height}{\includegraphics[height=0.1\textwidth,trim=0 -5 0 -5]{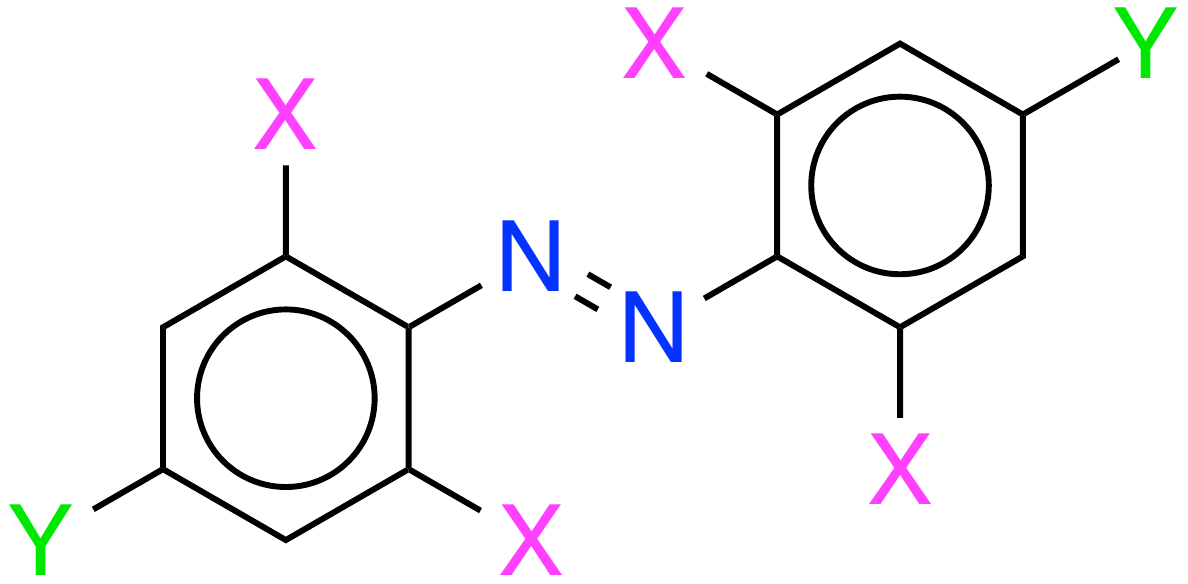}}  \raisebox{-.45\height}{\includegraphics[height=0.11\textwidth,trim=0 -5 0 -5]{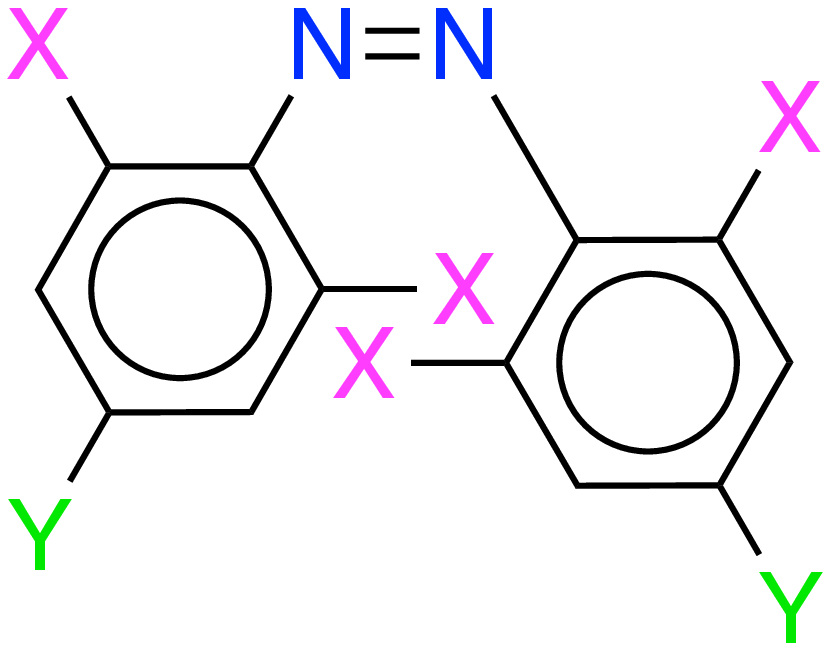}} &  9, 10, 13, 14, 15, 16, 17, 18, 21, 22, 95, 96, 131, 132, 133, 134, 135, 136, 137, 138, 139, 140, 161, 162 \\

\hline
\end{xltabular}

\newpage
\begin{table}[h]
\caption{Substituents used for combinatorial species generation. ``Azo'' denotes the azobenzene attachment site. } 
\label{sm_tab:substituents}
\end{table}

\renewcommand\tabularxcolumn[1]{m{#1}}
\begin{xltabular}{\textwidth}{|>{\centering\arraybackslash}m{.4\linewidth}|>{\centering\arraybackslash}m{.5\linewidth}|}
% \begin{xltabular}
\toprule
SMILES & Graph \\
\toprule

\hline 
[Azo]N(C)Cc1ccccn1 & \raisebox{-.45\height}{\includegraphics[height=0.055\textwidth,trim=0 -5 0 -5]{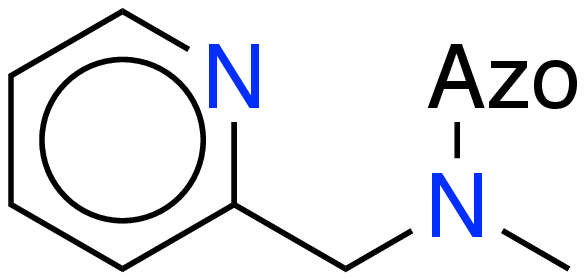}} \\ 

\hline 
[Azo]N & \raisebox{-.45\height}{\includegraphics[height=0.025\textwidth,trim=0 -5 0 -5]{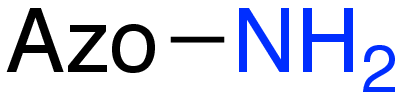}} \\ 

\hline 
[Azo]N(CC)Cc1ccccc1 & \raisebox{-.45\height}{\includegraphics[height=0.055\textwidth,trim=0 -5 0 -5]{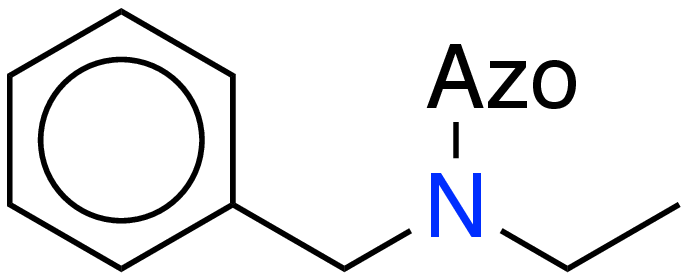}} \\ 

\hline 
[Azo]C(N)=O & \raisebox{-.45\height}{\includegraphics[height= 0.06\textwidth,trim=0 -5 0 -5]{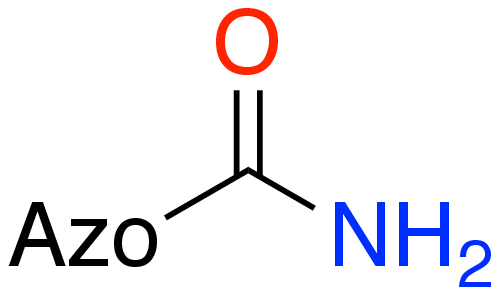}} \\ 

\hline 
[Azo]OCc1cc(OCc2ccccc2)cc(OCc2ccccc2)c1 & \raisebox{-.45\height}{\includegraphics[height=0.185\textwidth,trim=0 -5 0 -5]{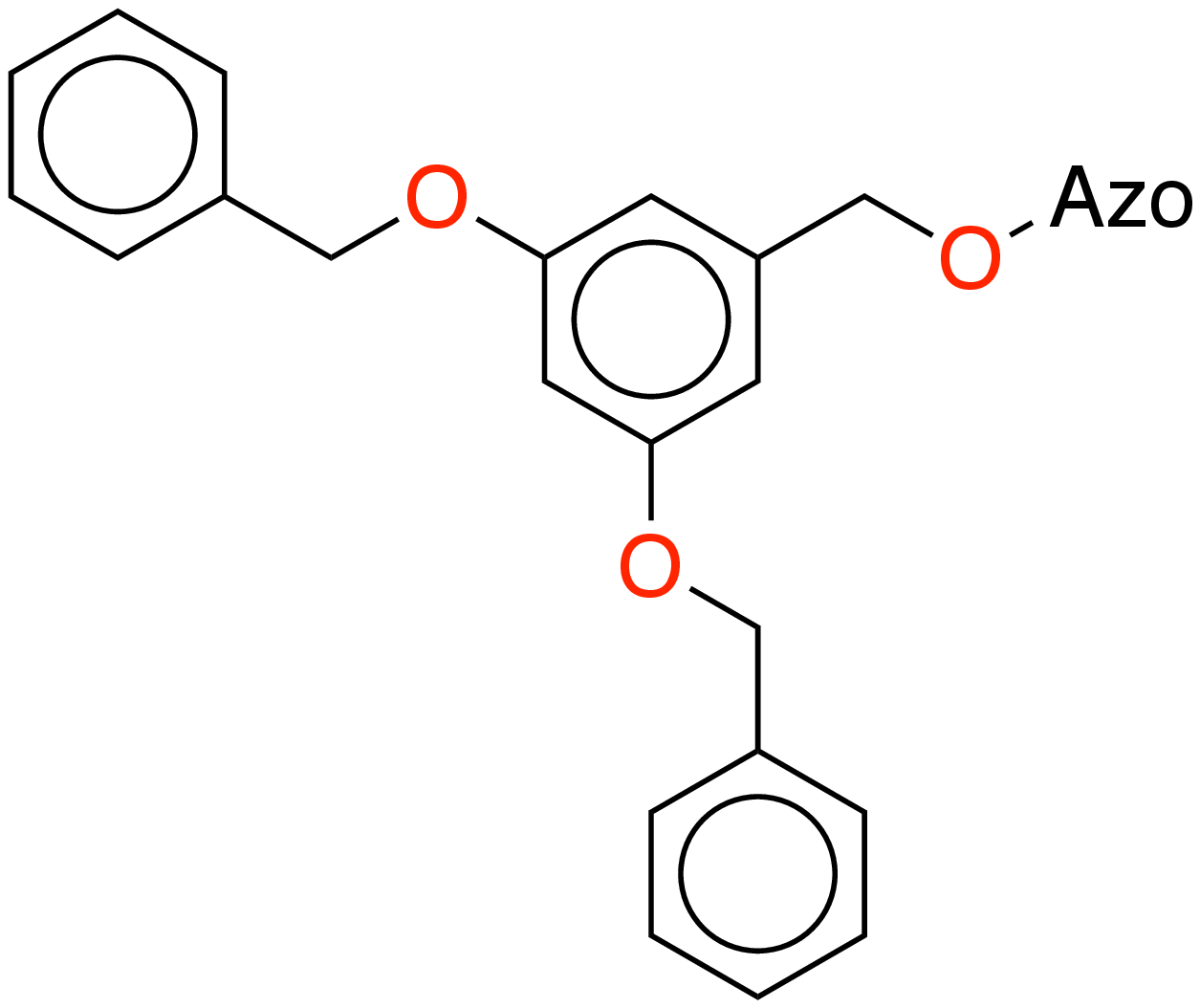}} \\ 

\hline 
[Azo]N1CCCCC1 & \raisebox{-.45\height}{\includegraphics[height=0.06\textwidth,trim=0 -5 0 -5]{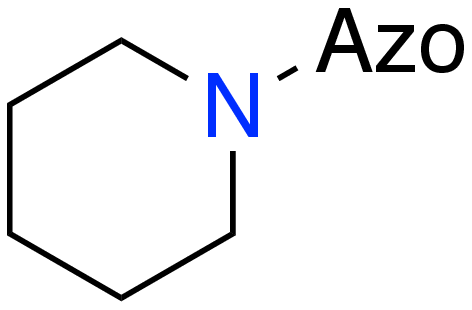}} \\ 

\hline 
[Azo]c1ccccc1 & \raisebox{-.45\height}{\includegraphics[height=0.06\textwidth,trim=0 -5 0 -5]{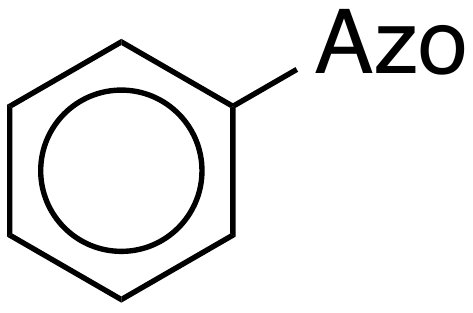}} \\ 

\hline 
[Azo]NC(=O)CCl & \raisebox{-.45\height}{\includegraphics[height=0.07\textwidth,trim=0 -5 0 -5]{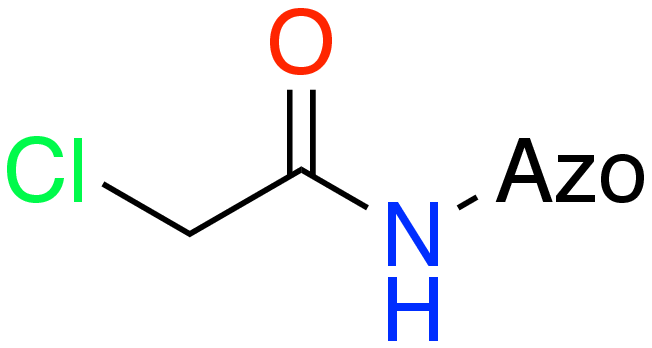}} \\ 

\hline 
[Azo]OC(C)=O & \raisebox{-.45\height}{\includegraphics[height=0.055\textwidth,trim=0 -5 0 -5]{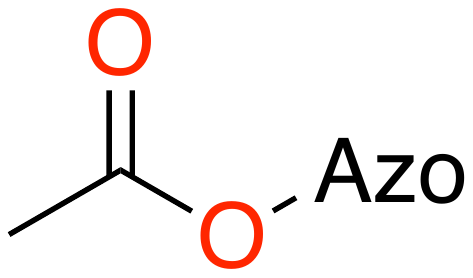}} \\ 

\hline 
[Azo]N1CCN(C(C)=O)CC1 & \raisebox{-.45\height}{\includegraphics[height=0.085\textwidth,trim=0 -5 0 -5]{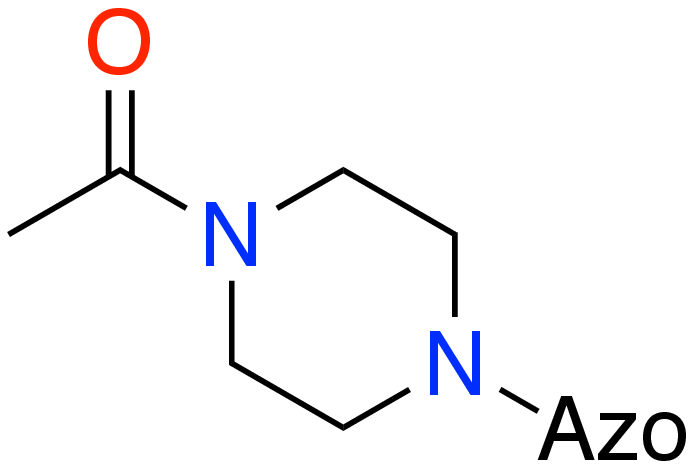}} \\ 

\hline 
[Azo]NC(=O)C[N+](CC)(CC)CC & \raisebox{-.45\height}{\includegraphics[height=0.075\textwidth,trim=0 -5 0 -5]{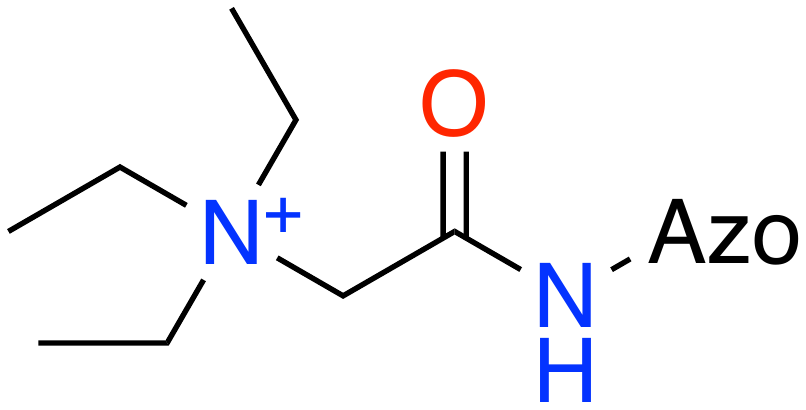}} \\ 

\hline 
[Azo]Cl & \raisebox{-.45\height}{\includegraphics[height=0.021\textwidth,trim=0 -5 0 -5]{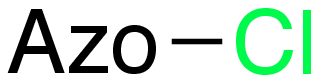}} \\ 

\hline 
[Azo]CO & \raisebox{-.45\height}{\includegraphics[height=0.027\textwidth,trim=0 -5 0 -5]{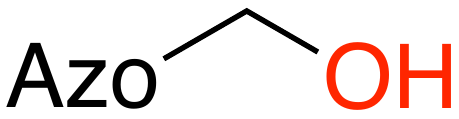}} \\ 

\hline 
[Azo]C(=O)OCC & \raisebox{-.45\height}{\includegraphics[height=0.055\textwidth,trim=0 -5 0 -5]{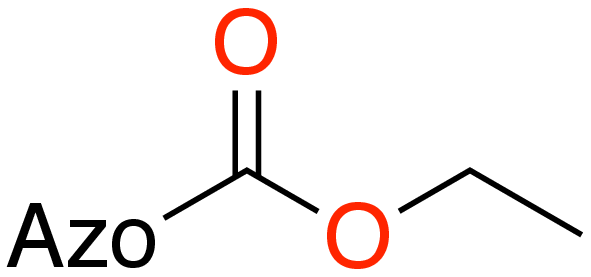}} \\ 

\hline 
[Azo]CC & \raisebox{-.45\height}{\includegraphics[height=0.028\textwidth,trim=0 -5 0 -5]{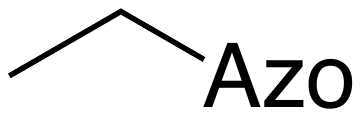}} \\ 

\hline 
[Azo]C(=O)O & \raisebox{-.45\height}{\includegraphics[height=0.055\textwidth,trim=0 -5 0 -5]{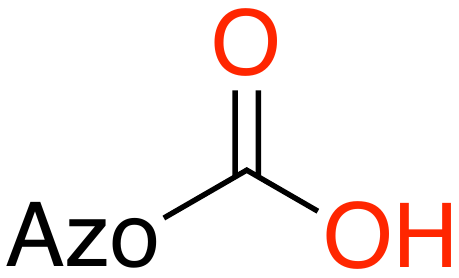}} \\ 

\hline 
[Azo]C\#C & \raisebox{-.45\height}{\includegraphics[height=0.02\textwidth,trim=0 -5 0 -5]{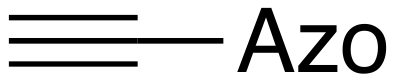}} \\ 

\hline 
[Azo]NC(=O)CN1C(=O)C=CC1=O & \raisebox{-.45\height}{\includegraphics[height=0.09\textwidth,trim=0 -5 0 -5]{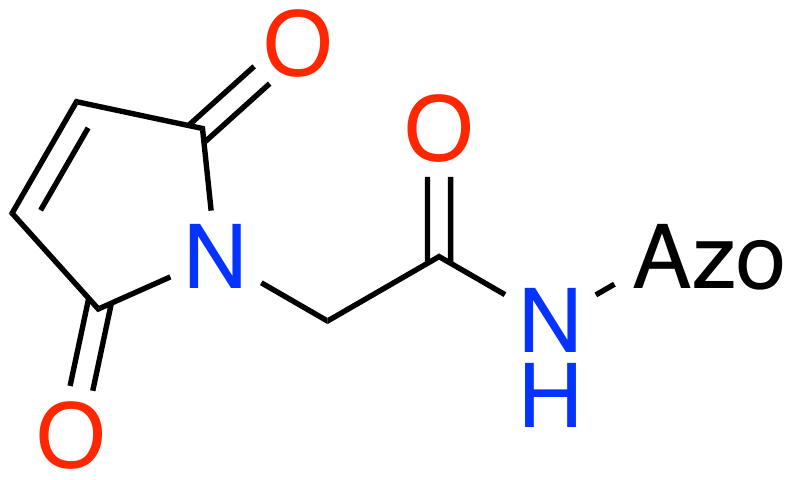}} \\ 

\hline 
[Azo][N+](=O)[O-] & \raisebox{-.45\height}{\includegraphics[height=0.055\textwidth,trim=0 -5 0 -5]{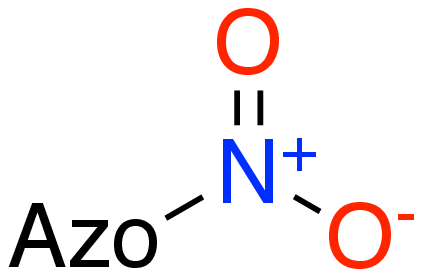}} \\ 

\hline 
[Azo][N+](C)(C)C & \raisebox{-.45\height}{\includegraphics[height=0.06\textwidth,trim=0 -5 0 -5]{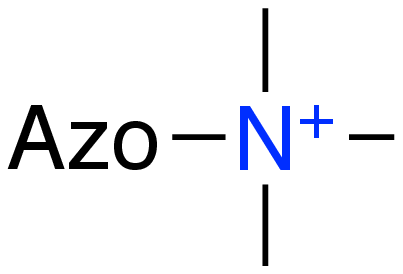}} \\ 

\hline 
[Azo]O & \raisebox{-.45\height}{\includegraphics[height=0.022\textwidth,trim=0 -5 0 -5]{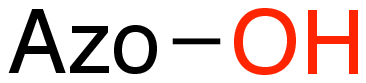}} \\ 

\hline 
[Azo]Oc1c(N)cccc1CC(C)C & \raisebox{-.45\height}{\includegraphics[height=0.08\textwidth,trim=0 -5 0 -5]{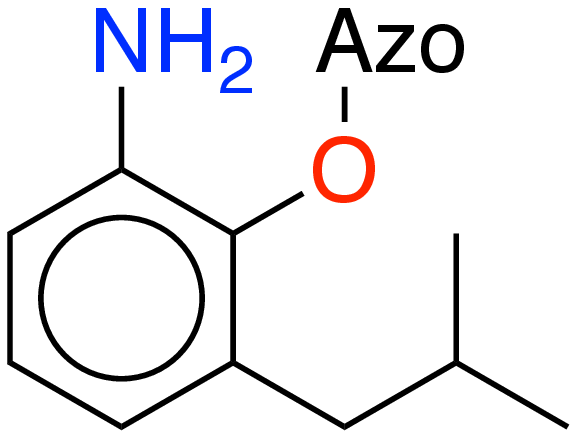}} \\ 

\hline 
[Azo]F & \raisebox{-.45\height}{\includegraphics[height=0.022\textwidth,trim=0 -5 0 -5]{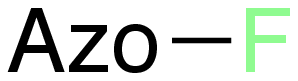}} \\ 

\hline 
[Azo]N1CCN(C(=O)CCl)CC1  & \raisebox{-.45\height}{\includegraphics[height=0.09\textwidth,trim=0 -5 0 -5]{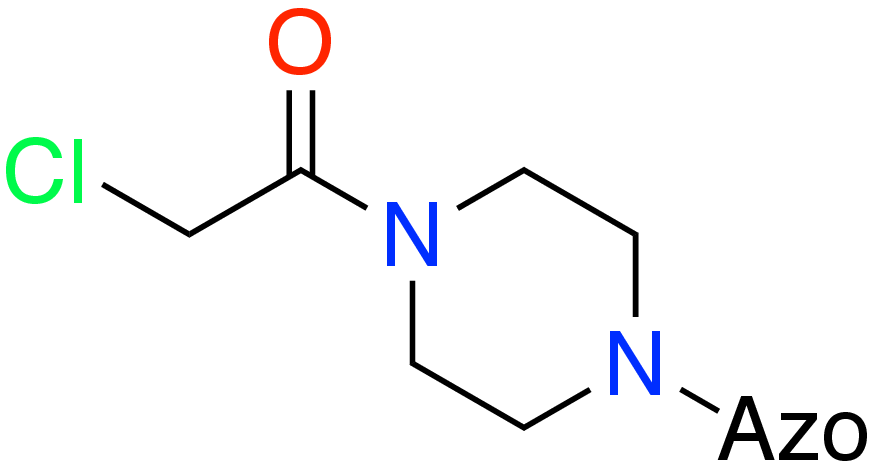}} \\ 

\hline 
[Azo]Oc1c(N)cccc1C(C)C & \raisebox{-.45\height}{\includegraphics[height=0.11\textwidth,trim=0 -5 0 -5]{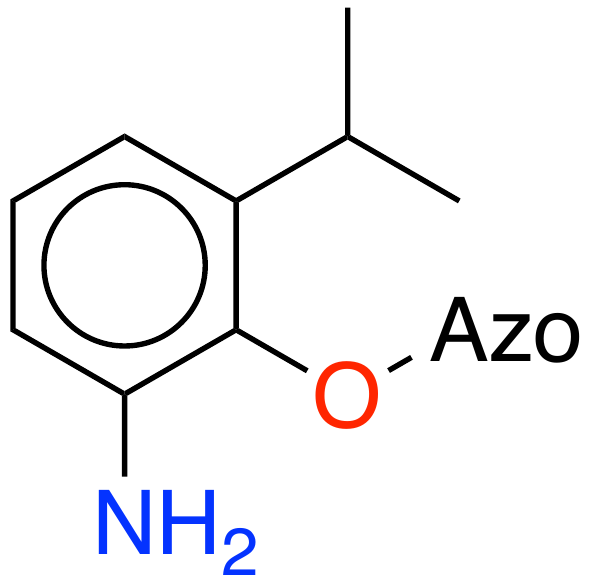}} \\ 

\hline 
[Azo]NCc1ccccn1 & \raisebox{-.45\height}{\includegraphics[height=0.06\textwidth,trim=0 -5 0 -5]{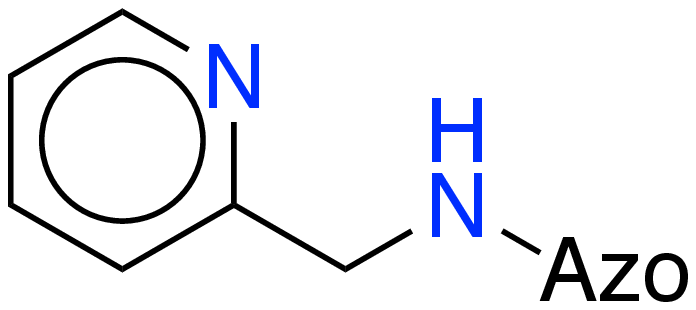}} \\ 

\hline 
[Azo]OCc1ccccc1 & \raisebox{-.45\height}{\includegraphics[height=0.06\textwidth,trim=0 -5 0 -5]{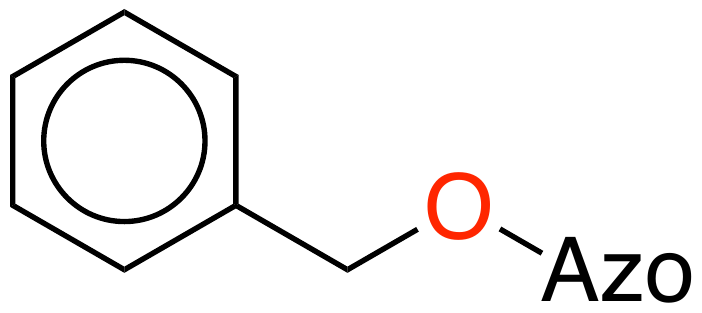}} \\ 

\hline 
[Azo]NC(C)=O & \raisebox{-.45\height}{\includegraphics[height=0.07\textwidth,trim=0 -5 0 -5]{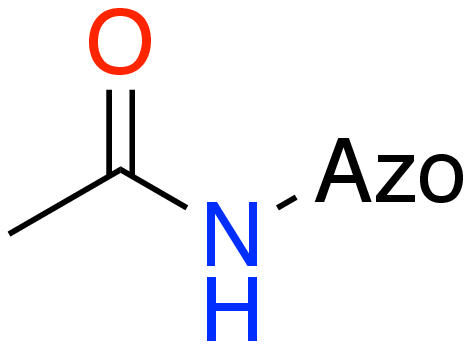}} \\ 

\hline 
[Azo]c1cn(-c2ccc3cc4ccccc4cc3c2)nn1 & \raisebox{-.45\height}{\includegraphics[height=0.085\textwidth,trim=0 -5 0 -5]{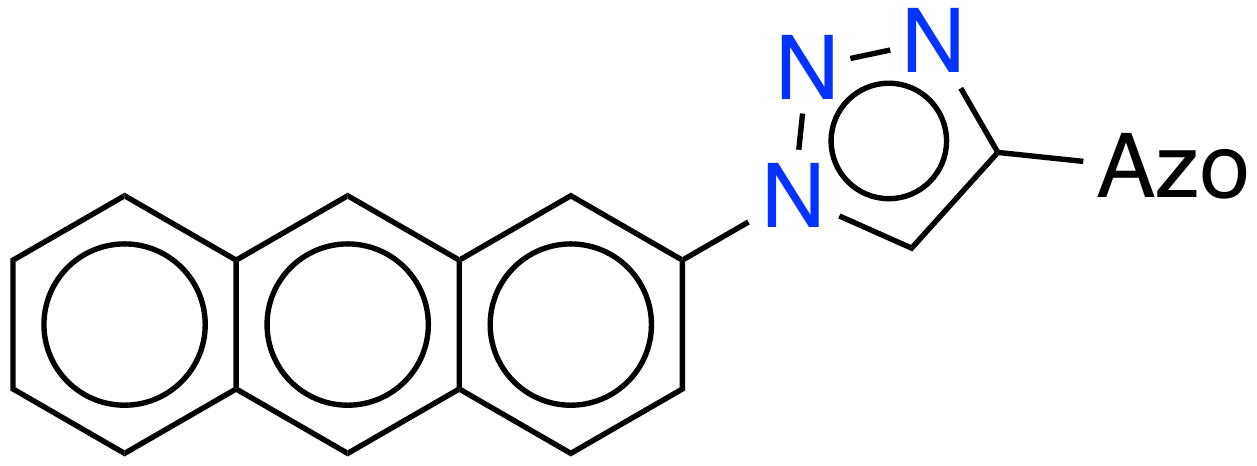}} \\ 

\hline 
[Azo]C & \raisebox{-.45\height}{\includegraphics[height=0.022\textwidth,trim=0 -5 0 -5]{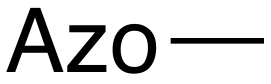}} \\ 

\hline 
[Azo]C(=O)C(F)(F)F & \raisebox{-.45\height}{\includegraphics[height=0.08\textwidth,trim=0 -5 0 -5]{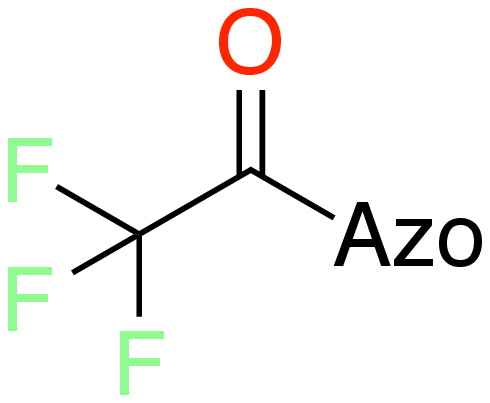}} \\ 

\hline 
[Azo]N1CCN(C)CC1 & \raisebox{-.45\height}{\includegraphics[height=0.06\textwidth,trim=0 -5 0 -5]{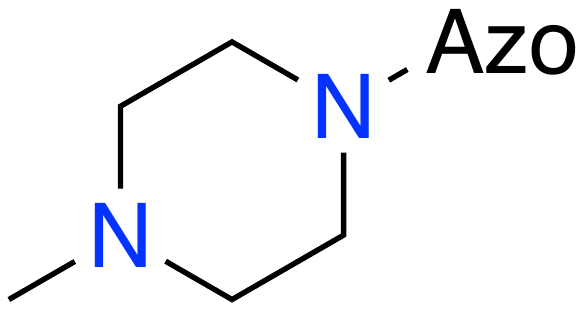}} \\ 

\hline 
[Azo]c1cn(-c2cccc3ccccc23)nn1 & \raisebox{-.45\height}{\includegraphics[height=0.13\textwidth,trim=0 -5 0 -5]{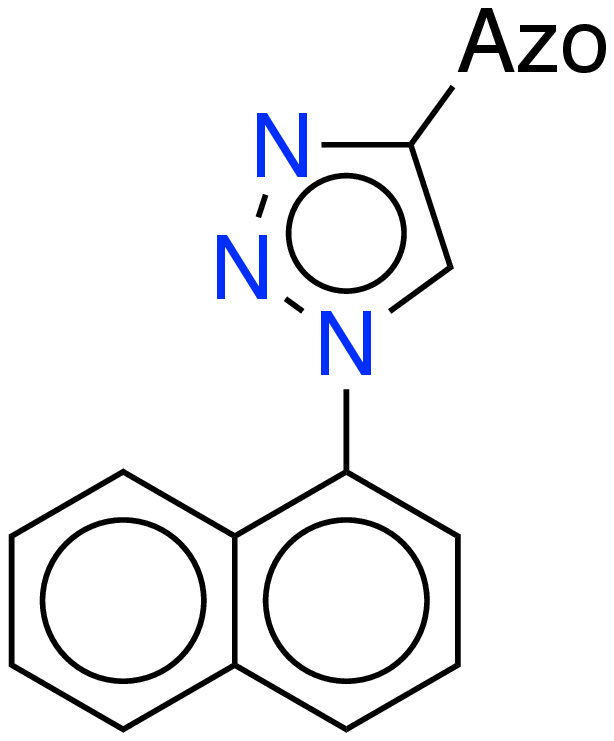}} \\ 

\hline 
[Azo]c1cn(-c2ccccc2)nn1 & \raisebox{-.45\height}{\includegraphics[height=0.085\textwidth,trim=0 -5 0 -5]{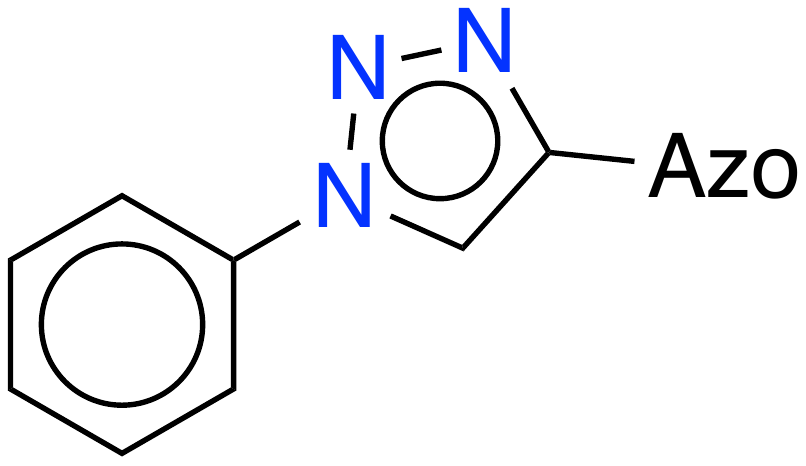}} \\ 

\hline 
[Azo]C(C)C & \raisebox{-.45\height}{\includegraphics[height=0.05\textwidth,trim=0 -5 0 -5]{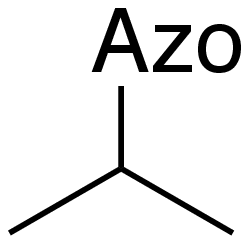}} \\ 

\hline 
[Azo]C(=O)OC & \raisebox{-.45\height}{\includegraphics[height=0.055\textwidth,trim=0 -5 0 -5]{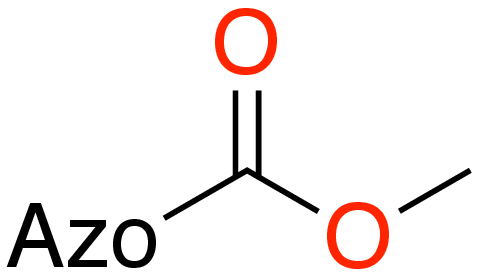}} \\ 

\hline 
[Azo]N1CCOCC1 & \raisebox{-.45\height}{\includegraphics[height=0.06\textwidth,trim=0 -5 0 -5]{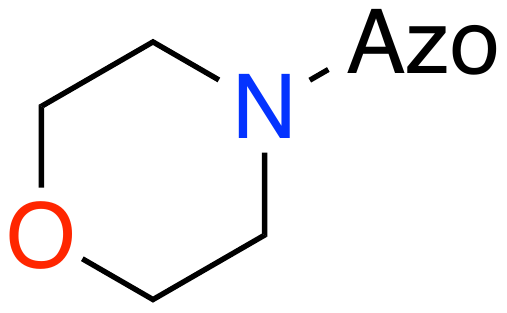}} \\ 

\hline 
[Azo]N(CC)c1ccccc1 & \raisebox{-.45\height}{\includegraphics[height=0.085\textwidth,trim=0 -5 0 -5]{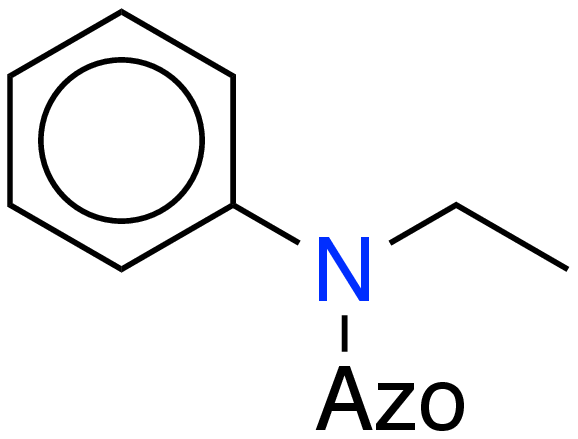}} \\ 

\hline 
[Azo]N(C)C & \raisebox{-.45\height}{\includegraphics[height=0.055\textwidth,trim=0 -5 0 -5]{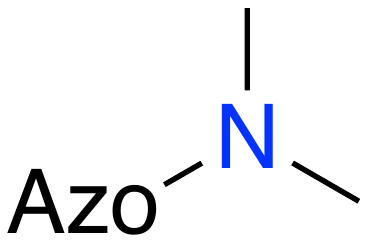}} \\ 

\hline 
[Azo]N1CCCC1 & \raisebox{-.45\height}{\includegraphics[height=0.05\textwidth,trim=0 -5 0 -5]{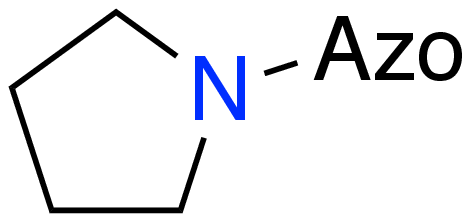}} \\ 

\hline 
[Azo]N1CC2CN(C(C)=O)CC2C1 & \raisebox{-.45\height}{\includegraphics[height=0.057\textwidth,trim=0 -5 0 -5]{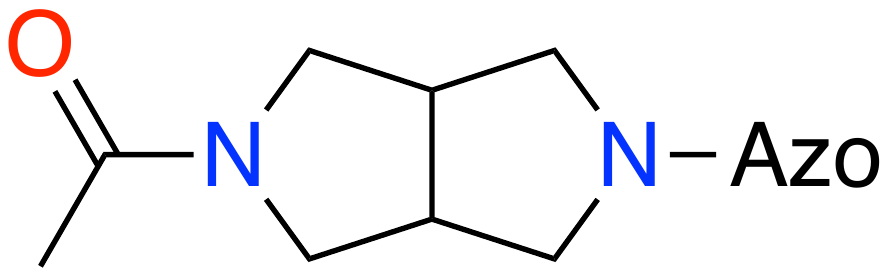}} \\ 

\hline 
[Azo]C(C)(C)C & \raisebox{-.45\height}{\includegraphics[height=0.057\textwidth,trim=0 -5 0 -5]{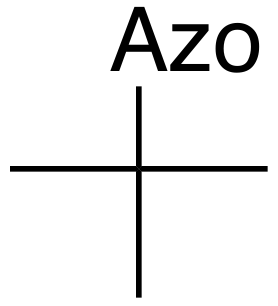}} \\ 

\hline 
[Azo]NC(=O)C=C & \raisebox{-.45\height}{\includegraphics[height=0.07\textwidth,trim=0 -5 0 -5]{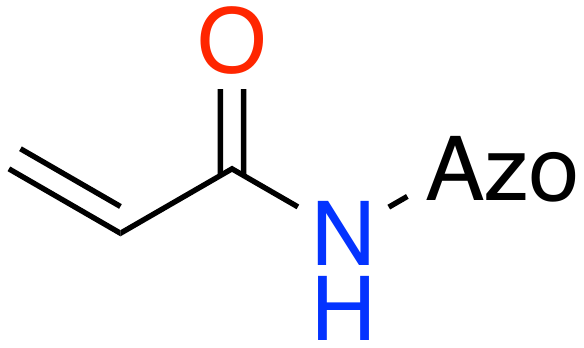}} \\ 

\hline 
[Azo]N(CCCCCCCC)CCCCCCCC & \raisebox{-.45\height}{\includegraphics[height=0.32\textwidth,trim=0 -5 0 -5]{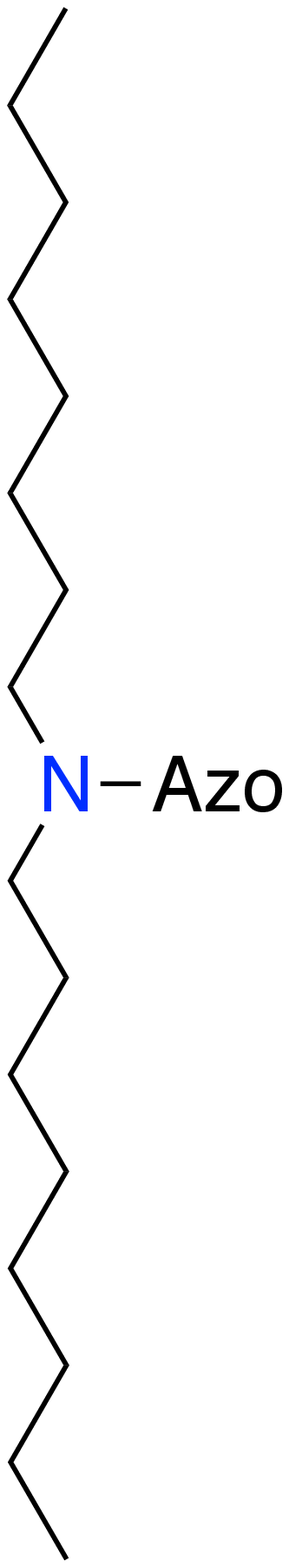}} \\ 

\hline 
[Azo]OCC & \raisebox{-.45\height}{\includegraphics[height=0.032\textwidth,trim=0 -5 0 -5]{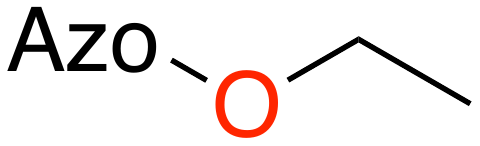}} \\ 

\hline 
[Azo]N(CC)CC & \raisebox{-.45\height}{\includegraphics[height=0.05\textwidth,trim=0 -5 0 -5]{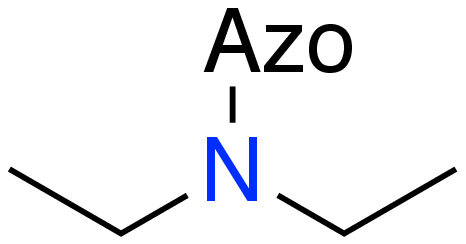}} \\ 

\hline 
[Azo]N(c1ccc2c(c1)C(C)(C)c1ccccc1-2)c1ccc2c(c1)C(C)(C)c1ccccc1-2 & \raisebox{-.45\height}{\includegraphics[height=0.14\textwidth,trim=0 -5 0 -5]{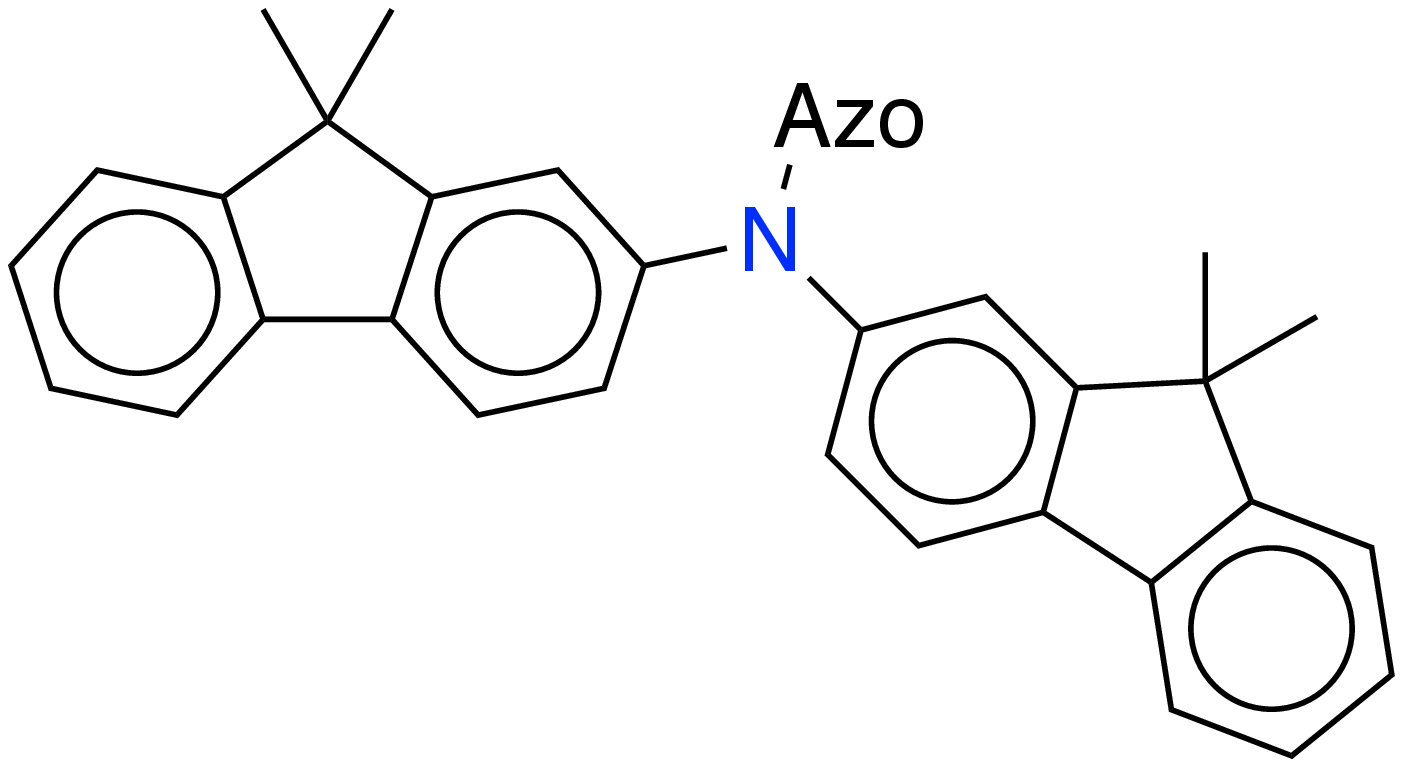}} \\ 

\hline 
[Azo]NC(=O)CCC[C@H](C[C@H] (N)C(=O)O)C(=O)O & \raisebox{-.45\height}{\includegraphics[height=0.09\textwidth,trim=0 -5 0 -5]{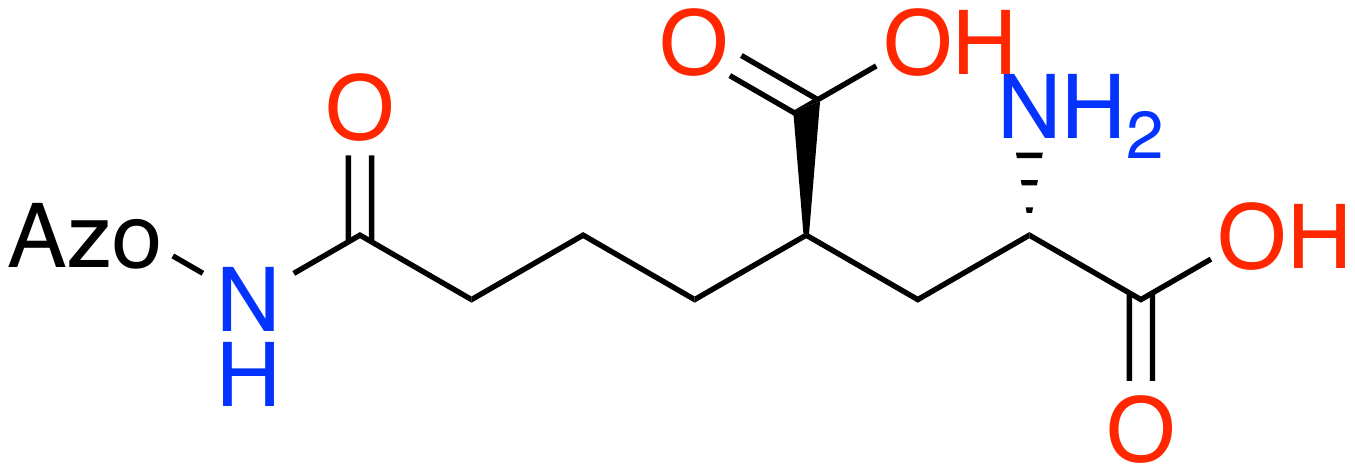}} \\

\hline\hline 

\end{xltabular}

\newpage

\begin{table}[h]
\caption{Test set species and literature quantum yields. For papers that reported the yield at different wavelengths, we chose the wavelength closest to the $n-\pi^*$ ($S_1$) absorption maximum. Quantum yields computed with FS surface hopping using the diabatic model are also shown.
} 
\label{sm_tab:test_lit_specs}
\end{table}

\renewcommand\tabularxcolumn[1]{m{#1}}
\begin{xltabular}{\textwidth}{|>{\centering\arraybackslash}m{.05\linewidth}|>{\centering\arraybackslash}m{.35\linewidth}|>{\centering\arraybackslash}m{.35\linewidth}|>{\centering\arraybackslash}m{.05\linewidth}|>{\centering\arraybackslash}m{.2\linewidth}|}
% \begin{xltabular}

\hline\endfoot

\toprule
\# & SMILES & Graph & Ref. & ($S_1$ yield, solvent) \\

\toprule
\hline

\textbf{1} & c1ccc(CNc2ccccc2/N=N/ c2ccccc2NCc2ccccn2)nc1 & \raisebox{-.45\height}{\includegraphics[height=0.14\textwidth,trim=0 -5 0 -5]{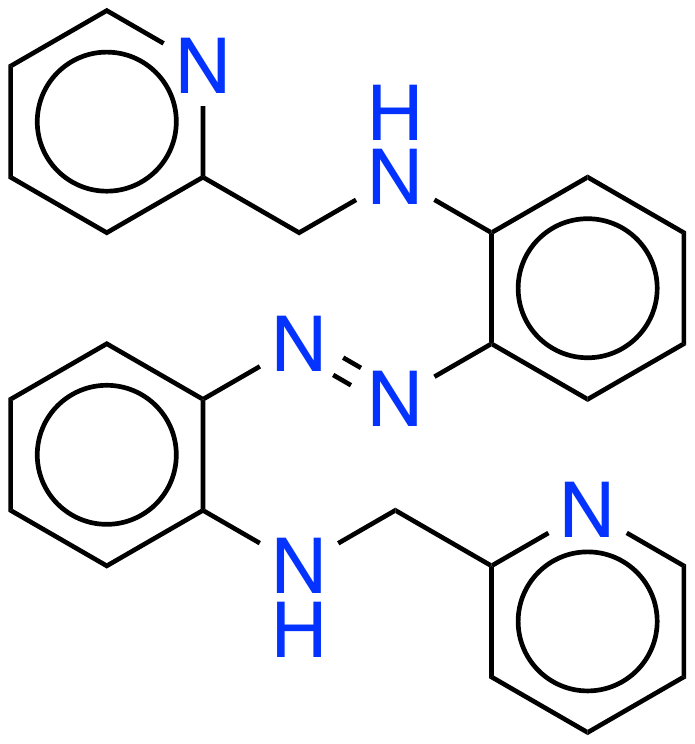}} & \cite{bandara2010proof} & \textbf{Exp}: (0\footnotemark, many) \newline \textbf{Calc}:  0 $\pm$ 0 \\

\hline
\textbf{2} & c1ccc(CNc2ccccc2/N=N{\textbackslash} c2ccccc2NCc2ccccn2)nc1 & \raisebox{-.45\height}{\includegraphics[height=0.21\textwidth,trim=0 -5 0 -5]{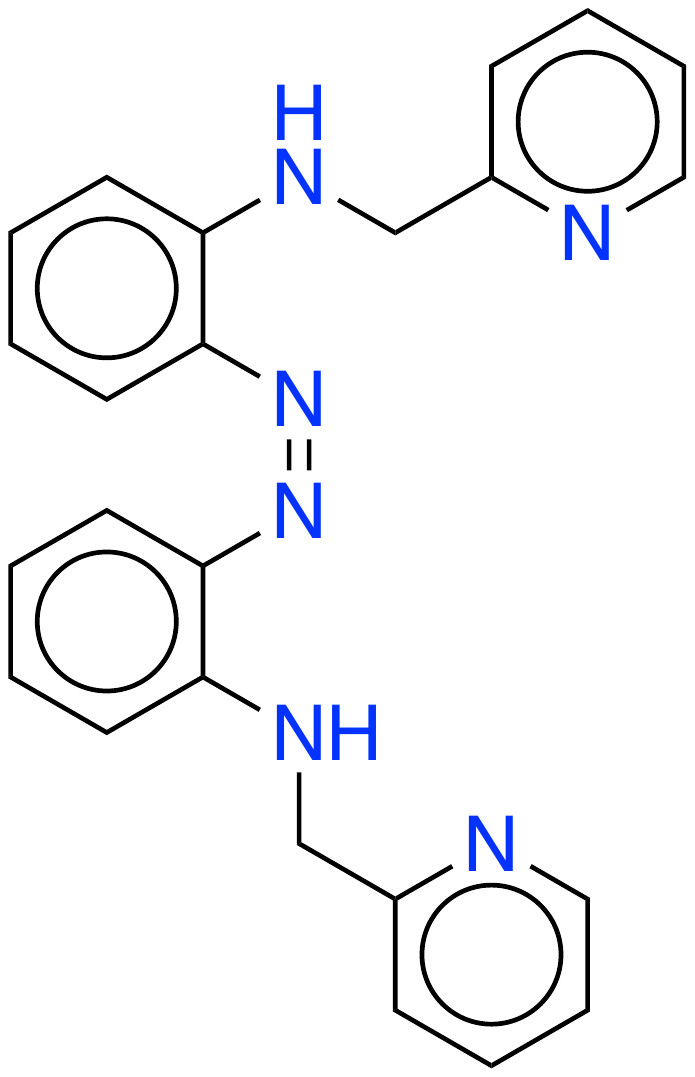}} & \cite{bandara2010proof} & \textbf{Exp}: None \newline \textbf{Calc}: 0.51 $\pm$ 0.02 \\

\hline
\textbf{3} & c1ccc(/N=N/c2cccc3ccccc23)cc1  & \raisebox{-.45\height}{\includegraphics[height=0.14\textwidth,trim=0 -5 0 -5]{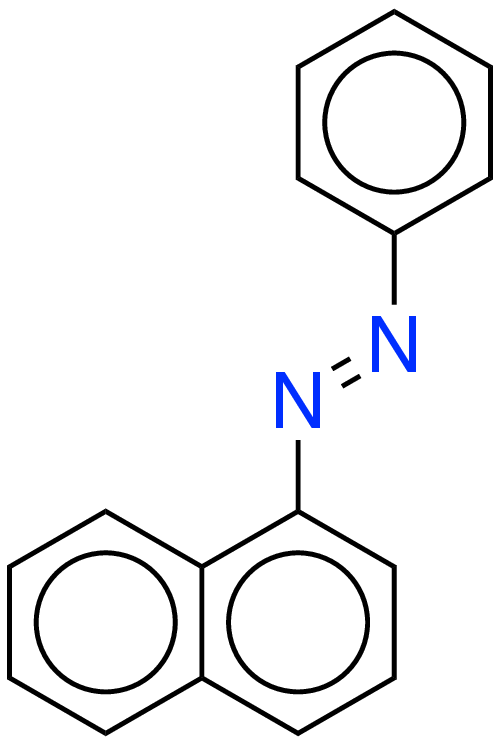}} & \cite{malkin1962temperature} & \textbf{Exp}: (0.25, methyl cyclopentane)\newline \textbf{Calc}: 0.43 $\pm$ 0.02 \\

\hline
\textbf{4} & c1ccc(/N=N{\textbackslash}c2cccc3ccccc23)cc1  & \raisebox{-.45\height}{\includegraphics[height=0.09\textwidth,trim=0 -5 0 -5]{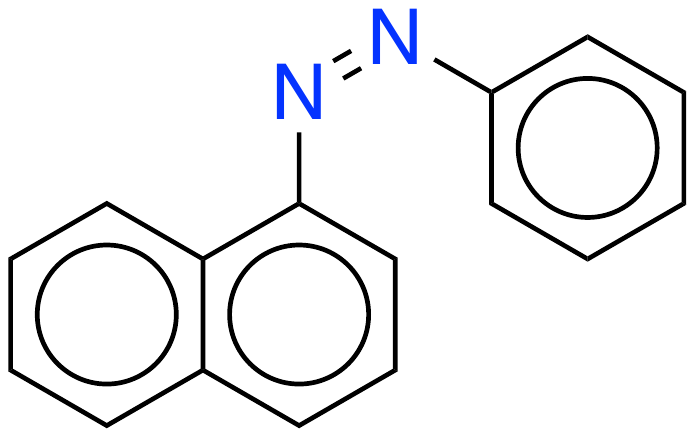}} & \cite{malkin1962temperature} & \textbf{Exp}: (0.49, methyl cyclopentane)\newline \textbf{Calc}: 0.57 $\pm$ 0.02 \\

\hline
\textbf{5} & CC1(C)c2ccccc2-c2ccc(N(c3ccc(/N=N/
c4ccccc4)cc3)c3ccc4c(c3)C(C)(C) c3ccccc3-4)cc21 & \raisebox{-.45\height}{\includegraphics[height=0.23\textwidth,trim=0 -5 0 -5]{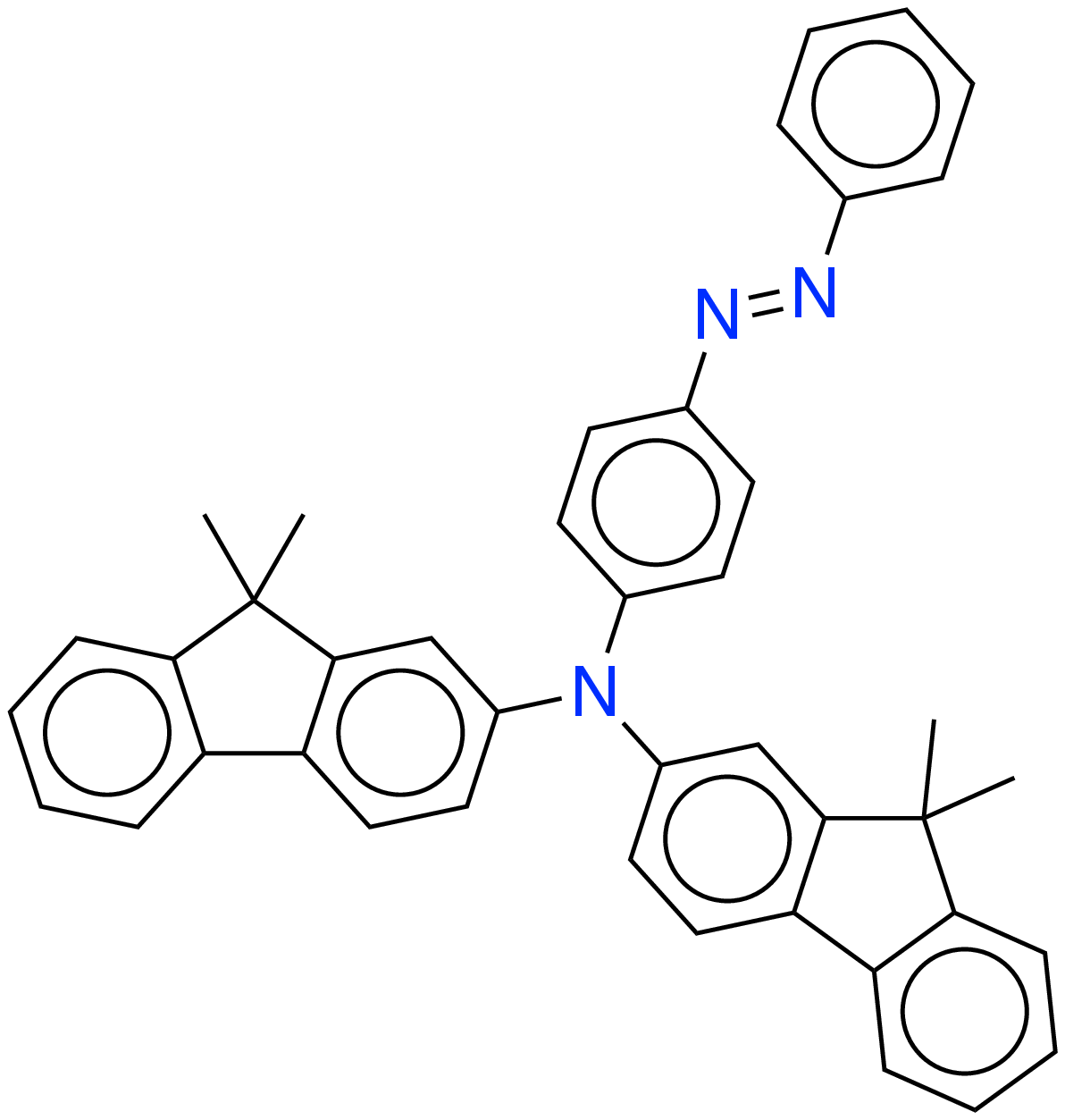}} & \cite{kurita2014photochromic} & \textbf{Exp}: (0.29, toluene)\newline \textbf{Calc}: 0.19 $\pm$ 0.05  \\

\hline
\textbf{6} & CC1(C)c2ccccc2-c2ccc(N(c3ccc(/N=N{\textbackslash} c4ccccc4)cc3)c3ccc4c(c3)C(C)(C) c3ccccc3-4)cc21 & \raisebox{-.45\height}{\includegraphics[height=0.18\textwidth,trim=0 -5 0 -5]{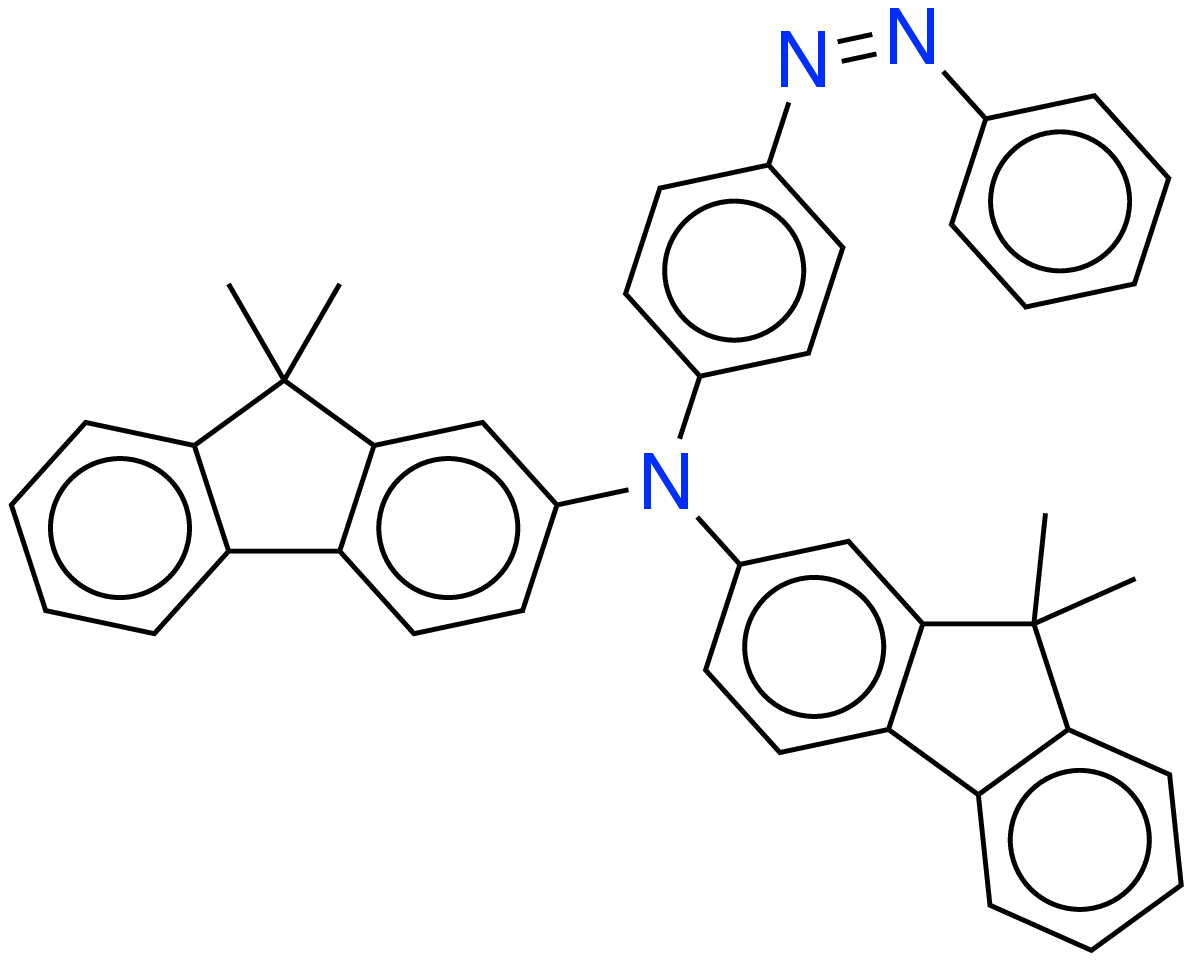}} & \cite{kurita2014photochromic} & \textbf{Exp}: (0.31, toluene)\newline \textbf{Calc}: 0.55 $\pm$ 0.02 \\
\hline

\hline

\textbf{7} & Cc1cc(C)c(/N=N/c2ccccc2)c(C)c1 & \raisebox{-.45\height}{\includegraphics[height=0.08\textwidth,trim=0 -5 0 -5]{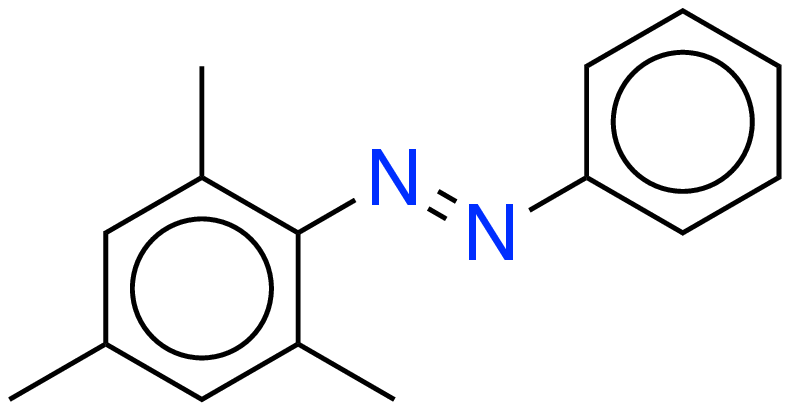}} & \cite{gegiou1968temperature} & \textbf{Exp}: (0.16, methylcyclohexane and isohexane)\newline \textbf{Calc}: 0.34 $\pm$ 0.03 \\
\hline

\textbf{8} & Cc1cc(C)c(/N=N{\textbackslash}c2ccccc2)c(C)c1 & \raisebox{-.45\height}{\includegraphics[height=0.11\textwidth,trim=0 -5 0 -5]{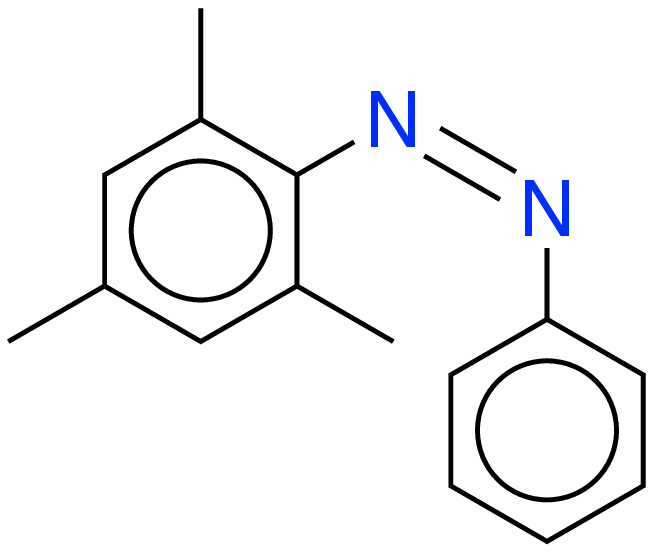}} & \cite{gegiou1968temperature} & \textbf{Exp}: (0.38, methylcyclohexane and isohexane)\newline \textbf{Calc}: 0.60 $\pm$ 0.02 \\
\hline

\textbf{9} & Cc1cc(C)c(/N=N/c2c(C) cc(C)cc2C)c(C)c1 & \raisebox{-.45\height}{\includegraphics[height=0.08\textwidth,trim=0 -5 0 -5]{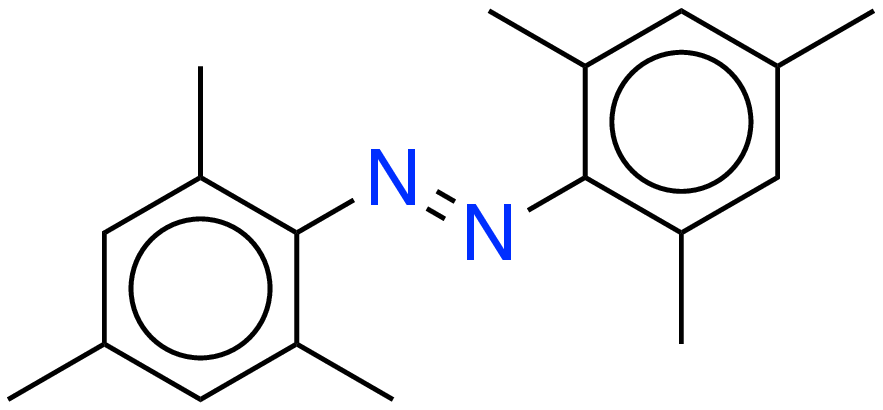}} & \cite{gegiou1968temperature,rau1988photoisomerization} & \textbf{Exp}: (0.16-0.22, methylcyclohexane and isohexane) \cite{gegiou1968temperature}, (0.24, n-hexane) \cite{rau1988photoisomerization}\newline \textbf{Calc}: 0.41 $\pm$ 0.02 \\
\hline

\textbf{10} & Cc1cc(C)c(/N=N{\textbackslash} c2c(C)cc(C)cc2C)c(C)c1 & \raisebox{-.45\height}{\includegraphics[height=0.12\textwidth,trim=0 -5 0 -5]{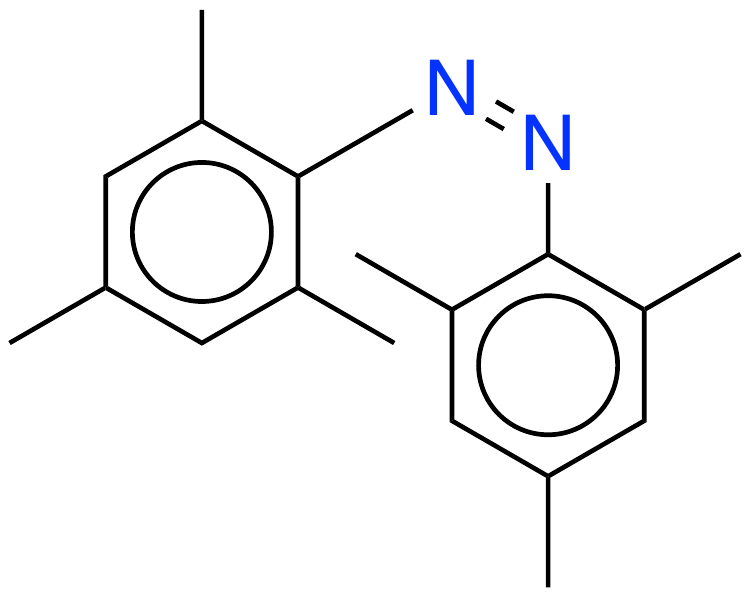}} & \cite{gegiou1968temperature,rau1988photoisomerization} & \textbf{Exp}: (0.44, methylcyclohexane and isohexane) \cite{gegiou1968temperature}, (0.5, n-hexane) \cite{rau1988photoisomerization}\newline \textbf{Calc}: 0.59 $\pm$ 0.02  \\
\hline

\textbf{11} & Cc1cc(/N=N/c2ccccc2N (C)Cc2ccccn2)c(N(C)Cc2ccccn2)cc1O & \raisebox{-.45\height}{\includegraphics[height=0.14\textwidth,trim=0 -5 0 -5]{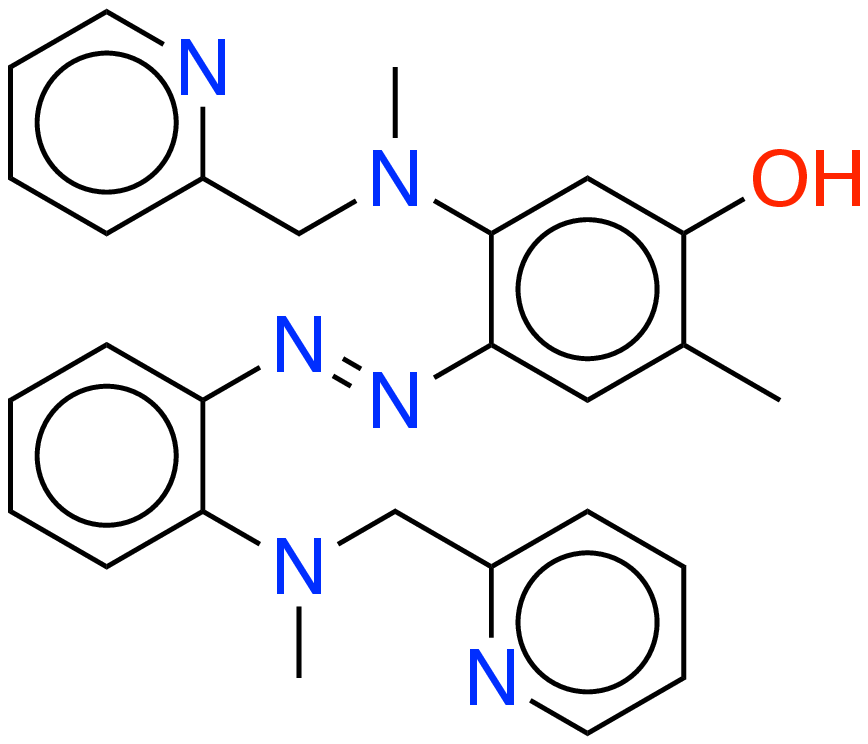}} & \cite{bandara2011short} & \textbf{Exp}: (0\footnotemark[\value{footnote}], many)\newline \textbf{Calc}: 0.33 $\pm$ 0.04 \\
\hline

\textbf{12} & Cc1cc(/N=N{\textbackslash}c2ccccc2N (C)Cc2ccccn2)c(N(C)Cc2ccccn2)cc1O & \raisebox{-.45\height}{\includegraphics[height=0.17\textwidth,trim=0 -5 0 -5]{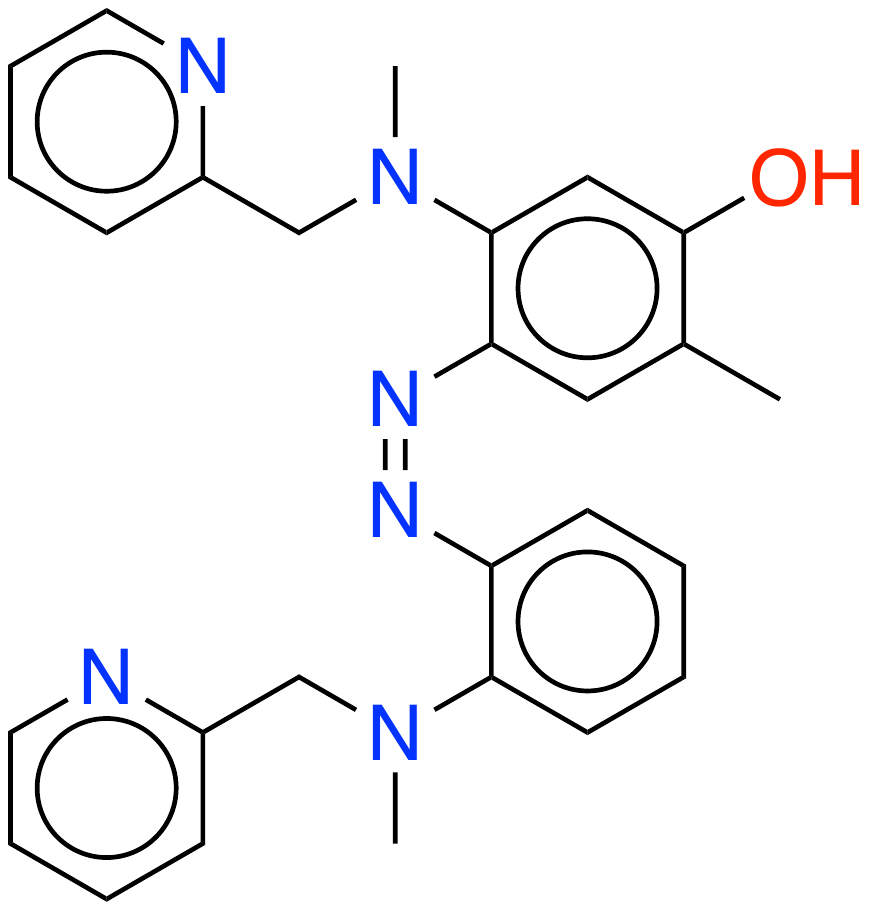}} & \cite{bandara2011short} & \textbf{Exp}: None\newline \textbf{Calc}: 0.53 $\pm$ 0.02 \\
\hline

\textbf{13} & CCc1cc(CC)c(/N=N/ c2c(CC)cc(CC)cc2CC)c(CC)c1 & \raisebox{-.45\height}{\includegraphics[height=0.08\textwidth,trim=0 -5 0 -5]{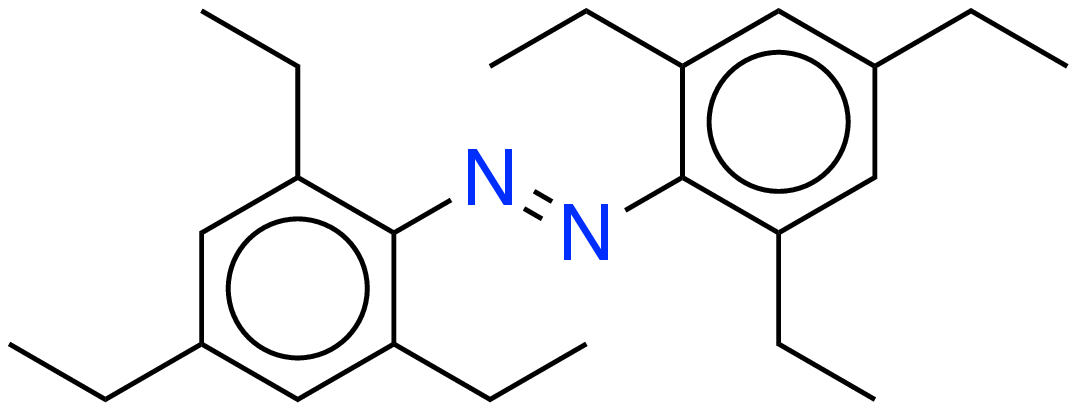}} & \cite{rau1988photoisomerization} & \textbf{Exp}: (0.25, n-hexane)\newline \textbf{Calc}: 0.32 $\pm$ 0.02 \\
\hline

\textbf{14} & CCc1cc(CC)c(/N=N{\textbackslash} c2c(CC)cc(CC)cc2CC)c(CC)c1 & \raisebox{-.45\height}{\includegraphics[height=0.15\textwidth,trim=0 -5 0 -5]{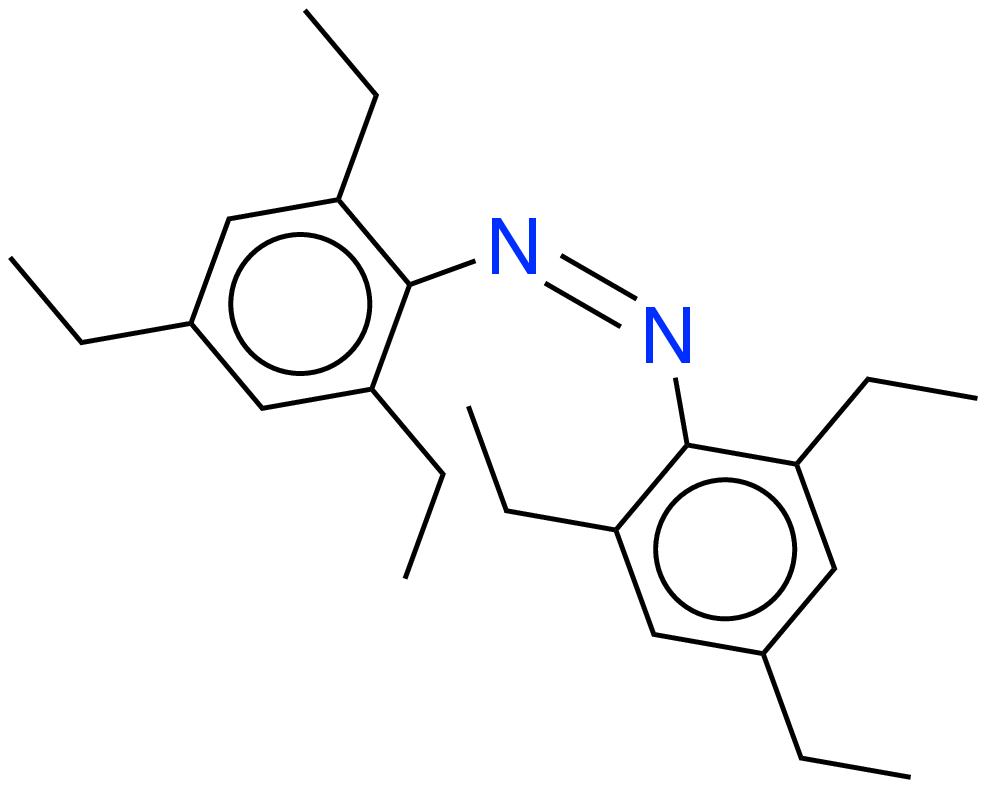}} & \cite{rau1988photoisomerization} & \textbf{Exp}: (0.5, n-hexane)\newline \textbf{Calc}: 0.49 $\pm$ 0.02 \\
\hline

\textbf{15} & CC(C)c1cc(C(C)C)c(/N=N/c2c(C (C)C)cc(C(C)C)cc2C(C)C)c(C(C)C)c1 & \raisebox{-.45\height}{\includegraphics[height=0.12\textwidth,trim=0 -5 0 -5]{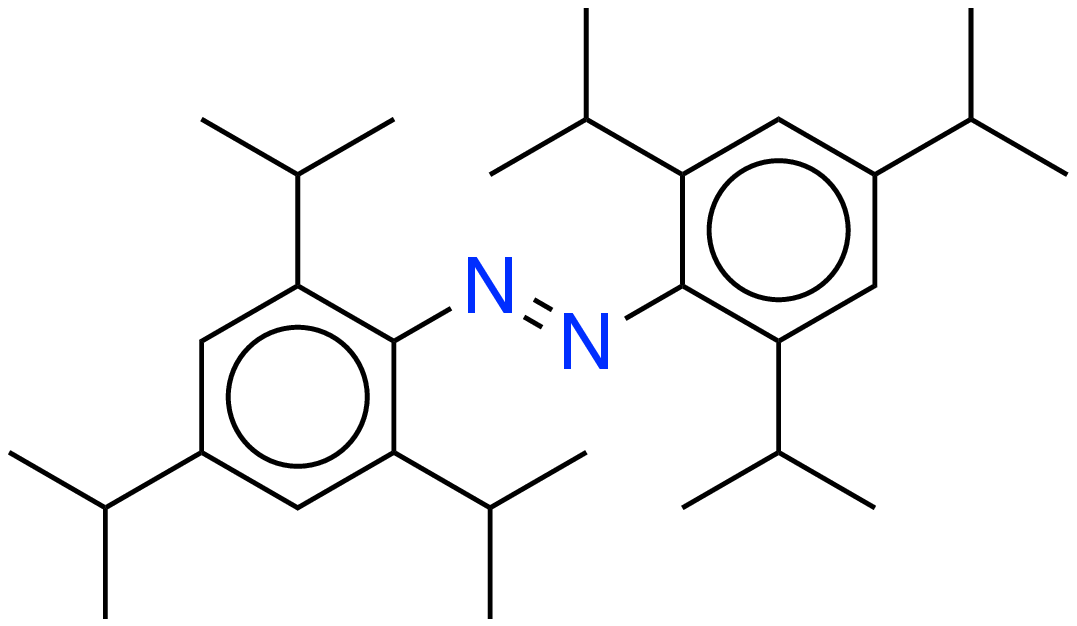}} & \cite{rau1988photoisomerization} & \textbf{Exp}: (0.19, n-hexane)\newline \textbf{Calc}: 0.27 $\pm$ 0.02 \\
\hline

\textbf{16} & CC(C)c1cc(C(C)C)c(/N=N{\textbackslash}c2c(C (C)C)cc(C(C)C)cc2C(C)C)c(C(C)C)c1 & \raisebox{-.45\height}{\includegraphics[height=0.17\textwidth,trim=0 -5 0 -5]{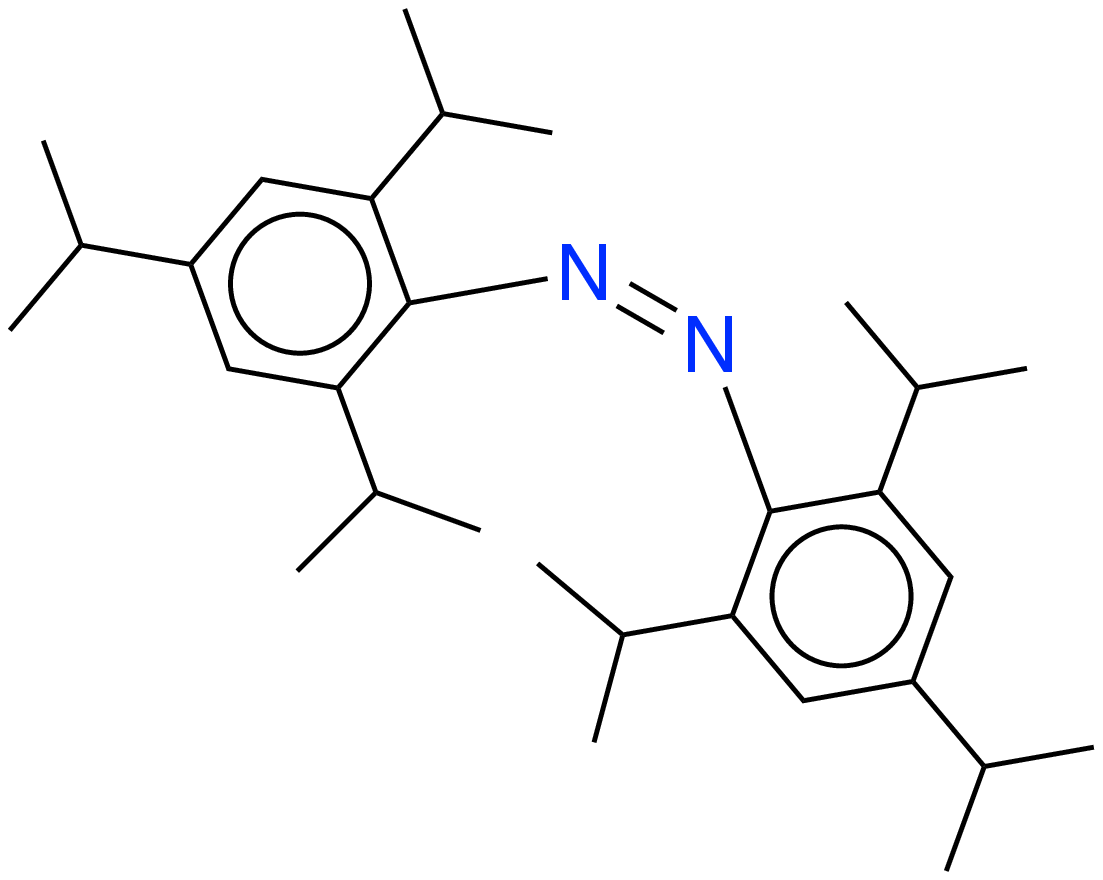}} & \cite{rau1988photoisomerization} & \textbf{Exp}: (0.55, n-hexane)\newline \textbf{Calc}: 0.53 $\pm$ 0.02 \\
\hline

\textbf{17} & CC(C)(C)c1cc(C(C)(C)C)c(/N=N/ c2c(C(C)(C)C)cc(C(C)(C)C)cc2C (C)(C)C)c(C(C)(C)C)c1 & \raisebox{-.45\height}{\includegraphics[height=0.12\textwidth,trim=0 -5 0 -5]{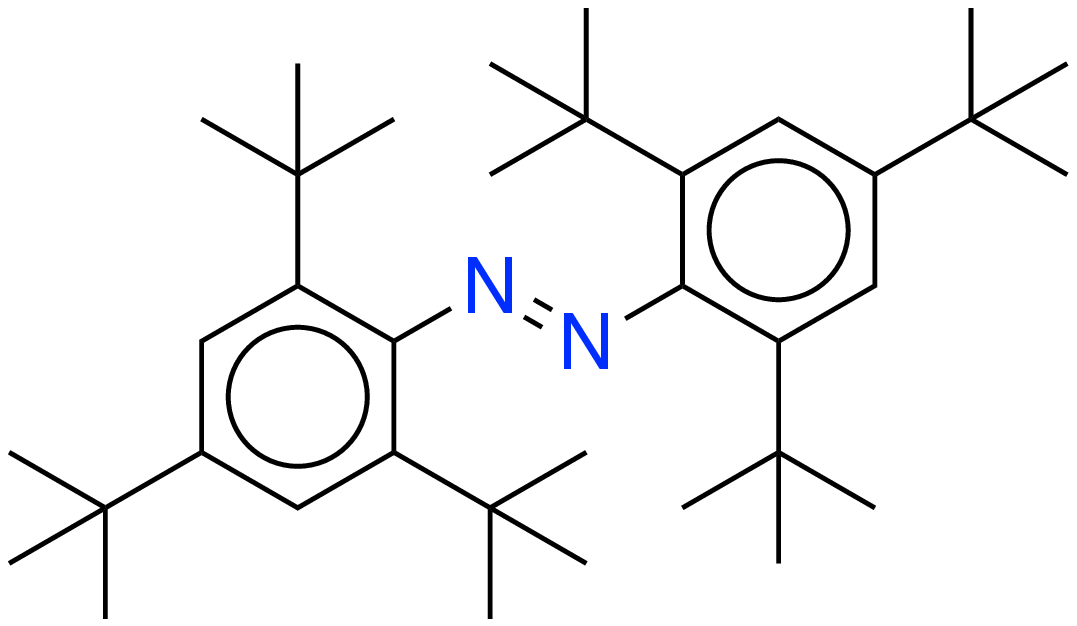}} & \cite{rau1988photoisomerization} & \textbf{Exp}: (0, n-hexane)\newline \textbf{Calc}: 0.01 $\pm$ 0.00 \\
\hline

\textbf{18} & CC(C)(C)c1cc(C(C)(C)C)c(/N=N{\textbackslash} c2c(C(C)(C)C)cc(C(C)(C)C)cc2C (C)(C)C)c(C(C)(C)C)c1 & \raisebox{-.45\height}{\includegraphics[height=0.19\textwidth,trim=0 -5 0 -5]{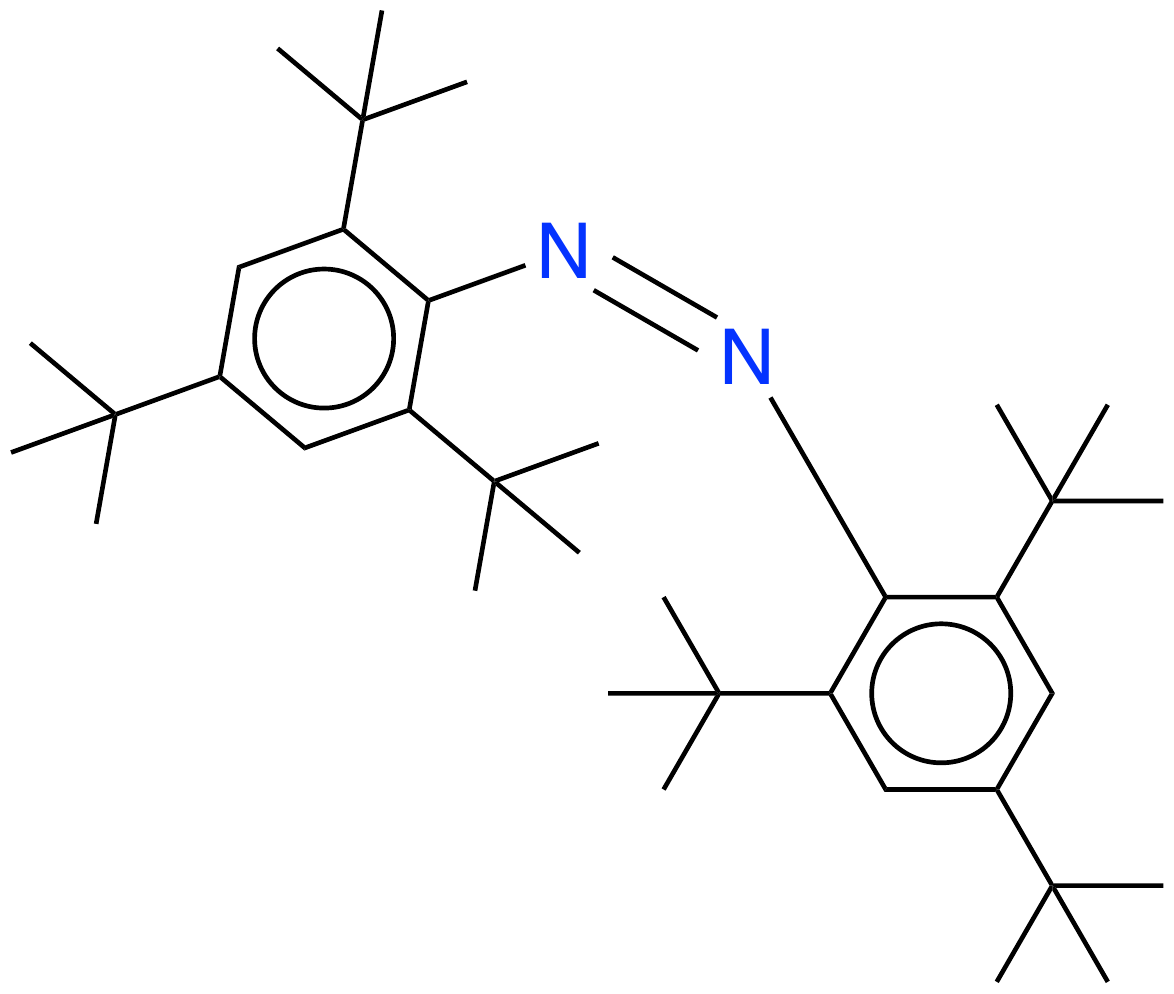}} & \cite{rau1988photoisomerization} & \textbf{Exp}: None\newline \textbf{Calc}: 0.78 $\pm$ 0.02 \\
\hline

\textbf{19} & CCCCCCCCN(CCCCCCCC)c1ccc (/N=N/c2ccc(C(=O)C(F)(F)F)cc2)cc1 & \raisebox{-.45\height}{\includegraphics[height=0.23\textwidth,trim=0 -5 0 -5]{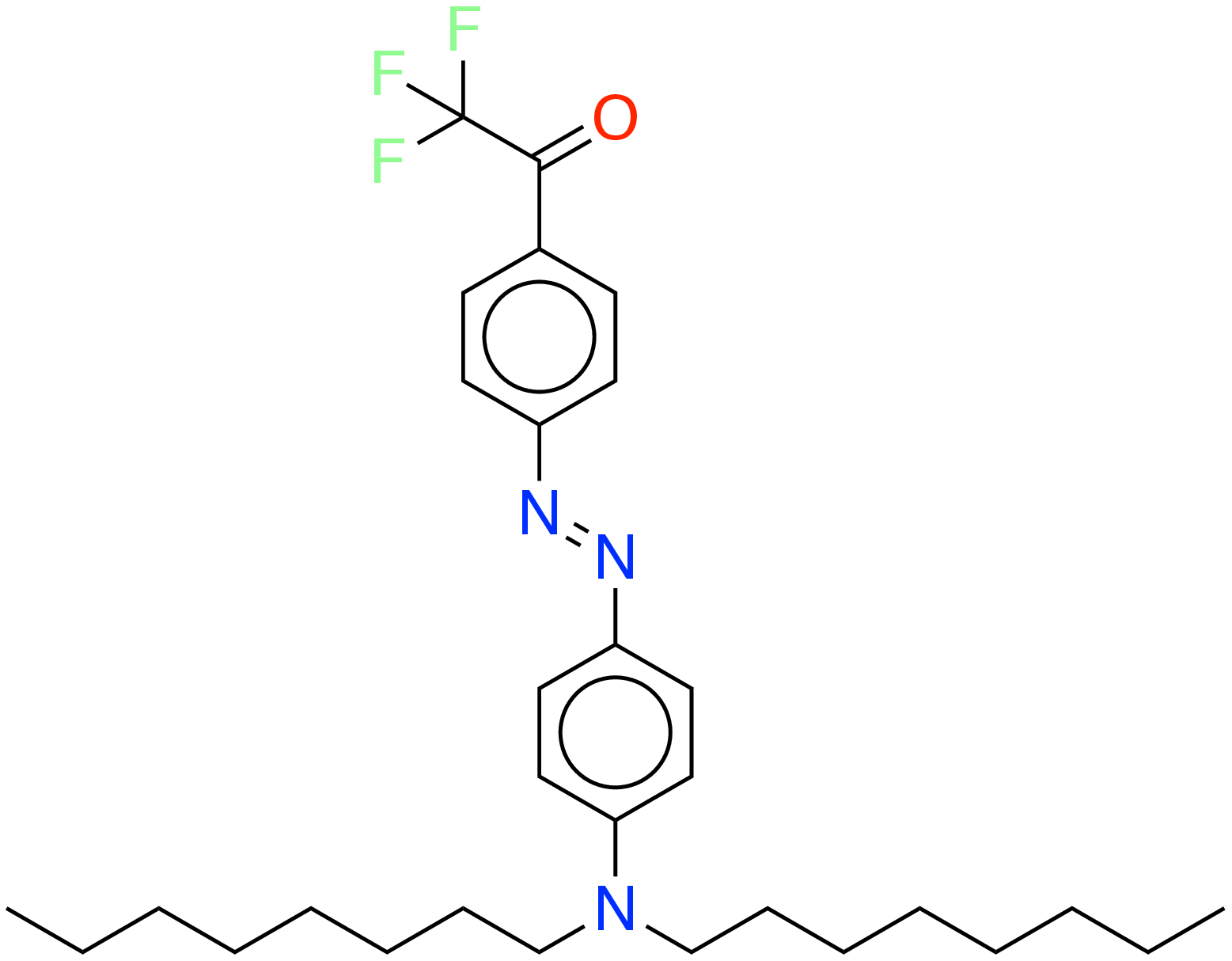}} & \cite{mohr2004photochemistry} & \textbf{Exp}: (0.16\footnotemark[\the\numexpr\value{footnote}+1\relax],  toluene)\newline \textbf{Calc}: 0.08 $\pm$ 0.04 \\ 
\hline

\textbf{20} & CCCCCCCCN(CCCCCCCC)c1ccc (/N=N{\textbackslash}c2ccc(C(=O)C(F)(F)F)cc2)cc1 & \raisebox{-.45\height}{\includegraphics[height=0.12\textwidth,trim=0 -5 0 -5]{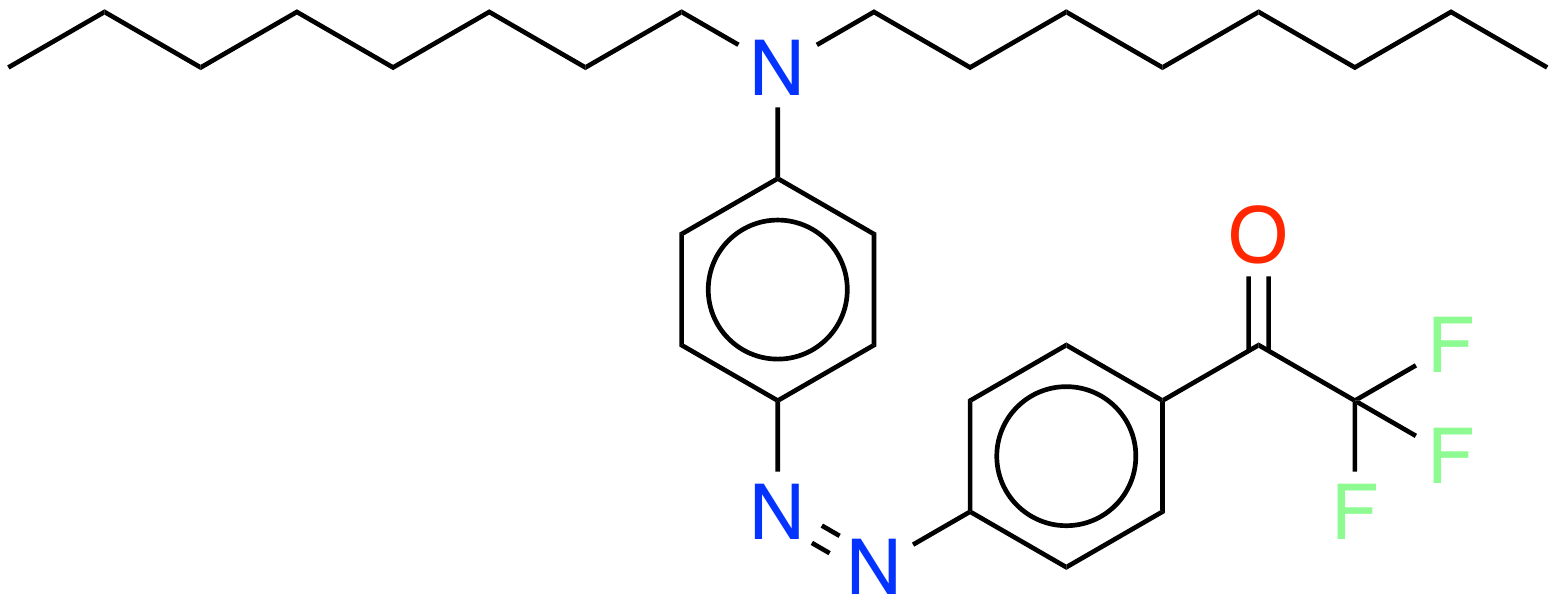}} & \cite{mohr2004photochemistry} & \textbf{Exp}: (0.7-1\footnotemark[\the\numexpr\value{footnote}+1\relax], toluene)\newline \textbf{Calc}: 0.66 $\pm$ 0.02  \\ 
\hline

\textbf{21} & CCOC(=O)c1cc(F)c(/N=N/c2c (F)cc(C(=O)OCC)cc2F)c(F)c1 & \raisebox{-.45\height}{\includegraphics[height=0.13\textwidth,trim=0 -5 0 -5]{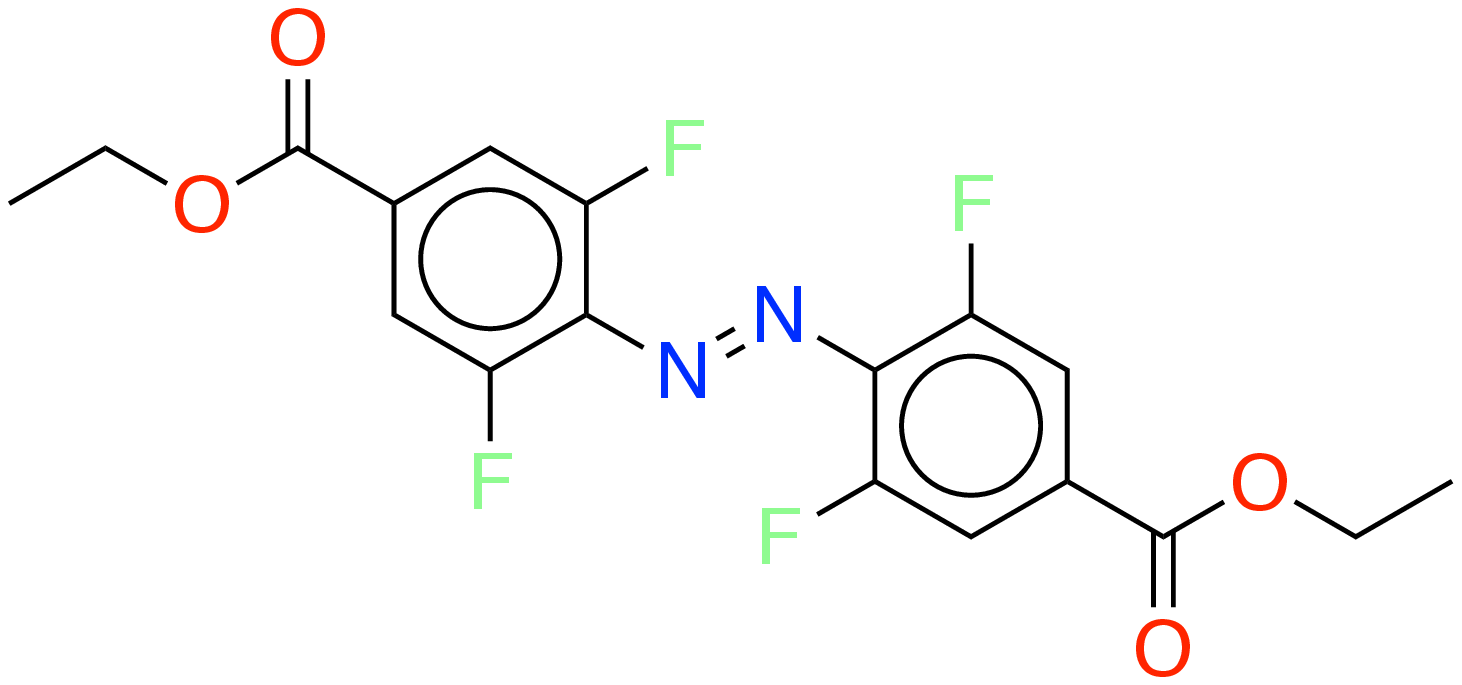}} & \cite{knie2014ortho} & \textbf{Exp}: (0.1, hexane)\newline \textbf{Calc}: 0.08 $\pm$ 0.02 \\ 
\hline

\textbf{22} & CCOC(=O)c1cc(F)c(/N=N{\textbackslash}c2c (F)cc(C(=O)OCC)cc2F)c(F)c1 & \raisebox{-.45\height}{\includegraphics[height=0.18\textwidth,trim=0 -5 0 -5]{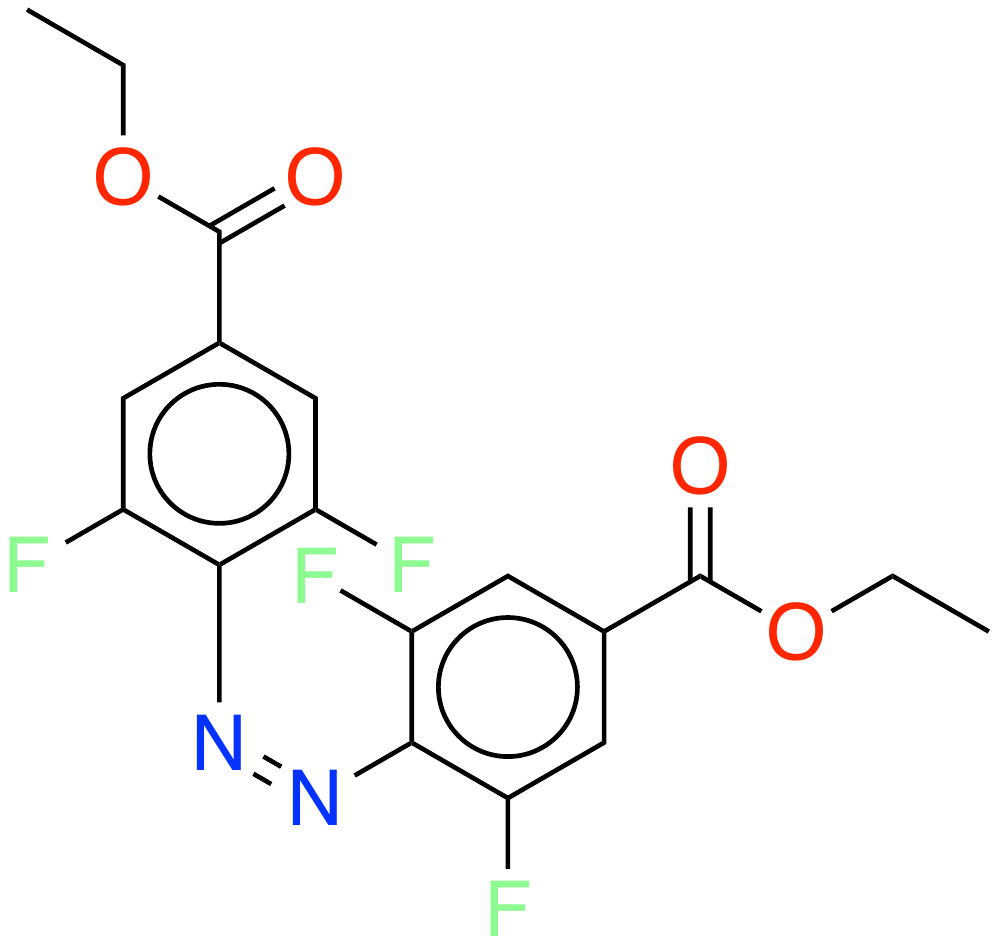}} & \cite{knie2014ortho} & \textbf{Exp}: (0.53, hexane)\newline \textbf{Calc}: 0.44 $\pm$ 0.02 \\ 
\hline

\textbf{23} & CC(=O)Oc1cc(N(C)Cc2ccccn2)c
(/N=N/c2ccccc2N(C)Cc2ccccn2)cc1C & \raisebox{-.45\height}{\includegraphics[height=0.14\textwidth,trim=0 -5 0 -5]{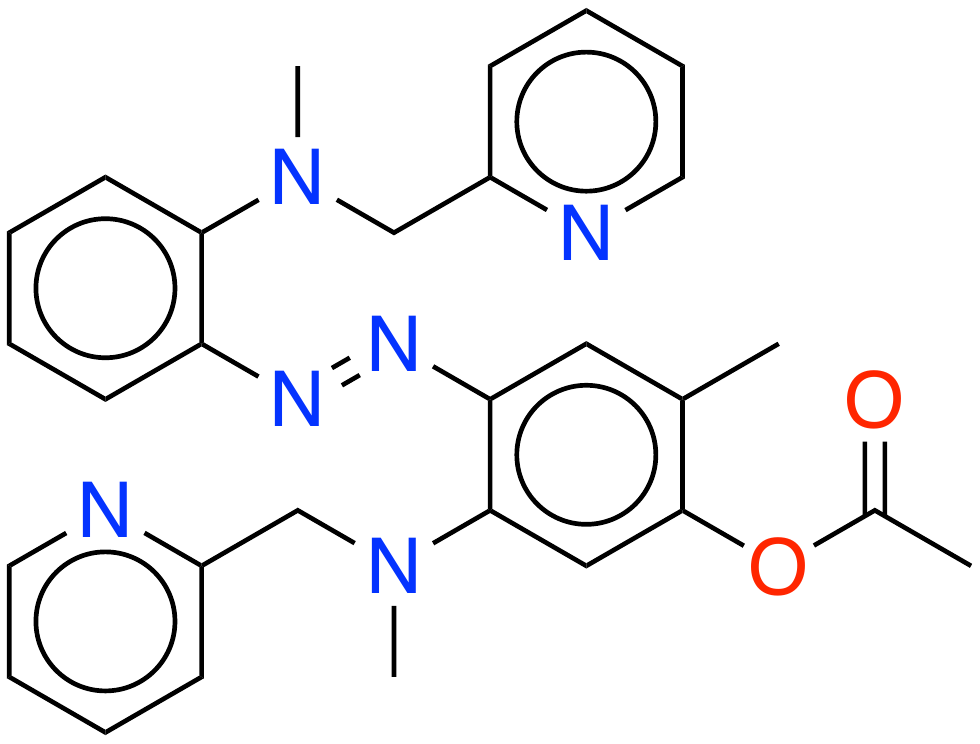}} & \cite{bandara2011short} & \textbf{Exp}: (0.17, C6D6)\newline \textbf{Calc}: 0.30 $\pm$ 0.04 \\ 
\hline

\textbf{24} & CC(=O)Oc1cc(N(C)Cc2ccccn2)c
(/N=N{\textbackslash}c2ccccc2N(C)Cc2ccccn2)cc1C & \raisebox{-.45\height}{\includegraphics[height=0.17\textwidth,trim=0 -5 0 -5]{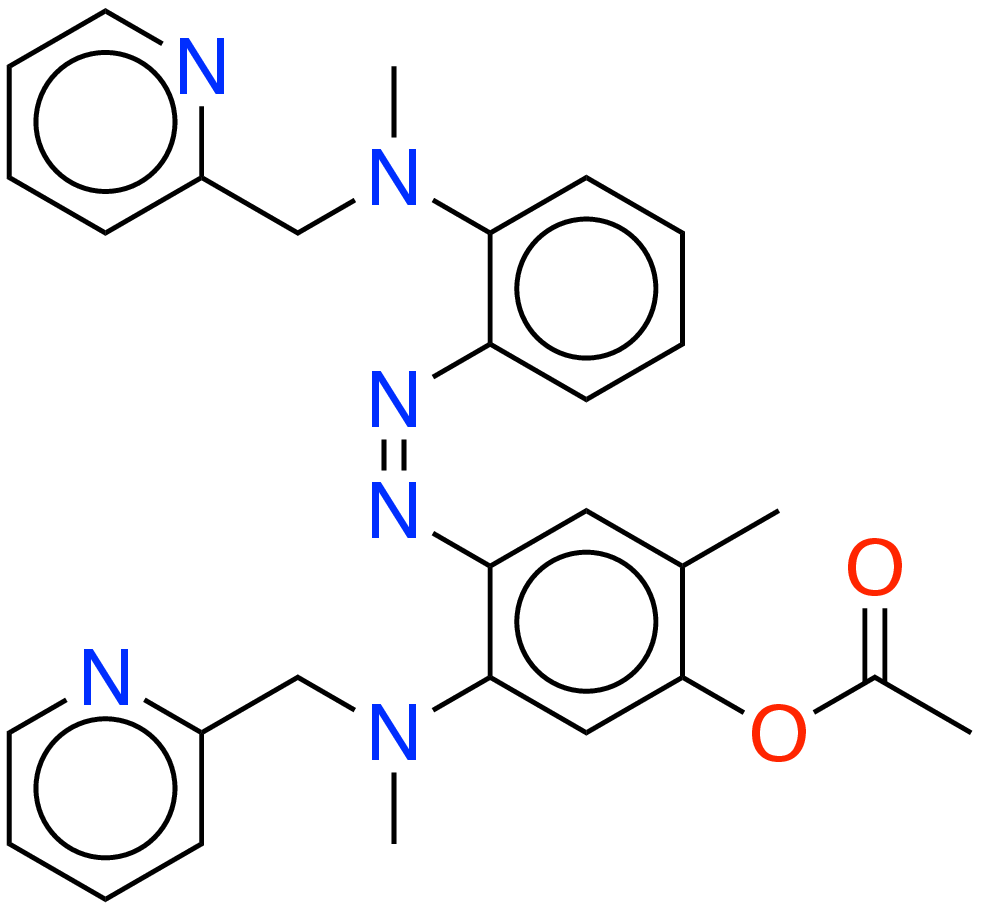}} & \cite{bandara2011short} & \textbf{Exp}: None\newline \textbf{Calc}: 0.51 $\pm$ 0.02 \\ 
\hline

\textbf{25} & CN(Cc1ccccn1)c1ccccc1 /N=N/c1ccccc1N(C)Cc1ccccn1 & \raisebox{-.45\height}{\includegraphics[height=0.14\textwidth,trim=0 -5 0 -5]{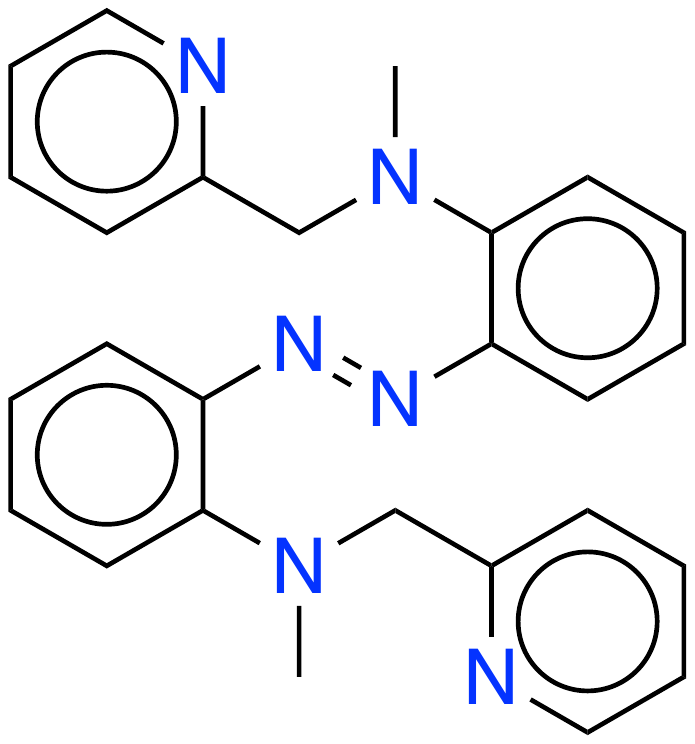}} & \cite{bandara2010proof} & \textbf{Exp}: (0.19, CDCl3)\newline \textbf{Calc}: 0.37 $\pm$ 0.03 \\ 
\hline

\textbf{26} & CN(Cc1ccccn1)c1ccccc1 /N=N{\textbackslash}c1ccccc1N(C)Cc1ccccn1 & \raisebox{-.45\height}{\includegraphics[height=0.18\textwidth,trim=0 -5 0 -5]{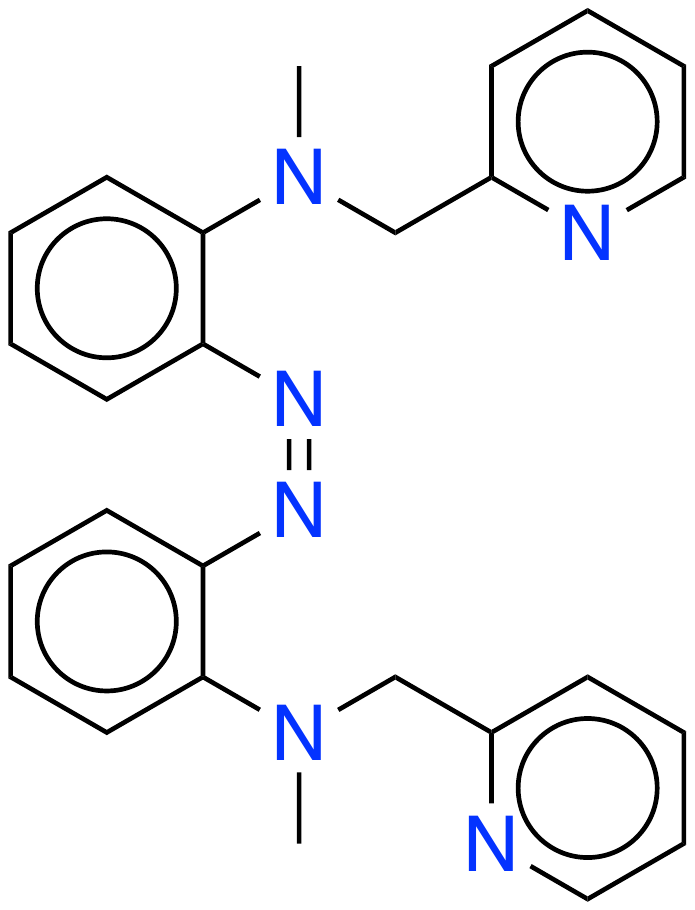}} & \cite{bandara2010proof} & \textbf{Exp}: None\newline \textbf{Calc}: 0.48 $\pm$ 0.02 \\ 
\hline

\textbf{27} & COc1ccc(/N=N/c2ccc ([N+](=O)[O-])cc2)cc1 & \raisebox{-.45\height}{\includegraphics[height=0.11\textwidth,trim=0 -5 0 -5]{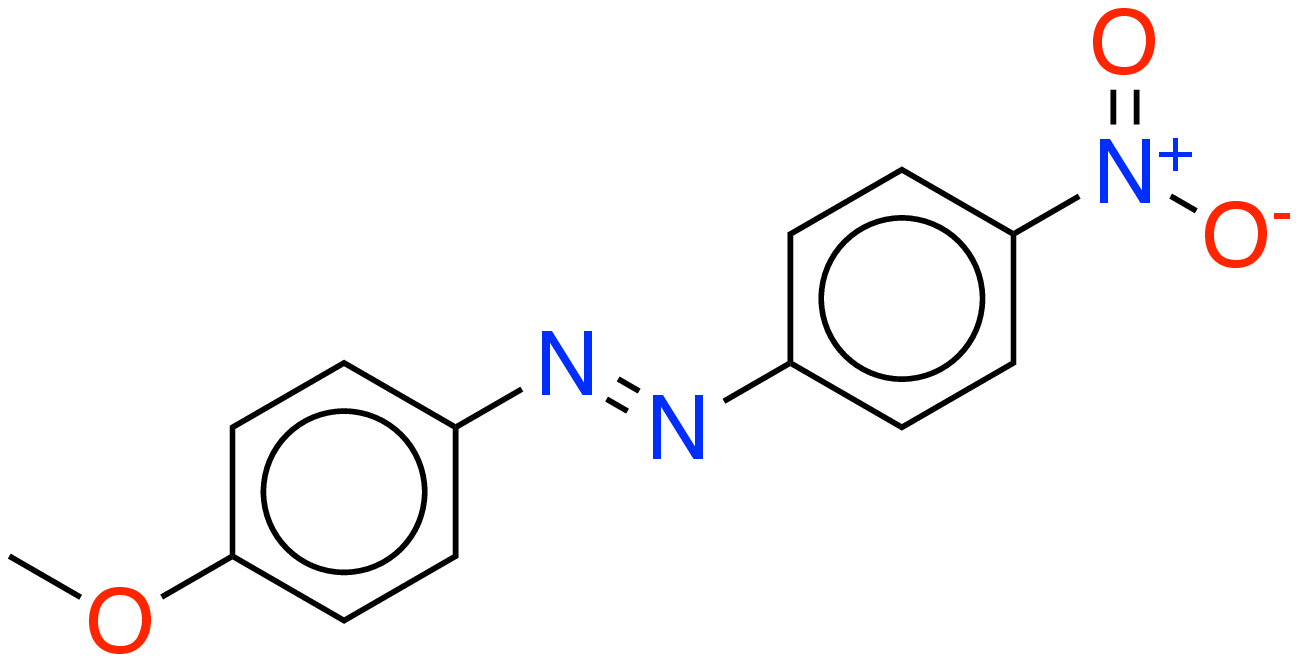}} & \cite{gegiou1968temperature} & \textbf{Exp}: (0.17, methylcyclohexane and isohexane)\newline \textbf{Calc}: 0.05 $\pm$ 0.02 \\ 

\hline

\textbf{28} & COc1ccc(/N=N{\textbackslash}c2ccc ([N+](=O)[O-])cc2)cc1 & \raisebox{-.45\height}{\includegraphics[height=0.115\textwidth,trim=0 -5 0 -5]{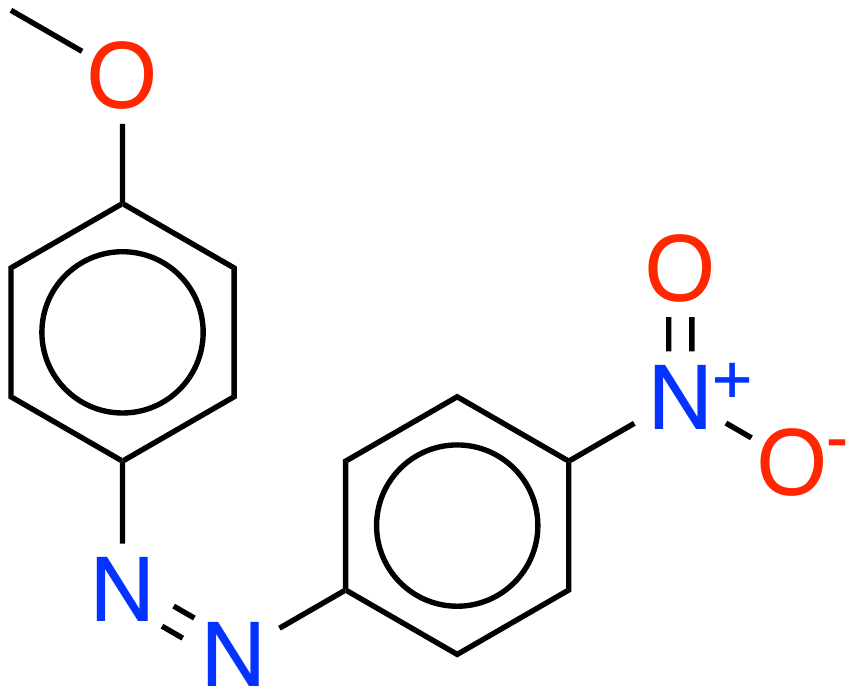}} & \cite{gegiou1968temperature} & \textbf{Exp}: (0.55, methylcyclohexane and isohexane)\newline \textbf{Calc}: 0.46 $\pm$ 0.02 \\ 
\hline

\textbf{29} & COc1cc(N(C)Cc2ccccn2) c(/N=N/c2ccccc2N(C)Cc2ccccn2)cc1C & \raisebox{-.45\height}{\includegraphics[height=0.15\textwidth,trim=0 -5 0 -5]{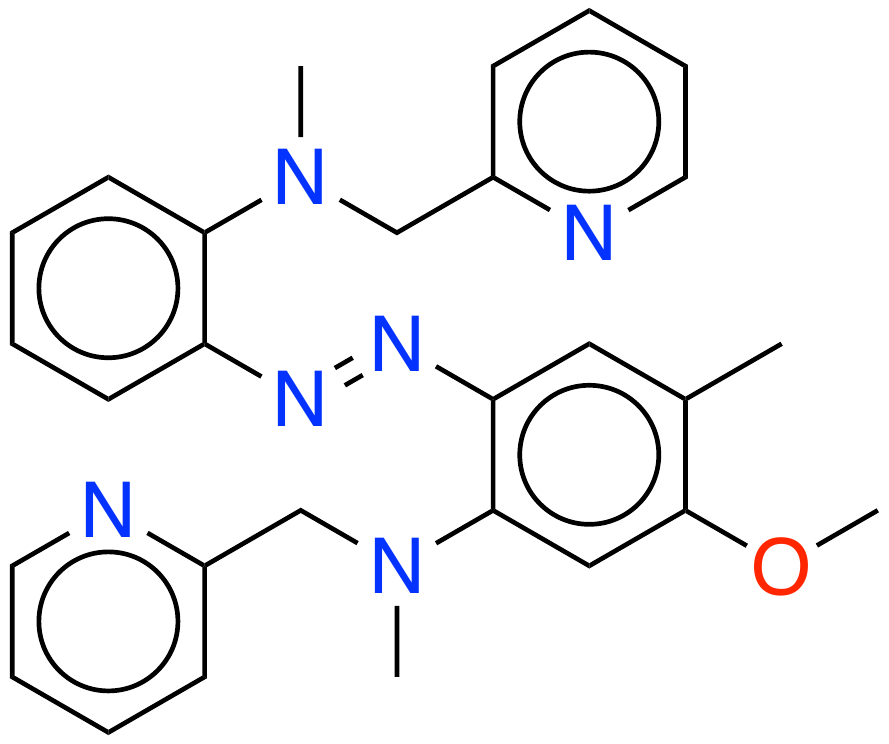}} & \cite{bandara2011short} & \textbf{Exp}: (0\footnotemark[\value{footnote}], many)\newline \textbf{Calc}: 0.35 $\pm$ 0.03 \\ 
\hline

\textbf{30} & COc1cc(N(C)Cc2ccccn2) c(/N=N{\textbackslash}c2ccccc2N(C)Cc2ccccn2)cc1C & \raisebox{-.45\height}{\includegraphics[height=0.18\textwidth,trim=0 -5 0 -5]{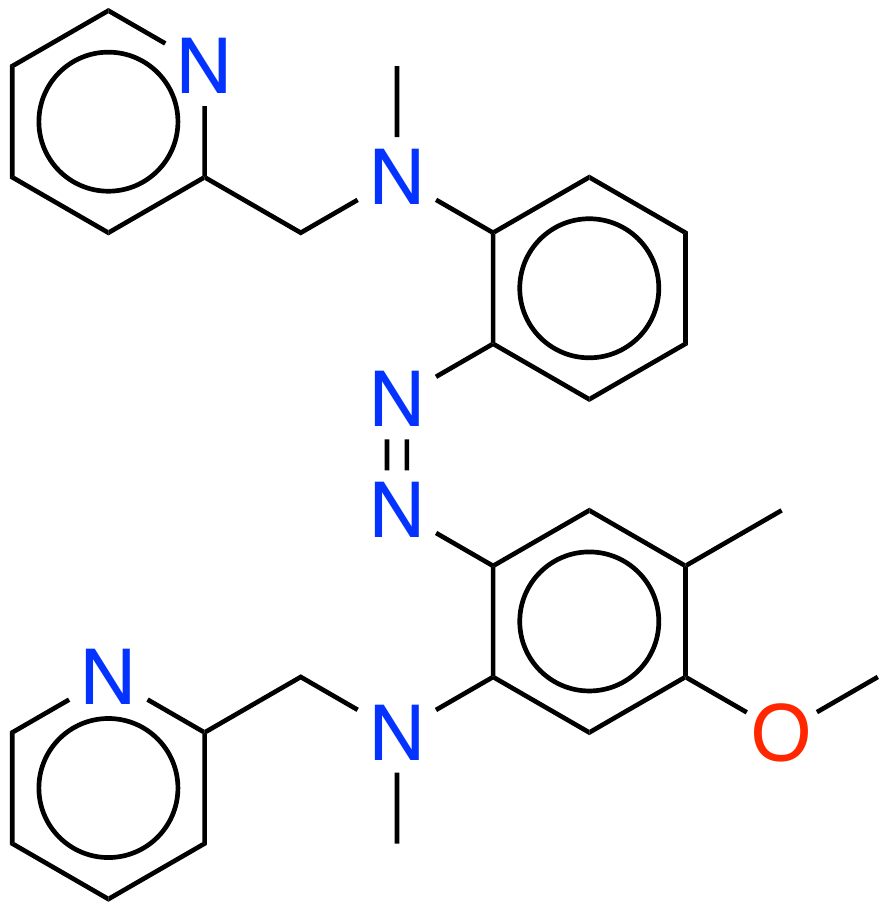}} & \cite{bandara2011short} & \textbf{Exp}: None\newline \textbf{Calc}: 0.54 $\pm$ 0.02 \\ 
\hline

\textbf{31} & Fc1cccc(F)c1/N=N/c1ccccc1 & \raisebox{-.45\height}{\includegraphics[height=0.09\textwidth,trim=0 -5 0 -5]{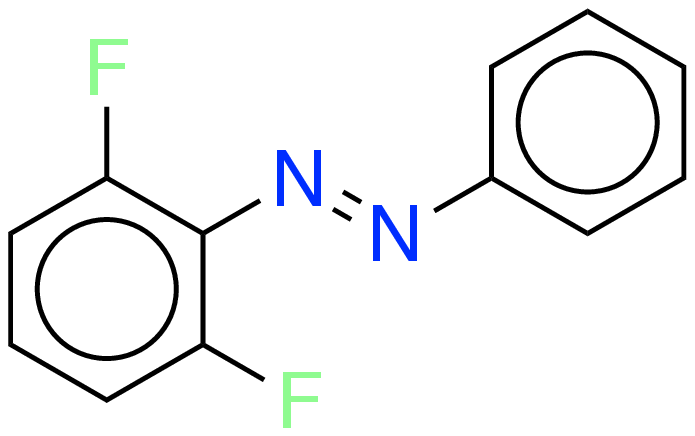}} & \cite{knie2014ortho} & \textbf{Exp}: (0.32, hexane)\newline \textbf{Calc}: 0.36 $\pm$ 0.02 \\ 
\hline

\textbf{32} & Fc1cccc(F)c1/N=N{\textbackslash}c1ccccc1 & \raisebox{-.45\height}{\includegraphics[height=0.12\textwidth,trim=0 -5 0 -5]{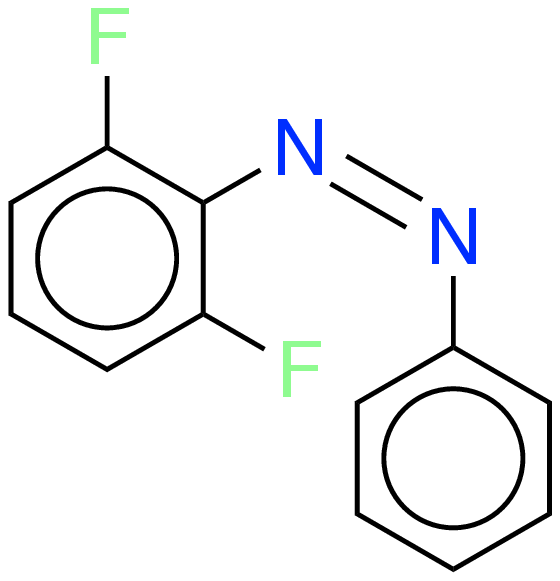}} & \cite{knie2014ortho} & \textbf{Exp}: (0.55, hexane)\newline \textbf{Calc}: 0.53 $\pm$ 0.02 \\ 
\hline

\textbf{33} & Fc1cccc(F)c1/N=N/c1c(F)cc(-c2ccccc2)cc1F & \raisebox{-.45\height}{\includegraphics[height=0.12\textwidth,trim=0 -5 0 -5]{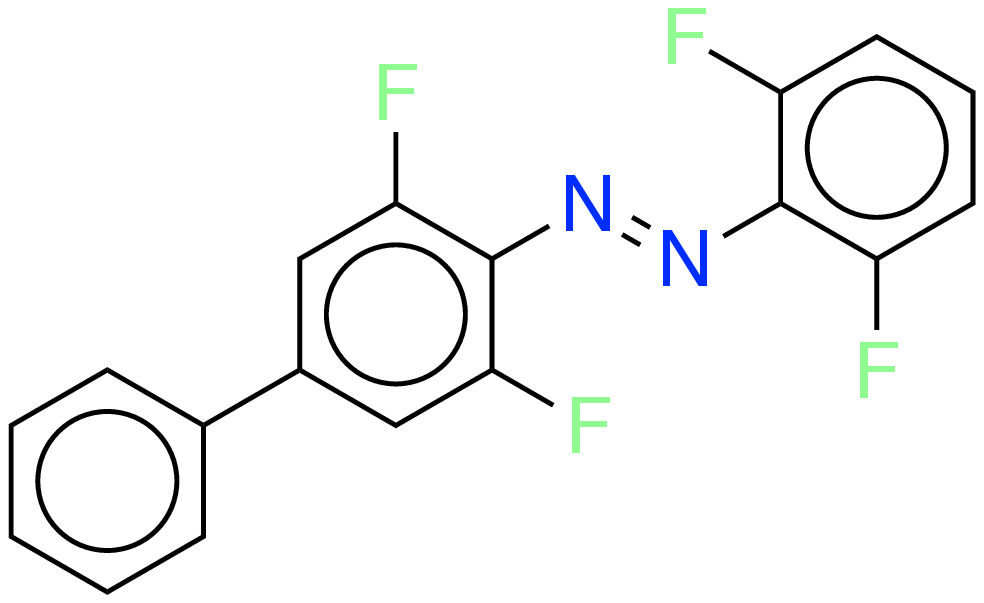}} & \cite{moreno2015two} & \textbf{Exp}: (0.15 $\pm$ 0.015, n-hexane)\newline \textbf{Calc}: 0.41 $\pm$ 0.02  \\ 
\hline

\textbf{34} & Fc1cccc(F)c1/N=N{\textbackslash}c1c(F)cc(-c2ccccc2)cc1F & \raisebox{-.45\height}{\includegraphics[height=0.12\textwidth,trim=0 -5 0 -5]{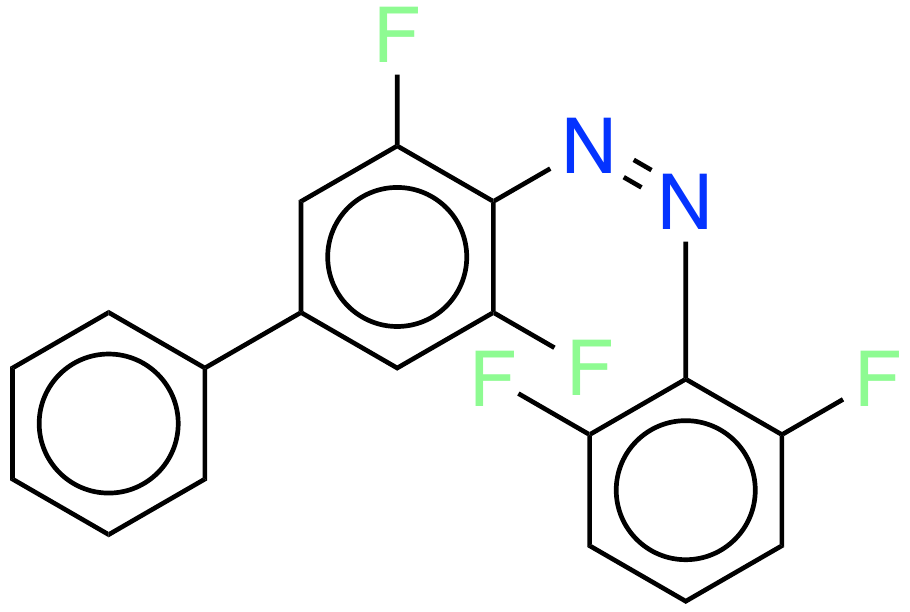}} & \cite{moreno2015two} & \textbf{Exp}: (0.28 $\pm$ 0.028, n-hexane)\newline \textbf{Calc}: 0.51 $\pm$ 0.02 \\ 
\hline

\textbf{35} & Nc1ccccc1/N=N/c1ccccc1NCc1ccccn1 & \raisebox{-.45\height}{\includegraphics[height=0.12\textwidth,trim=0 -5 0 -5]{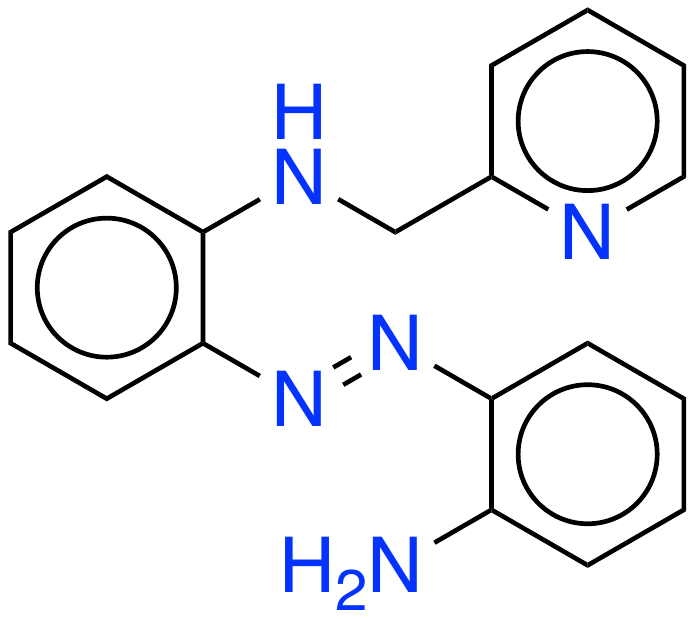}} & \cite{bandara2010proof} & \textbf{Exp}: (0\footnotemark[\value{footnote}], many)\newline \textbf{Calc}: 0.04 $\pm$ 0.04  \\ 
\hline

\textbf{36} & Nc1ccccc1/N=N{\textbackslash}c1ccccc1NCc1ccccn1 & \raisebox{-.45\height}{\includegraphics[height=0.18\textwidth,trim=0 -5 0 -5]{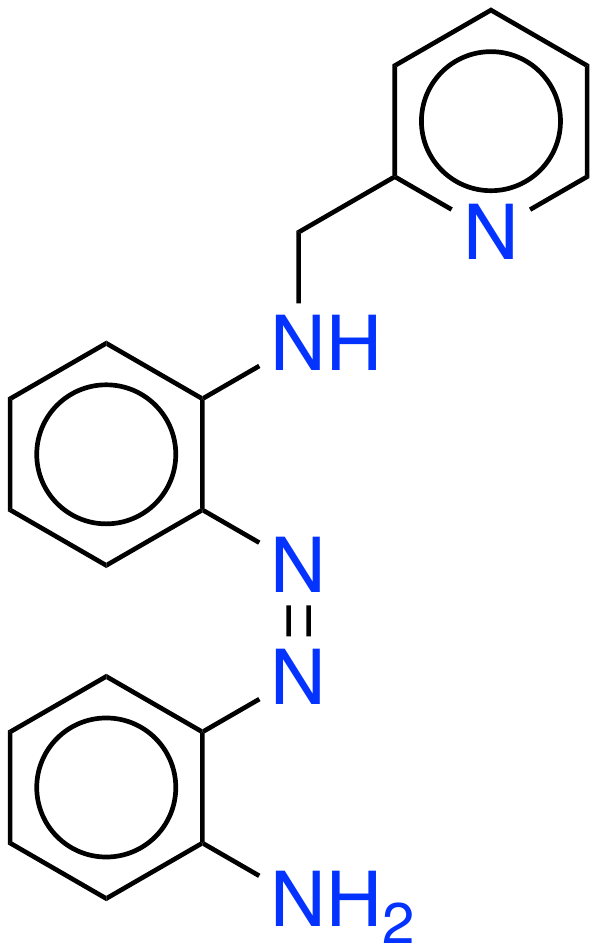}} & \cite{bandara2010proof} & \textbf{Exp}:  None\newline \textbf{Calc}: 0.54 $\pm$ 0.02 \\ 
\hline

\textbf{37} & OCc1ccc(/N=N/c2ccc (OCc3ccccc3)cc2)cc1 & \raisebox{-.45\height}{\includegraphics[height=0.11\textwidth,trim=0 -5 0 -5]{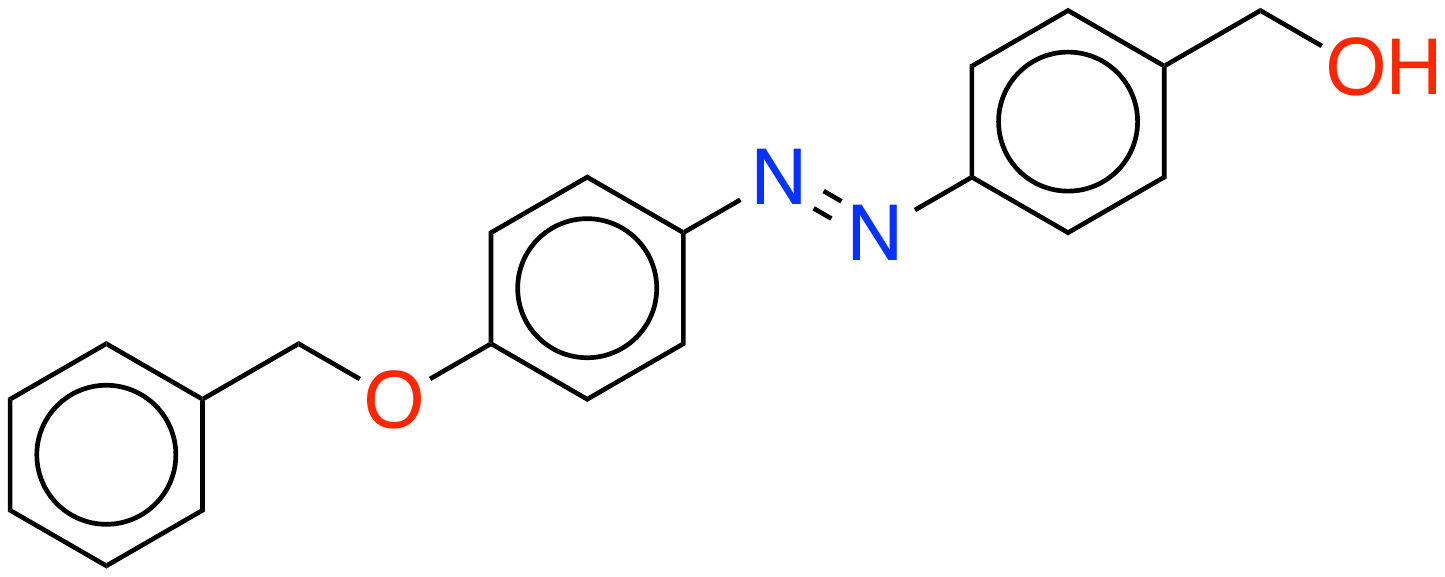}} & \cite{sierocki2006photoisomerization} & \textbf{Exp}: (0.4, dichloromethane)\newline \textbf{Calc}: 0.13 $\pm$ 0.03 \\ 
\hline

\textbf{38} & OCc1ccc(/N=N{\textbackslash}c2ccc (OCc3ccccc3)cc2)cc1 & \raisebox{-.45\height}{\includegraphics[height=0.12\textwidth,trim=0 -5 0 -5]{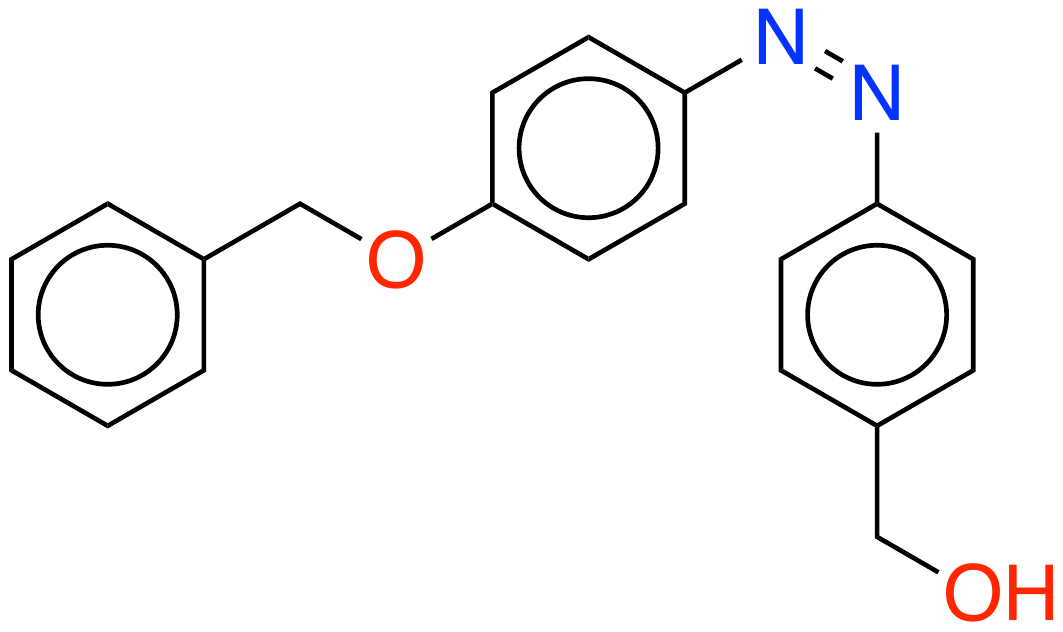}} & \cite{sierocki2006photoisomerization} & \textbf{Exp}: (0.61, dichloromethane)\newline \textbf{Calc}: 0.53 $\pm$ 0.02 \\ 
\hline

\textbf{39} & OCc1ccc(/N=N/c2ccc (OCc3cc(OCc4ccccc4)cc (OCc4ccccc4)c3)cc2)cc1 & \raisebox{-.45\height}{\includegraphics[height=0.19\textwidth,trim=0 -5 0 -5]{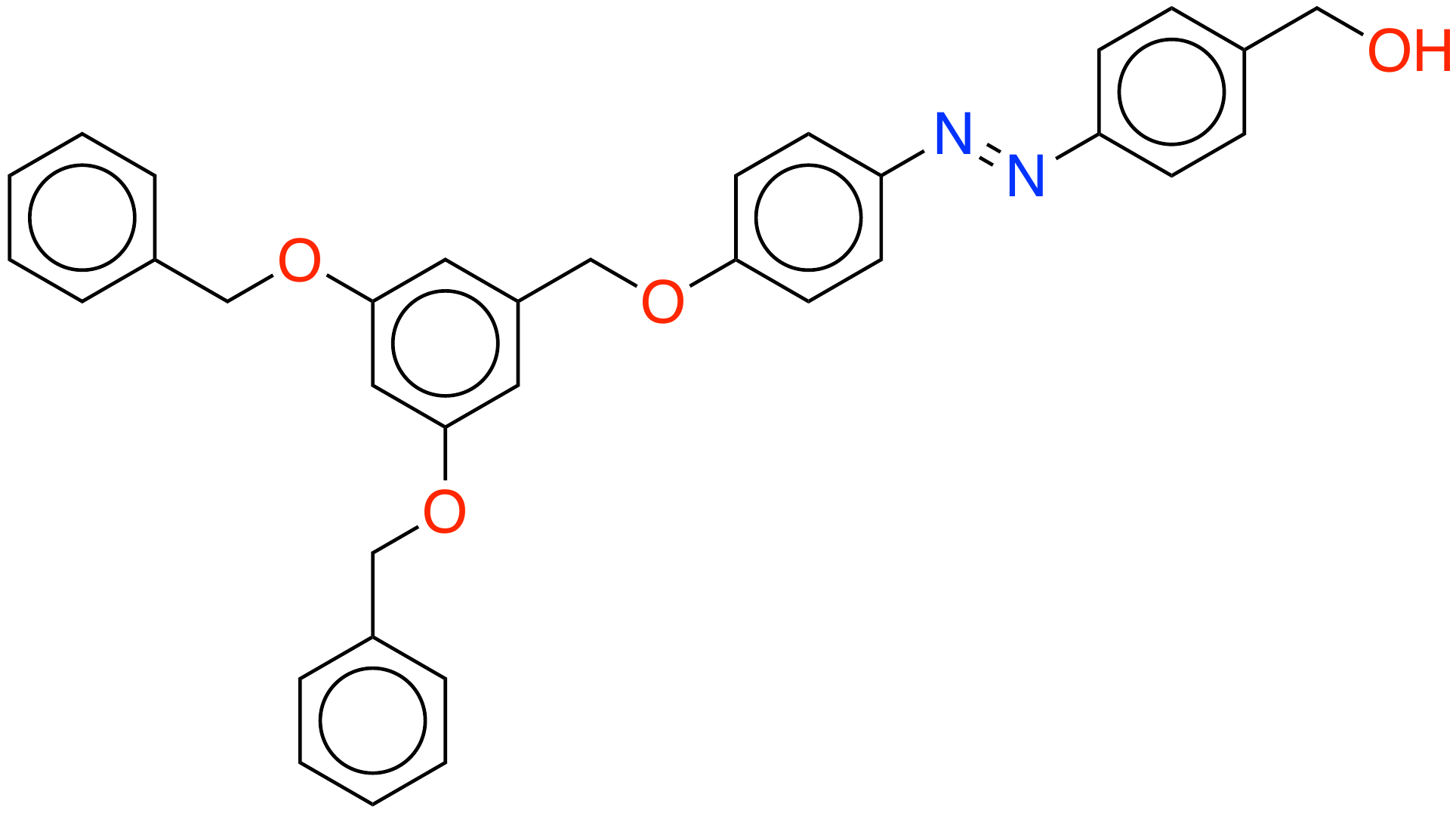}} & \cite{sierocki2006photoisomerization} & \textbf{Exp}: (0.36, dichloromethane)\newline \textbf{Calc}: 0.13 $\pm$ 0.03 \\ 
\hline

\textbf{40} & OCc1ccc(/N=N{\textbackslash}c2ccc (OCc3cc(OCc4ccccc4)cc (OCc4ccccc4)c3)cc2)cc1 & \raisebox{-.45\height}{\includegraphics[height=0.17\textwidth,trim=0 -5 0 -5]{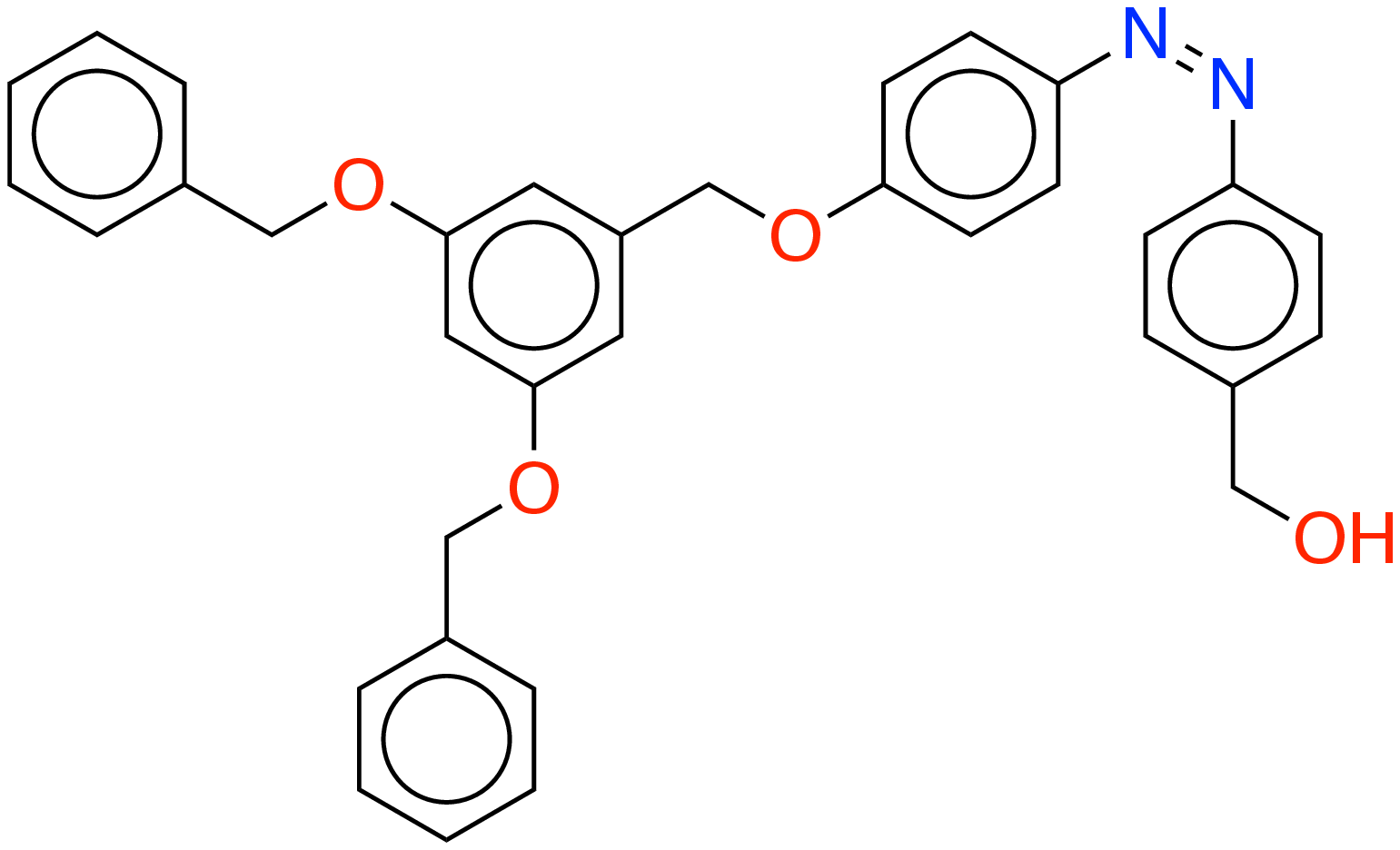}} & \cite{sierocki2006photoisomerization} & \textbf{Exp}: (0.64, dichloromethane)\newline \textbf{Calc}: 0.54 $\pm$ 0.02 \\ 

\hline\hline
\end{xltabular}
\addtocounter{footnote}{-1}\footnotetext{In Refs. \cite{bandara2010proof, bandara2011short}, the non-reactive azobenzene derivatives were irradiated from 300 to 600 nm. This range covered both $S_0\to S_1$ and $S_0\to S_2$ excitation. Only small changes in absorbance were observed, indicating minimal \textit{trans}$\to$\textit{cis} isomerization. However, the $S_2$ transition was highly redshifted, and thus had significant overlap with the $S_1$ transition. The $S_2$ transition also had a much higher oscillator strength. Hence the small absorbance changes could have been due to the $S_1$ transition, and so the $S_1$ yield may not have been precisely 0. The $S_1$ yield should therefore be interpreted as ``small'', rather than exactly zero. }
\addtocounter{footnote}{0}\footnotetext{Averaged over excitations at 313, 436, and 546 nm.}

\newpage
\begin{table}[h]
\caption{Training and validation set literature species used for dense configurational sampling.} 
\label{sm_tab:train_lit_specs}
\end{table}

\renewcommand\tabularxcolumn[1]{m{#1}}
\begin{xltabular}{\textwidth}{|>{\centering\arraybackslash}m{.1\linewidth}|>{\centering\arraybackslash}m{.45\linewidth}|>{\centering\arraybackslash}m{.35\linewidth}|>{\centering\arraybackslash}m{.1\linewidth}| }
% \begin{xltabular}

\hline\endfoot

\toprule

\# & SMILES & Graph & Ref. \\
\toprule
\hline

\textbf{41} & c1ccc2c(c1)CCc1ccccc1/N=N/2 & \raisebox{-.45\height}{\includegraphics[height=0.065\textwidth,trim=0 -5 0 -5]{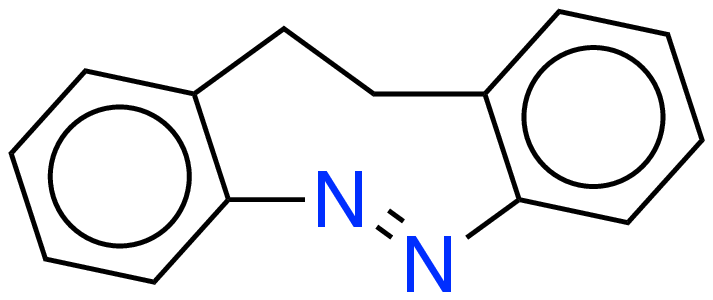}} & \cite{siewertsen2009highly}  \\ 
\hline

\textbf{42} & c1ccc2c(c1)CCc1ccccc1/N=N{\textbackslash}2 & \raisebox{-.45\height}{\includegraphics[height=0.065\textwidth,trim=0 -5 0 -5]{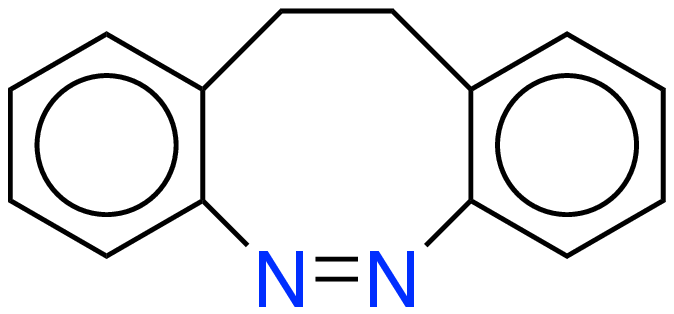}} & \cite{siewertsen2009highly}  \\ 
\hline

\textbf{43} & c1ccc2c(c1)CNc1ccccc1/N=N/2 & \raisebox{-.45\height}{\includegraphics[height=0.09\textwidth,trim=0 -5 0 -5]{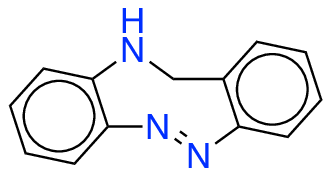}} & \cite{lentes2019nitrogen}  \\ 
\hline

\textbf{44} & c1ccc2c(c1)CNc1ccccc1/N=N{\textbackslash}2 & \raisebox{-.45\height}{\includegraphics[height=0.095\textwidth,trim=0 -5 0 -5]{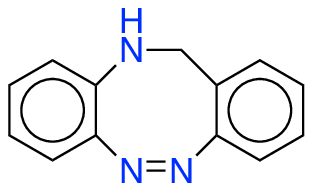}} & \cite{lentes2019nitrogen}  \\ 
\hline

\textbf{45} & c1ccc2c(c1)COc1ccccc1/N=N/2 & \raisebox{-.45\height}{\includegraphics[height=0.065\textwidth,trim=0 -5 0 -5]{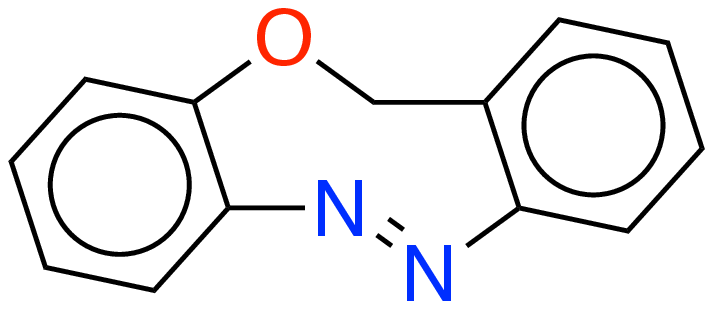}} & \cite{lentes2019nitrogen}  \\ 
\hline

\textbf{46} & c1ccc2c(c1)COc1ccccc1/N=N{\textbackslash}2 & \raisebox{-.45\height}{\includegraphics[height=0.07\textwidth,trim=0 -5 0 -5]{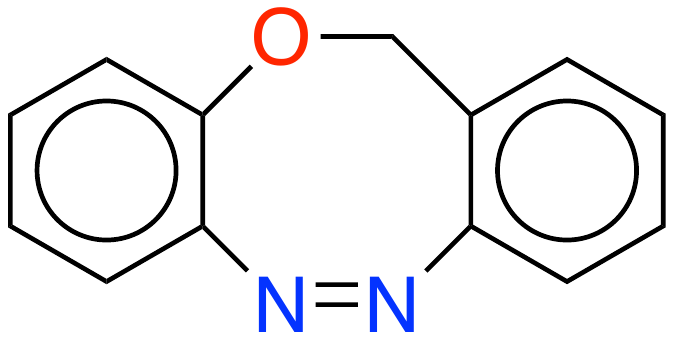}} & \cite{lentes2019nitrogen}  \\ 
\hline

\textbf{47} & c1ccc2c(c1)CSc1ccccc1/N=N/2 & \raisebox{-.45\height}{\includegraphics[height=0.065\textwidth,trim=0 -5 0 -5]{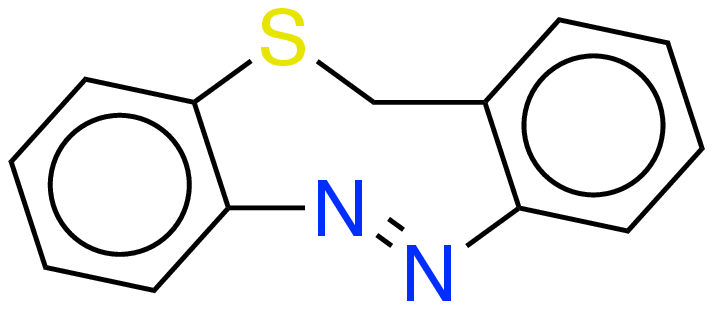}} & \cite{lentes2019nitrogen}  \\ 
\hline

\textbf{48} & c1ccc2c(c1)CSc1ccccc1/N=N{\textbackslash}2 & \raisebox{-.45\height}{\includegraphics[height=0.07\textwidth,trim=0 -5 0 -5]{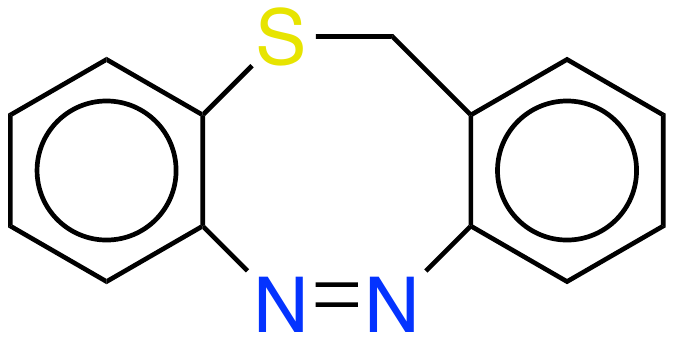}} & \cite{lentes2019nitrogen}  \\ 
\hline

\textbf{49} & c1ccc2cc(/N=N/c3ccc4ccccc4c3)ccc2c1 & \raisebox{-.45\height}{\includegraphics[height=0.085\textwidth,trim=0 -5 0 -5]{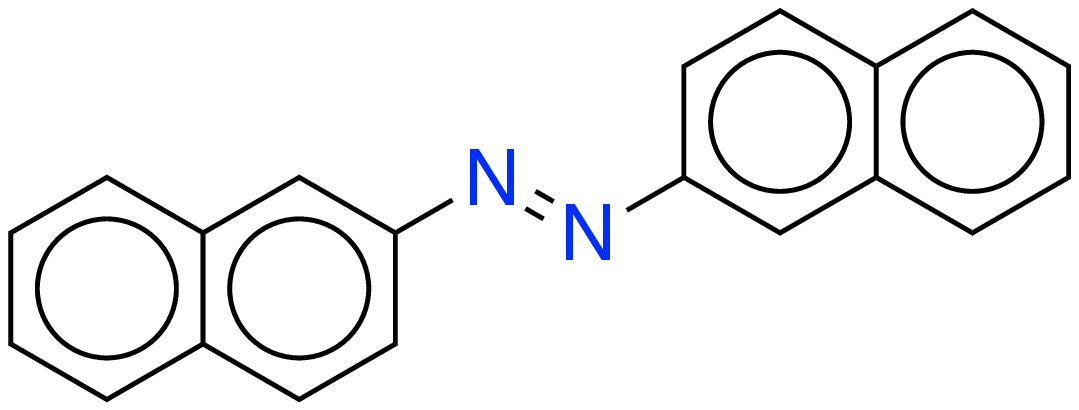}} & \cite{malkin1962temperature}  \\ 
\hline

\textbf{50} & c1ccc2cc(/N=N{\textbackslash}c3ccc4ccccc4c3)ccc2c1 & \raisebox{-.45\height}{\includegraphics[height=0.122\textwidth,trim=0 -5 0 -5]{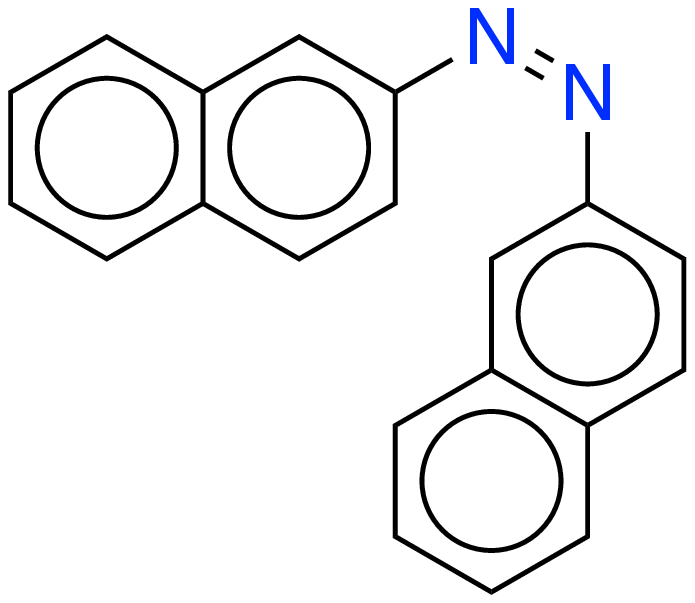}} & \cite{malkin1962temperature}  \\ 
\hline

\textbf{51} & c1ccc(/N=N/c2ccccc2)cc1 & \raisebox{-.45\height}{\includegraphics[height=0.085\textwidth,trim=0 -5 0 -5]{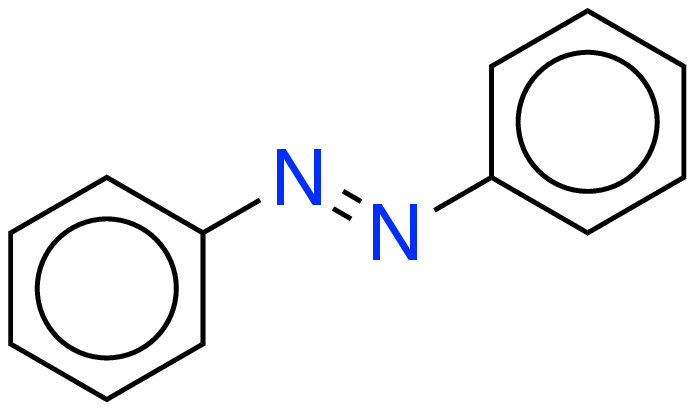}} & \cite{moreno2015two, birnbaum1954photo, zimmerman1958photochemical, bortolus1979cis, rau1988photoisomerization, gegiou1968temperature, rau1984further}  \\ 
\hline

\textbf{52} & c1ccc(/N=N{\textbackslash}c2ccccc2)cc1 & \raisebox{-.45\height}{\includegraphics[height=0.09\textwidth,trim=0 -5 0 -5]{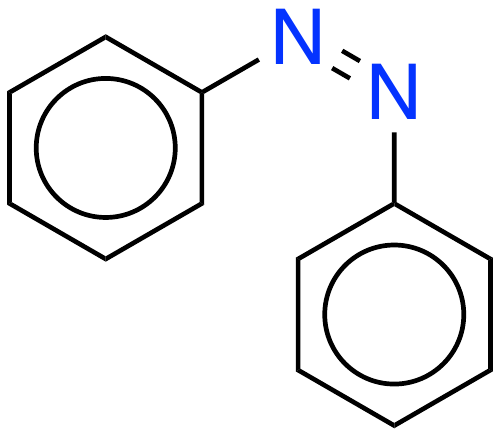}} & \cite{moreno2015two, birnbaum1954photo, zimmerman1958photochemical, bortolus1979cis, rau1988photoisomerization, gegiou1968temperature, rau1984further}  \\ 
\hline

\textbf{53} & CC1(C)c2cccc(N)c2Oc2c(/N=N/ c3cccc4c3Oc3c(N)cccc3C4(C)C)cccc21 & \raisebox{-.45\height}{\includegraphics[height=0.183\textwidth,trim=0 -5 0 -5]{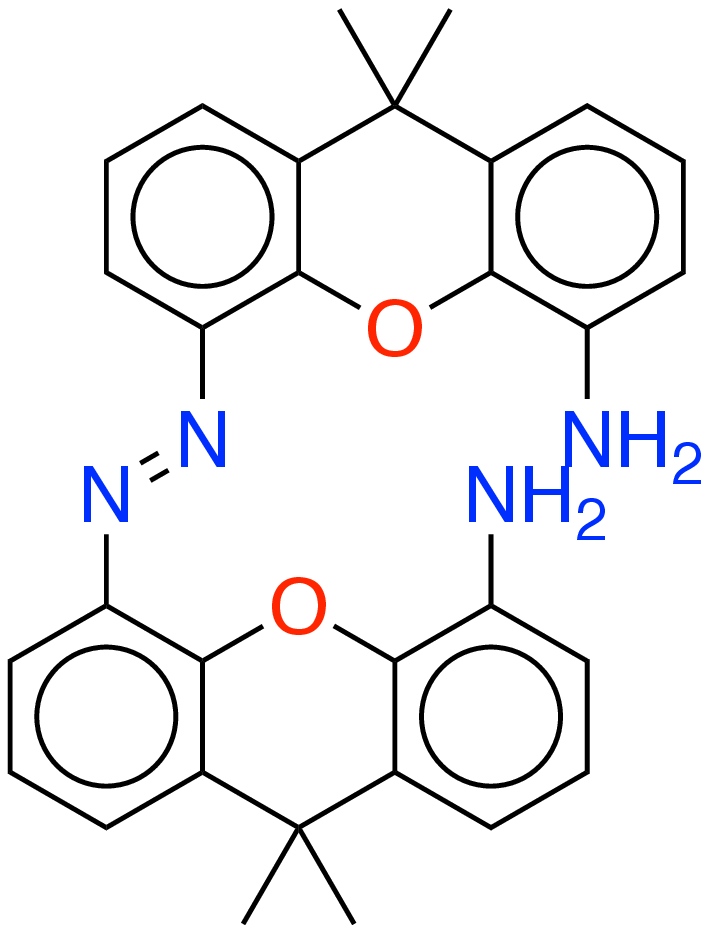}} & \cite{nagamani2005photoinduced}  \\ 
\hline

\textbf{54} & CC1(C)c2cccc(N)c2Oc2c(/N=N{\textbackslash} c3cccc4c3Oc3c(N)cccc3C4(C)C)cccc21 & \raisebox{-.45\height}{\includegraphics[height=0.167\textwidth,trim=0 -5 0 -5]{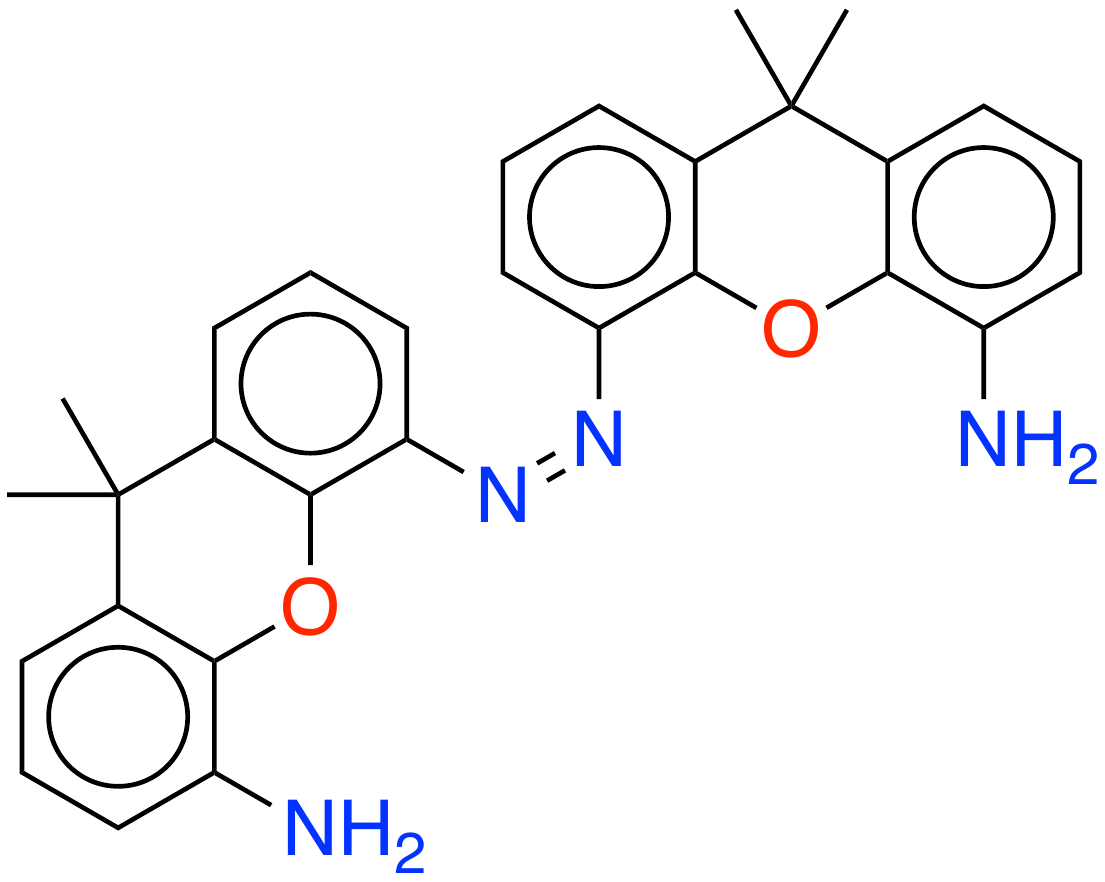}} & \cite{nagamani2005photoinduced}  \\ 
\hline

\textbf{55} & Cc1cccc(C)c1/N=N/c1c(C)cccc1C & \raisebox{-.45\height}{\includegraphics[height=0.085\textwidth,trim=0 -5 0 -5]{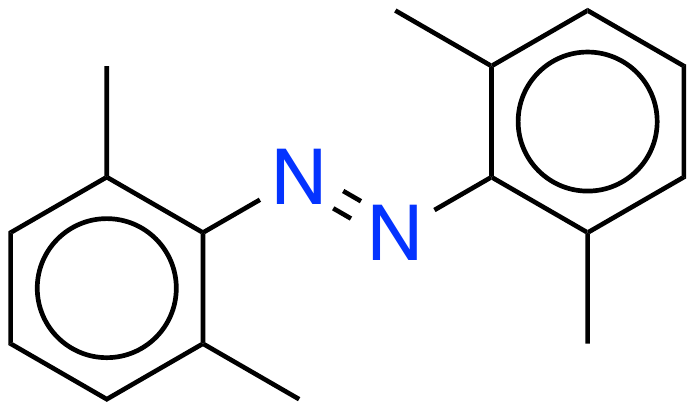}} & \cite{olmsted1983photochemical}  \\ 
\hline

\textbf{56} & Cc1cccc(C)c1/N=N{\textbackslash}c1c(C)cccc1C & \raisebox{-.45\height}{\includegraphics[height=0.123\textwidth,trim=0 -5 0 -5]{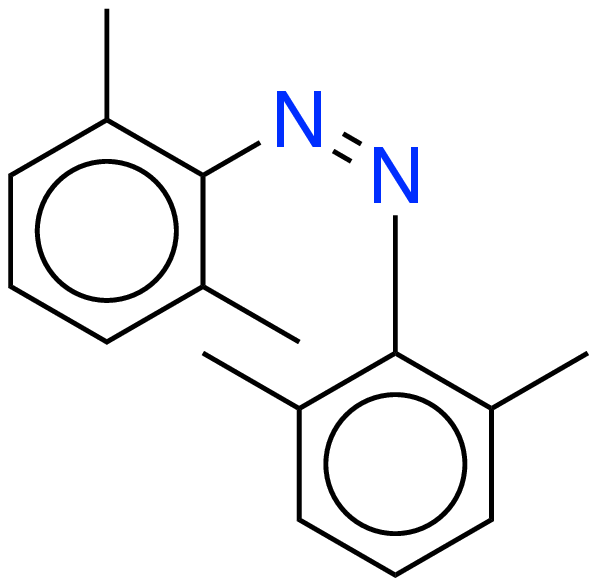}} & \cite{olmsted1983photochemical}  \\ 
\hline

\textbf{57} & Cc1cccc(C)c1/N=N/c1ccc(C(=O)O)cc1 & \raisebox{-.45\height}{\includegraphics[height=0.11\textwidth,trim=0 -5 0 -5]{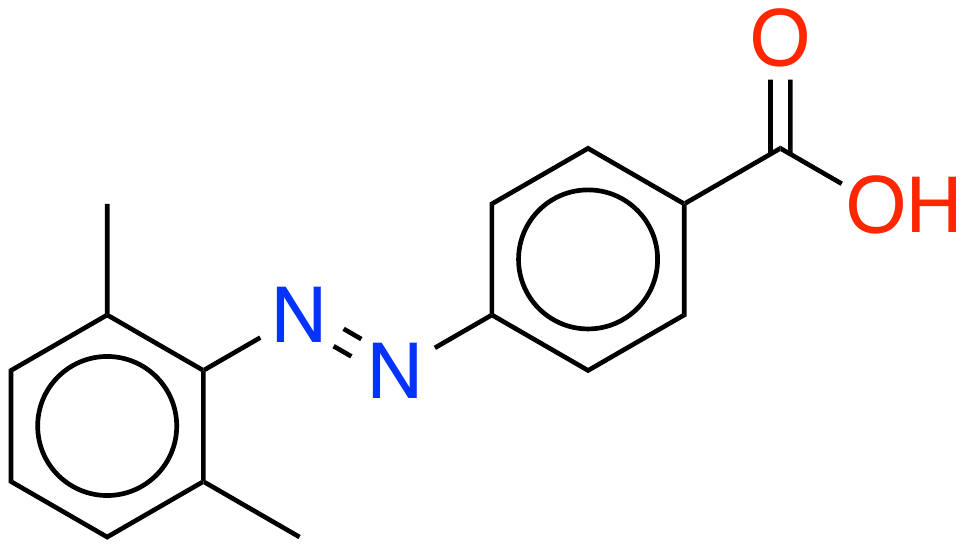}} & \cite{chen2012ultrafast}  \\ 
\hline

\textbf{58} & Cc1cccc(C)c1/N=N{\textbackslash}c1ccc(C(=O)O)cc1 & \raisebox{-.45\height}{\includegraphics[height=0.15\textwidth,trim=0 -5 0 -5]{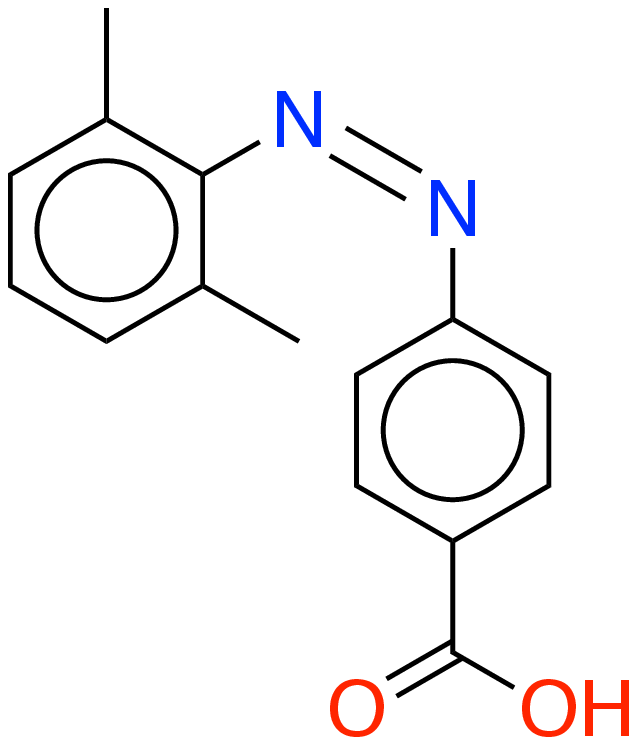}} & \cite{chen2012ultrafast}  \\ 
\hline

\textbf{59} & Cc1ccc(/N=N/c2ccccc2)cc1 & \raisebox{-.45\height}{\includegraphics[height=0.085\textwidth,trim=0 -5 0 -5]{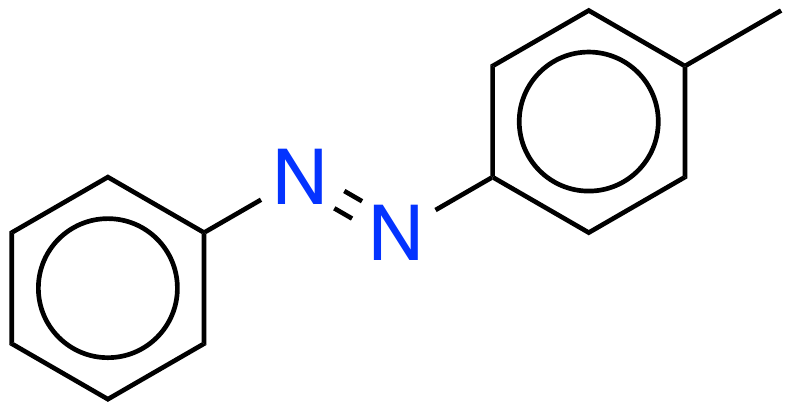}} & \cite{birnbaum1954photo}  \\ 
\hline

\textbf{60} & Cc1ccc(/N=N{\textbackslash}c2ccccc2)cc1 & \raisebox{-.45\height}{\includegraphics[height=0.09\textwidth,trim=0 -5 0 -5]{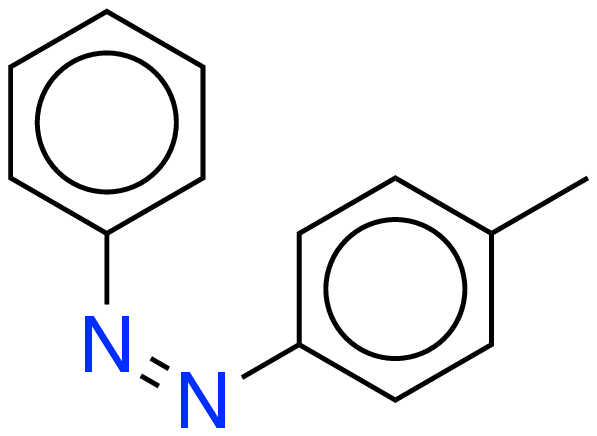}} & \cite{birnbaum1954photo}  \\ 
\hline

\textbf{61} & Cc1ccc(/N=N/c2ccc(C)cc2)cc1 & \raisebox{-.45\height}{\includegraphics[height=0.09\textwidth,trim=0 -5 0 -5]{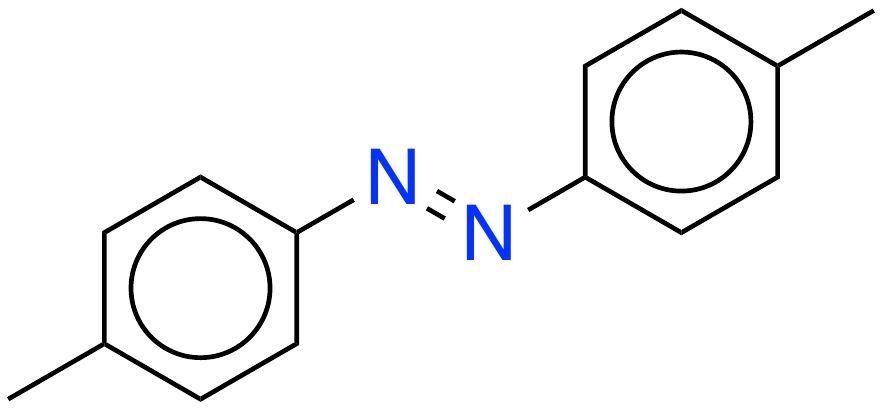}} & \cite{birnbaum1954photo}  \\ 
\hline

\textbf{62} & Cc1ccc(/N=N{\textbackslash}c2ccc(C)cc2)cc1 & \raisebox{-.45\height}{\includegraphics[height=0.11\textwidth,trim=0 -5 0 -5]{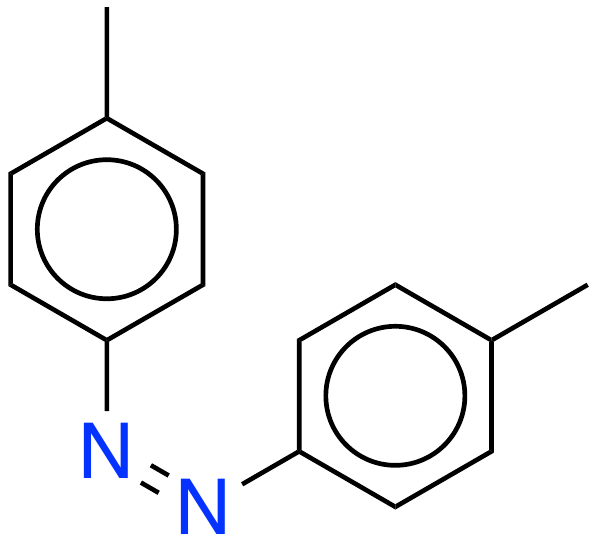}} & \cite{birnbaum1954photo}  \\ 
\hline

\textbf{63} & CCc1ccccc1/N=N/c1ccccc1CC & \raisebox{-.45\height}{\includegraphics[height=0.09\textwidth,trim=0 -5 0 -5]{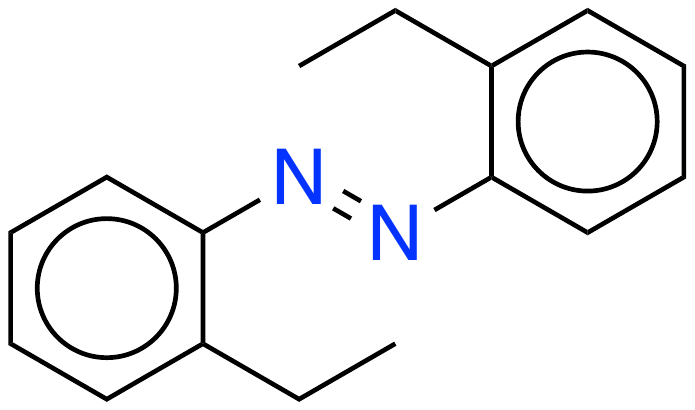}} & \cite{olmsted1983photochemical}  \\ 
\hline

\textbf{64} & CCc1ccccc1/N=N{\textbackslash}c1ccccc1CC & \raisebox{-.45\height}{\includegraphics[height=0.125\textwidth,trim=0 -5 0 -5]{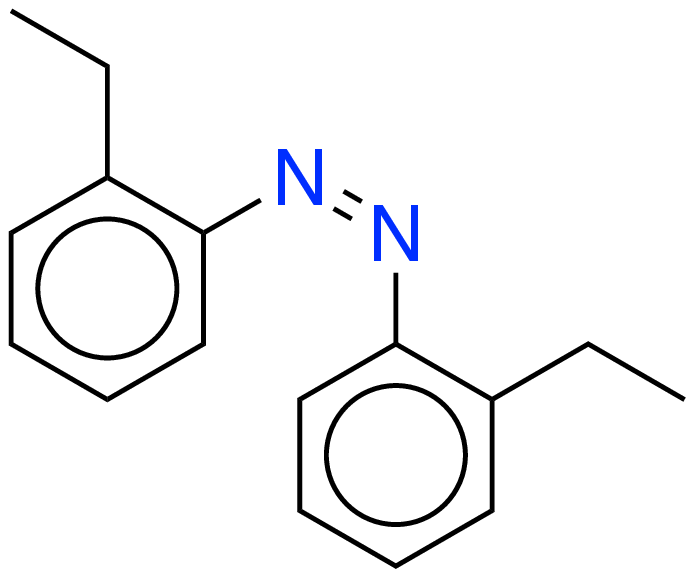}} & \cite{olmsted1983photochemical}  \\ 
\hline

\textbf{65} & C\#Cc1ccc(/N=N/c2ccc (N(CC)CC)cc2)c([N+](=O)[O-])c1 & \raisebox{-.45\height}{\includegraphics[height=0.23\textwidth,trim=0 -5 0 -5]{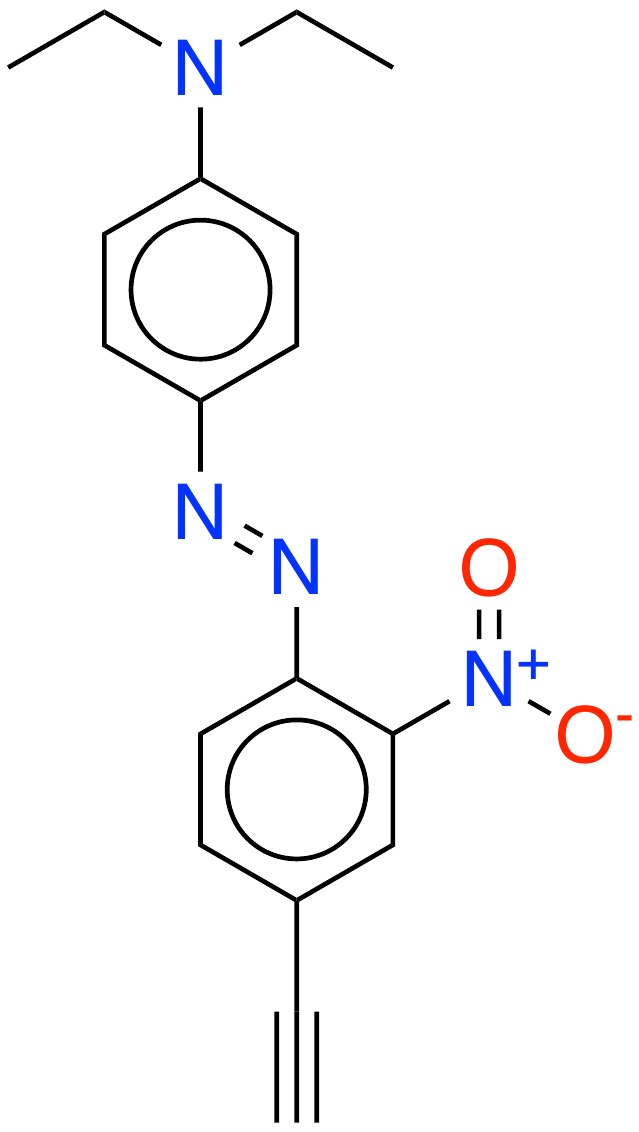}} & \cite{goulet2014effect}  \\ 
\hline

\textbf{66} & C\#Cc1ccc(/N=N{\textbackslash}c2ccc (N(CC)CC)cc2)c([N+](=O)[O-])c1 & \raisebox{-.45\height}{\includegraphics[height=0.175\textwidth,trim=0 -5 0 -5]{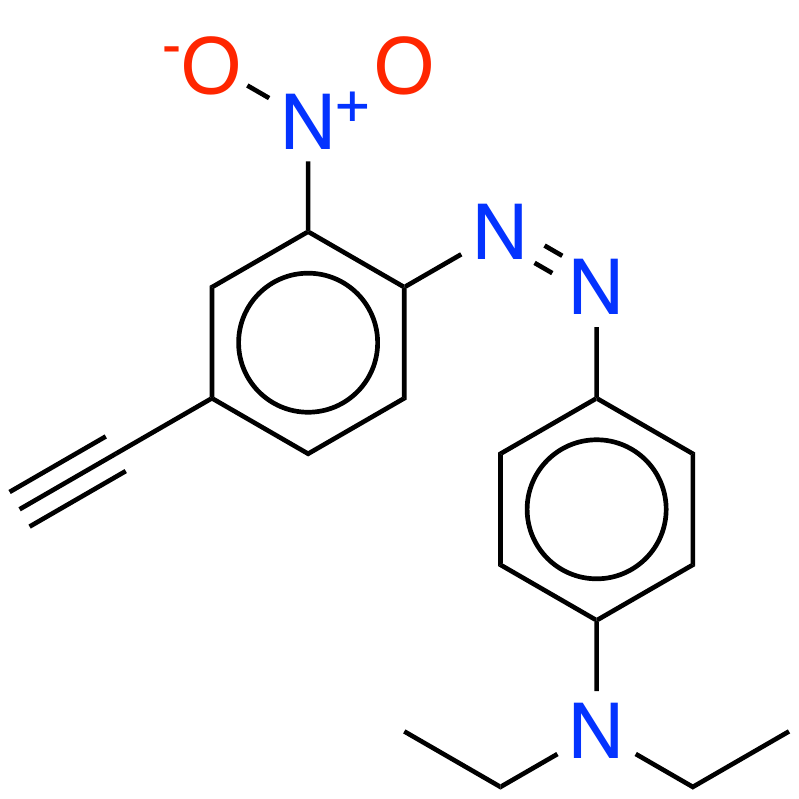}} & \cite{goulet2014effect}  \\ 
\hline

\textbf{67} & CC(C)c1ccc(/N=N/c2ccc (C(C)C)cc2C(C)C)c(C(C)C)c1 & \raisebox{-.45\height}{\includegraphics[height=0.14\textwidth,trim=0 -5 0 -5]{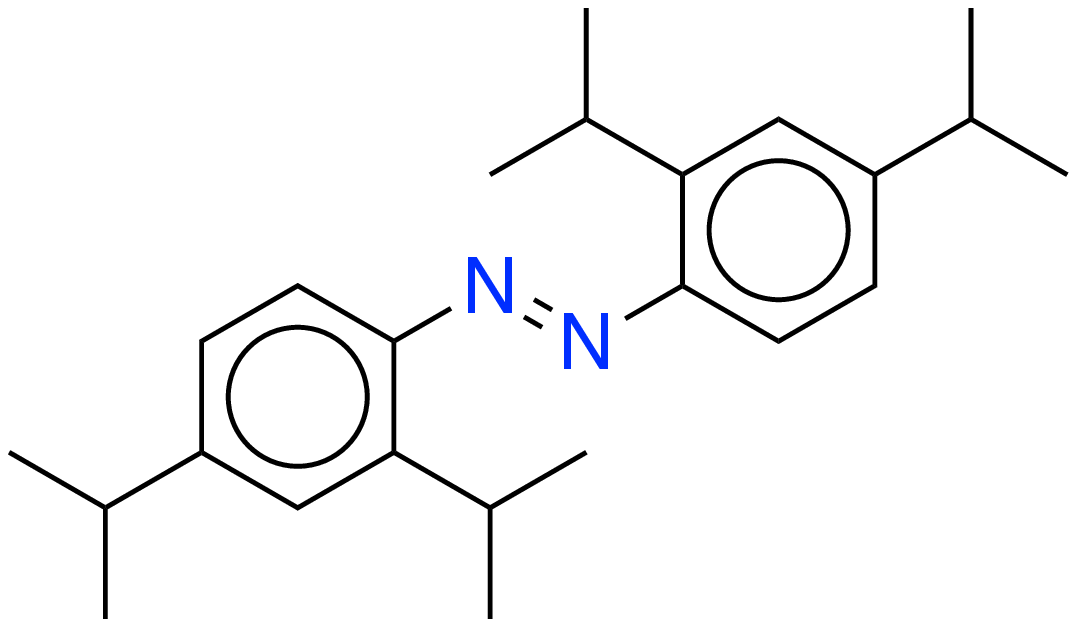}} & \cite{rau1988photoisomerization}  \\ 
\hline

\textbf{68} & CC(C)c1ccc(/N=N{\textbackslash}c2ccc (C(C)C)cc2C(C)C)c(C(C)C)c1 & \raisebox{-.45\height}{\includegraphics[height=0.16\textwidth,trim=0 -5 0 -5]{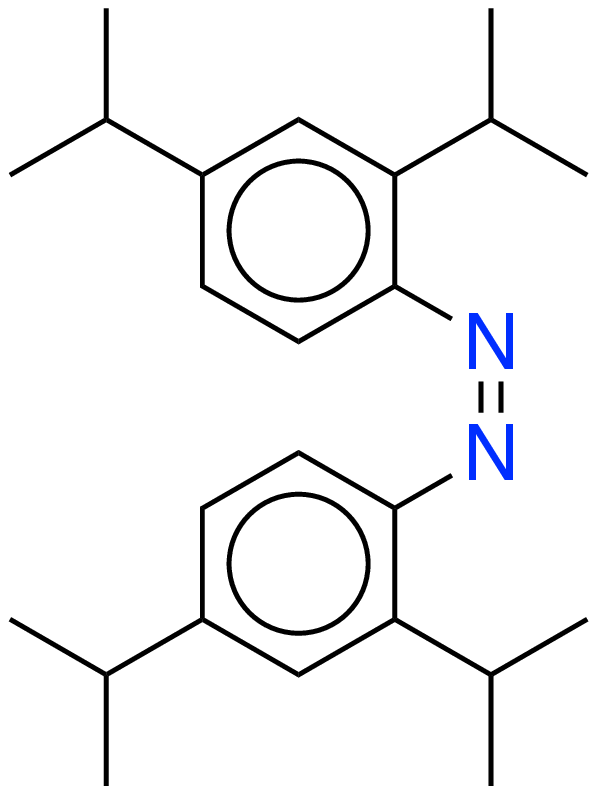}} & \cite{rau1988photoisomerization}  \\ 
\hline

\textbf{69} & C=CC(=O)Nc1ccc(/N=N/c2ccc (NC(=O)C[N+](CC)(CC)CC)cc2)cc1

& \raisebox{-.45\height}{\includegraphics[height=0.12\textwidth,trim=0 -5 0 -5]{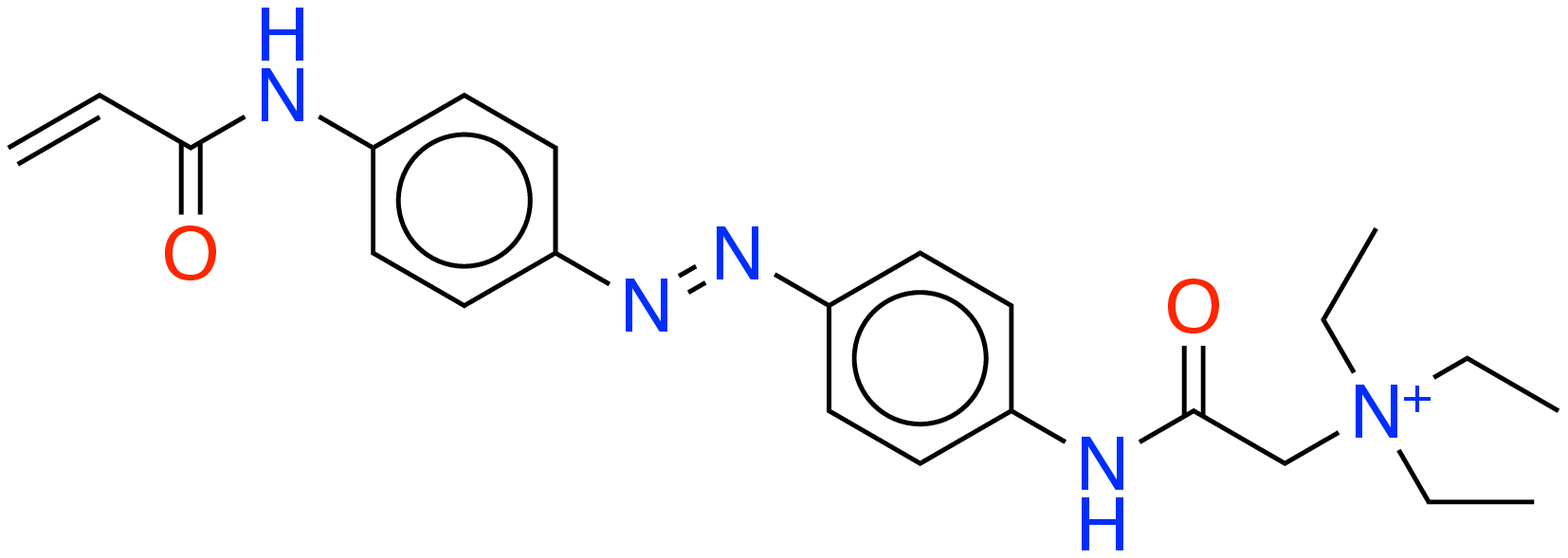}} & \cite{mourot2011tuning}  \\ 
\hline

\textbf{70} & C=CC(=O)Nc1ccc(/N=N{\textbackslash}c2ccc (NC(=O)C[N+](CC)(CC)CC)cc2)cc1 & \raisebox{-.45\height}{\includegraphics[height=0.16\textwidth,trim=0 -5 0 -5]{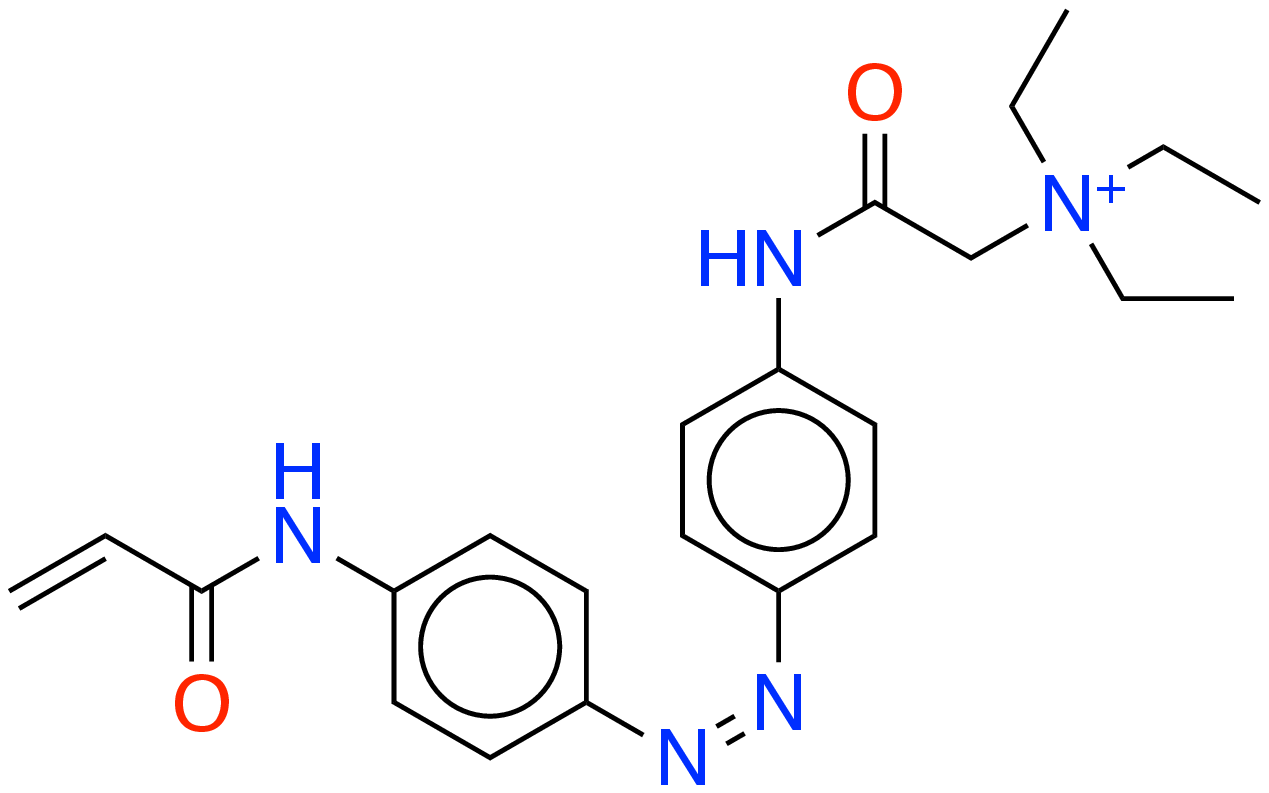}} & \cite{mourot2011tuning}  \\ 
\hline

\textbf{71} & CCN1Cc2ccccc2/N=N/c2ccccc21 & \raisebox{-.45\height}{\includegraphics[height=0.1 \textwidth,trim=0 -5 01 -5]{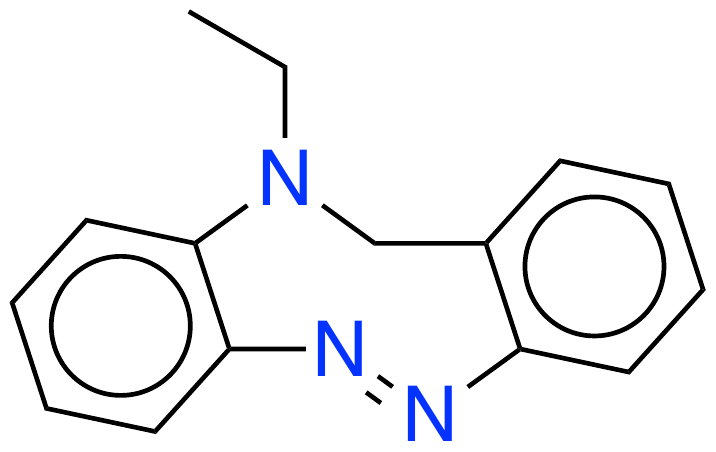}} & \cite{lentes2019nitrogen}  \\ 
\hline

\textbf{72} & CCN1Cc2ccccc2/N=N{\textbackslash}c2ccccc21 & \raisebox{-.45\height}
{\includegraphics[height=0.09 \textwidth,trim=0 -5 01 -5]{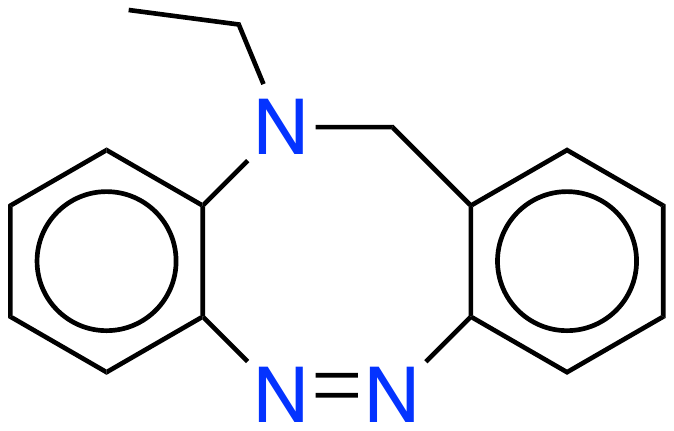}} 
& \cite{lentes2019nitrogen}  \\ 
\hline

\textbf{73} & CCN(c1ccccc1)c1ccc(/N=N/ c2ccc(NC(=O)C[N+](CC)(CC)CC)cc2)cc1 & \raisebox{-.45\height}
{\includegraphics[height=0.17 \textwidth,trim=0 -5 01 -5]{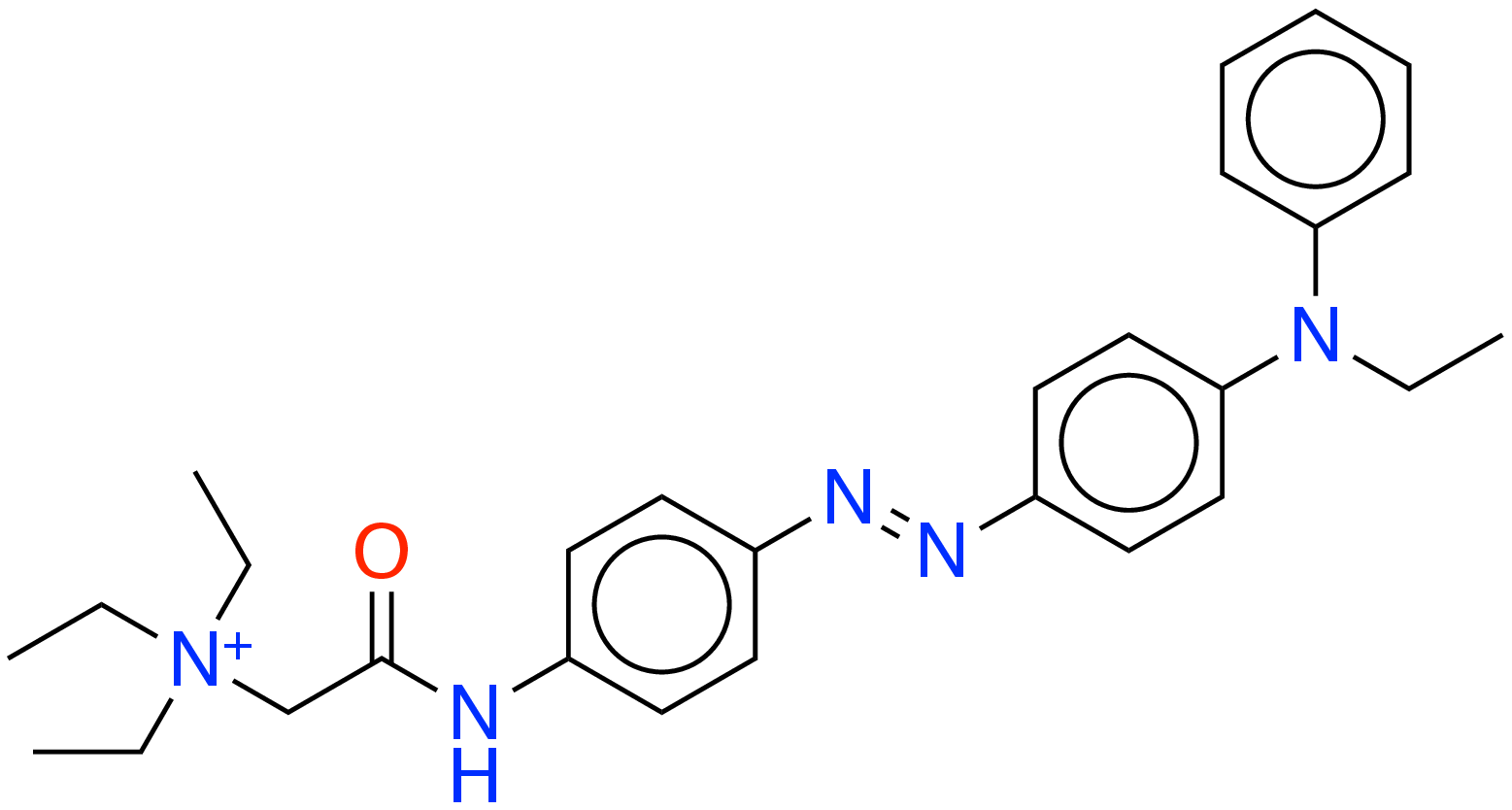}} 
& \cite{mourot2011tuning}  \\ 
\hline

\textbf{74} & CCN(c1ccccc1)c1ccc(/N=N{\textbackslash} c2ccc(NC(=O)C[N+](CC)(CC)CC)cc2)cc1 & \raisebox{-.45\height}{\includegraphics[height=0.16 \textwidth,trim=0 -5 01 -5]{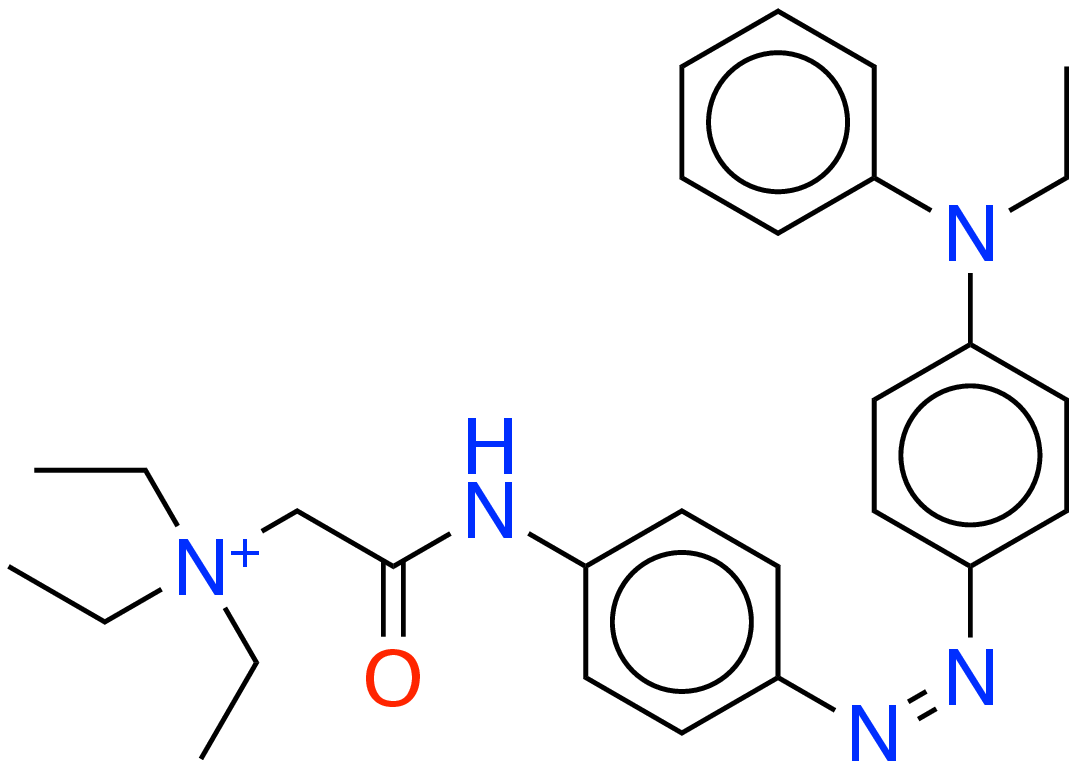}} 
& \cite{mourot2011tuning}  \\ 
\hline

\textbf{75} & CCN(Cc1ccccc1)c1ccc(/N=N/ c2ccc(NC(=O)C[N+](CC)(CC)CC)cc2)cc1 & \raisebox{-.45\height}{\includegraphics[height=0.25 \textwidth,trim=0 -5 01 -5]{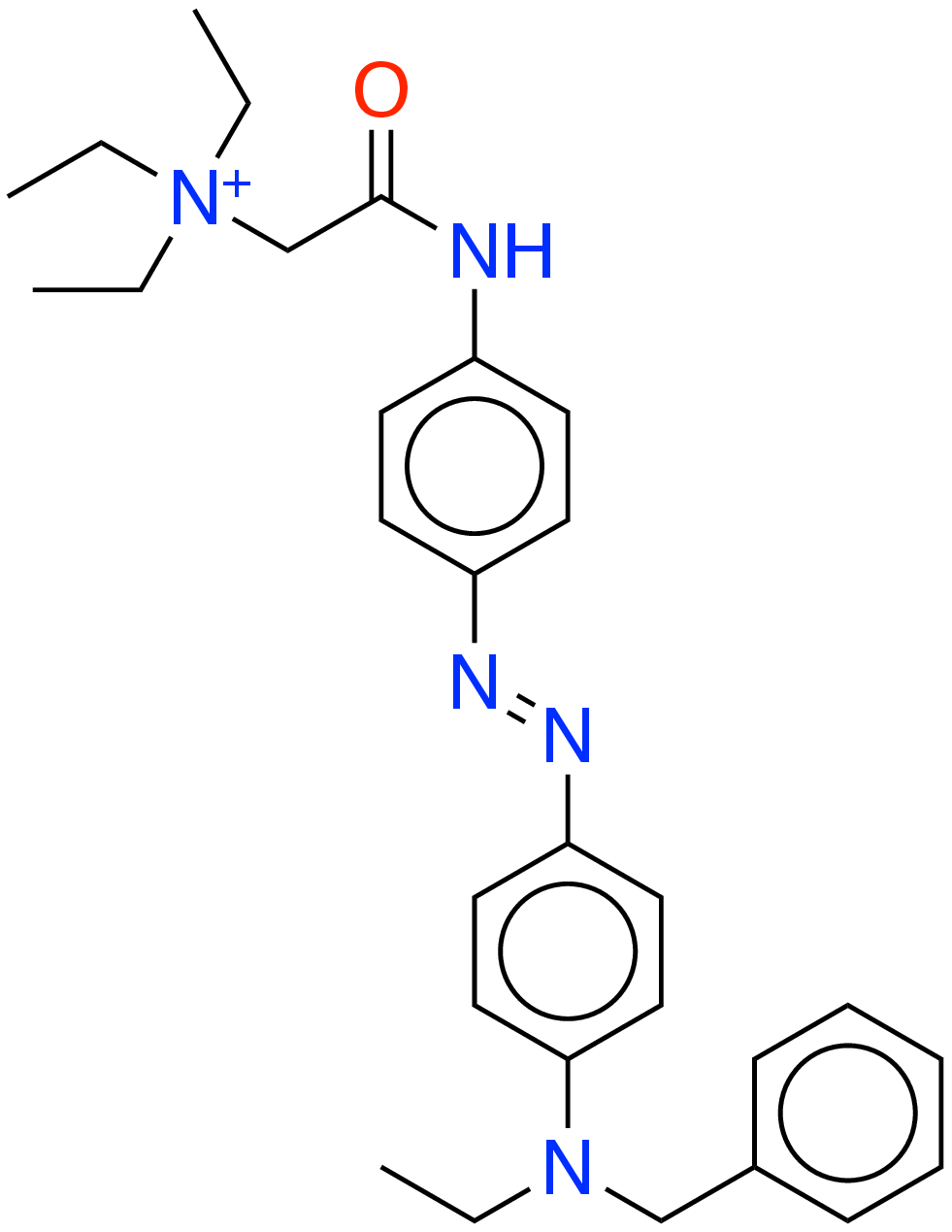}} & \cite{mourot2011tuning}  \\ 
\hline

\textbf{76} & CCN(Cc1ccccc1)c1ccc(/N=N{\textbackslash} c2ccc(NC(=O)C[N+](CC)(CC)CC)cc2)cc1 & \raisebox{-.45\height}{\includegraphics[height=0.12 \textwidth,trim=0 -5 01 -5]{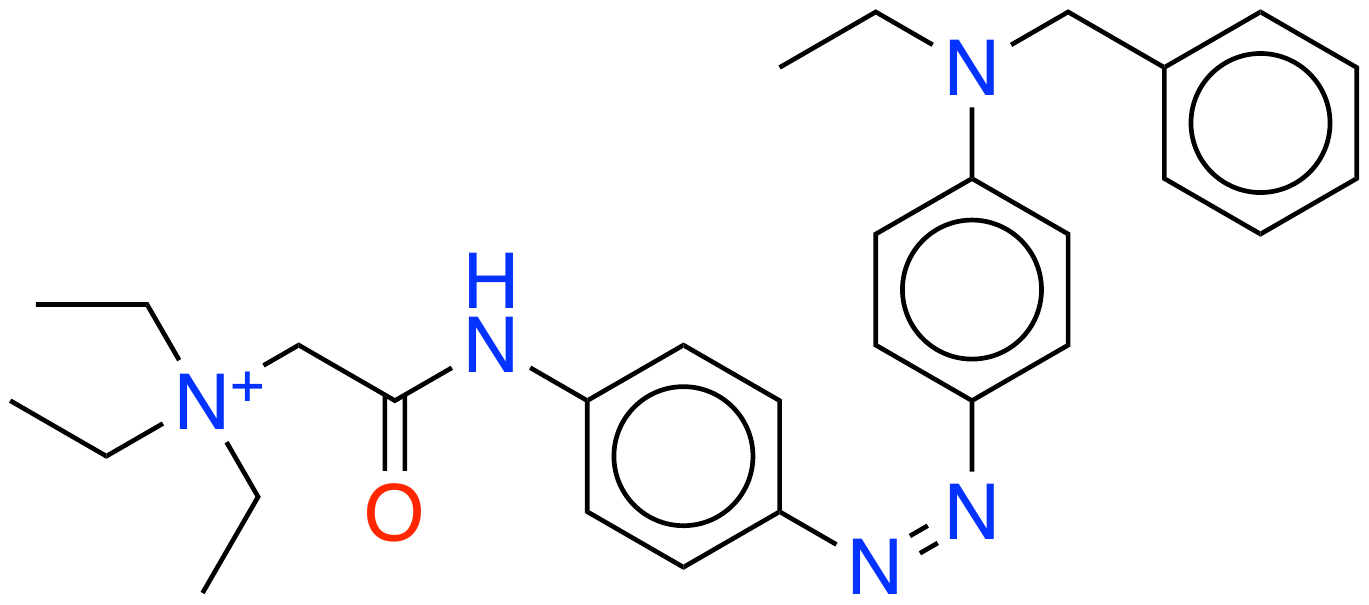}} & \cite{mourot2011tuning}  \\ 
\hline

\textbf{77} & CCN(CC)c1ccc(/N=N/c2ccc (-c3cn(-c4ccc5cc6ccccc6cc5c4)nn3)cc2 [N+](=O)[O-])cc1 & \raisebox{-.45\height}{\includegraphics[height=0.38 \textwidth,trim=0 -5 01 -5]{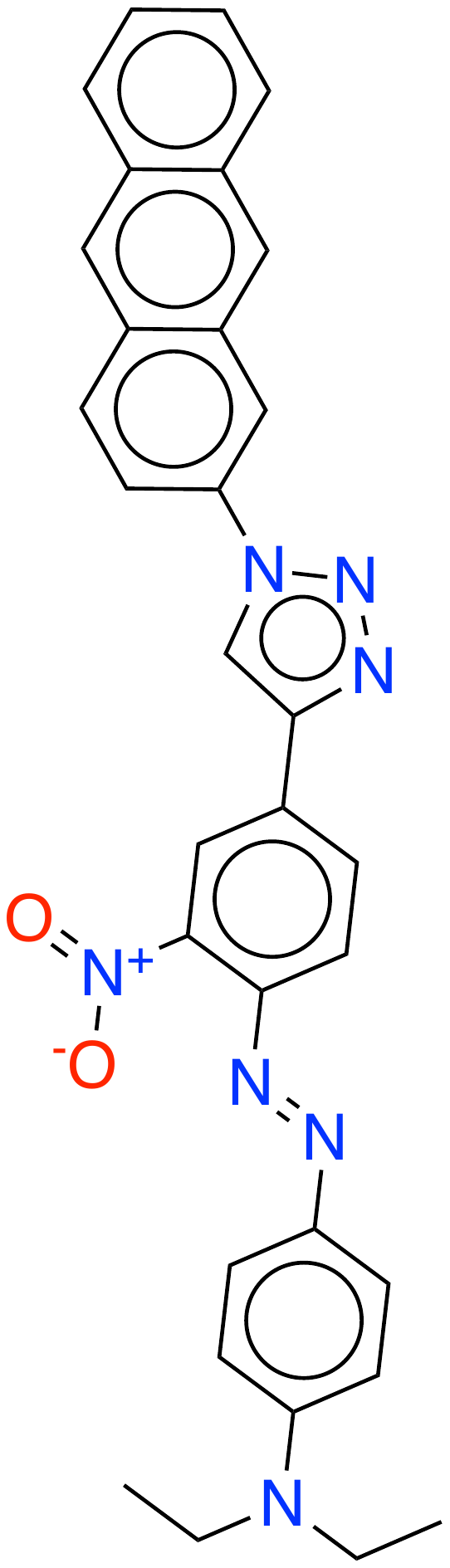}} & \cite{goulet2014effect}  \\ 
\hline

\textbf{78} & CCN(CC)c1ccc(/N=N{\textbackslash}c2ccc (-c3cn(-c4ccc5cc6ccccc6cc5c4)nn3)cc2 [N+](=O)[O-])cc1 & \raisebox{-.45\height}{\includegraphics[height=0.18 \textwidth,trim=0 -5 01 -5]{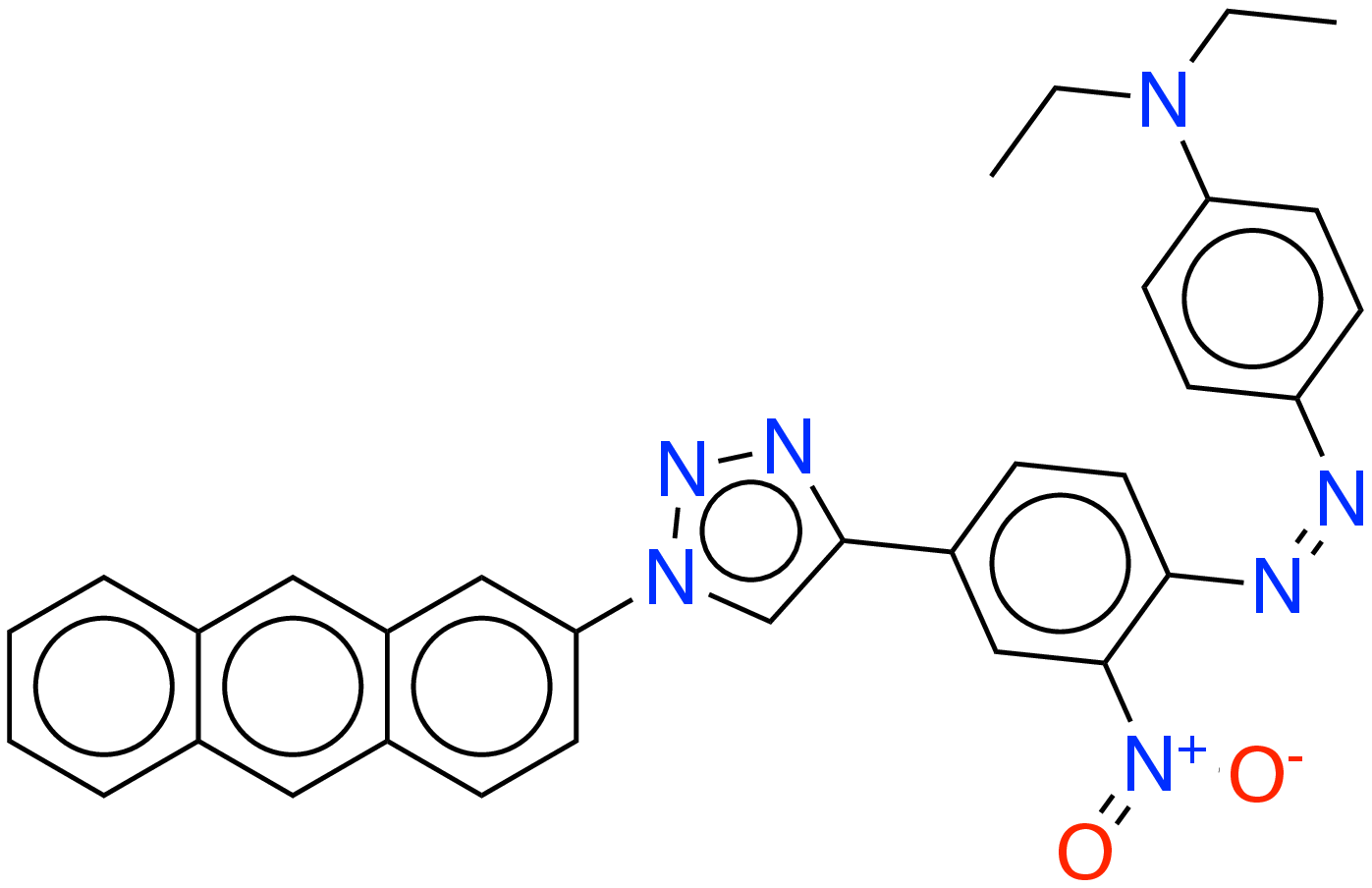}} & \cite{goulet2014effect}  \\ 
\hline

\textbf{79} & CCN(CC)c1ccc(/N=N/c2ccc(-c3cn (-c4cccc5ccccc45)nn3)cc2[N+](=O)[O-])cc1 & \raisebox{-.45\height}{\includegraphics[height=0.26 \textwidth,trim=0 -5 01 -5]{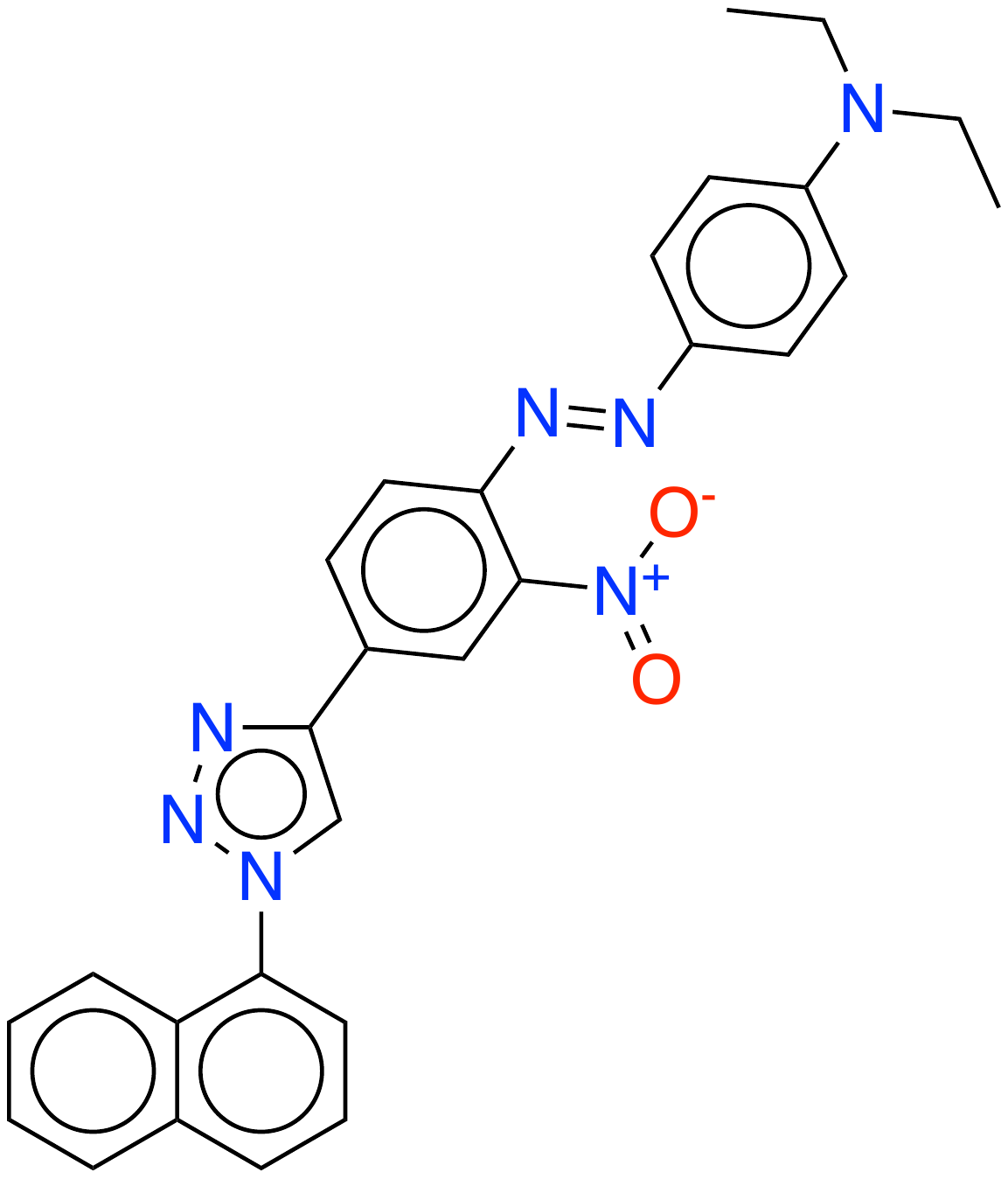}} & \cite{goulet2014effect}  \\ 
\hline

\textbf{80} & CCN(CC)c1ccc(/N=N{\textbackslash}c2ccc(-c3cn (-c4cccc5ccccc45)nn3)cc2[N+](=O)[O-])cc1 & \raisebox{-.45\height}{\includegraphics[height=0.24 \textwidth,trim=0 -5 01 -5]{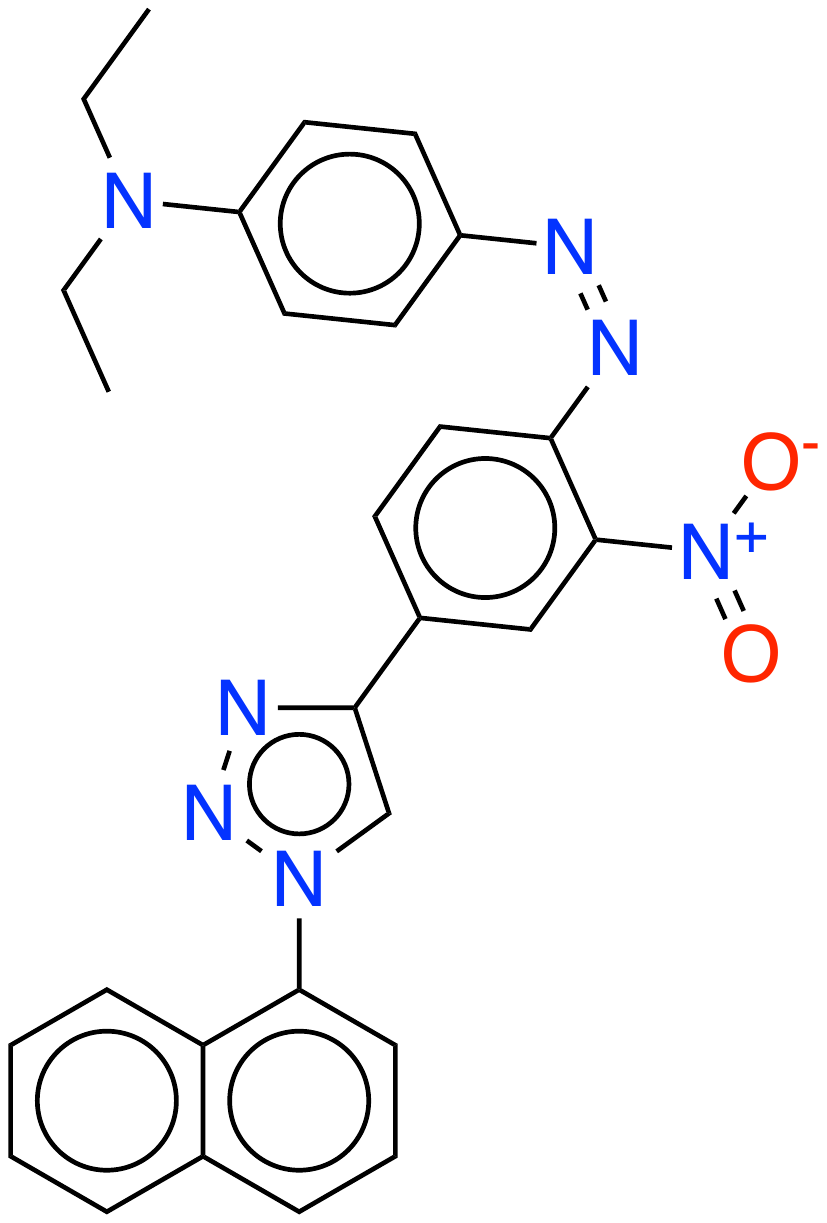}} & \cite{goulet2014effect}  \\ 
\hline

\textbf{81} & CCN(CC)c1ccc(/N=N/c2ccc (-c3cn(-c4ccccc4)nn3)cc2[N+](=O)[O-])cc1 & \raisebox{-.45\height}{\includegraphics[height=0.29 \textwidth,trim=0 -5 01 -5]{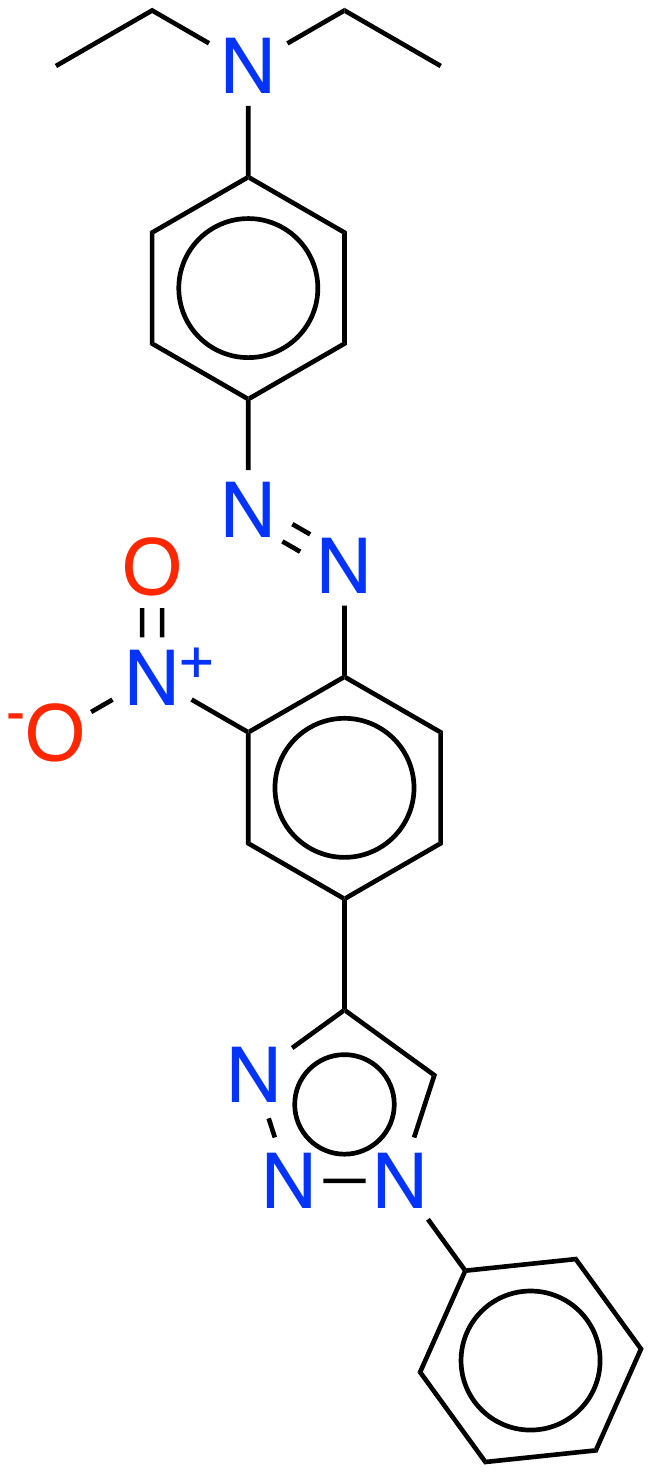}} & \cite{goulet2014effect}  \\ 
\hline

\textbf{82} & CCN(CC)c1ccc(/N=N{\textbackslash}c2ccc (-c3cn(-c4ccccc4)nn3)cc2[N+](=O)[O-])cc1 & \raisebox{-.45\height}{\includegraphics[height=0.16 \textwidth,trim=0 -5 01 -5]{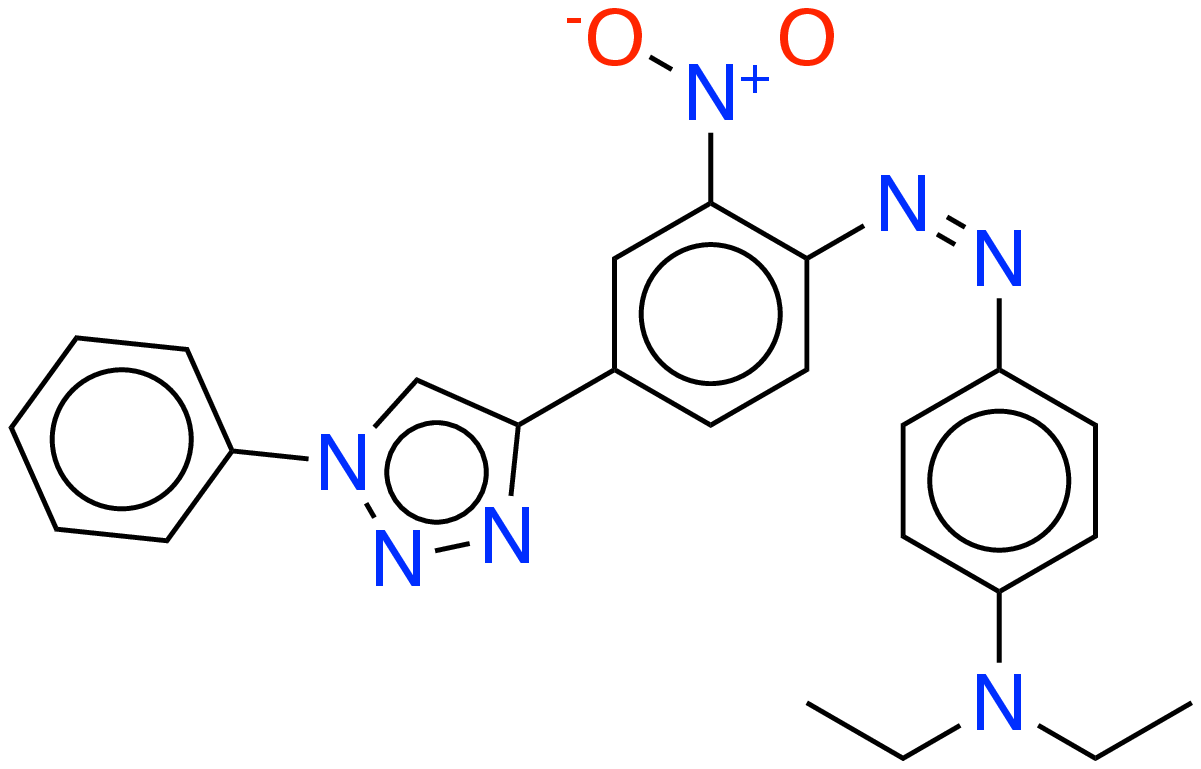}} & \cite{goulet2014effect}  \\ 
\hline

\textbf{83} & CCN(CC)c1ccc(/N=N/c2ccc (NC(=O)C[N+](CC)(CC)CC)cc2)cc1 & \raisebox{-.45\height}{\includegraphics[height=0.26 \textwidth,trim=0 -5 01 -5]{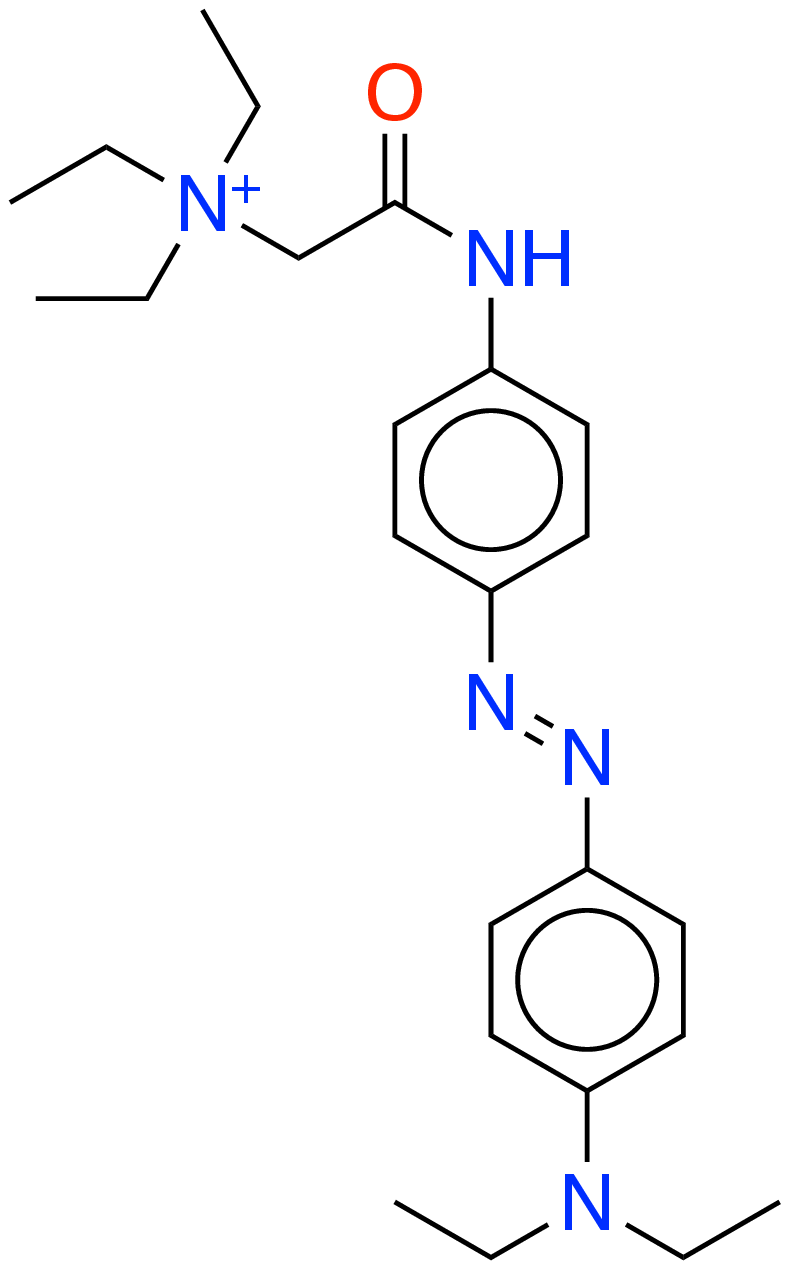}} & \cite{mourot2011tuning}  \\ 
\hline

\textbf{84} & CCN(CC)c1ccc(/N=N{\textbackslash}c2ccc (NC(=O)C[N+](CC)(CC)CC)cc2)cc1 & \raisebox{-.45\height}{\includegraphics[height=0.12 \textwidth,trim=0 -5 01 -5]{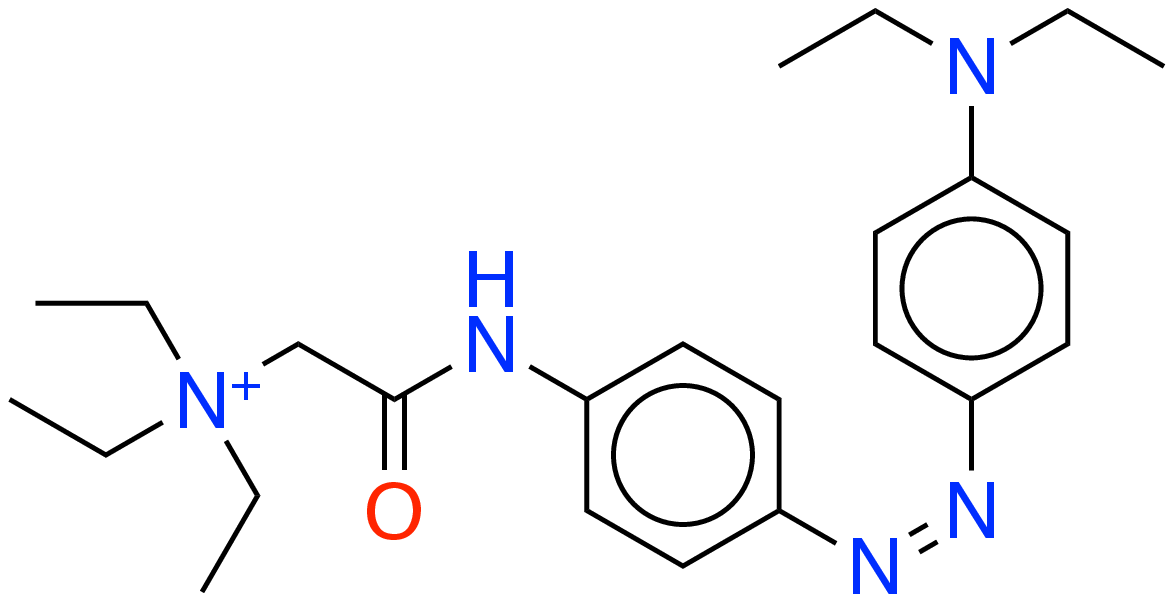}} & \cite{mourot2011tuning}  \\ 
\hline

\textbf{85} & CCN(CC)c1ccc(/N=N/c2c(F) c(F)c(NC(=O)C[N+](CC)(CC)CC)c(F)c2F)cc1 & \raisebox{-.45\height}{\includegraphics[height=0.25 \textwidth,trim=0 -5 01 -5]{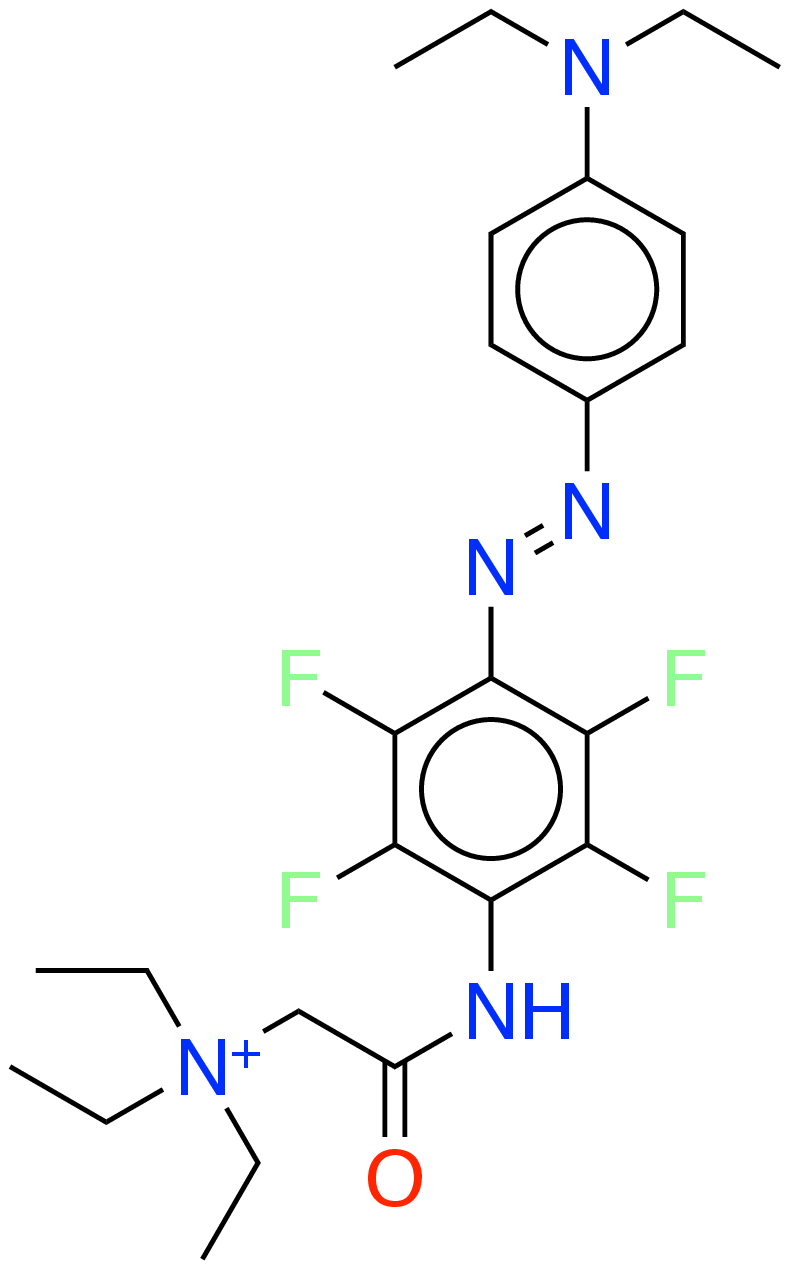}} & \cite{mourot2011tuning}  \\ 
\hline

\textbf{86} & CCN(CC)c1ccc(/N=N{\textbackslash}c2c(F) c(F)c(NC(=O)C[N+](CC)(CC)CC)c(F)c2F)cc1 & \raisebox{-.45\height}{\includegraphics[height=0.25 \textwidth,trim=0 -5 01 -5]{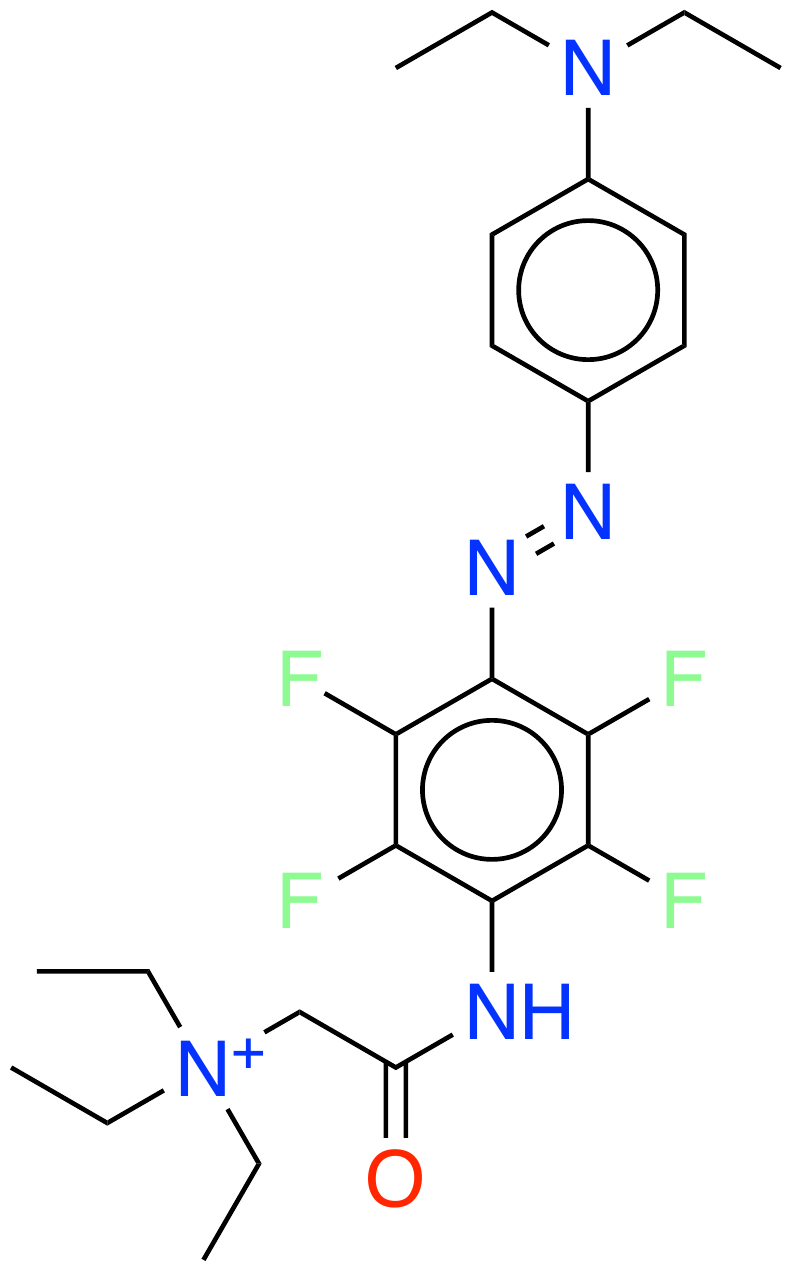}} & \cite{mourot2011tuning}  \\ 
\hline

\textbf{87} & CCN(CC)c1cc(NC(C)=O)ccc1/N=N/ c1ccc(NC(C)=O)cc1N(CC)CC & \raisebox{-.45\height}{\includegraphics[height=0.12 \textwidth,trim=0 -5 01 -5]{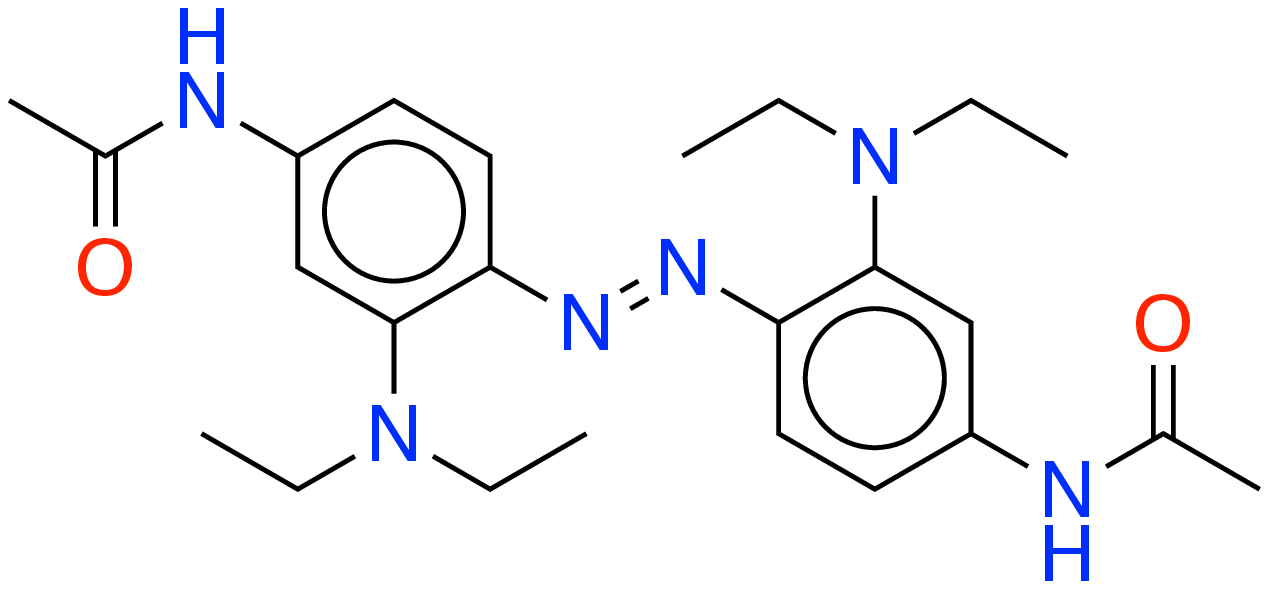}} & \cite{sadovski2009spectral}  \\ 
\hline

\textbf{88} & CCN(CC)c1cc(NC(C)=O)ccc1/N=N{\textbackslash} c1ccc(NC(C)=O)cc1N(CC)CC & \raisebox{-.45\height}{\includegraphics[height=0.18 \textwidth,trim=0 -5 01 -5]{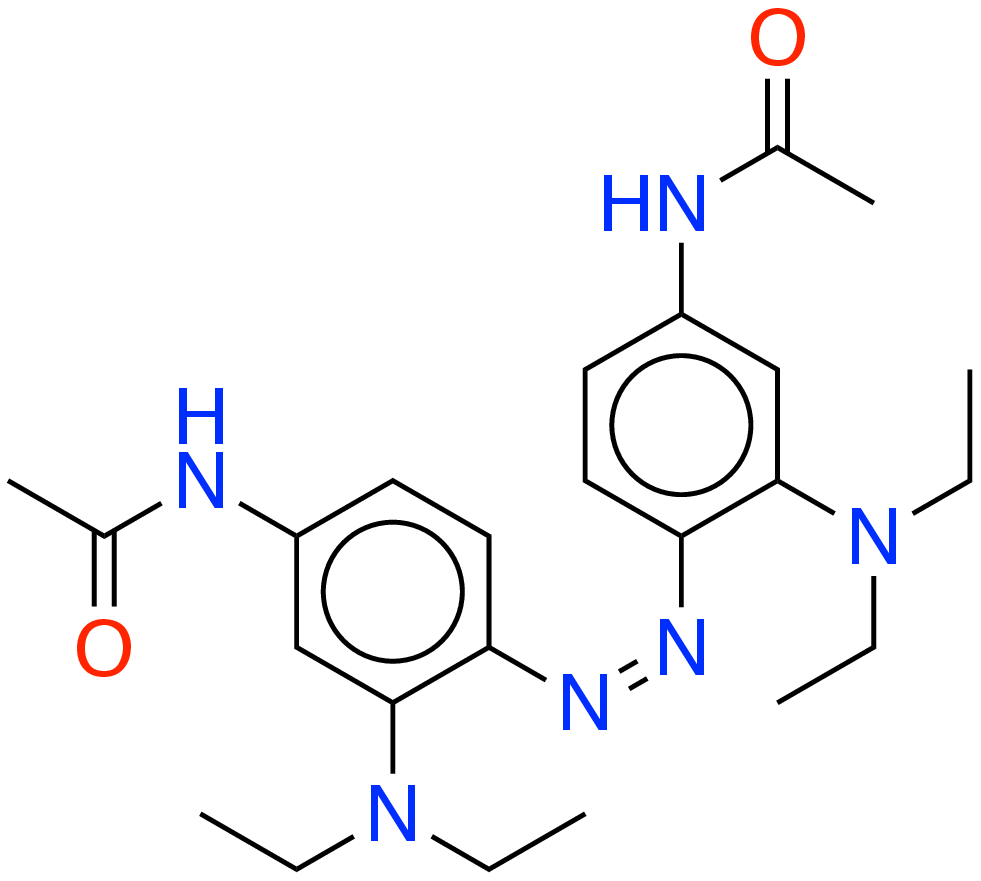}} & \cite{sadovski2009spectral}  \\ 
\hline

\textbf{89} & CCOc1ccc(/N=N/c2ccccc2)cc1 & \raisebox{-.45\height}{\includegraphics[height=0.09 \textwidth,trim=0 -5 01 -5]{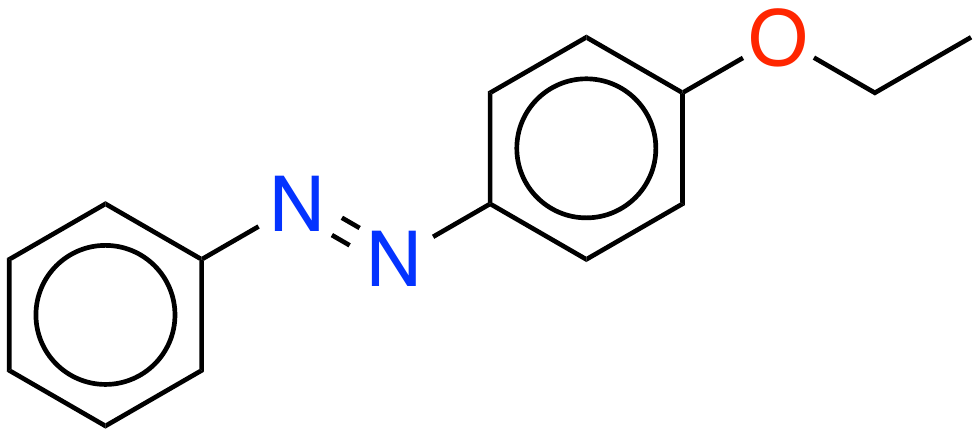}} & \cite{birnbaum1954photo}  \\ 
\hline

\textbf{90} & CCOc1ccc(/N=N{\textbackslash}c2ccccc2)cc1 & \raisebox{-.45\height}{\includegraphics[height=0.09 \textwidth,trim=0 -5 01 -5]{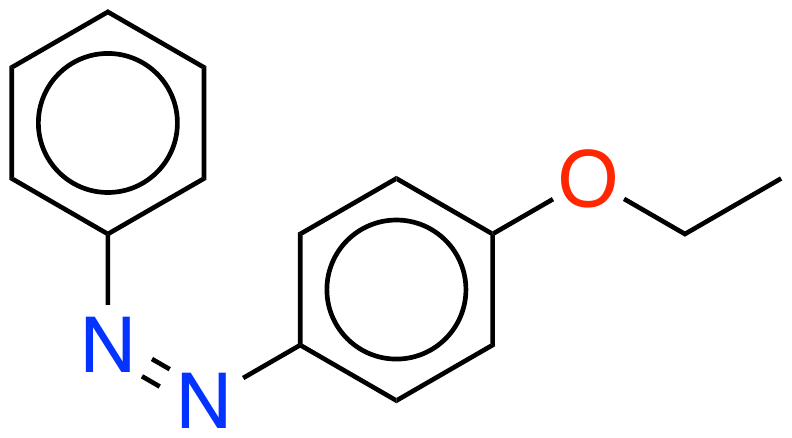}} & \cite{birnbaum1954photo}  \\ 
\hline

\textbf{91} & CC(=O)N1Cc2ccccc2/N=N/c2ccccc21 & \raisebox{-.45\height}{\includegraphics[height=0.1 \textwidth,trim=0 -5 01 -5]{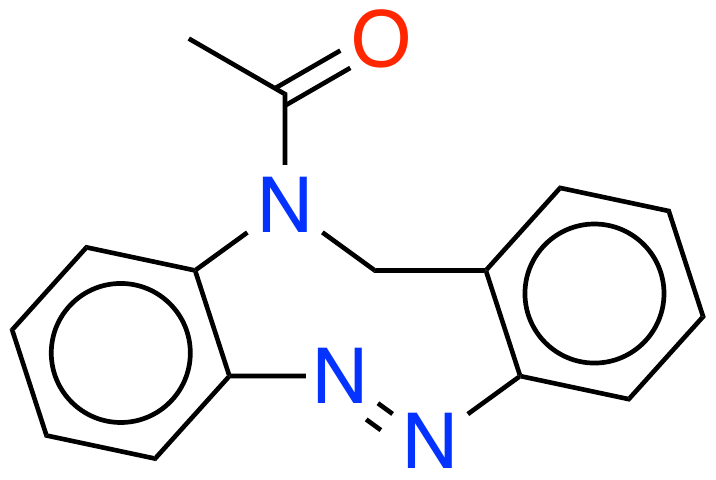}} & \cite{lentes2019nitrogen}  \\ 
\hline

\textbf{92} & CC(=O)N1Cc2ccccc2/N=N{\textbackslash}c2ccccc21 & \raisebox{-.45\height}{\includegraphics[height=0.1 \textwidth,trim=0 -5 01 -5]{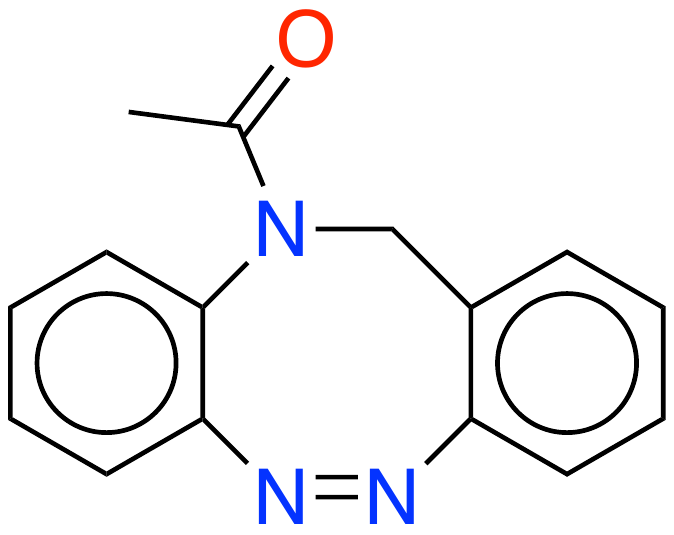}} & \cite{lentes2019nitrogen}  \\ 
\hline

\textbf{93} & CC(=O)Nc1ccc2c(c1)CCc1cc (NC(C)=O)ccc1/N=N/2 & \raisebox{-.45\height}{\includegraphics[height=0.1 \textwidth,trim=0 -5 01 -5]{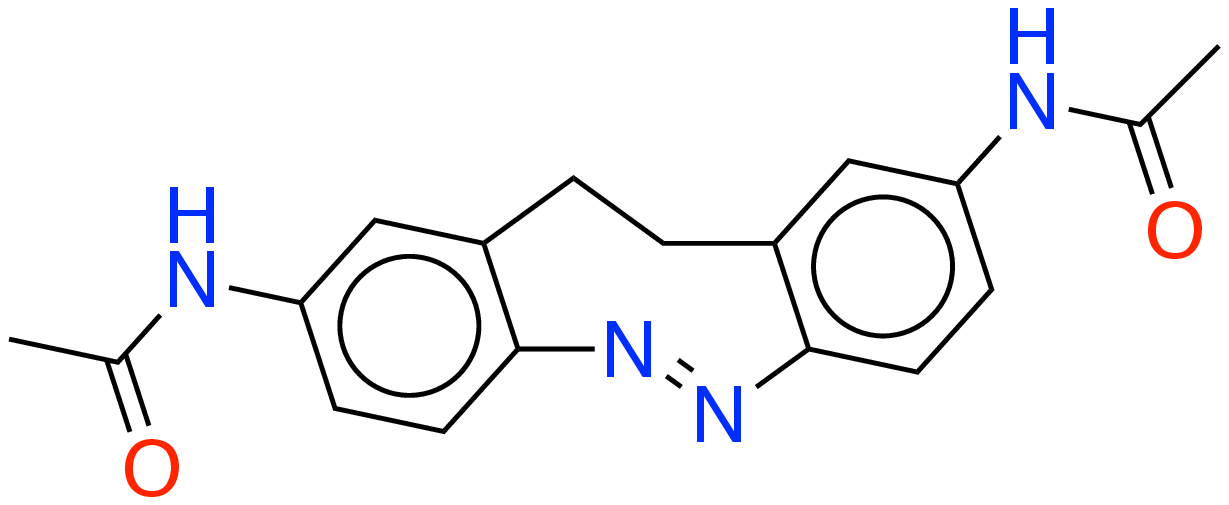}} & \cite{samanta2012bidirectional}  \\ 
\hline

\textbf{94} & CC(=O)Nc1ccc2c(c1)CCc1cc (NC(C)=O)ccc1/N=N{\textbackslash}2 & \raisebox{-.45\height}{\includegraphics[height=0.085 \textwidth,trim=0 -5 01 -5]{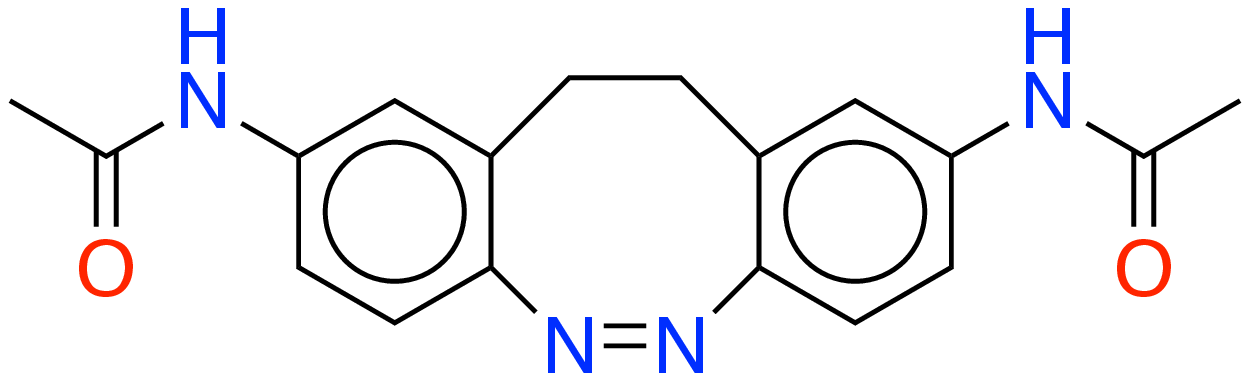}} & \cite{samanta2012bidirectional}  \\ 
\hline

\textbf{95} & CC(=O)Nc1cc(Cl)c(/N=N/c2c (Cl)cc(NC(C)=O)cc2Cl)c(Cl)c1 & \raisebox{-.45\height}{\includegraphics[height=0.125 \textwidth,trim=0 -5 01 -5]{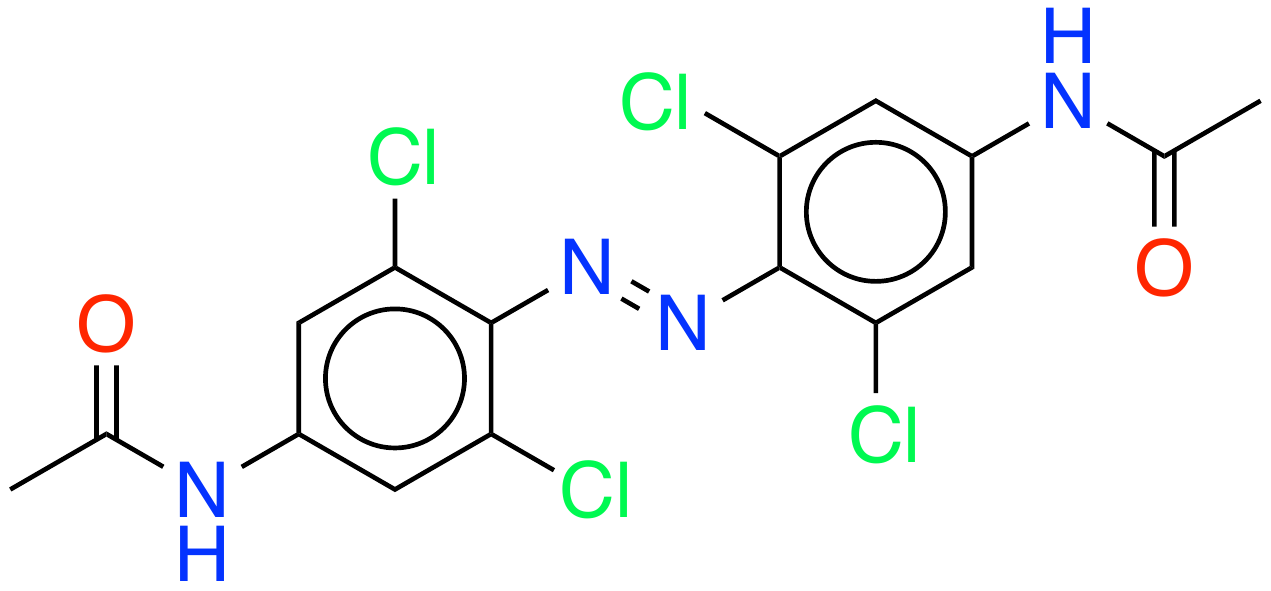}} & \cite{samanta2013photoswitching}  \\ 
\hline

\textbf{96} & CC(=O)Nc1cc(Cl)c(/N=N{\textbackslash}c2c (Cl)cc(NC(C)=O)cc2Cl)c(Cl)c1 & \raisebox{-.45\height}{\includegraphics[height=0.17 \textwidth,trim=0 -5 01 -5]{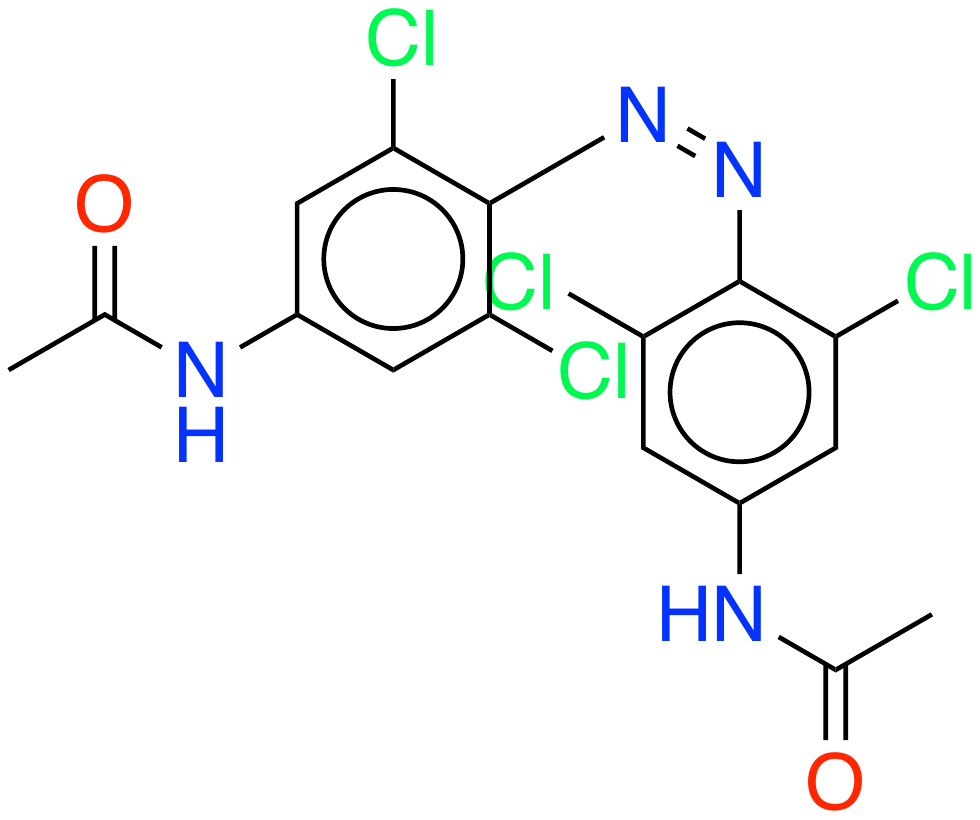}} & \cite{samanta2013photoswitching}  \\ 
\hline

\textbf{97} & CC(=O)Nc1ccc(/N=N/c2ccc(NC(C)=O)cc2)cc1 & \raisebox{-.45\height}{\includegraphics[height=0.125 \textwidth,trim=0 -5 01 -5]{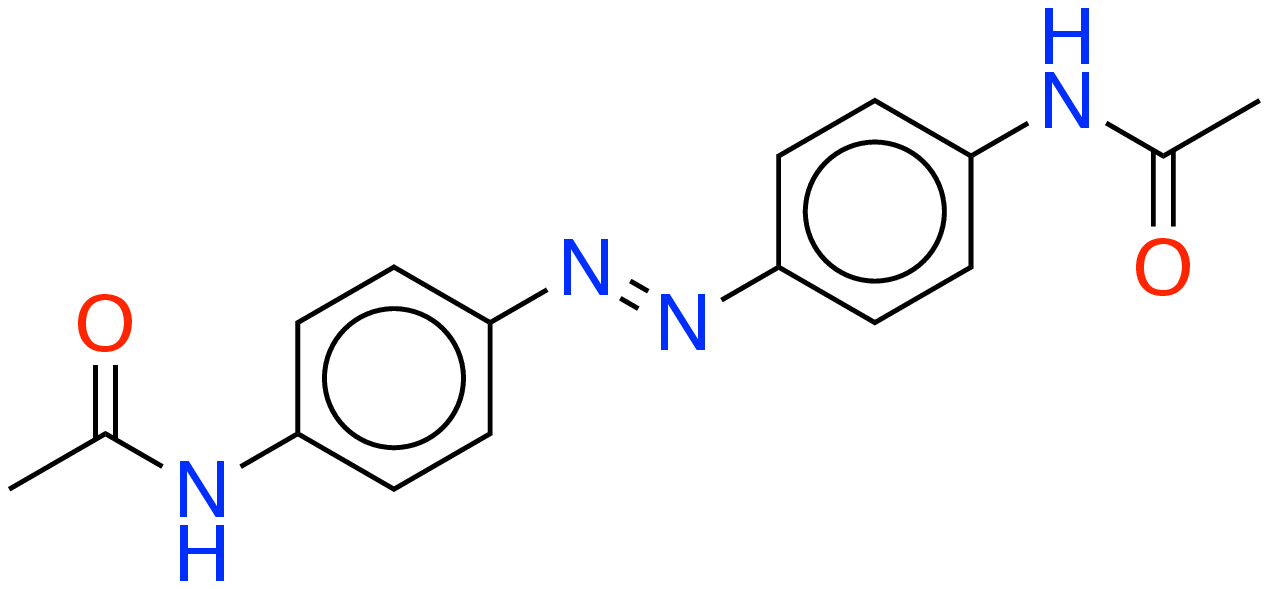}} & \cite{gascon2015optimized}  \\ 
\hline

\textbf{98} & CC(=O)Nc1ccc(/N=N{\textbackslash}c2ccc(NC(C)=O)cc2)cc1 & \raisebox{-.45\height}{\includegraphics[height=0.16 \textwidth,trim=0 -5 01 -5]{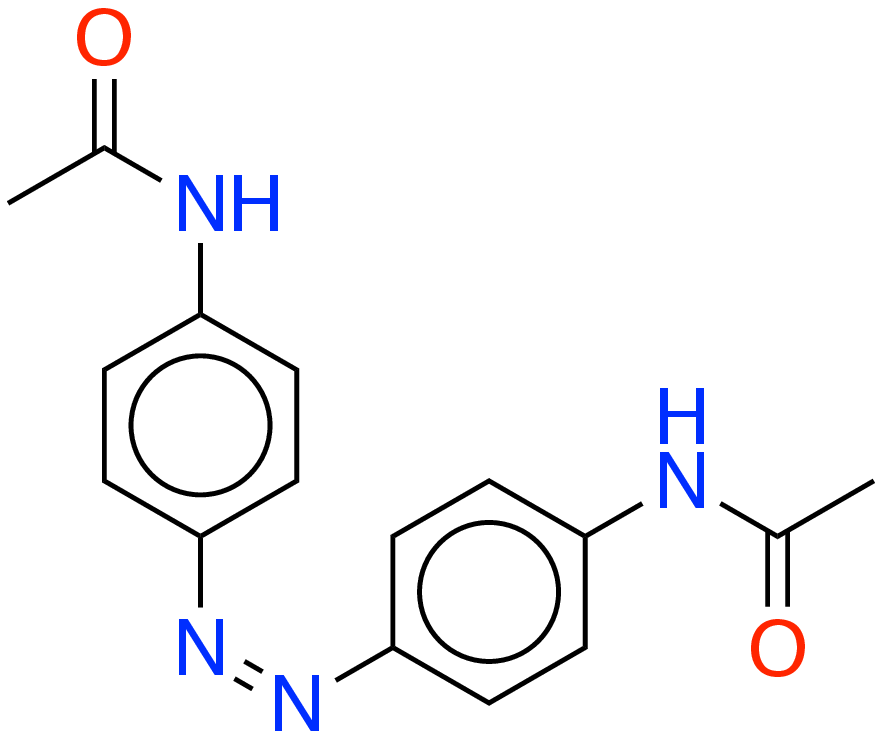}} & \cite{gascon2015optimized}  \\ 
\hline

\textbf{99} & CC(=O)Nc1ccc(/N=N/c2ccc (NC(C)=O)cc2N2CCCC2)c(N2CCCC2)c1 & \raisebox{-.45\height}{\includegraphics[height=0.15 \textwidth,trim=0 -5 01 -5]{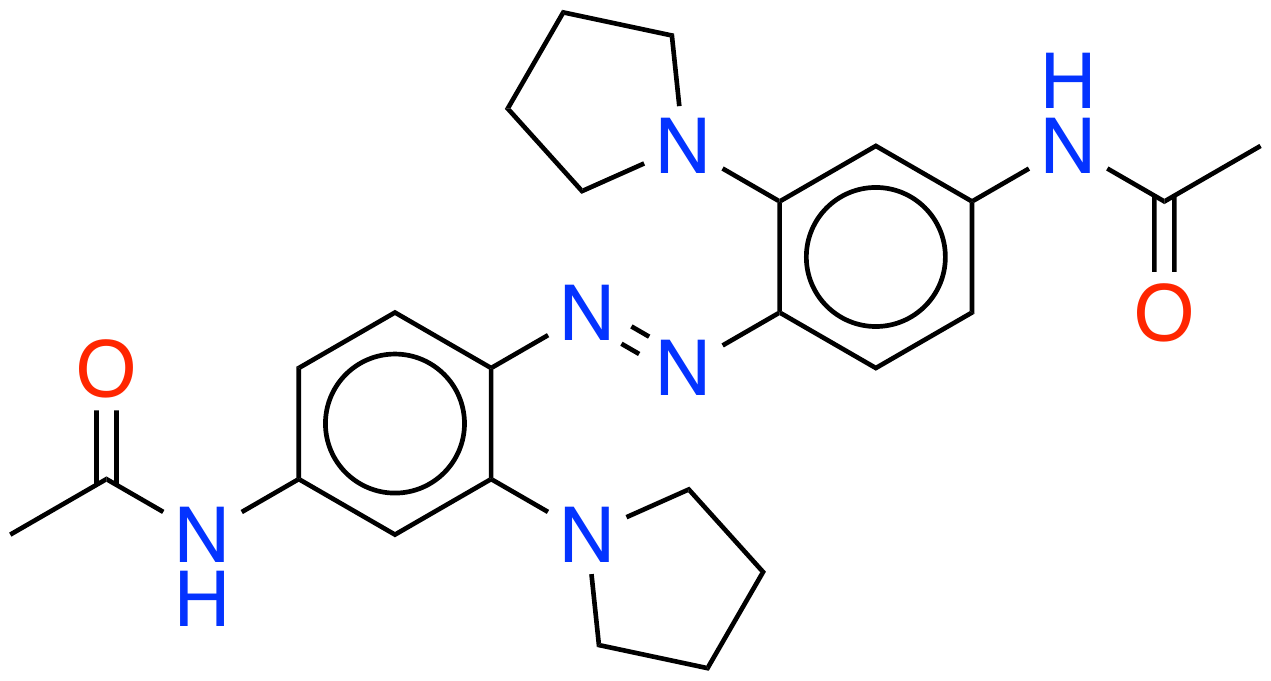}} & \cite{sadovski2009spectral}  \\ 
\hline

\textbf{100} & CC(=O)Nc1ccc(/N=N{\textbackslash}c2ccc (NC(C)=O)cc2N2CCCC2)c(N2CCCC2)c1 & \raisebox{-.45\height}{\includegraphics[height=0.18 \textwidth,trim=0 -5 01 -5]{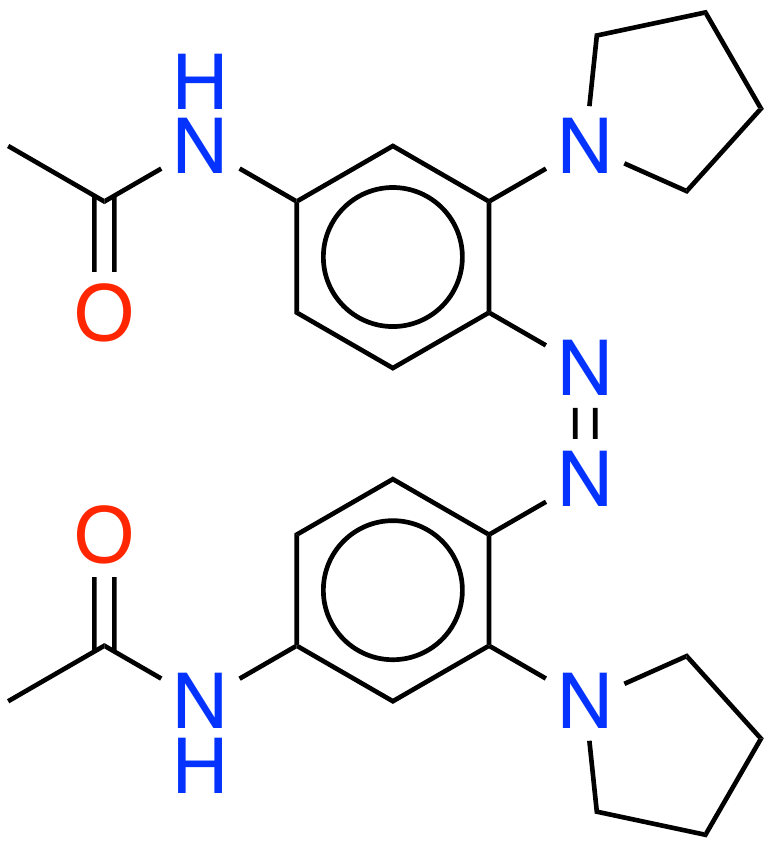}} & \cite{sadovski2009spectral}  \\ 
\hline

\textbf{101} & CC(=O)Nc1ccc(/N=N/c2ccc (NC(C)=O)cc2N2CCCCC2)c(N2CCCCC2)c1 & \raisebox{-.45\height}{\includegraphics[height=0.16 \textwidth,trim=0 -5 01 -5]{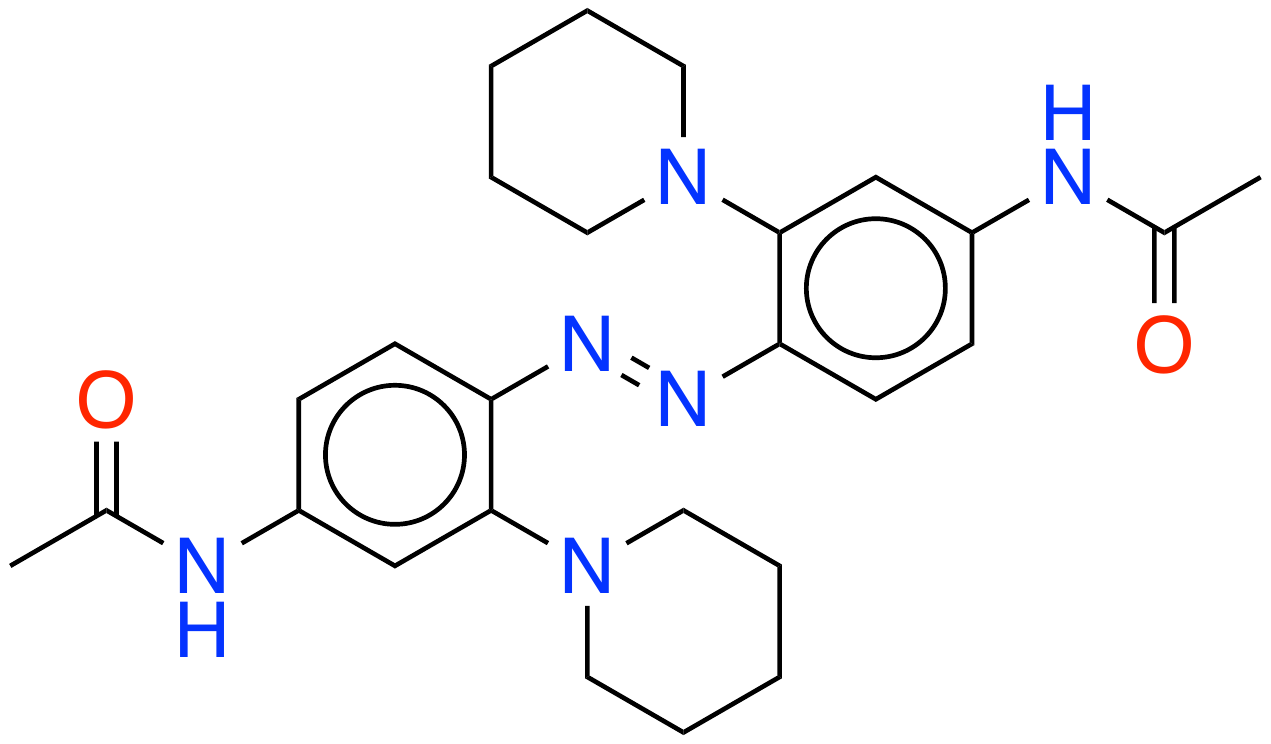}} & \cite{sadovski2009spectral}  \\ 
\hline

\textbf{102} & CC(=O)Nc1ccc(/N=N{\textbackslash}c2ccc (NC(C)=O)cc2N2CCCCC2)c(N2CCCCC2)c1 & \raisebox{-.45\height}{\includegraphics[height=0.18 \textwidth,trim=0 -5 01 -5]{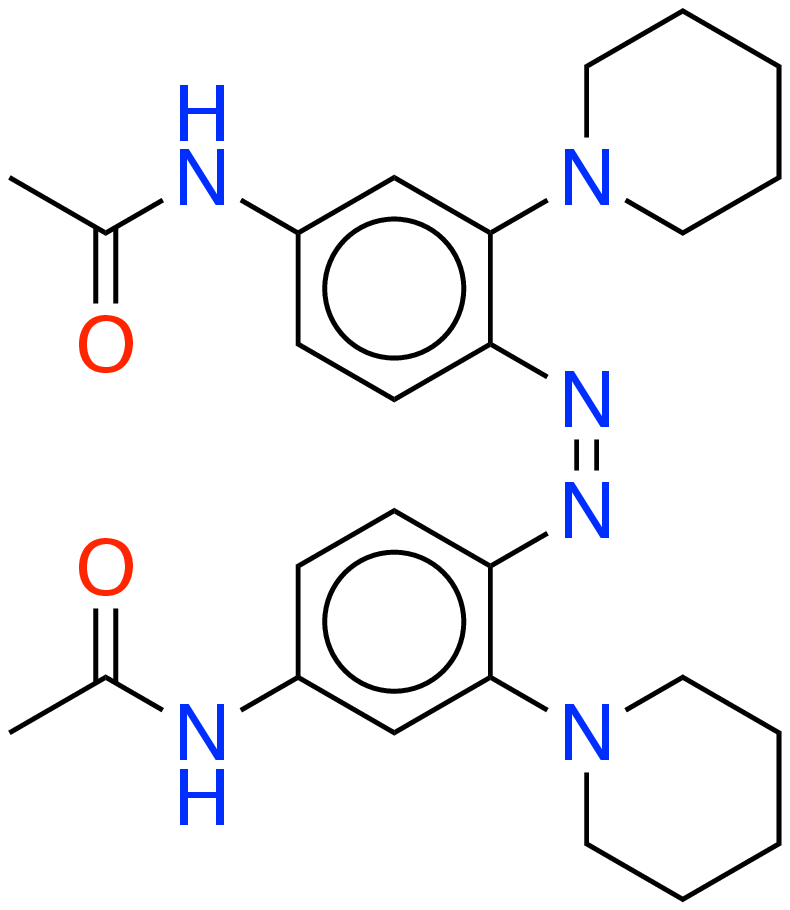}} & \cite{sadovski2009spectral}  \\ 
\hline

\textbf{103} & CC(=O)Nc1ccc(/N=N/c2ccc (NC(C)=O)cc2N2CCN(C)CC2) c(N2CCN(C)CC2)c1 & \raisebox{-.45\height}{\includegraphics[height=0.15 \textwidth,trim=0 -5 01 -5]{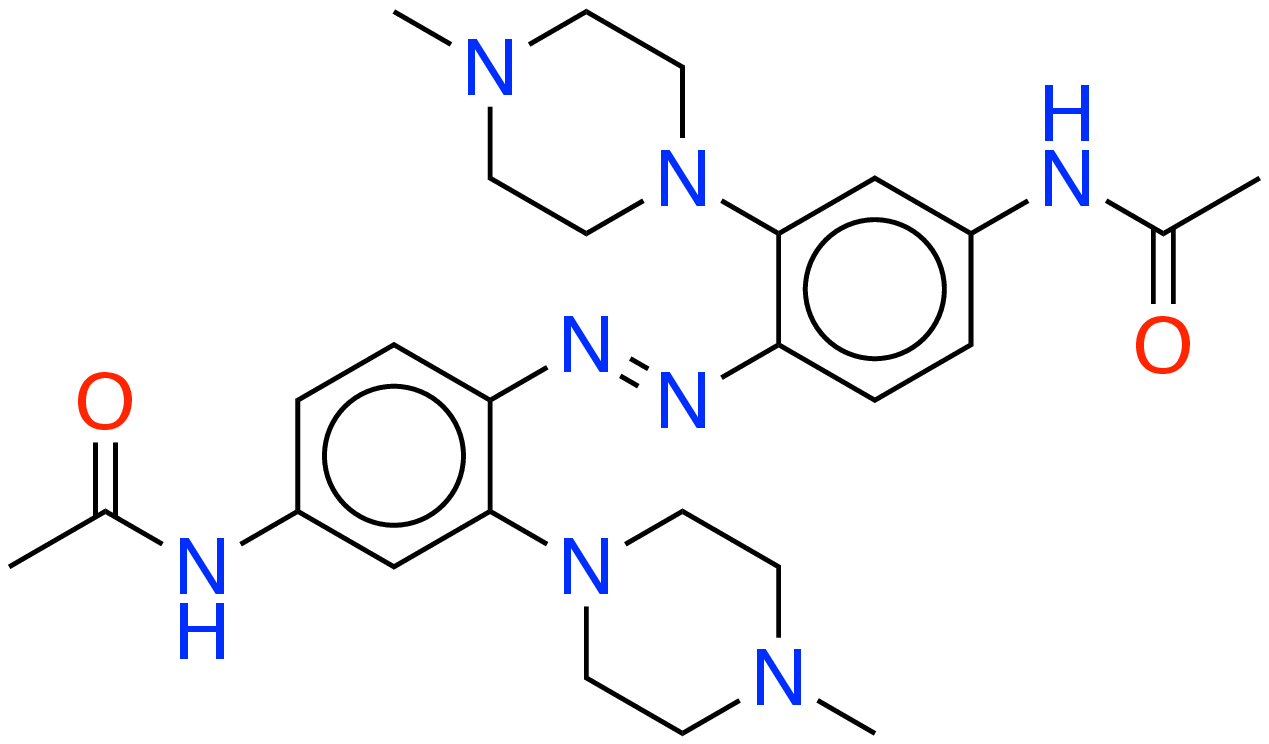}} & \cite{sadovski2009spectral}  \\ 
\hline

\textbf{104} & CC(=O)Nc1ccc(/N=N{\textbackslash}c2ccc (NC(C)=O)cc2N2CCN(C)CC2) c(N2CCN(C)CC2)c1 & \raisebox{-.45\height}{\includegraphics[height=0.18 \textwidth,trim=0 -5 01 -5]{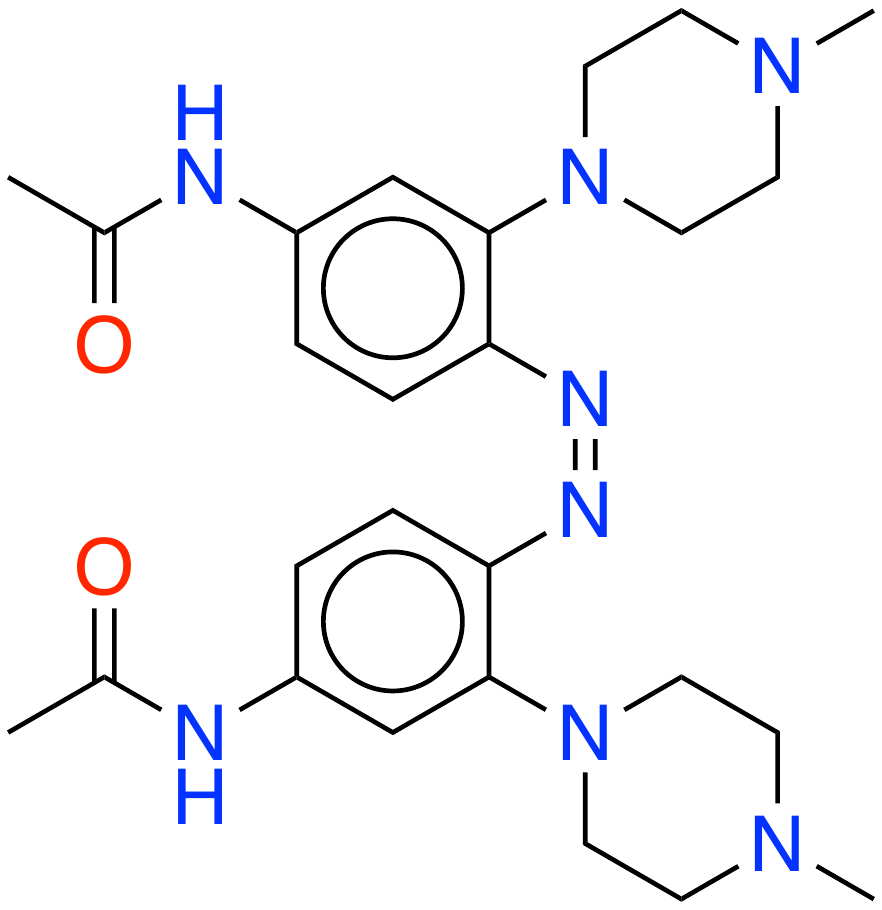}} & \cite{sadovski2009spectral}  \\ 
\hline

\textbf{105} & CC(=O)Nc1ccc(/N=N/c2ccc (NC(C)=O)cc2N2CCOCC2)c(N2CCOCC2)c1 & \raisebox{-.45\height}{\includegraphics[height=0.15 \textwidth,trim=0 -5 01 -5]{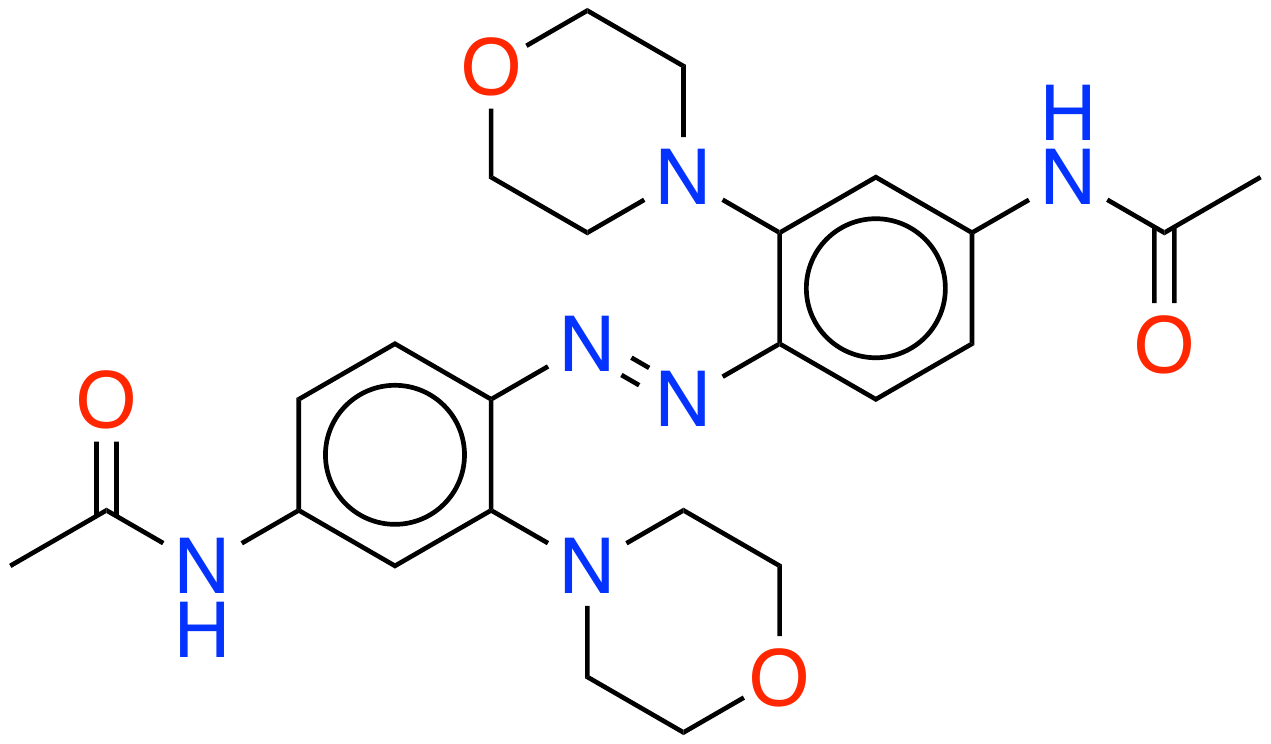}} & \cite{sadovski2009spectral}  \\ 
\hline

\textbf{106} & CC(=O)Nc1ccc(/N=N{\textbackslash}c2ccc (NC(C)=O)cc2N2CCOCC2)c(N2CCOCC2)c1 & \raisebox{-.45\height}{\includegraphics[height=0.185 \textwidth,trim=0 -5 01 -5]{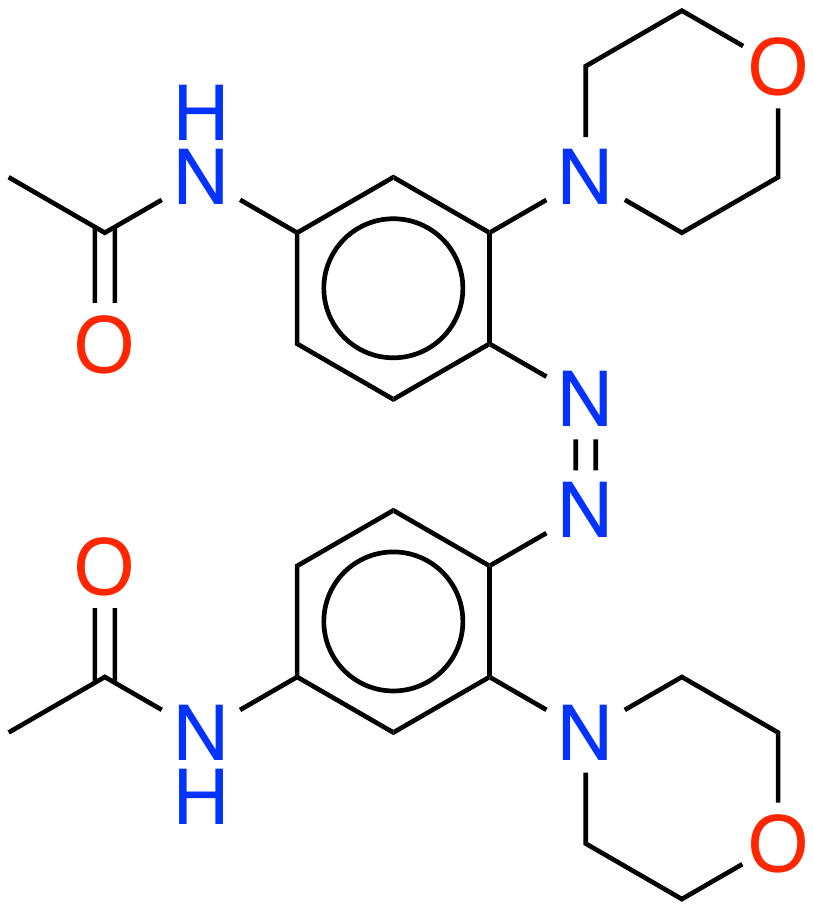}} & \cite{sadovski2009spectral}  \\ 
\hline

\textbf{107} & CC(=O)Nc1ccc(/N=N/c2ccc(NC (C)=O)cc2N(C)C)c(N(C)C)c1
& \raisebox{-.45\height}{\includegraphics[height=0.13 \textwidth,trim=0 -5 01 -5]{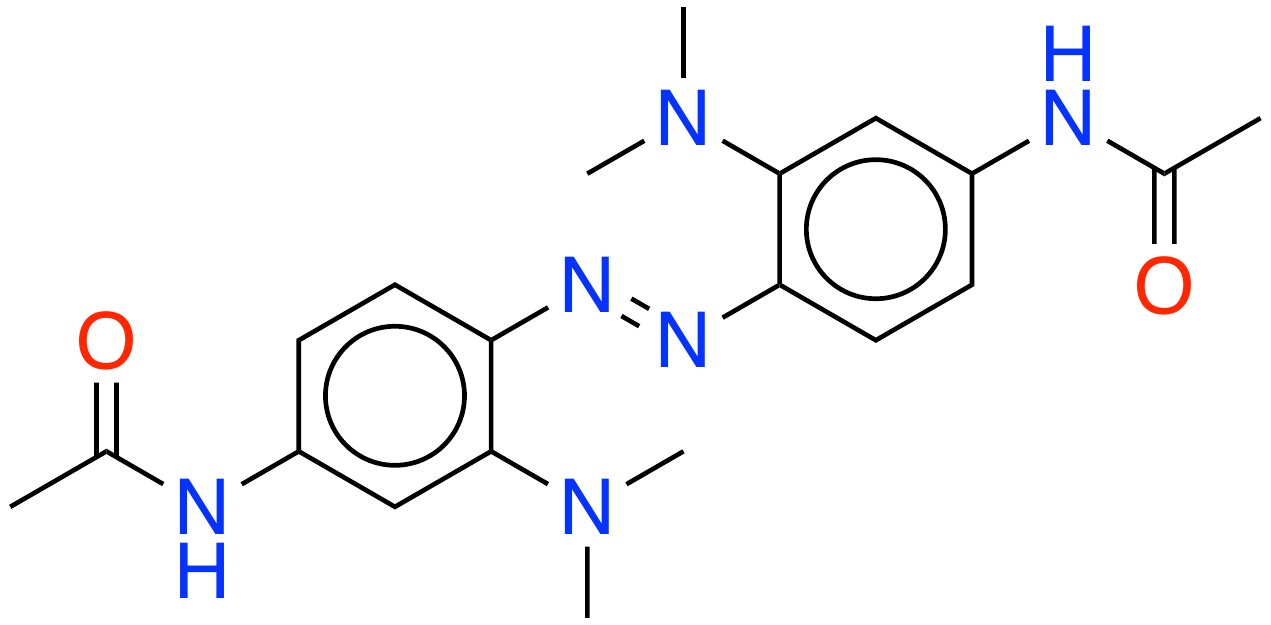}} & \cite{sadovski2009spectral}  \\ 
\hline

\textbf{108} & CC(=O)Nc1ccc(/N=N{\textbackslash}c2ccc(NC (C)=O)cc2N(C)C)c(N(C)C)c1 & \raisebox{-.45\height}{\includegraphics[height=0.16 \textwidth,trim=0 -5 01 -5]{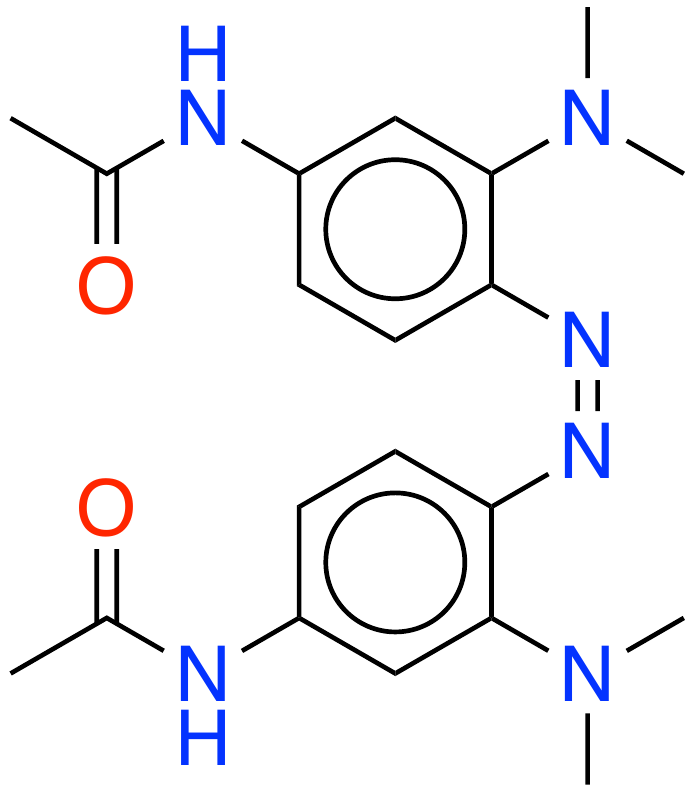}} & \cite{sadovski2009spectral}  \\ 
\hline

\textbf{109} & Clc1ccc(/N=N/c2ccccc2)cc1 & \raisebox{-.45\height}{\includegraphics[height=0.09 \textwidth,trim=0 -5 01 -5]{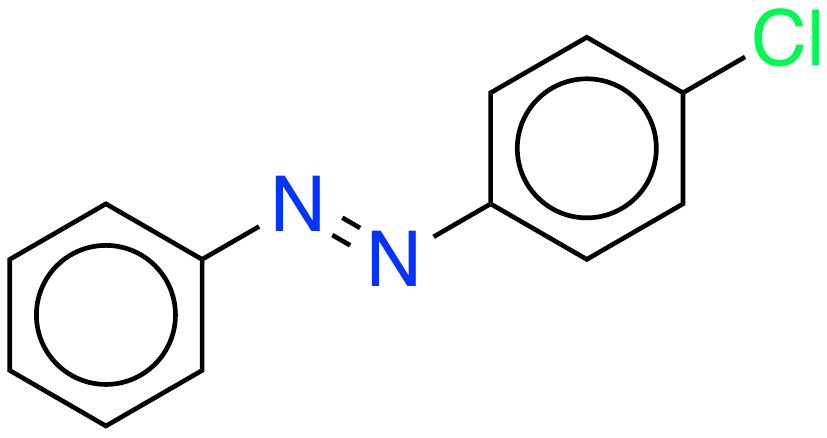}} & \cite{birnbaum1954photo}  \\ 
\hline

\textbf{110} & Clc1ccc(/N=N{\textbackslash}c2ccccc2)cc1 & \raisebox{-.45\height}{\includegraphics[height=0.09 \textwidth,trim=0 -5 01 -5]{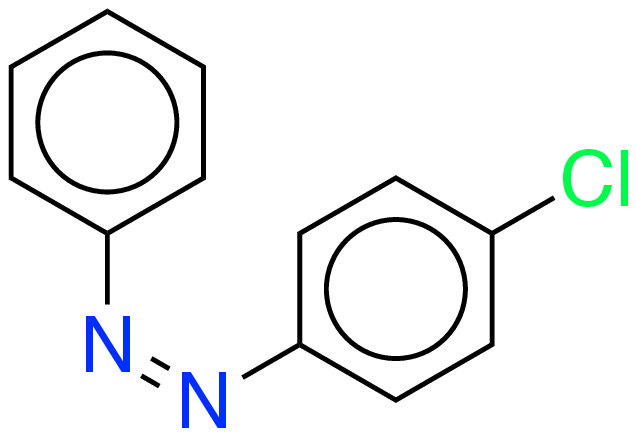}} & \cite{birnbaum1954photo}  \\ 
\hline

\textbf{111} & CN1Cc2ccccc2/N=N/c2ccccc21 & \raisebox{-.45\height}{\includegraphics[height=0.085 \textwidth,trim=0 -5 01 -5]{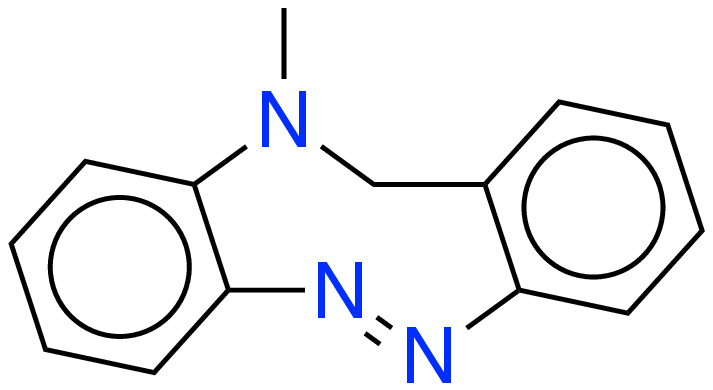}} & \cite{lentes2019nitrogen}  \\ 
\hline

\textbf{112} & CN1Cc2ccccc2/N=N{\textbackslash}c2ccccc21 & \raisebox{-.45\height}{\includegraphics[height=0.085 \textwidth,trim=0 -5 01 -5]{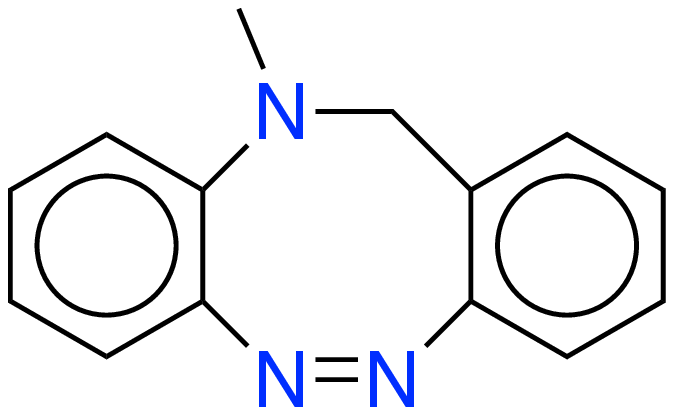}} & \cite{lentes2019nitrogen}  \\ 
\hline

\textbf{113} & CN(C)c1ccc(/N=N/c2ccccc2)cc1 & \raisebox{-.45\height}{\includegraphics[height=0.11 \textwidth,trim=0 -5 01 -5]{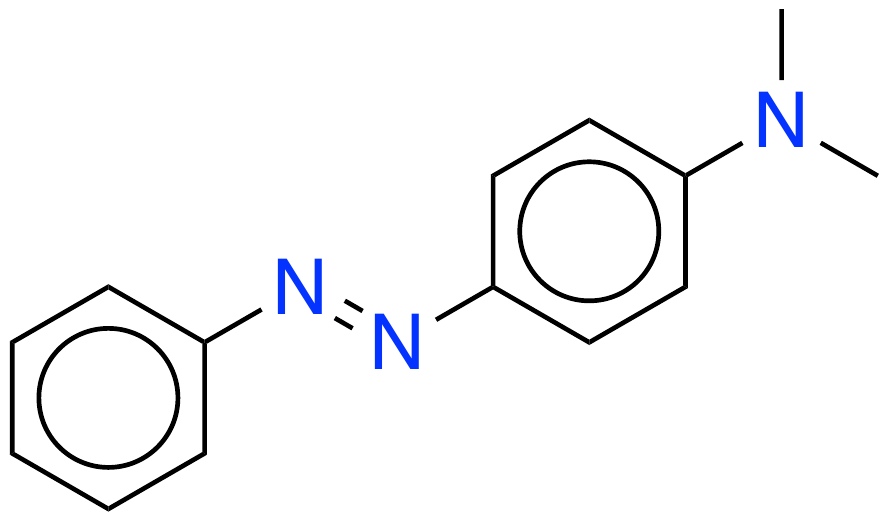}} & \cite{rau1990photochemical,albini1983photochemistry}  \\ 
\hline

\textbf{114} & CN(C)c1ccc(/N=N{\textbackslash}c2ccccc2)cc1 & \raisebox{-.45\height}{\includegraphics[height=0.09 \textwidth,trim=0 -5 01 -5]{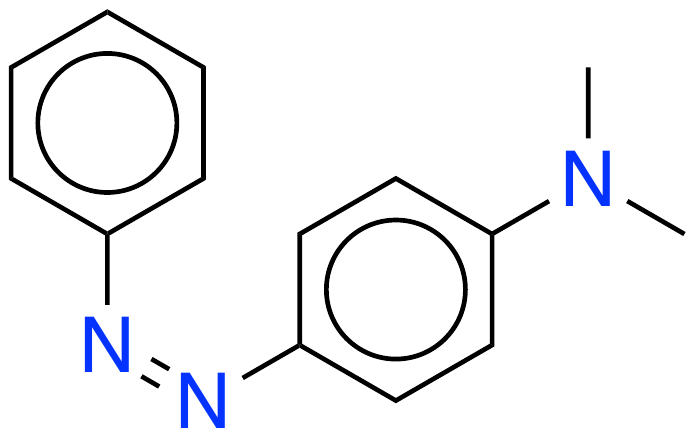}} & \cite{rau1990photochemical,albini1983photochemistry}  \\ 
\hline

\textbf{115} & CN(C)c1ccc(/N=N/c2cccc([N+](=O)[O-])c2)cc1 & \raisebox{-.45\height}{\includegraphics[height=0.11 \textwidth,trim=0 -5 01 -5]{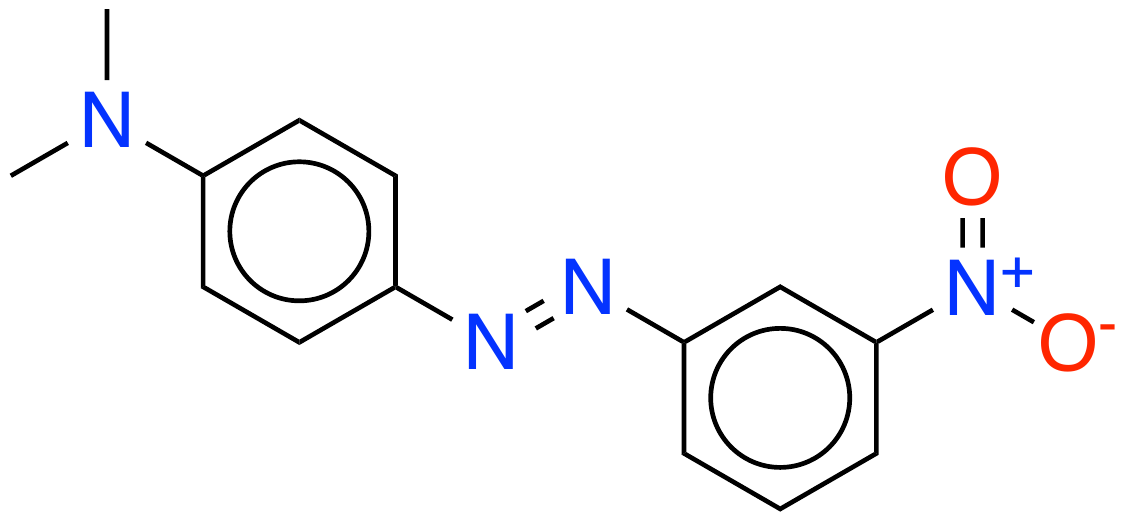}} & \cite{olmsted1983photochemical}  \\ 
\hline

\textbf{116} & CN(C)c1ccc(/N=N{\textbackslash}c2cccc([N+](=O)[O-])c2)cc1 & \raisebox{-.45\height}{\includegraphics[height=0.14 \textwidth,trim=0 -5 01 -5]{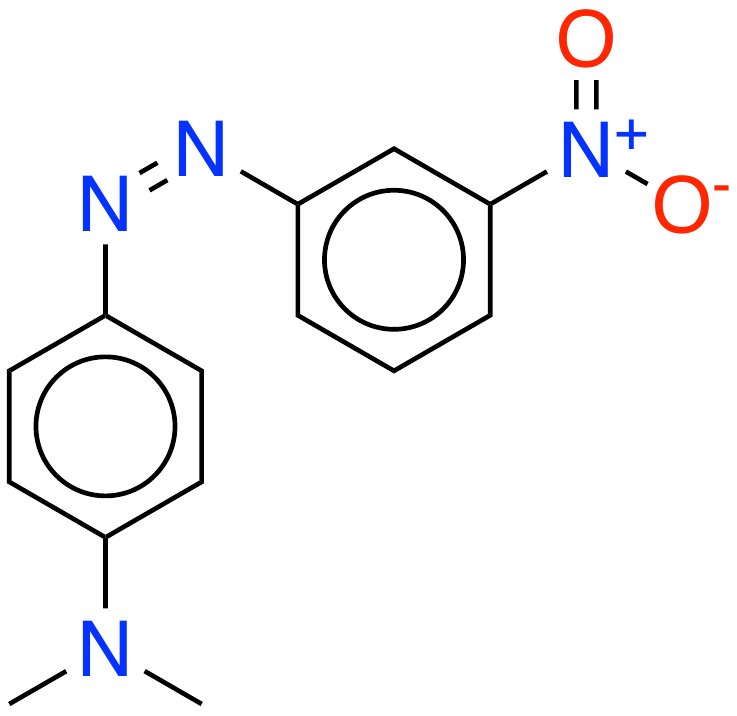}} & \cite{olmsted1983photochemical}  \\ 
\hline

\textbf{117} & CN(C)c1ccc(/N=N/c2ccc(C(=O)O)cc2)cc1 & \raisebox{-.45\height}{\includegraphics[height=0.13 \textwidth,trim=0 -5 01 -5]{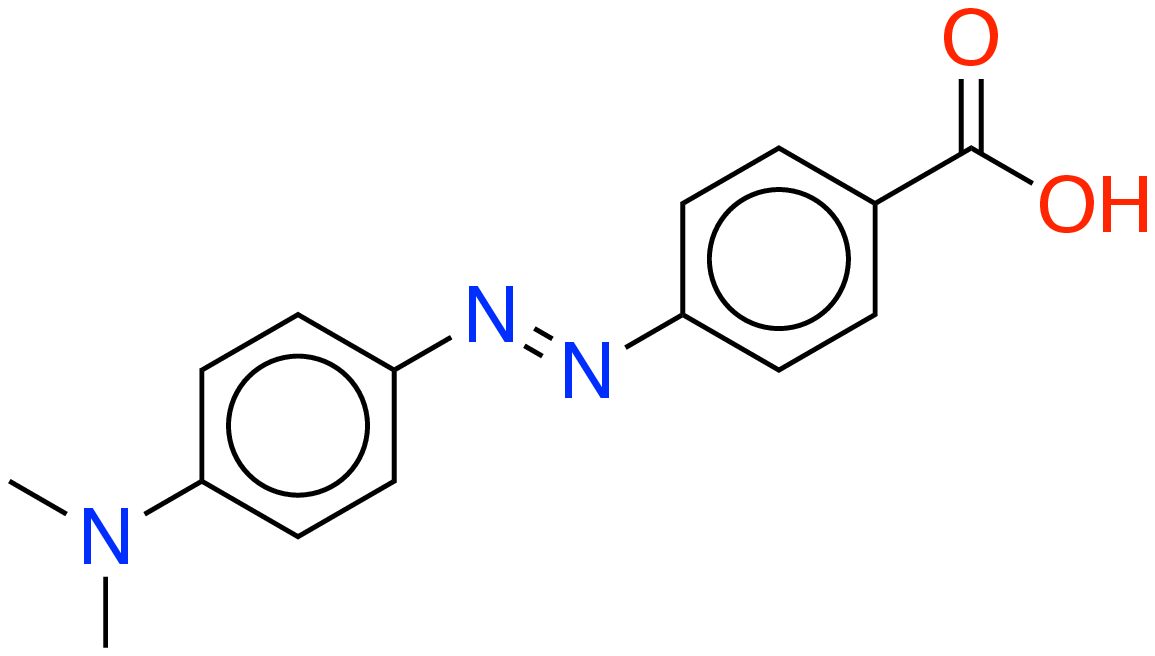}} & \cite{lee2009solvent}  \\ 
\hline

\textbf{118} & CN(C)c1ccc(/N=N{\textbackslash}c2ccc(C(=O)O)cc2)cc1 & \raisebox{-.45\height}{\includegraphics[height=0.12 \textwidth,trim=0 -5 01 -5]{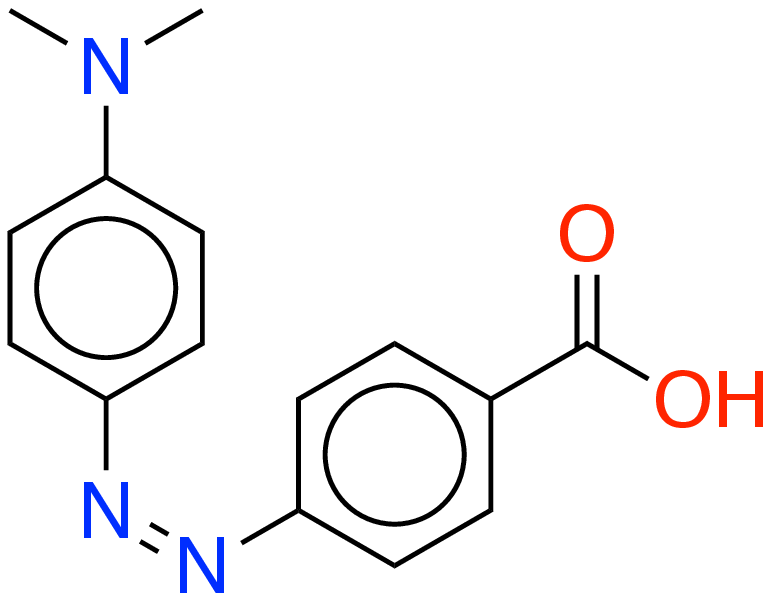}} & \cite{lee2009solvent}  \\ 
\hline

\textbf{119} & CN(C)c1ccc(/N=N/c2ccc([N+](=O)[O-])cc2)cc1 & \raisebox{-.45\height}{\includegraphics[height=0.13 \textwidth,trim=0 -5 01 -5]{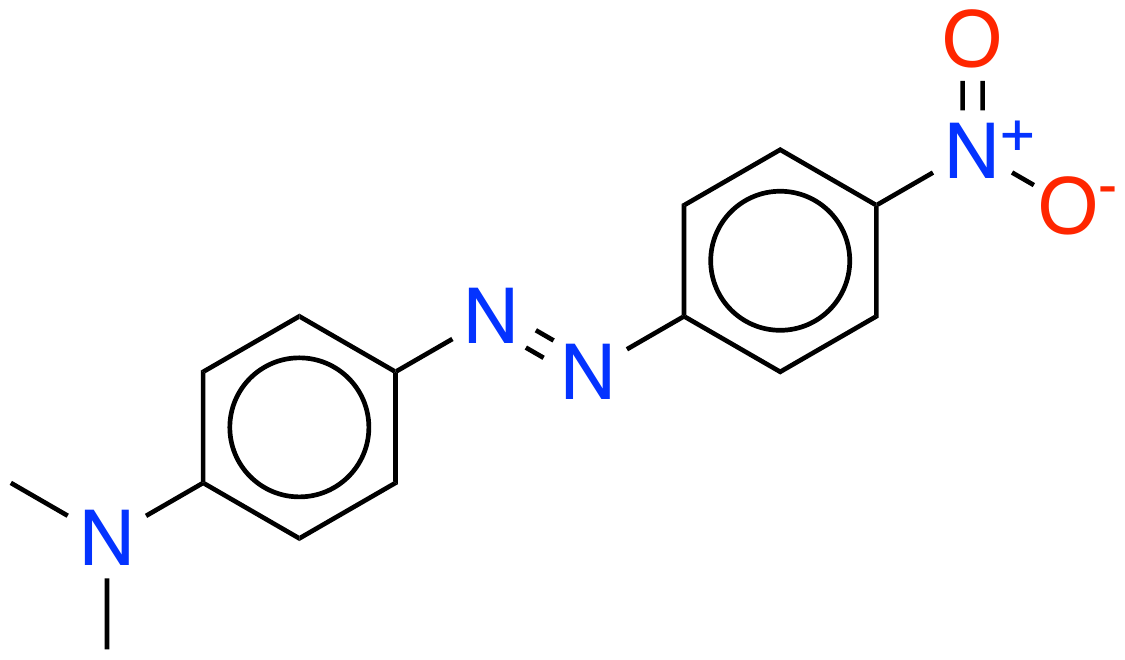}} & \cite{koller2007slower}  \\ 
\hline

\textbf{120} & CN(C)c1ccc(/N=N{\textbackslash}c2ccc([N+](=O)[O-])cc2)cc1 & \raisebox{-.45\height}{\includegraphics[height=0.12 \textwidth,trim=0 -5 01 -5]{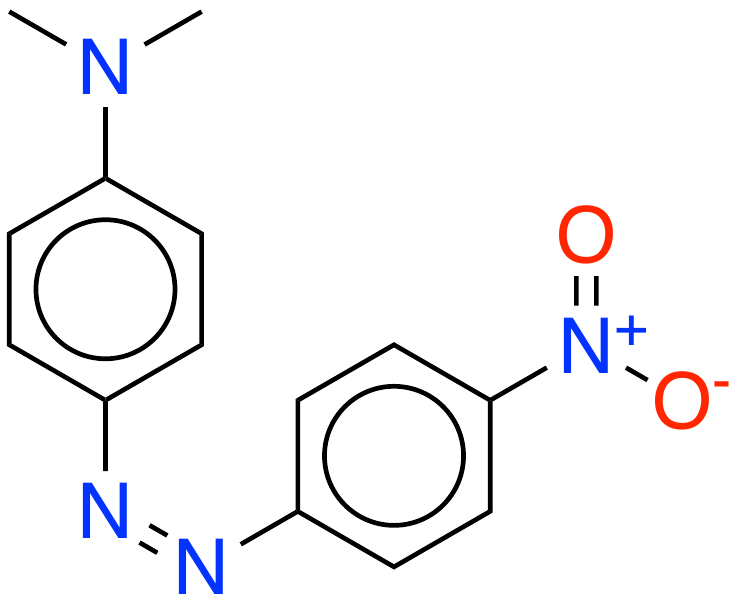}} & \cite{koller2007slower}  \\ 
\hline

\textbf{121} & CN(C)c1ccc(/N=N/c2ccc(S(=O)(=O)O)cc2)cc1 & \raisebox{-.45\height}{\includegraphics[height=0.13 \textwidth,trim=0 -5 01 -5]{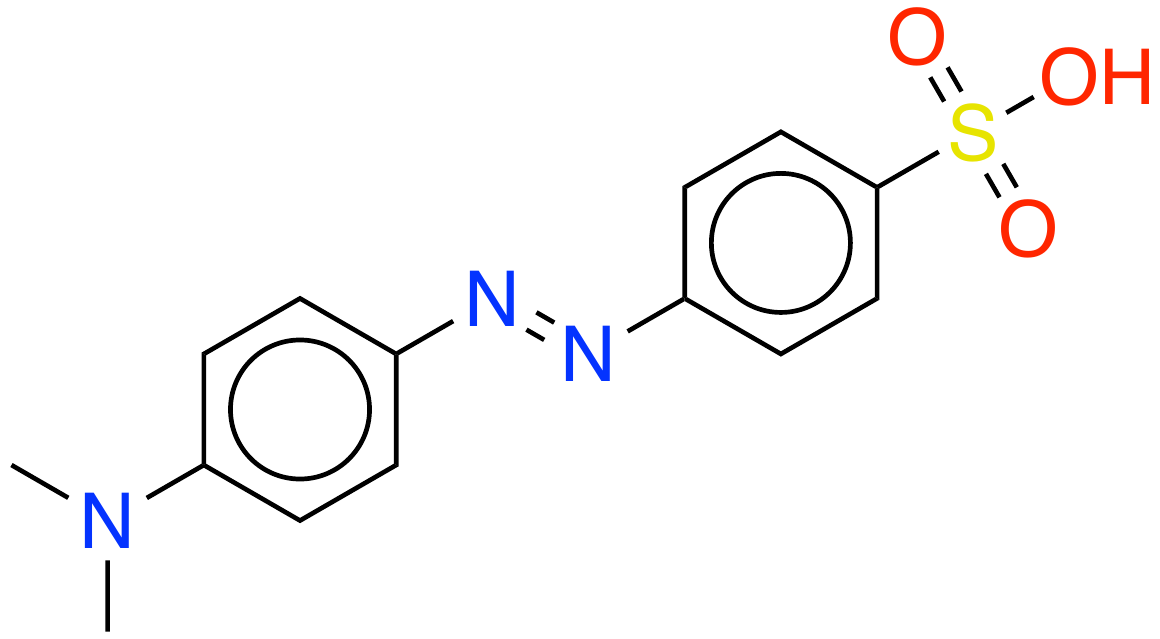}} & \cite{olmsted1983photochemical}  \\ 
\hline

\textbf{122} & CN(C)c1ccc(/N=N{\textbackslash}c2ccc(S(=O)(=O)O)cc2)cc1 & \raisebox{-.45\height}{\includegraphics[height=0.12 \textwidth,trim=0 -5 01 -5]{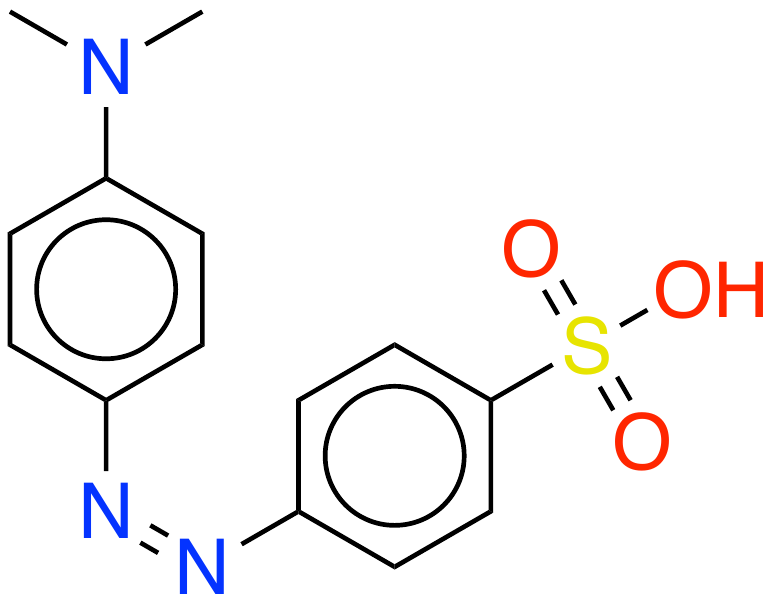}} & \cite{olmsted1983photochemical}  \\ 
\hline

\textbf{123} & C[N+](C)(C)c1ccc(/N=N/c2ccc ([N+](C)(C)C)cc2)cc1 & \raisebox{-.45\height}{\includegraphics[height=0.12 \textwidth,trim=0 -5 01 -5]{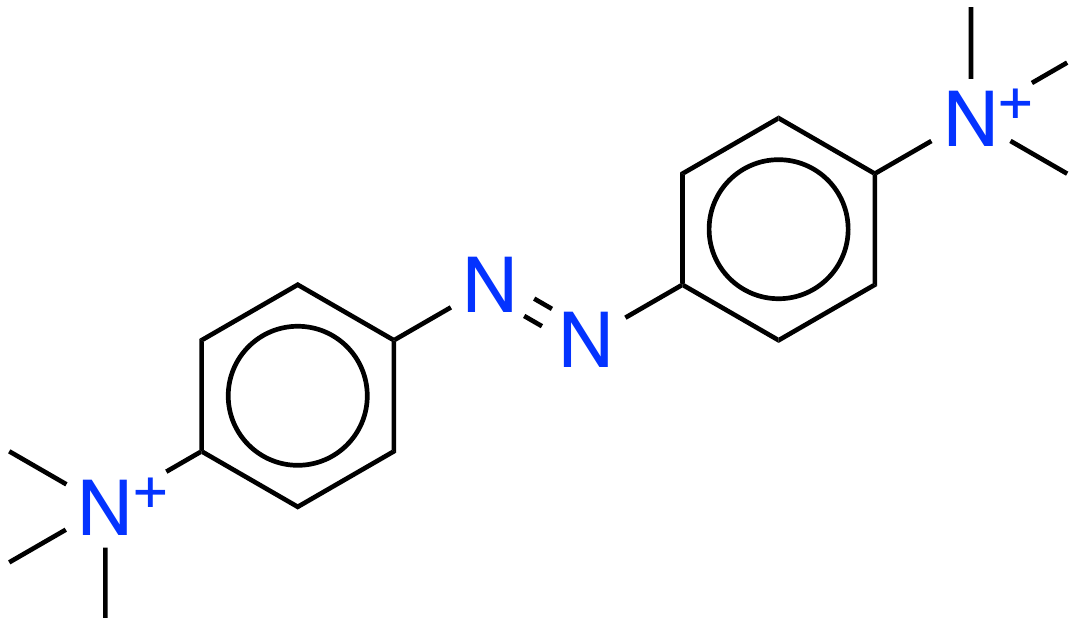}} & \cite{umemoto2014pinning}  \\ 
\hline

\textbf{124} & C[N+](C)(C)c1ccc(/N=N{\textbackslash}c2ccc ([N+](C)(C)C)cc2)cc1 & \raisebox{-.45\height}{\includegraphics[height=0.12 \textwidth,trim=0 -5 01 -5]{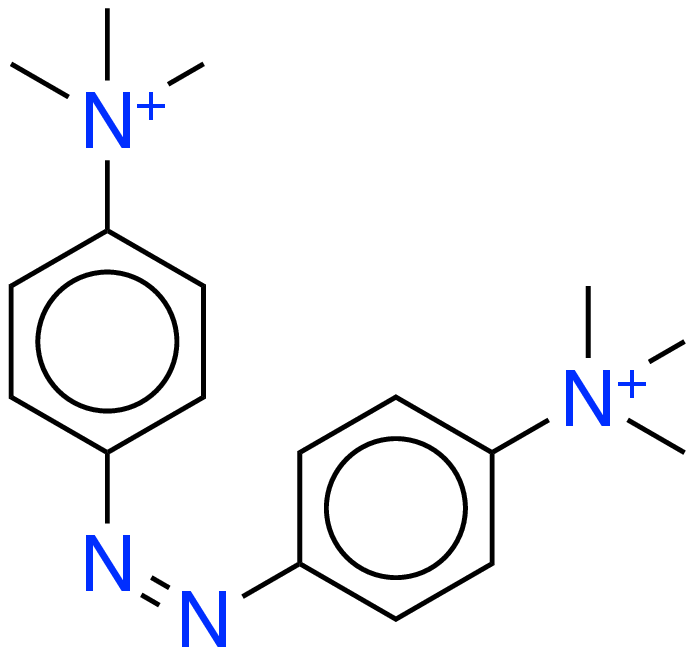}} & \cite{umemoto2014pinning}  \\ 
\hline

\textbf{125} & COc1ccccc1/N=N/c1ccccc1OC & \raisebox{-.45\height}{\includegraphics[height=0.095 \textwidth,trim=0 -5 01 -5]{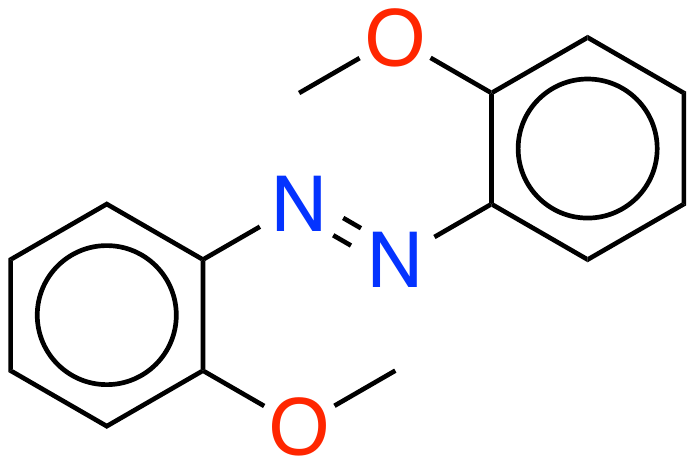}} & \cite{olmsted1983photochemical, tahara1997photoisomerization}  \\ 
\hline

\textbf{126} & COc1ccccc1/N=N{\textbackslash}c1ccccc1OC & \raisebox{-.45\height}{\includegraphics[height=0.115 \textwidth,trim=0 -5 01 -5]{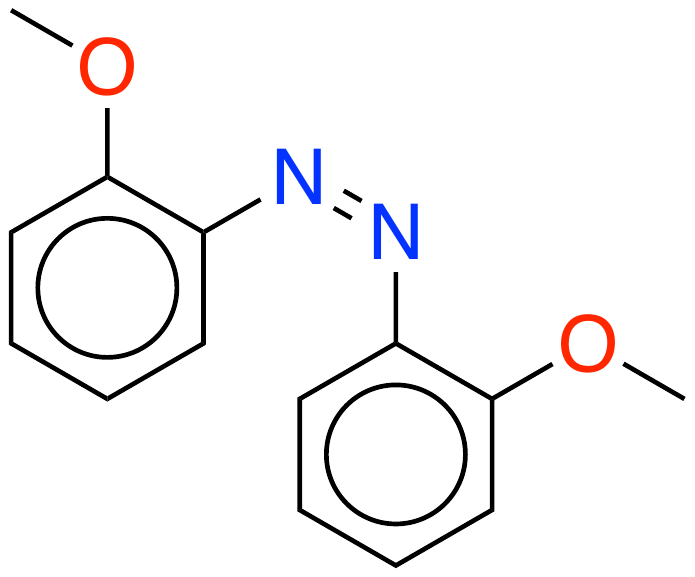}} & \cite{olmsted1983photochemical, tahara1997photoisomerization}  \\ 
\hline

\textbf{127} & COc1ccc(/N=N/c2ccc(CO)cc2)cc1 & \raisebox{-.45\height}{\includegraphics[height=0.09 \textwidth,trim=0 -5 01 -5]{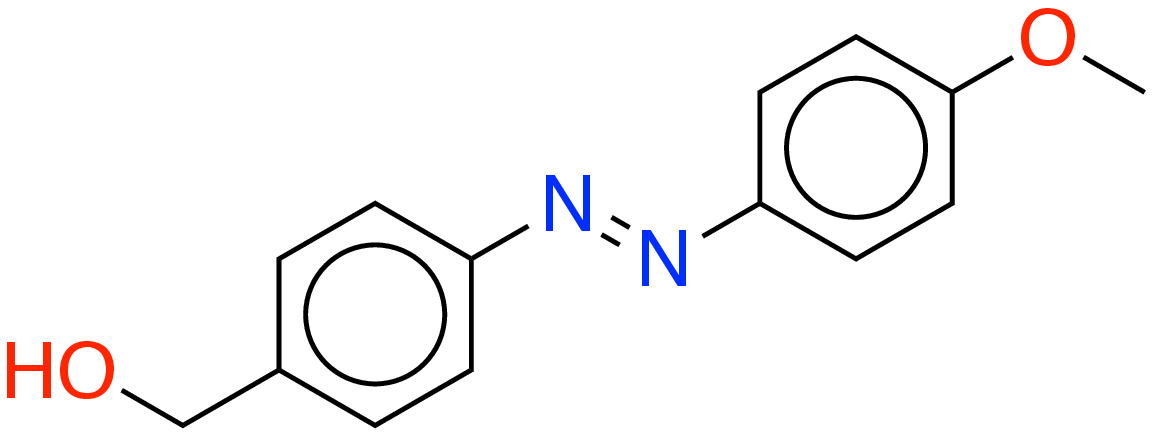}} & \cite{sierocki2006photoisomerization}  \\ 
\hline

\textbf{128} & COc1ccc(/N=N{\textbackslash}c2ccc(CO)cc2)cc1 & \raisebox{-.45\height}{\includegraphics[height=0.125 \textwidth,trim=0 -5 01 -5]{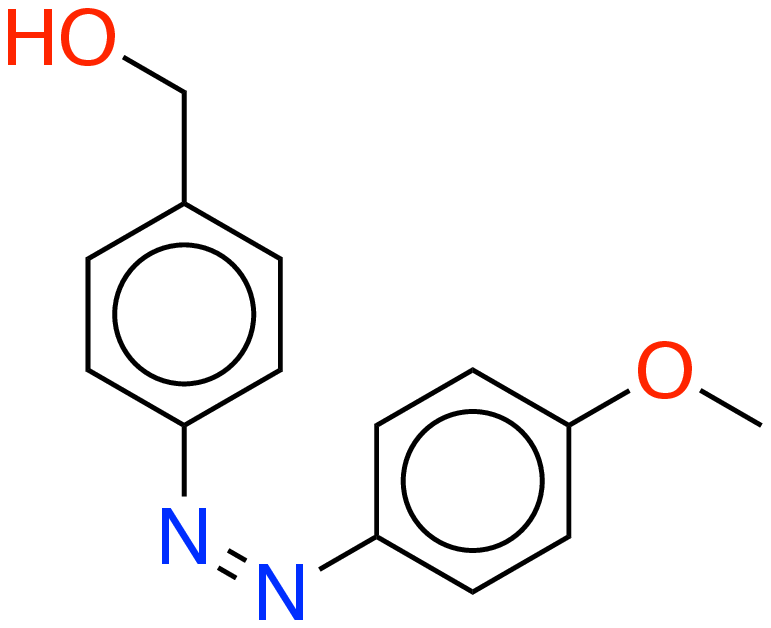}} & \cite{sierocki2006photoisomerization}  \\ 
\hline

\textbf{129} & COc1ccc(/N=N/c2ccc(OC)cc2)cc1 & \raisebox{-.45\height}{\includegraphics[height=0.1 \textwidth,trim=0 -5 01 -5]{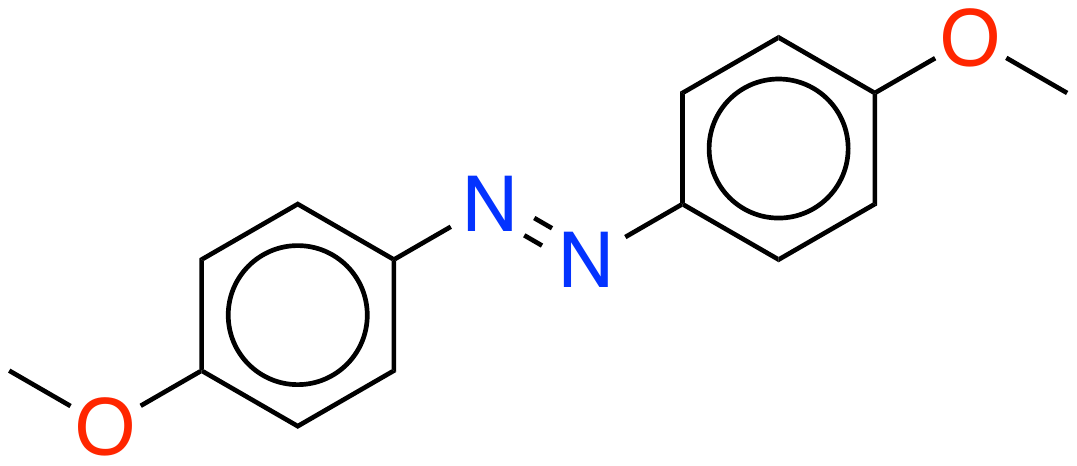}} & \cite{olmsted1983photochemical}  \\ 
\hline

\textbf{130} & COc1ccc(/N=N{\textbackslash}c2ccc(OC)cc2)cc1 & \raisebox{-.45\height}{\includegraphics[height=0.12 \textwidth,trim=0 -5 01 -5]{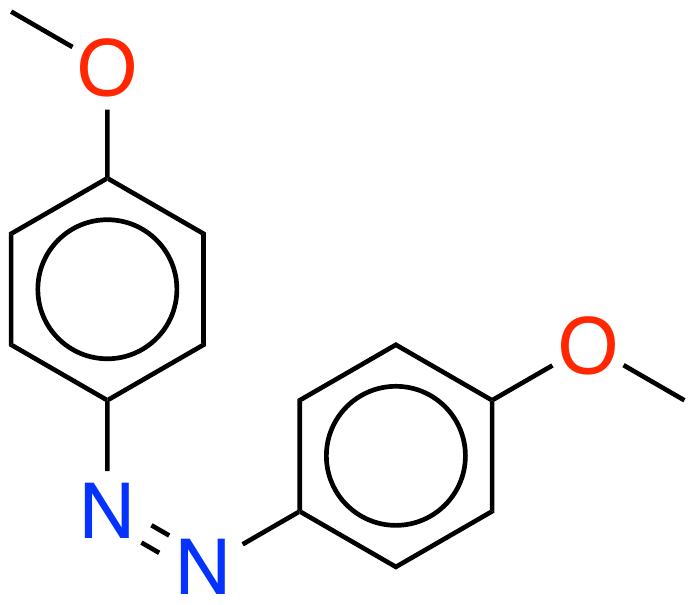}} & \cite{olmsted1983photochemical}  \\ 
\hline

\textbf{131} & COc1cc(N2CC3CN(C(C)=O)CC3C2) c(OC)cc1/N=[NH+]/c1cc(OC)c (N2CC3CN(C(C)=O)CC3C2)cc1OC & \raisebox{-.45\height}{\includegraphics[height=0.4 \textwidth,trim=0 -5 01 -5]{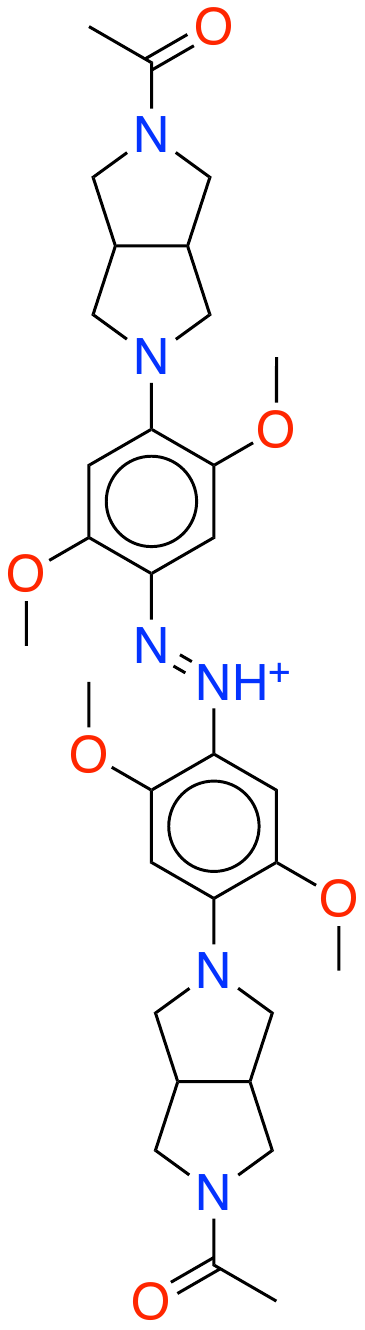}} & \cite{dong2015red}  \\ 
\hline

\textbf{132} & COc1cc(N2CC3CN(C(C)=O)CC3C2) c(OC)cc1/N=[NH+]{\textbackslash}c1cc(OC)c (N2CC3CN(C(C)=O)CC3C2)cc1OC & \raisebox{-.45\height}{\includegraphics[height=0.25 \textwidth,trim=0 -5 01 -5]{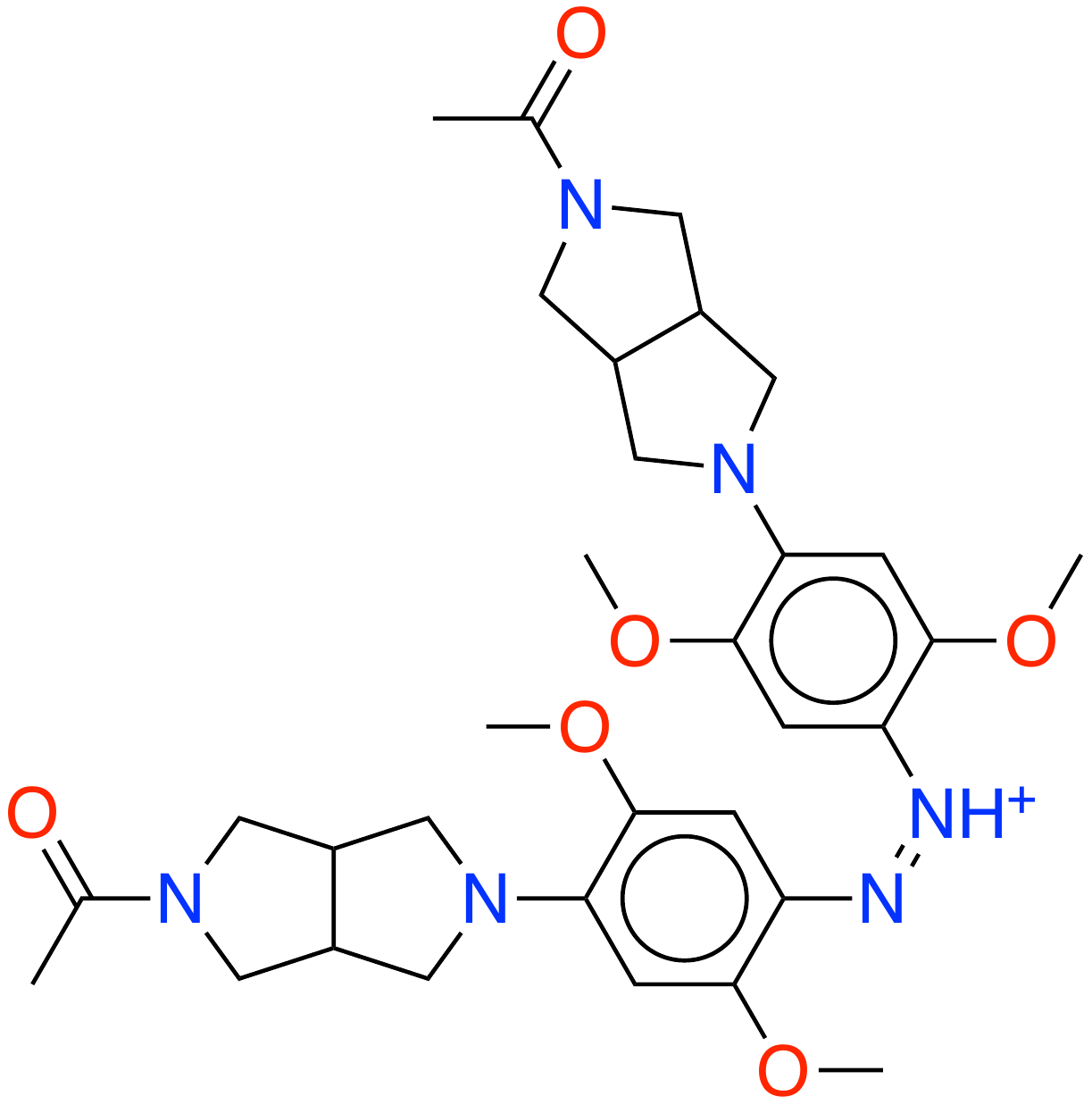}} & \cite{dong2015red}  \\ 
\hline

\textbf{133} & COc1cc(N2CCN(C(C)=O)CC2)c(OC) cc1/N=[NH+]/c1cc(OC)c(N2CCN(C (C)=O)CC2)cc1OC & \raisebox{-.45\height}{\includegraphics[height=0.21 \textwidth,trim=0 -5 01 -5]{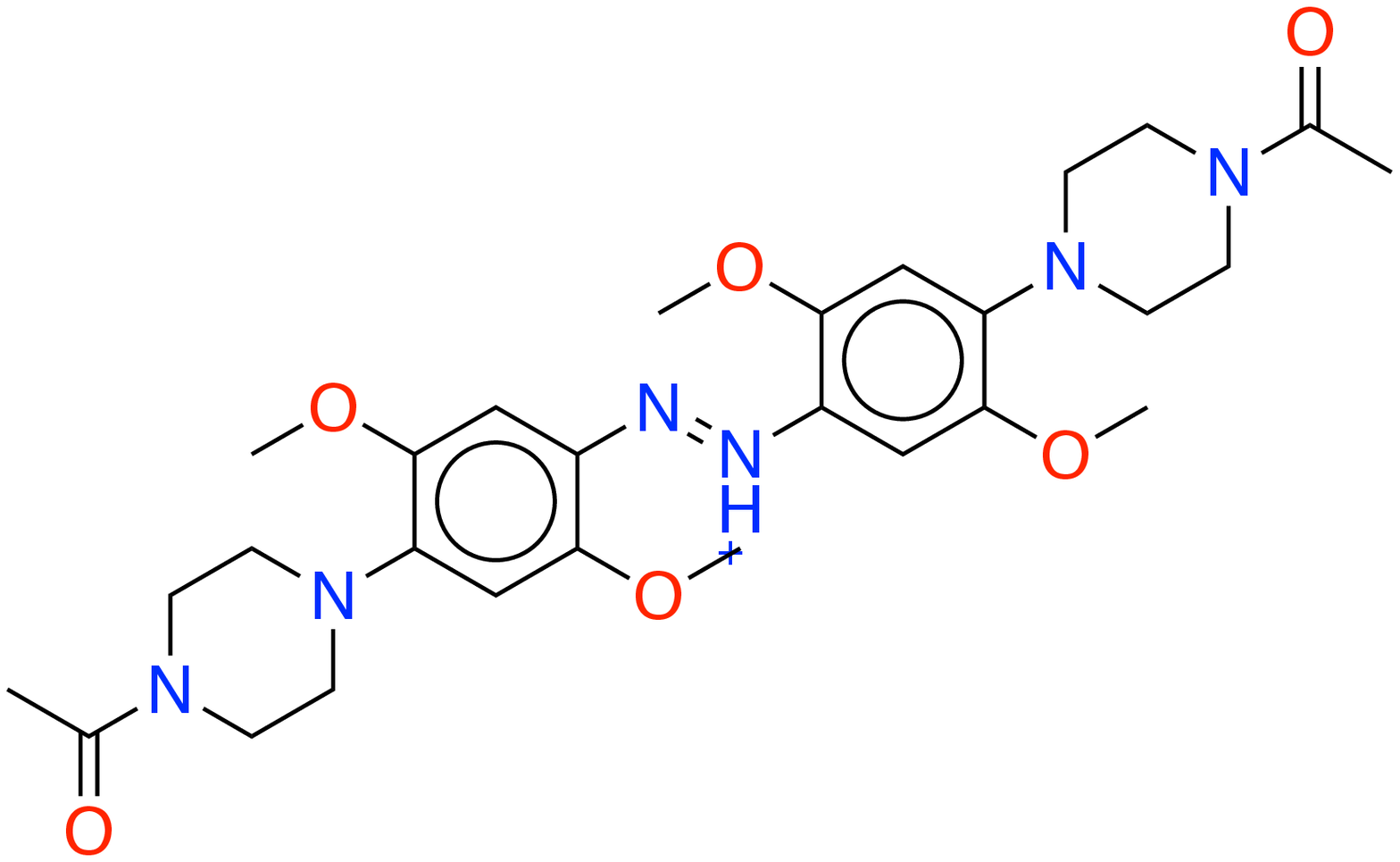}} & \cite{dong2015red}  \\ 
\hline

\textbf{134} & COc1cc(N2CCN(C(C)=O)CC2)c(OC) cc1/N=[NH+]{\textbackslash}c1cc(OC)c(N2CCN(C (C)=O)CC2)cc1OC & \raisebox{-.45\height}{\includegraphics[height=0.24 \textwidth,trim=0 -5 01 -5]{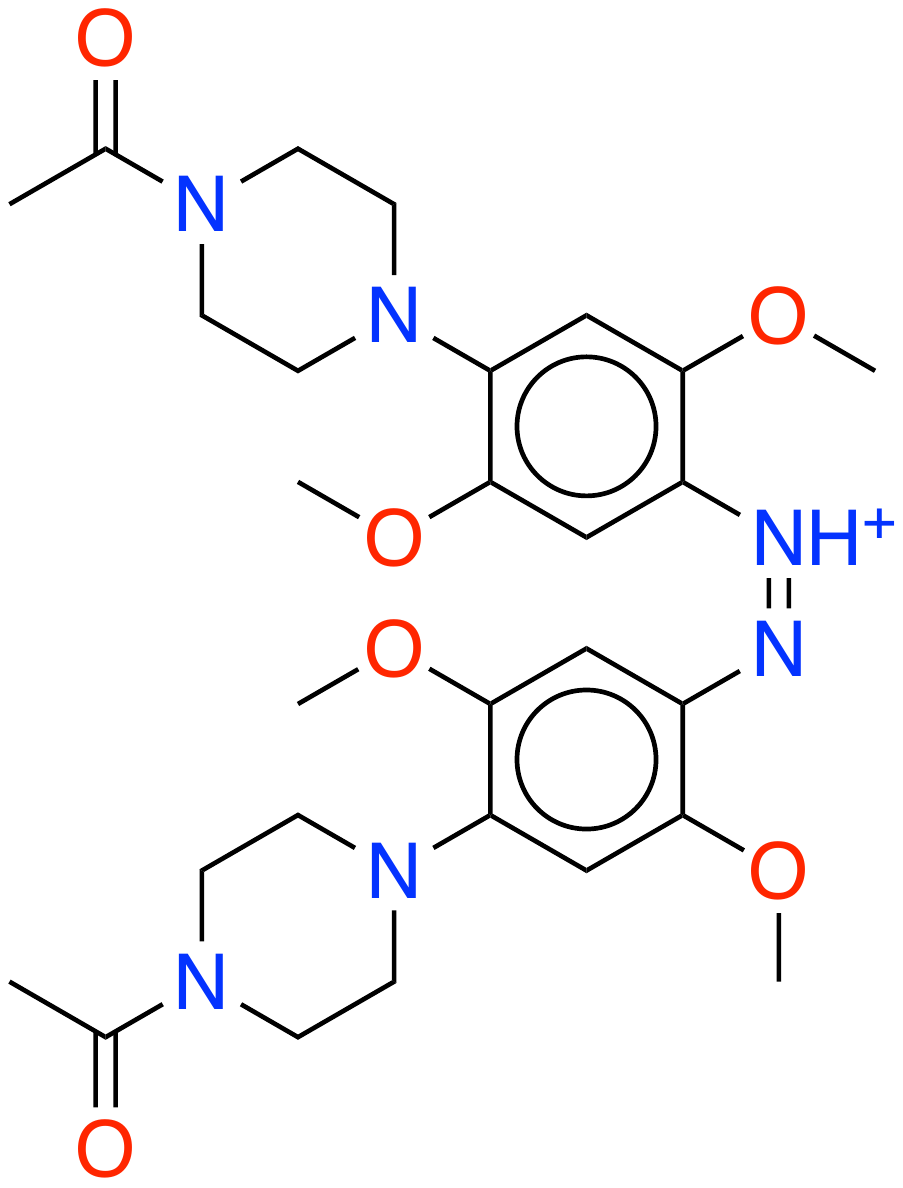}} & \cite{dong2015red}  \\ 
\hline

\textbf{135} & COc1cc(N2CCN(C(=O)CCl)CC2) cc(OC)c1/N=N/c1c(OC)cc(N2CCN (C(=O)CCl)CC2)cc1OC & \raisebox{-.45\height}{\includegraphics[height=0.4 \textwidth,trim=0 -5 01 -5]{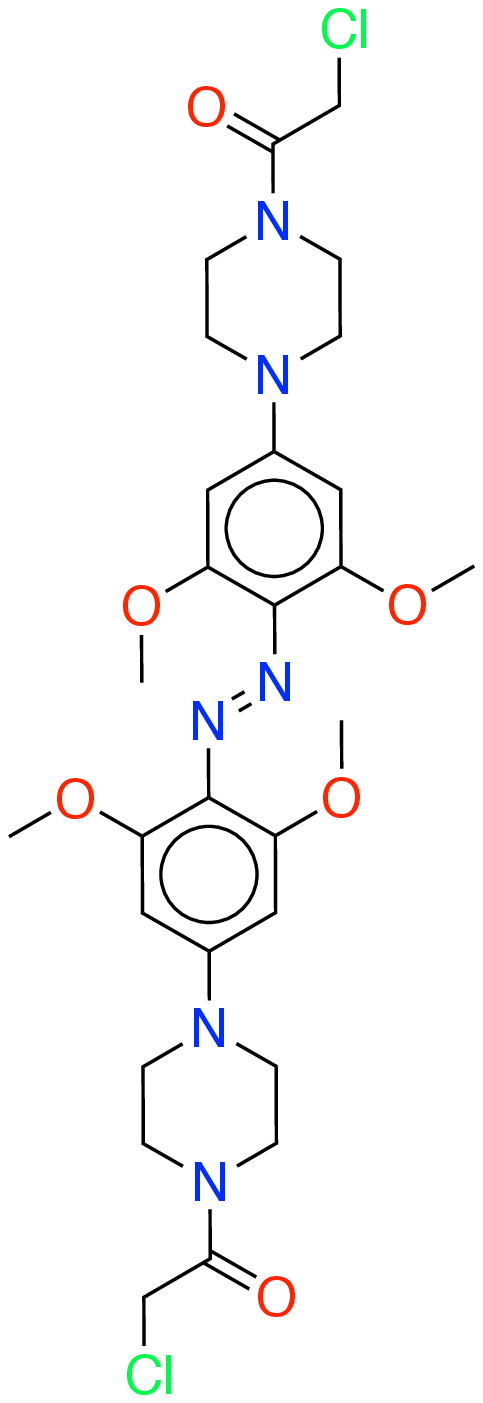}} & \cite{samanta2013photoswitching2}  \\ 
\hline

\textbf{136} & COc1cc(N2CCN(C(=O)CCl)CC2) cc(OC)c1/N=N{\textbackslash}c1c(OC)cc(N2CCN (C(=O)CCl)CC2)cc1OC & \raisebox{-.45\height}{\includegraphics[height=0.25 \textwidth,trim=0 -5 01 -5]{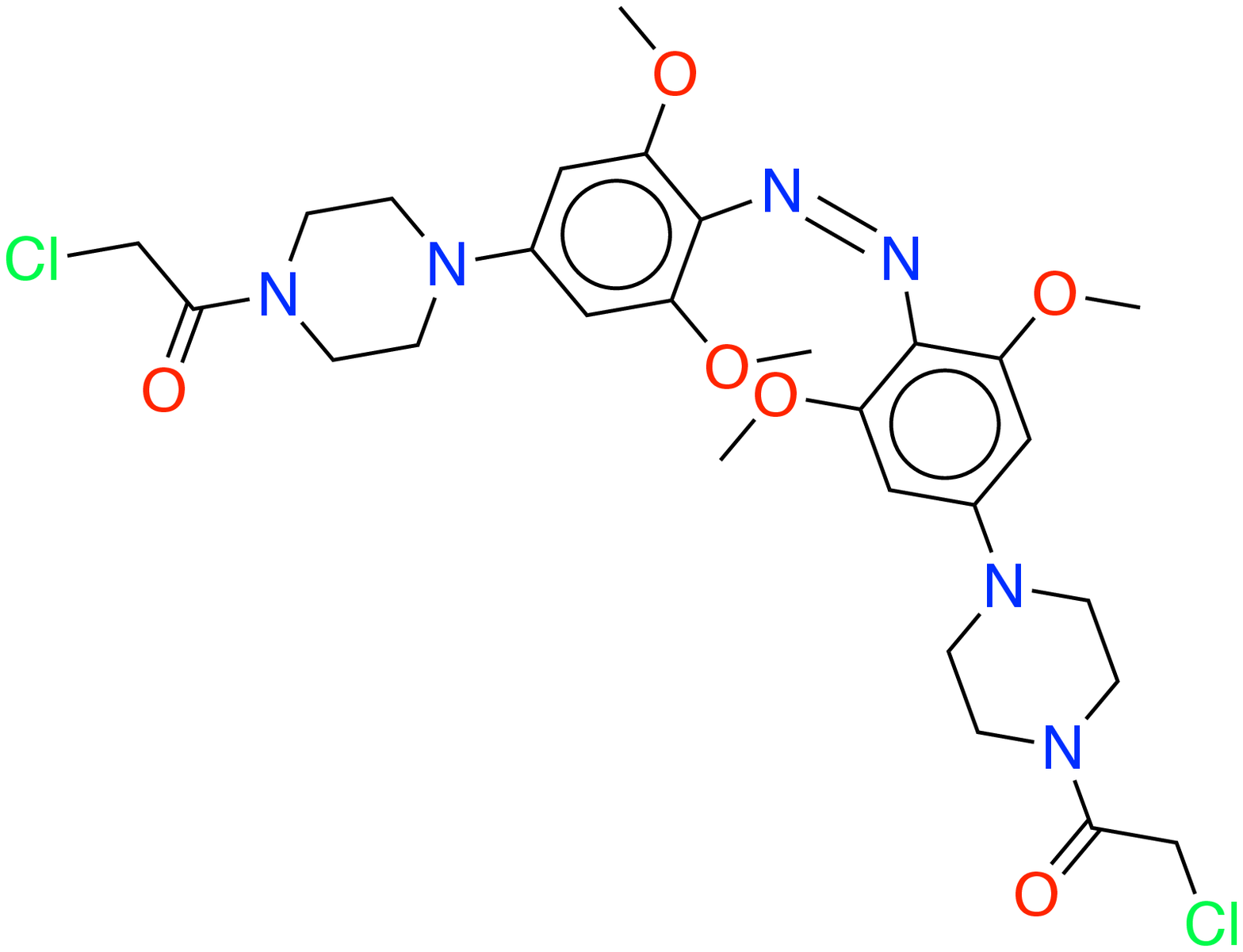}} & \cite{samanta2013photoswitching2}  \\ 
\hline

\textbf{137} & COc1cc(NC(C)=O)cc(OC)c1/N=N/ c1c(OC)cc(NC(C)=O)cc1OC & \raisebox{-.45\height}{\includegraphics[height=0.12 \textwidth,trim=0 -5 01 -5]{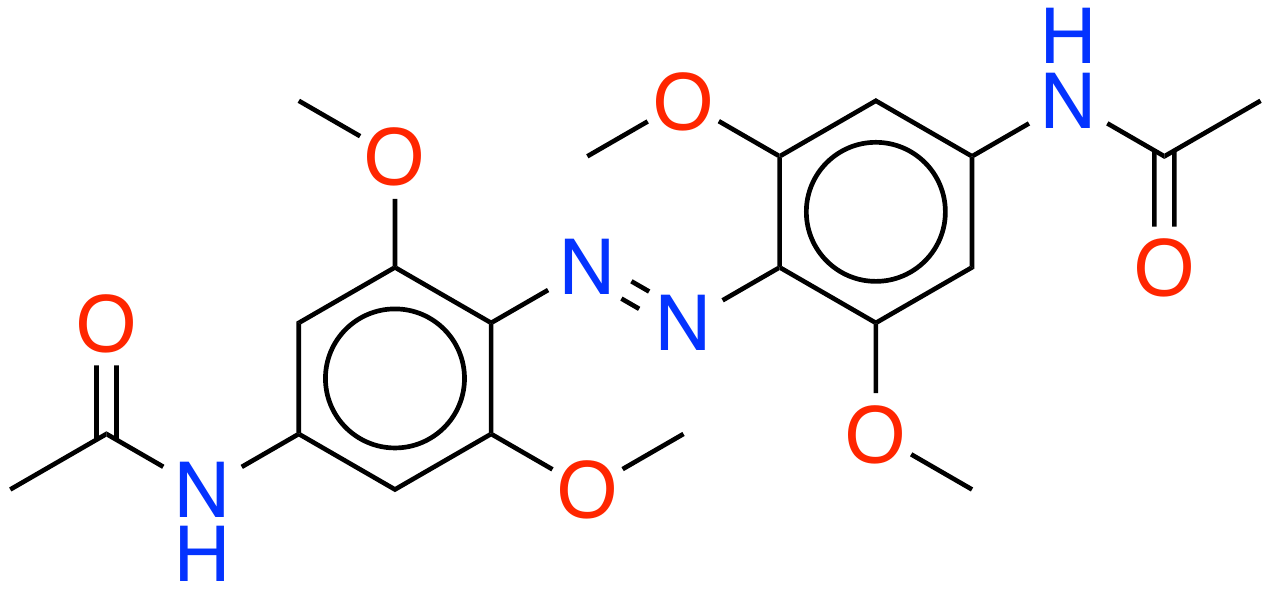}} & \cite{samanta2013photoswitching}  \\ 
\hline

\textbf{138} & COc1cc(NC(C)=O)cc(OC)c1/N=N{\textbackslash} c1c(OC)cc(NC(C)=O)cc1OC & \raisebox{-.45\height}{\includegraphics[height=0.185 \textwidth,trim=0 -5 01 -5]{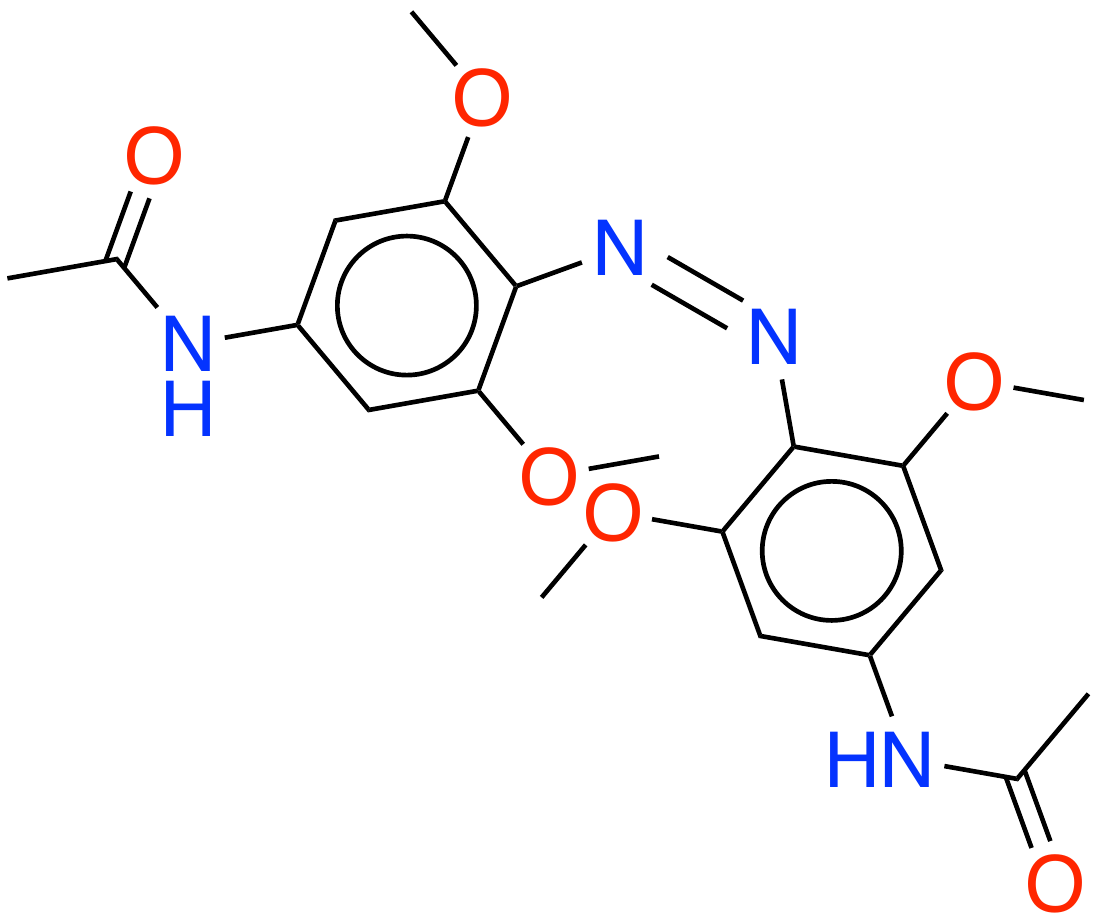}} & \cite{samanta2013photoswitching}  \\ 
\hline

\textbf{139} & COc1cc(NC(=O)CCl)cc(OC)c1 /N=N/c1c(OC)cc(NC(=O)CCl)cc1OC & \raisebox{-.45\height}{\includegraphics[height=0.12 \textwidth,trim=0 -5 01 -5]{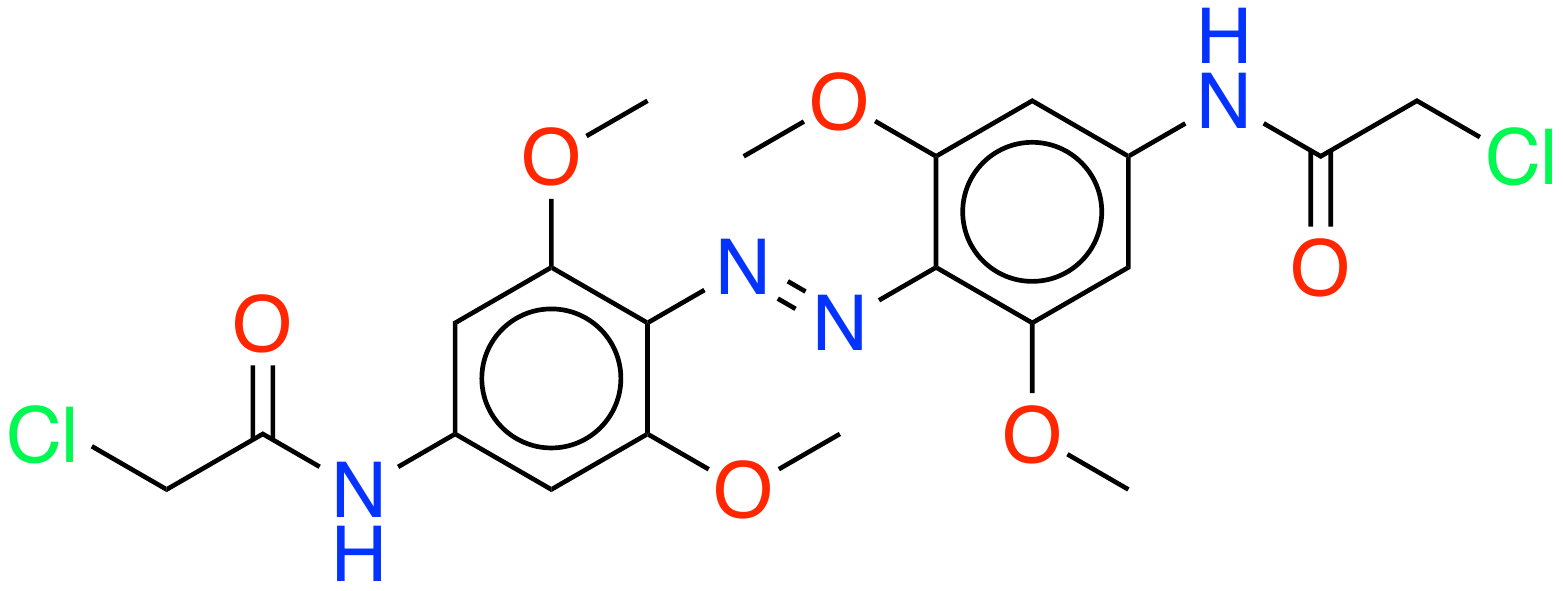}} & \cite{samanta2013photoswitching}  \\ 
\hline

\textbf{140} & COc1cc(NC(=O)CCl)cc(OC)c1 /N=N{\textbackslash}c1c(OC)cc(NC(=O)CCl)cc1OC & \raisebox{-.45\height}{\includegraphics[height=0.13 \textwidth,trim=0 -5 01 -5]{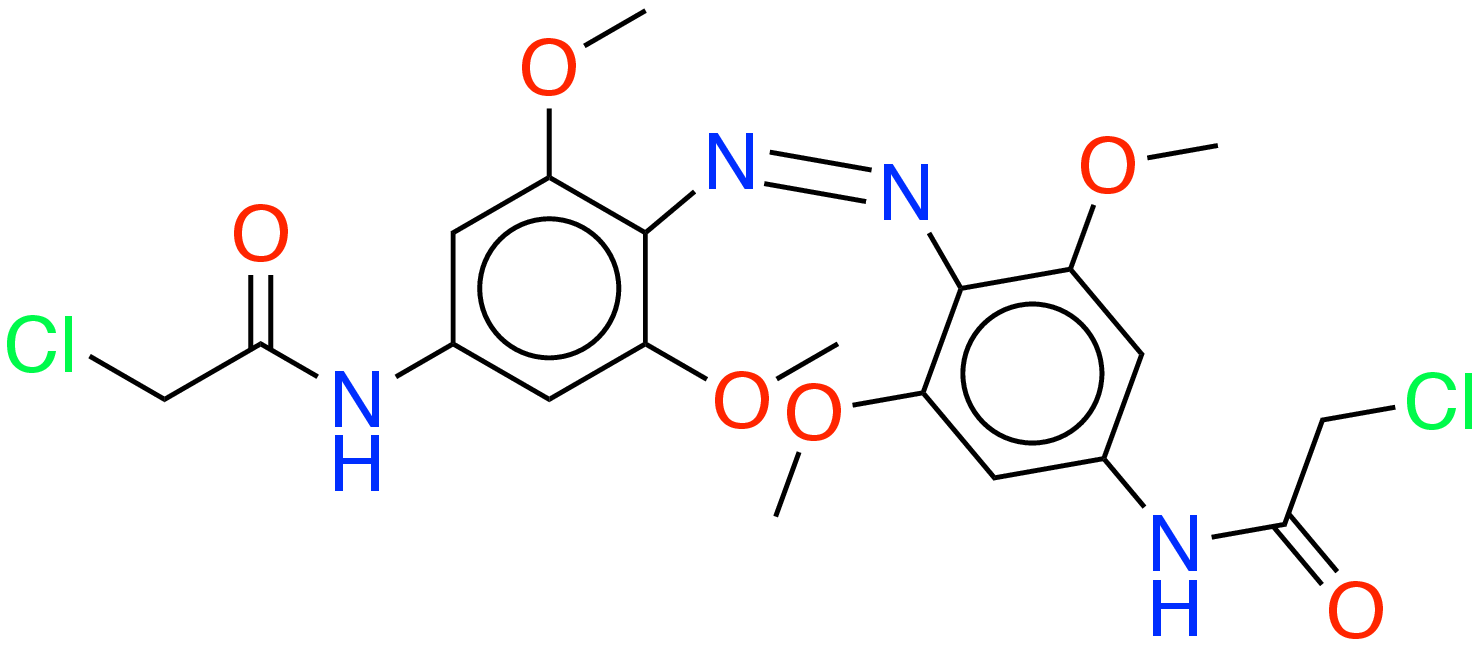}} & \cite{samanta2013photoswitching}  \\ 
\hline

\textbf{141} & COC(=O)c1ccc(/N=N/c2ccccc2)cc1 & \raisebox{-.45\height}{\includegraphics[height=0.11 \textwidth,trim=0 -5 01 -5]{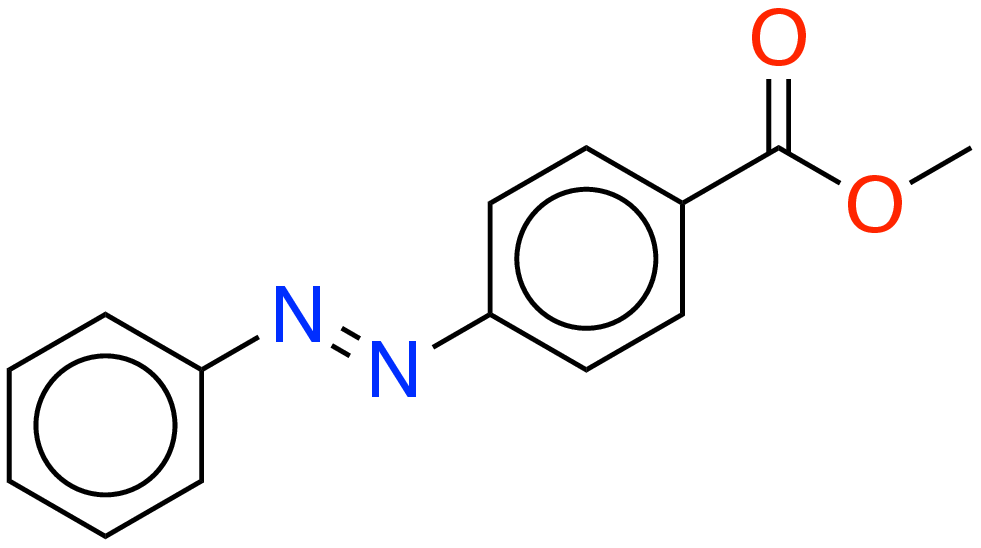}} & \cite{birnbaum1954photo}  \\ 
\hline

\textbf{142} & COC(=O)c1ccc(/N=N{\textbackslash}c2ccccc2)cc1 & \raisebox{-.45\height}{\includegraphics[height=0.09 \textwidth,trim=0 -5 01 -5]{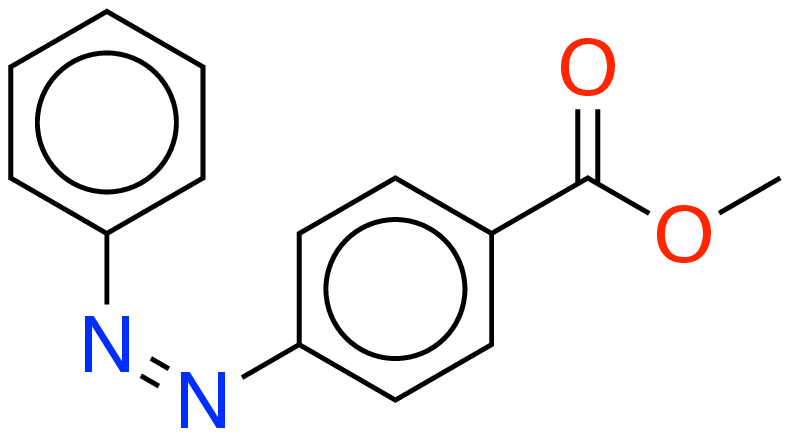}} & \cite{birnbaum1954photo}  \\ 
\hline

\textbf{143} & Fc1cccc(F)c1/N=N/c1c(F)cccc1F & \raisebox{-.45\height}{\includegraphics[height=0.095 \textwidth,trim=0 -5 01 -5]{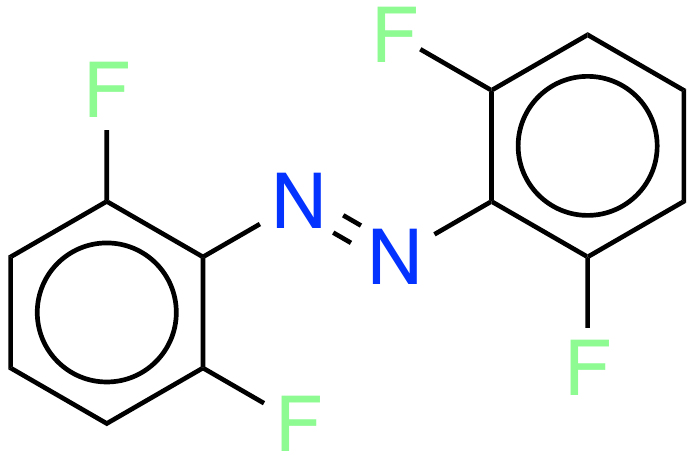}} & \cite{knie2014ortho}  \\ 
\hline

\textbf{144} & Fc1cccc(F)c1/N=N{\textbackslash}c1c(F)cccc1F & \raisebox{-.45\height}{\includegraphics[height=0.125 \textwidth,trim=0 -5 01 -5]{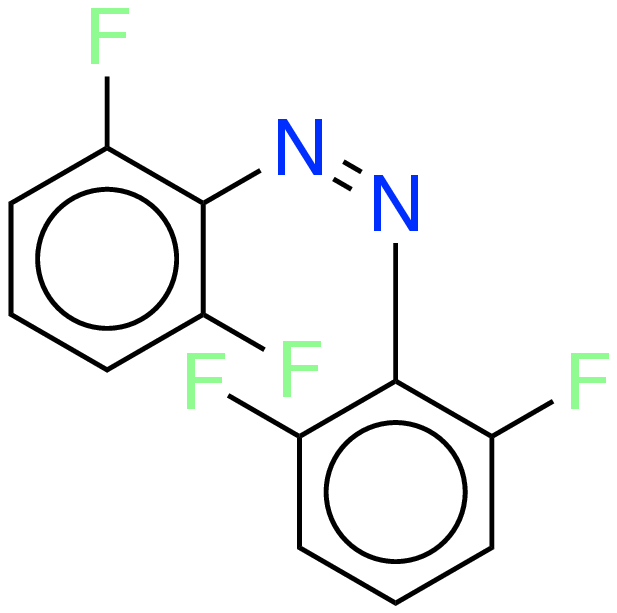}} & \cite{knie2014ortho}  \\ 
\hline

\textbf{145} & Fc1ccc(/N=N/c2ccccc2)cc1  & \raisebox{-.45\height}{\includegraphics[height=0.095 \textwidth,trim=0 -5 01 -5]{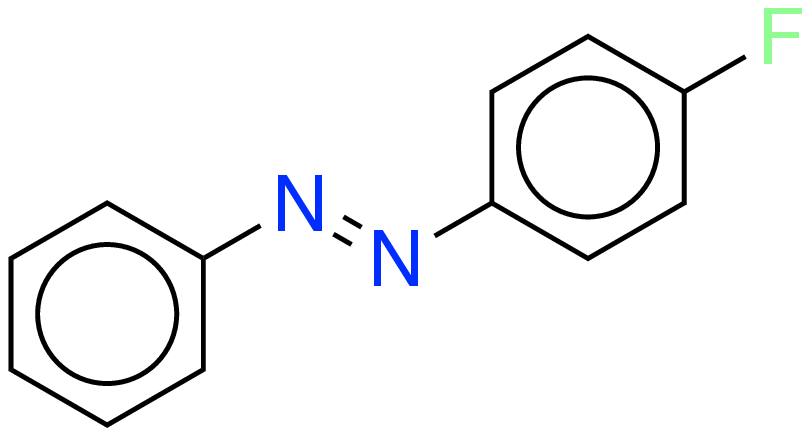}} & \cite{birnbaum1954photo}  \\ 
\hline

\textbf{146} & Fc1ccc(/N=N{\textbackslash}c2ccccc2)cc1  & \raisebox{-.45\height}{\includegraphics[height=0.095 \textwidth,trim=0 -5 01 -5]{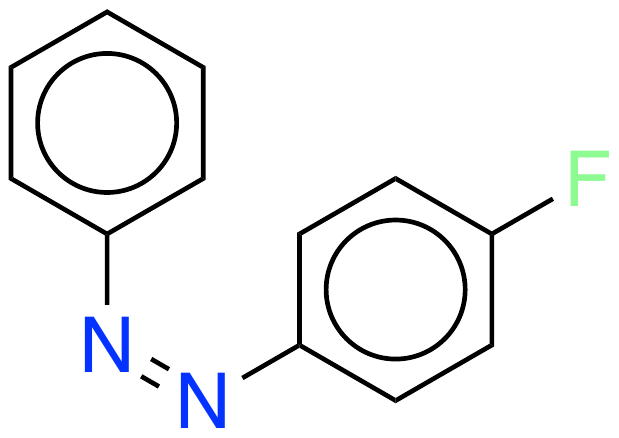}} & \cite{birnbaum1954photo}  \\ 
\hline

\textbf{147} & N[C@@H](C[C@@H](CCCC(=O) Nc1ccc(/N=N/c2ccc(NC(=O)CN3 C(=O)C=CC3=O)cc2)cc1) C(=O)O)C(=O)O & \raisebox{-.45\height}{\includegraphics[height=0.45 \textwidth,trim=0 -5 01 -5]{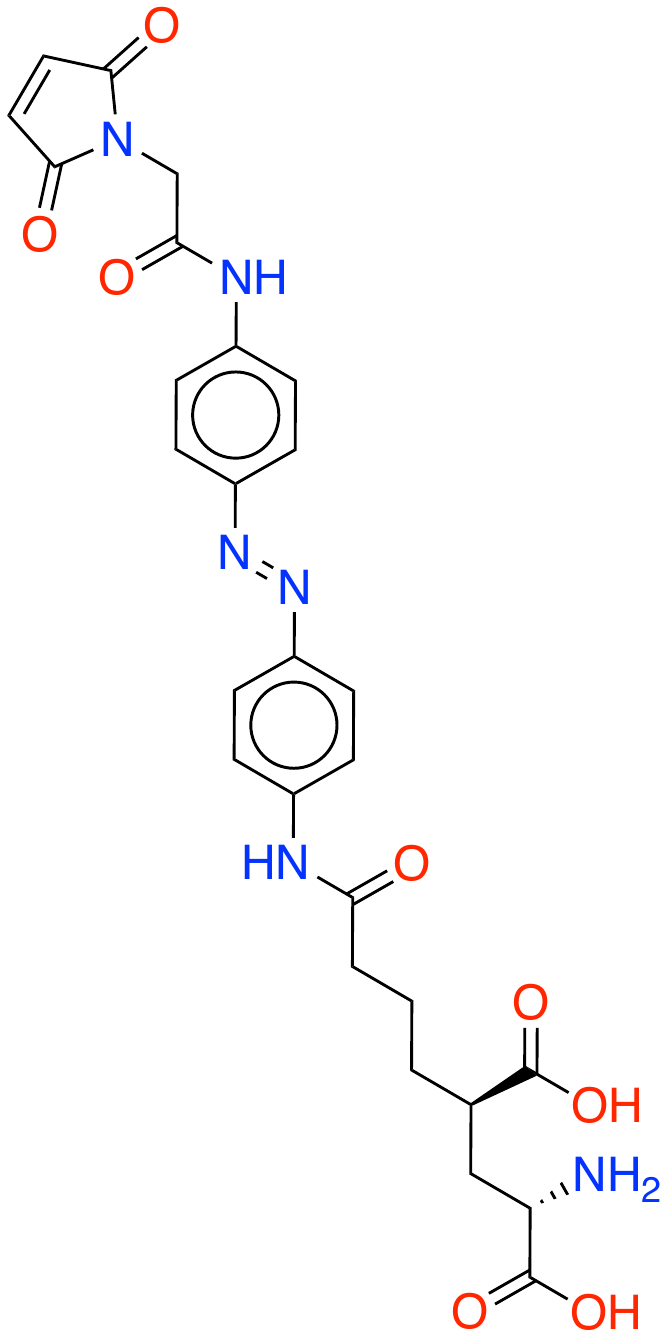}} & \cite{carroll2015two}  \\ 
\hline

\textbf{148} & N[C@@H](C[C@@H](CCCC(=O) Nc1ccc(/N=N{\textbackslash}c2ccc(NC(=O)CN3 C(=O)C=CC3=O)cc2)cc1) C(=O)O)C(=O)O & \raisebox{-.45\height}{\includegraphics[height=0.36 \textwidth,trim=0 -5 01 -5]{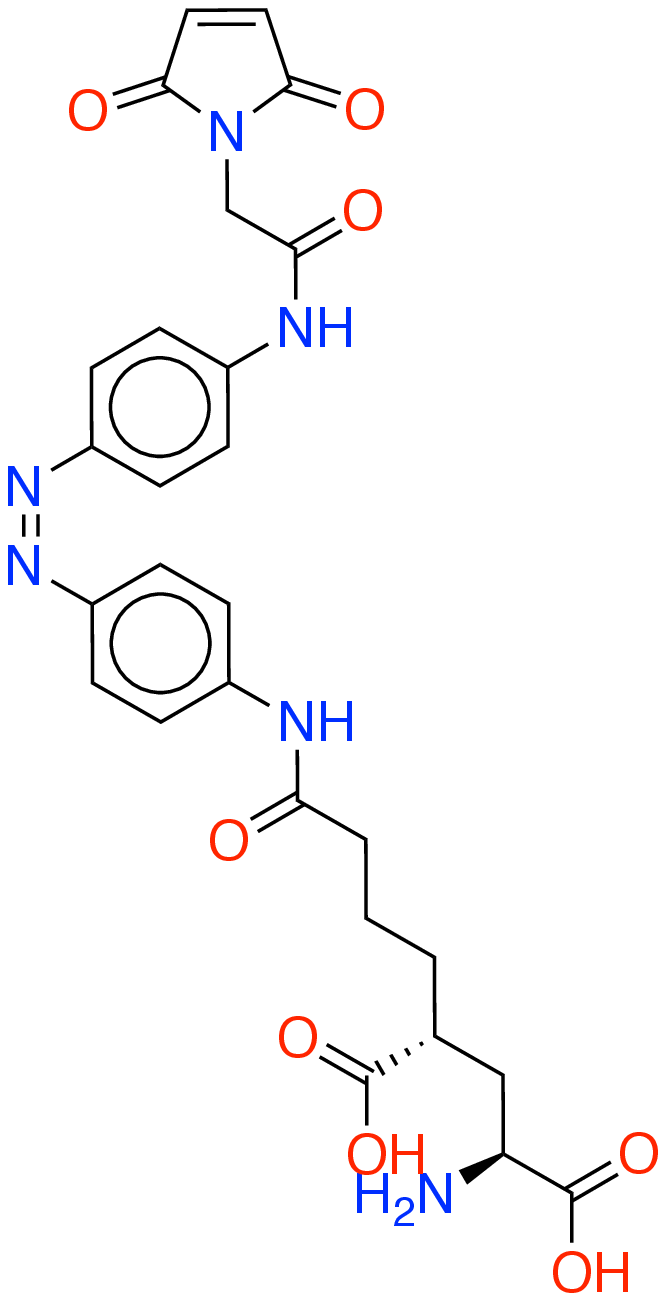}} & \cite{carroll2015two}  \\ 
\hline

\textbf{149} & NC(=O)c1cccc(/N=N/c2ccccc2)c1 & \raisebox{-.45\height}{\includegraphics[height=0.085 \textwidth,trim=0 -5 01 -5]{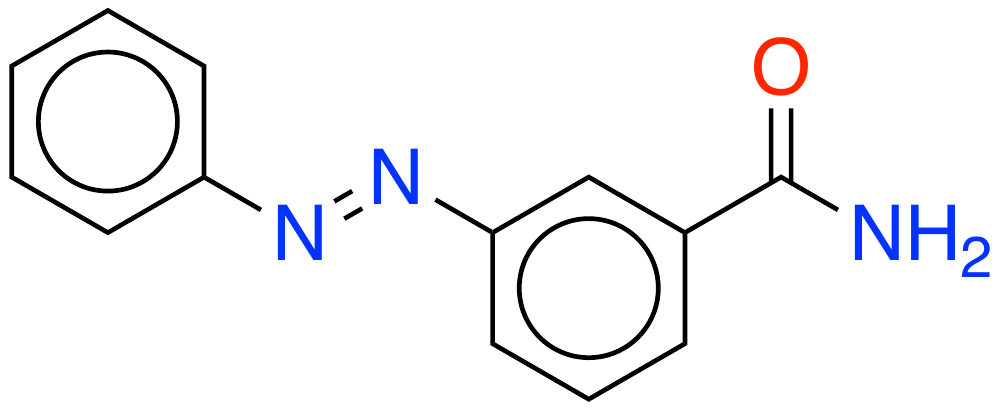}} & \cite{archut1998azobenzene}  \\ 
\hline

\textbf{150} & NC(=O)c1cccc(/N=N{\textbackslash}c2ccccc2)c1 & \raisebox{-.45\height}{\includegraphics[height=0.11 \textwidth,trim=0 -5 01 -5]{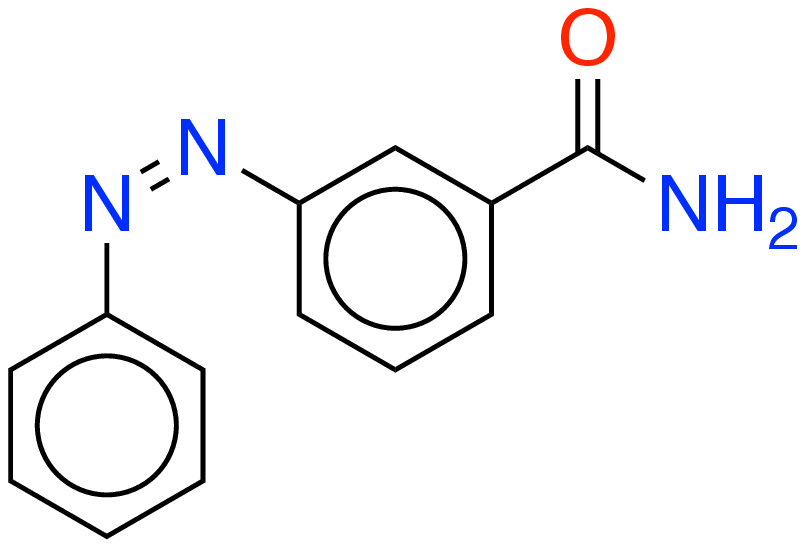}} & \cite{archut1998azobenzene}  \\ 
\hline

\textbf{151} & NC(=O)c1ccc(/N=N/c2ccccc2)cc1 & \raisebox{-.45\height}{\includegraphics[height=0.11 \textwidth,trim=0 -5 01 -5]{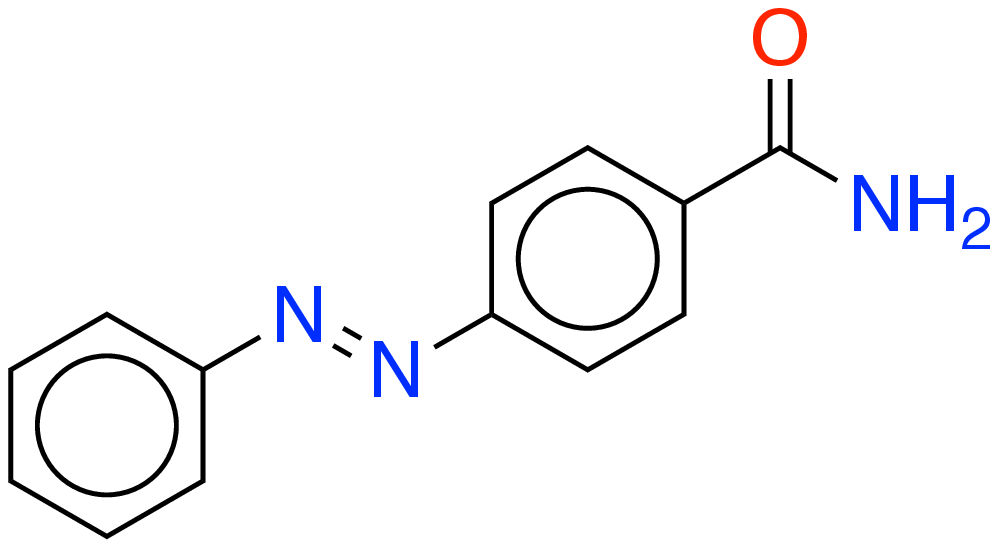}} & \cite{archut1998azobenzene}  \\ 
\hline

\textbf{152} & NC(=O)c1ccc(/N=N{\textbackslash}c2ccccc2)cc1 & \raisebox{-.45\height}{\includegraphics[height=0.085 \textwidth,trim=0 -5 01 -5]{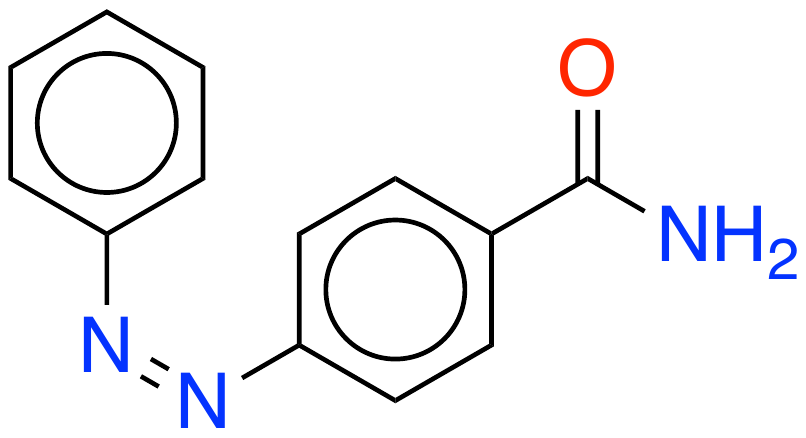}} & \cite{archut1998azobenzene}  \\ 
\hline

\textbf{153} & O=C1c2cccc(c2)/N=N/ c2cccc(c2)C(=O)N2CCOCCOCCN1CCOCCOCC2 & \raisebox{-.45\height}{\includegraphics[height=0.15 \textwidth,trim=0 -5 01 -5]{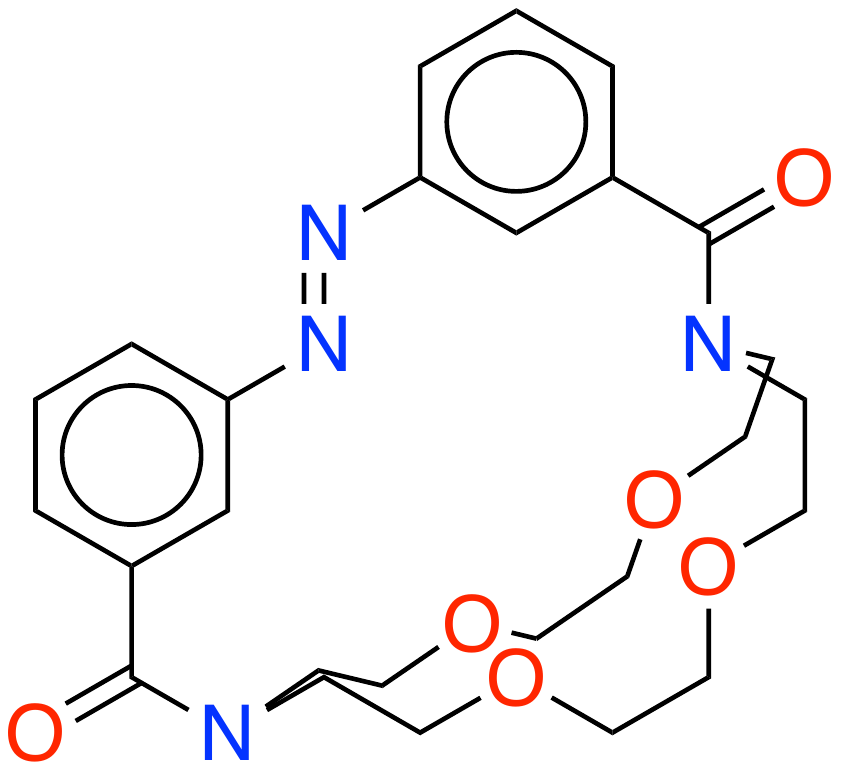}} & \cite{rau1984further, lednev1998photoisomerization}  \\ 
\hline

\textbf{154} & O=C1c2cccc(c2)/N=N{\textbackslash} c2cccc(c2)C(=O)N2CCOCCOCCN1CCOCCOCC2 & \raisebox{-.45\height}{\includegraphics[height=0.15 \textwidth,trim=0 -5 01 -5]{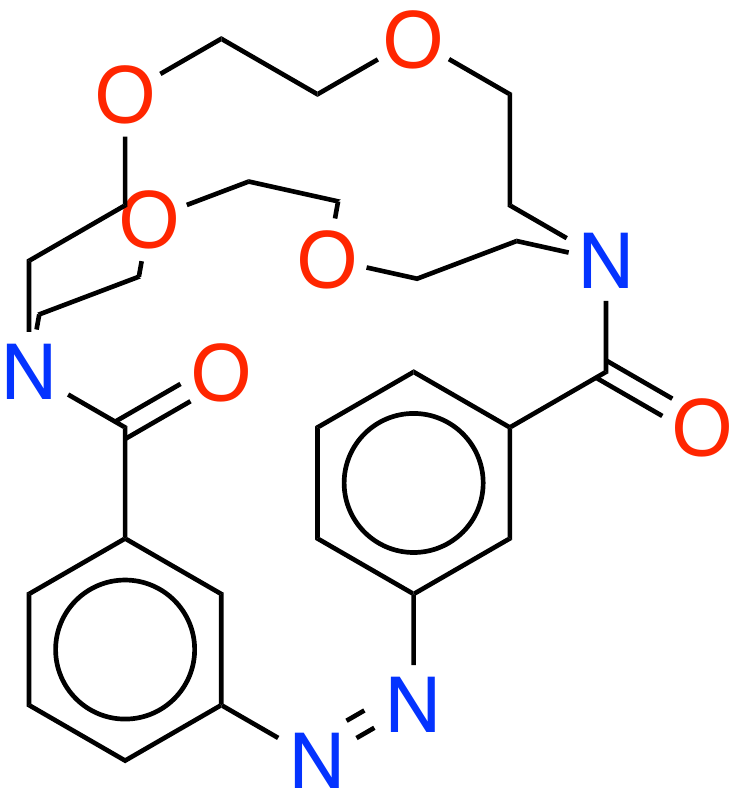}} & \cite{rau1984further, lednev1998photoisomerization}  \\ 
\hline

\textbf{155} & OCc1ccc(/N=N/c2ccc(O)cc2)cc1 & \raisebox{-.45\height}{\includegraphics[height=0.09 \textwidth,trim=0 -5 01 -5]{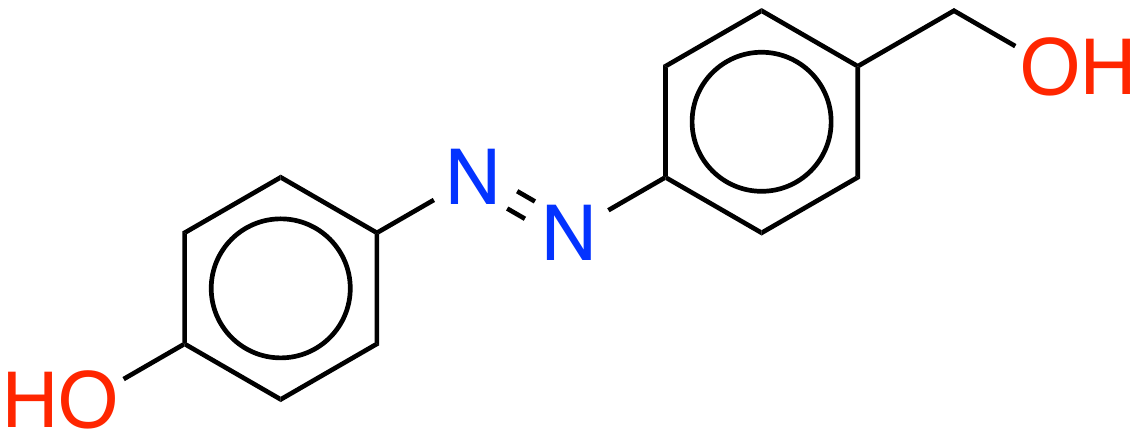}} & \cite{sierocki2006photoisomerization}  \\ 
\hline

\textbf{156} & OCc1ccc(/N=N{\textbackslash}c2ccc(O)cc2)cc1 & \raisebox{-.45\height}{\includegraphics[height=0.12 \textwidth,trim=0 -5 01 -5]{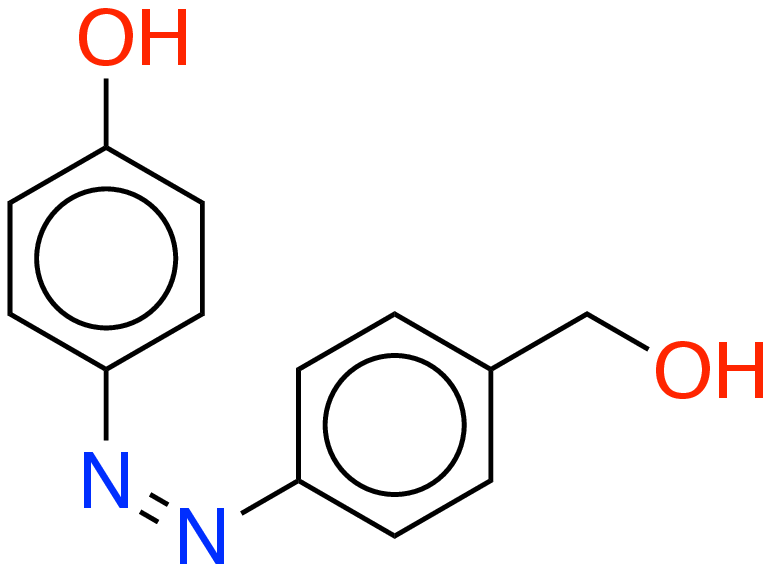}} & \cite{sierocki2006photoisomerization}  \\ 
\hline

\textbf{157} & O=C(CCl)N1CCN(c2ccc (/N=N/c3ccc(N4CCN(C(=O) CCl)CC4)cc3)cc2)CC1 & \raisebox{-.45\height}{\includegraphics[height=0.425 \textwidth,trim=0 -5 01 -5]{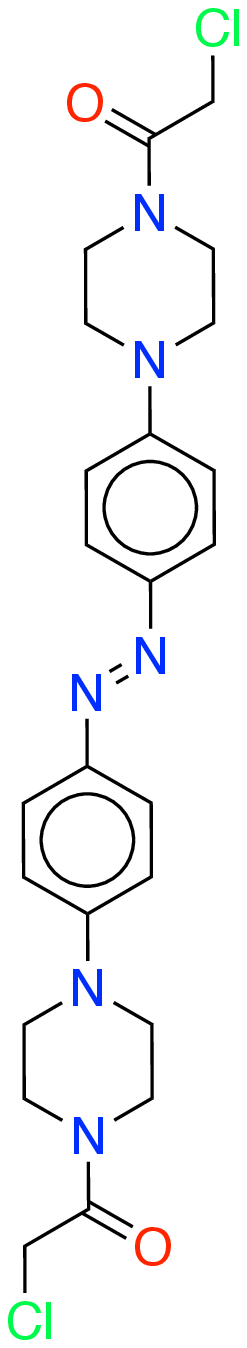}} & \cite{beharry2008photo}  \\ 
\hline

\textbf{158} & O=C(CCl)N1CCN(c2ccc (/N=N{\textbackslash}c3ccc(N4CCN(C(=O) CCl)CC4)cc3)cc2)CC1 & \raisebox{-.45\height}{\includegraphics[height=0.225 \textwidth,trim=0 -5 01 -5]{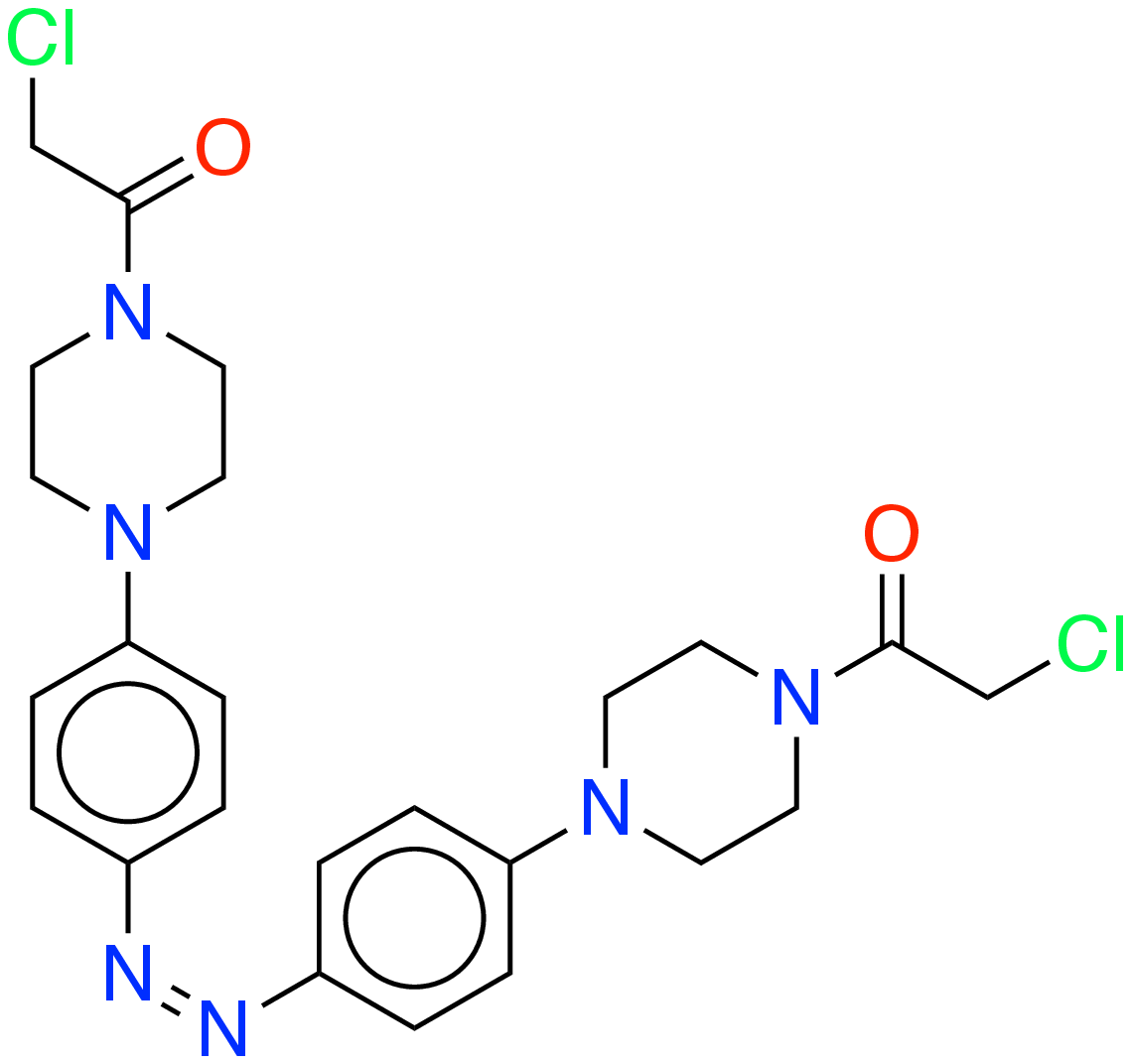}} & \cite{beharry2008photo}  \\ 
\hline

\textbf{159} & O=C(CCl)Nc1ccc2c(c1)CC c1cc(NC(=O)CCl)ccc1/N=N/2& \raisebox{-.45\height}{\includegraphics[height=0.11 \textwidth,trim=0 -5 01 -5]{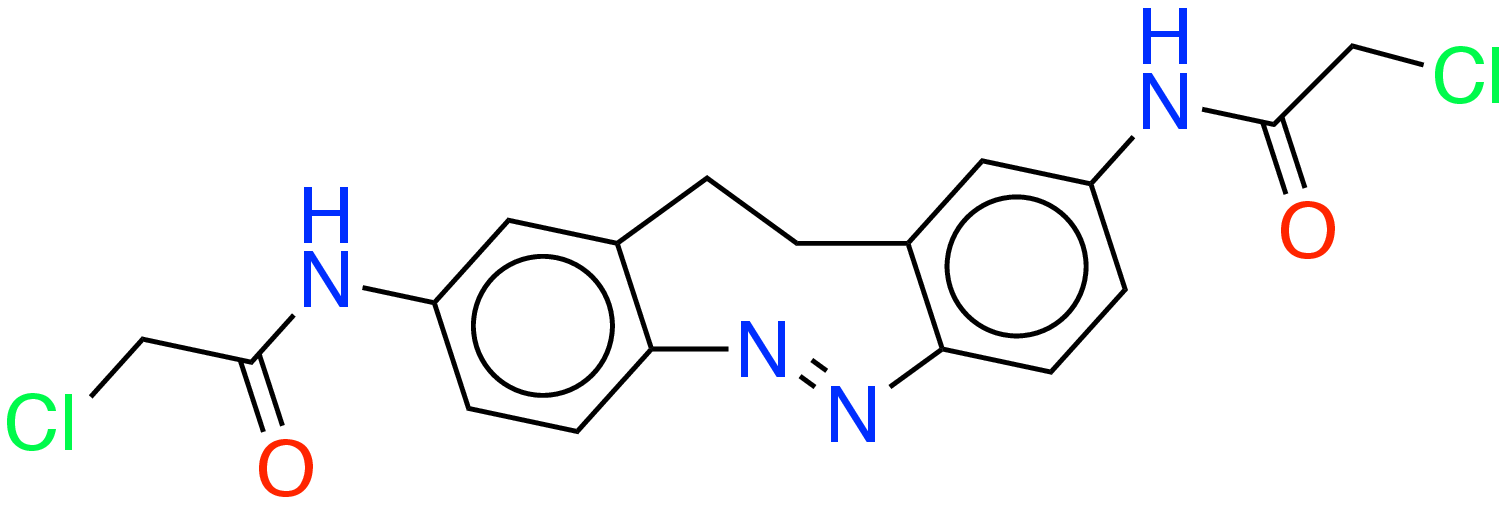}} & \cite{samanta2012bidirectional}  \\ 
\hline

\textbf{160} & O=C(CCl)Nc1ccc2c(c1)CC c1cc(NC(=O)CCl)ccc1/N=N{\textbackslash}2 & \raisebox{-.45\height}{\includegraphics[height=0.08 \textwidth,trim=0 -5 01 -5]{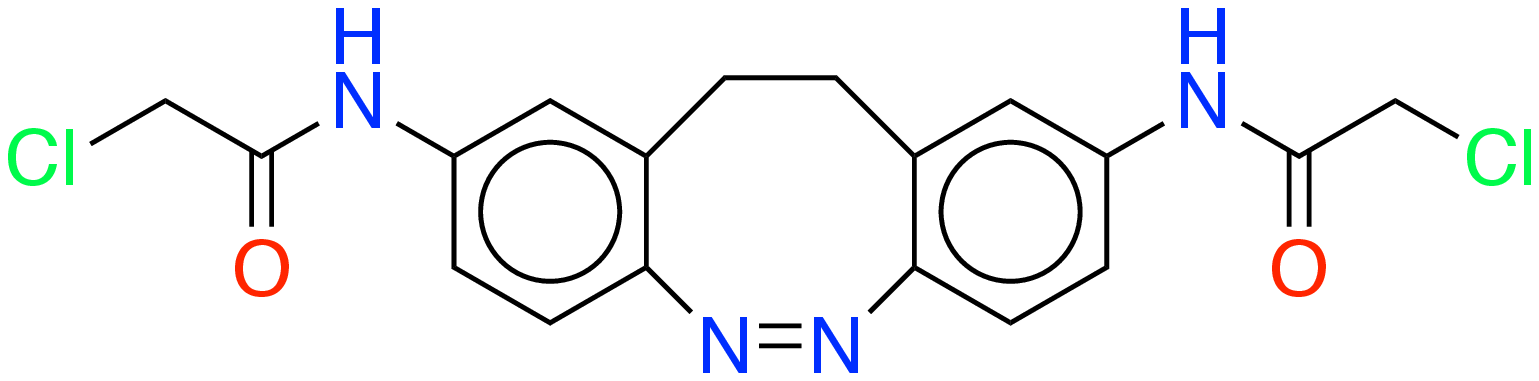}} & \cite{samanta2012bidirectional}  \\ 
\hline

\textbf{161} & O=C(CCl)Nc1cc(Cl)c(/N=N/ c2c(Cl)cc(NC(=O)CCl)cc2Cl)c(Cl)c1 & \raisebox{-.45\height}{\includegraphics[height=0.12 \textwidth,trim=0 -5 01 -5]{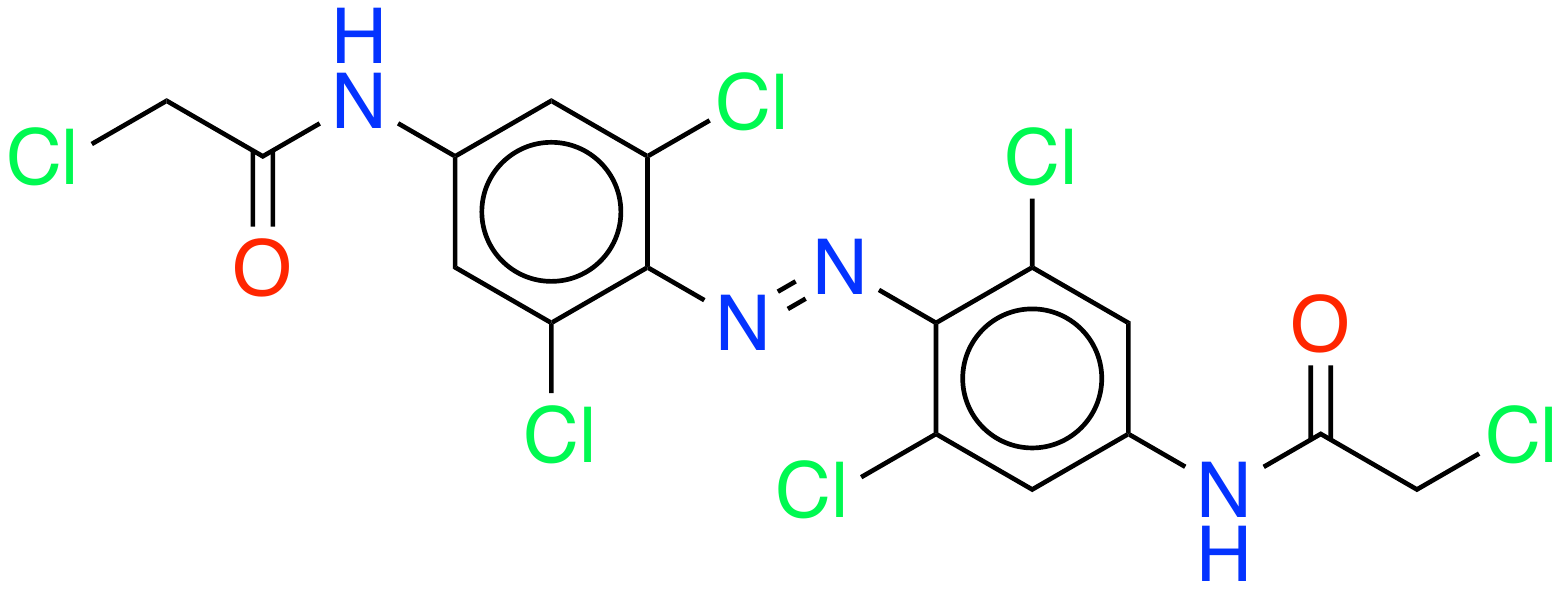}} & \cite{samanta2013photoswitching}  \\ 
\hline

\textbf{162} & O=C(CCl)Nc1cc(Cl)c(/N=N{\textbackslash} c2c(Cl)cc(NC(=O)CCl)cc2Cl)c(Cl)c1 & \raisebox{-.45\height}{\includegraphics[height=0.185 \textwidth,trim=0 -5 01 -5]{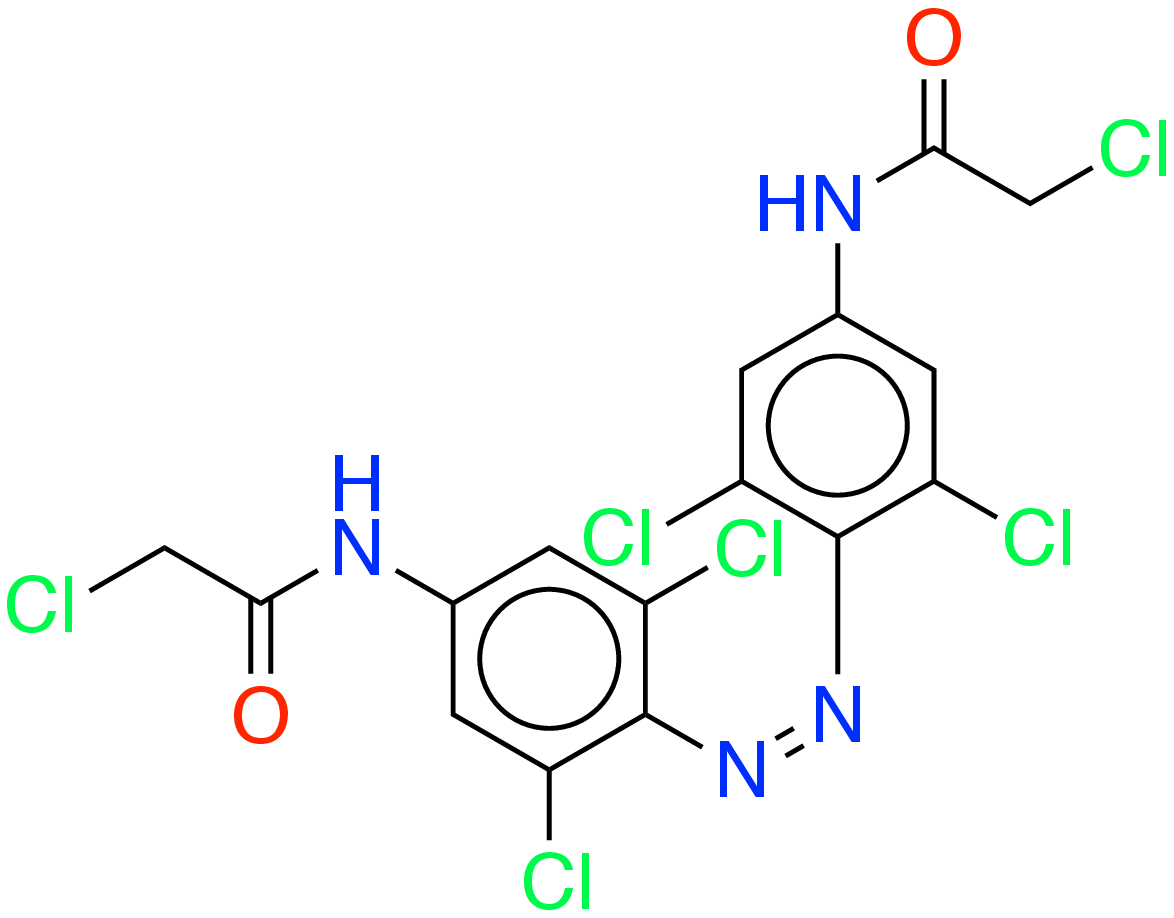}} & \cite{samanta2013photoswitching}   \\

\textbf{163} & CCN(CC)c1ccc(/N=N/c2ccc(OC)cc2)cc1 & \raisebox{-.45\height}{\includegraphics[height=0.22\textwidth,trim=0 -5 0 -5]{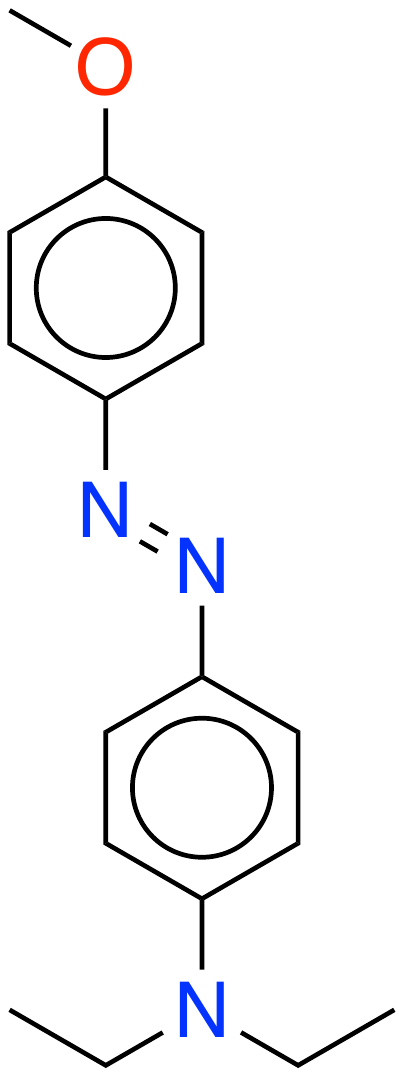}} & \cite{albini1983photochemistry}   \\ 
\hline

\textbf{164} & CCN(CC)c1ccc(/N=N{\textbackslash}c2ccc(OC)cc2)cc1 & \raisebox{-.45\height}{\includegraphics[height=0.12\textwidth,trim=0 -5 0 -5]{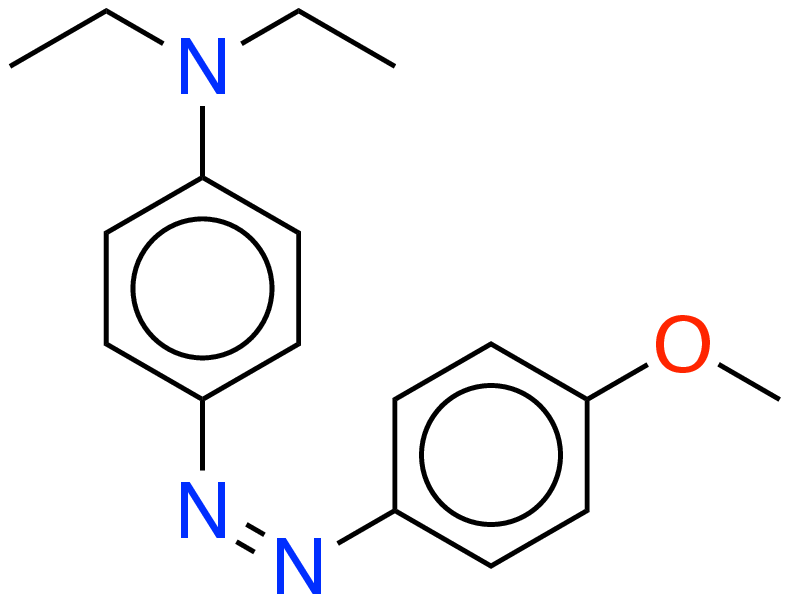}} & \cite{albini1983photochemistry}  \\ 
\hline

\hline
\end{xltabular}